 \documentclass[11pt]{article}

 \pdfoutput=1

 \usepackage[polutonikogreek,english]{babel}

 \usepackage{textgreek} 

\usepackage{bigints}

\usepackage[round]{natbib} 

\usepackage{graphicx,siunitx,xspace}
\usepackage{amsmath}
\usepackage{amssymb}
\usepackage{hyperref}
\usepackage{enumitem}
\usepackage{color}
\usepackage{enumitem}
\usepackage{float}
\usepackage{amsfonts,bm}
\usepackage{epstopdf}
\usepackage[mathscr]{euscript}

  \usepackage[normalem]{ulem}
  \usepackage{textcomp}

\usepackage[utf8]{inputenc}
\usepackage[T1]{fontenc}
\usepackage{pifont} 

\usepackage{chemformula}

\usepackage{float}
\usepackage{ifthen}
\begin{document}


\begin{center}
{\bf \Large
An English translation of the monograph of Max 
\citet[][]{Planck_Das_Prinzip_der_Erhaltung_der_Energie_1887}}
\\ \vspace*{2mm}
\hspace*{-0mm}
{\bf \Large 
``\,\underline{Das Princip der Erhaltung der Energie}\,''}
\\ \vspace*{2mm}
{\bf \Large or: ``\,\underline{The principle of the conservation of energy}\,''}
\\ \vspace*{2mm}
{\bf \Large
to provide a readable version of the German content.
}
\\ \vspace*{2mm}
{\bf \large\color{red}
Translated by Dr. Hab. Pascal Marquet 
}
\\ \vspace*{2mm}
{\bf\bf\color{red}  \large Possible contact at: 
    pascalmarquet@yahoo.com}
    \vspace*{1mm}
    \\
{\bf\bf\color{red} 
    Web Google-sites:
    \url{https://sites.google.com/view/pascal-marquet}
    \\ ArXiv: 
    \url{https://arxiv.org/find/all/1/all:+AND+pascal+marquet/0/1/0/all/0/1}
    \\ Research-Gate:
    \url{https://www.researchgate.net/profile/Pascal-Marquet/research}
}
\\ \vspace*{1mm}
\end{center}

\hspace*{65mm} Version-1 / \today

\vspace*{-2mm} 
\begin{center}
--------------------------------------------------- 
\end{center}
\vspace*{-11mm}

\bibliographystyle{ametsoc2014}
\bibliography{Book_FAQ_Thetas_arXiv}

\vspace*{-2mm} 
\begin{center}
--------------------------------------------------- 
\end{center}
\vspace*{-2mm}

Uncertainties/alternatives in the translation are indicated {\color{red} (in red)} with {\it\color{red} italic terms}, together with some additional footnotes (indicated with {\it\color{red} P. Marquet)}. 
Moreover, I have added some highlight (shown as \dashuline{\,dashing text}), in particular about the \dashuline{energy} concept.

I have written {\bf in bold} those parts of the text that deal in particular with the problem of determining {\bf the zero state of energies} {\it\color{red}(\,``\,Nullzustand der Energie\,''\,)}, in particular in the Section~\ref{Section-1} (R. Clausius, W. Thomson), in the Section~\ref{Section-2}, and at the end of the Section~\ref{Subsection-3-2}.
It is indeed for this aspect that I felt the need to translate the 2 thesis memoirs (1879 for the Doctor dissertation; 1880 for the Habilitation) and the 2 articles (1887 about the conservation of energy; 1943 about the discovery of quanta in physics), all written in German by Max Planck.

Max Planck's 1887 contribution should be seen as nothing more than a synthesis of the late 1880s, with many concepts that would be strongly modified (or even invalidated) by the coming revolutions of relativistic and quantum mechanics (at the beginning of the 20th century).
I have, of course, kept \dashuline{the original text of Planck (in black)} unchanged, while sometimes including {\it\color{red}\dashuline{additional notes (in red)}} to put into perspective the concepts that are sometimes outdated (and considered invalid) today.

Planck's text is organised almost linearly within each chapter, without any separation and most often with very long sentences, as is often the case in German. 
I therefore wanted to add a number of section-like headings (in red), in order to make Planck's chosen organisation clearer, with in the Section~\ref{Section-1} the names of the various scientists appearing just before the description of their contributions.

In addition, as Max Planck refers many times to the pagination of his text for cross-references, I have added {\it\color{red}\dashuline{(always in red)}} the beginning of each page (more or less one sentence, depending on the pages).  

Do not hesitate to contact me in case of mistakes or any trouble in the English translation from the German text.
\vspace*{-3mm}

\begin{center}
---------------------------------------------------
\end{center}
\vspace*{-8mm}

\newpage
  \tableofcontents

\vspace*{4mm}
\begin{center}
========================================================
\end{center}
\vspace*{-2mm}


{\Large\bf\underline{{\it\color{red}Preface of P. Marquet}}}
\vspace*{1mm}

{\color{purple}
According to the German page on 
Wikipedia$\,$\footnote{\color{red}\it$\:$See: 
\url{https://de.wikipedia.org/wiki/Max_Planck}}$\,$:
``{\it\,During his time in Kiel, Planck took part in a competition ``\,On the nature of energy\,'' organised by the Faculty of Philosophy at the University of G\"ottingen in 1884 for the year 1887.\,}''
}

{\color{purple}
As a result, \citet[][]{Planck_Das_Prinzip_der_Erhaltung_der_Energie_1887}:
``{\it\,was awarded second prize for his monograph ``\,The Principle of the Conservation of Energy'' (``Das Princip der Erhaltung der Energie'' in German), and as the first prize was not awarded, Planck unofficially emerged as the winner of the competition. The jury particularly emphasised ``\,the methodical way of thinking, the thorough mathematical-physical education of the author [and] the prudence of his judgement''. 
He was presumably denied the first prize because he favoured \dashuline{\,Helmholtz's} work over that of G\"ottingen professor \dashuline{\,Wilhelm Eduard Weber} in his essay. 
There was a fierce scientific dispute between the two physicists at the time.\,}''
}
\vspace*{2mm}

\begin{center}
========================================================
\end{center}
\vspace*{2mm}

{\Large\bf \underline{Foreword {\it\color{red}(of Max Planck)}} (p.III-XII)}
\vspace*{2mm}


The occasion for writing this book was a competition organised by the Faculty of Philosophy in Göttingen for the Beneke Prize Foundation for the year 1887. 
The task, which was set in 1884, was as follows: 
\vspace*{-4mm}

\begin{center}
---------------------------------------------------
\end{center}
\vspace*{-2mm}

``{\it Since \dashuline{\,Thomas Young} (Lectures on Natural Philosophy, London 1807, Lecture VIII) \dashuline{\,energy} has been attributed to bodies by many physicists, and since \dashuline{William Thomson\,} (Philosophical Magazine and Journal of Science, IV Series, London 1855, p.523) the \dashuline{\,principle of the conservation of energy} has frequently been pronounced as a principle valid for all bodies, by which the same principle seems to be understood which had already been pronounced earlier by \dashuline{Helmholtz\,} under the name of the \dashuline{principle of the conservation of force}.

A precise historical development of the meaning and \dashuline{use of the word energy in physics} is now required.
Then a thorough physical investigation as to whether \dashuline{\,different kinds of energy\,} are to be distinguished, and how each of them is to be defined, and finally in what way the \dashuline{\,principle of the conservation of energy\:} can be established and proved as a \dashuline{\,universally} valid \dashuline{\,law of nature}.}''
\vspace*{2mm}

\begin{center}
---------------------------------------------------
\end{center}
\vspace*{-2mm}

The train of ideas that guided me in working on this task, as well as the goal that I had in mind, will be most clearly illustrated if I share the essentials from the comments that accompanied the submission of my work to the jury: 

``{\it First of all, I would like to make a few introductory remarks about the plan and execution of the work. Having always been particularly interested in the theory of energy, the idea of making it the subject of a more detailed work by dealing with the prize task was all the more obvious to me, as I personally often felt the lack of a work which, calculated primarily for the physicist of the subject, unites under a common point of view the manifold forms and applications of which the concept of energy is capable, and which are presented by various authors in the most diverse ways.
For as much as has been written and spoken about the doctrine of energy in the last 40 years, with the sole exception of Helmholtz's treatise on the conservation of energy from 1847, all publications relating to it, insofar as they deal with the general concept of energy and not only special applications of it, for example to the theory of heat, are primarily intended for a wider circle of readers: from the writings of J. R. Mayer to the more extensive works of A. Secchi, G. Krebs, Balfour Stewart and others.

It is therefore essentially a practical-physical need that I am endeavouring to satisfy with this paper, and I only hope that this conception does not stray too far from the meaning of the task at hand to be considered a treatment of it. Incidentally, the limitation I have indicated also proved to be very appropriate in view of the nature of the material to be dealt with, as it swelled to such dimensions during the preparatory work that without a thorough review of all the material I had to fear jeopardising the unity of the presentation, which was my main concern.
Thus, above all, I was allowed to simply skip the discussion of all philosophical speculations that go beyond the purely physical field, which have often enough been linked to the concept of energy. Furthermore, in accordance with the physical conception, which I have also expressed in the choice of the title, I have always been able to place the main emphasis of the investigation on the \dashuline{\,principle of the conservation of energy}, while I have treated the concept of energy in detail only in so far as it can be brought into connection with the principle, starting from the idea that the concept of energy only gains its meaning for physics through the principle which contains it. On the other hand, in view of the fact that the traces of the existence of the \dashuline{concept of energy} go back considerably further into the past than the use of the word, it seemed essential to me to at least briefly commemorate the historical development of the concept, even before it was \dashuline{\,named by Th. Young}.

For the rest, I have naturally endeavoured to follow the wording of the problem as closely as possible. In particular, I have divided it into three main sections: \\
\hspace*{5mm} 1) {historical development}; \\
\hspace*{5mm} 3) {different kinds of energy}; \\\
\hspace*{5mm} 2) {formulation and proof of the principle of the conservation of energy}. \\
However, following considerations which in part only became apparent in the course of the work itself, I have taken the liberty of rearranging the titles insofar as I have placed the section mentioned third in the task before the second. On the one hand, this change of order seemed to me to be too insignificant to meet with serious reservations. On the other hand, I believed that I could better fulfil the intentions of the task setter by taking the greatest possible account of the clarity of the treatment than if I had impaired the lively context, as it seemed to my subjective feeling, by clinging too anxiously to the letter.

According to the starting-point of my view, already stated above, the centre of gravity of the whole work lies in the second and third sections. But I have also devoted much labour and care to the first. In particular, I believe I have completely exhausted the number of facts which are in any way remarkable, just as I believe I can vouch for the correctness of all statements and quotations, which I have always checked in detail as far as the works in question were accessible to me.

It is an old experience that almost every scientific discovery of some importance is made more than once, and by different researchers independently of each other. However, it is just as often the case that as soon as the discovery gains a certain degree of prestige, a whole number of applicants immediately arise to claim the fame of priority. These questions, which, as is well known, have been most vividly discussed with regard to the principle of the conservation of energy, and which have unfortunately led, among other things, to excessive attacks on personalities who are among the first to work on the development of the physical sciences, have only been touched upon in the present paper to the extent that the objective presentation of the historical development of the principle seemed to me to require it. I may feel all the less called upon to judge in such matters, as many of the men are still alive who are able to give the most competent judgement through their personal testimony. I may only add a brief general comment here.

Certainly, the person who expresses an important idea for the first time has gained a lasting merit. 
However, it will always be important to check whether he was also fully aware of the scope of this idea, and whether he was able to do something with this idea and develop it further. 
If this condition is applied when assessing priority disputes, the number of concurrents will certainly have to be significantly reduced. It has become almost fashionable nowadays to search the writings of older physicists and philosophers for statements reminiscent of the principle of the \dashuline{\,conservation of energy} or of the \dashuline{\,mechanical theory of heat}.
Much has already been found in this direction, and much more would undoubtedly be found if one were to search further. 
However, as important as it seems to establish the fact that certain ideas were quietly growing in the minds of individual outstanding minds long before they were handed over to mankind as the ripe fruit of the commons, one must not unilaterally attribute the credit for the discovery to those 
 who perhaps had no idea at all about the potential for development of the germ that an occasionally expressed thought contained.

When it is a question of making clear the meaning of a physical proposition for the investigation of the laws of the phenomenal world, it is above all necessary to compare the content of the proposition with such facts as appear to be established with complete certainty by experience, and the deeper and more comprehensive the meaning which one attaches to the proposition to be investigated, the more closely one must follow the direct results of observation, which after all form the only reliable starting point of all natural science. This applies in an outstanding degree to the principle of the \dashuline{\,conservation of energy}, a theorem of such universal significance, with such a profound effect on all scientific theories, that it cannot be purified carefully enough of all hypothetical ideas which are so easily formed in order to facilitate an overview of the legal connection between the various natural phenomena.
For if any doubtful premise, any unproved hypothesis, is introduced into the investigation, it is not the principle itself that is tested, but that hypothesis at the same time, and any difference between theory and experience will then have to be set down not only to the account of the principle, but equally well to that of the hypothesis. 

On the basis of this consideration, I believed that the main focus of my presentation should always be on basing the \dashuline{\,concept and principle of energy} primarily on pure empirical facts, avoiding all hypotheses as far as possible, including the various molecular hypotheses, even though some of them have gained a notable place in science in recent times. Thus I have also excluded \dashuline{\,Carnot-Clausius's Principle}: the so-called \dashuline{\,second law of the mechanical theory of heat}, with its corollaries, from the investigation, because it is itself developed from the principle of energy, by adding to it an entirely new element: the conditions of the \dashuline{\,transformation of the different kinds of energy into each other} (p. 129). I intend perhaps to make it the subject of a special work.

Only when the ground on which the \dashuline{\,doctrine of energy} rests is thus securely established, may one begin to apply it to more remote fields of research ; but then the principle is no longer to be investigated in and of itself, but conversely it serves as a guide to test other hypotheses against it. In this respect I have gone as far as the time at my disposal would permit. It could not occur to me to give a complete survey of all the individual applications that have ever been made of the \dashuline{\,Principle}, but I believe I have at least contributed something new, if not, of course, in the number of external facts, at least in the manner of conception.\,}''

Finally, I feel obliged to express my sincere thanks to the high faculty for the many-sided scientific encouragement and support that I have received through working on this rich and beautiful task.

Apart from my treatment of the task, two others were received which were not honoured. The Faculty's judgement on mine follows 
verbatim: 

``{\it In the first section the author combines the development of the concept of energy with a detailed historical theory of the \dashuline{equivalence theorem of mechanical heat\,}, which bears the most favourable testimony to his sound and independent judgement and his thorough acquaintance with the sources. The epoch-making achievements that prepared and founded the principle are presented with great clarity and the most precise expertise. The continuity of the development is preserved by an appreciation of the intermediate elements based on a fine scientific feeling.

The fact that the gradual spread of the principle and its application to the various fields of physics have also been given a very detailed account was not exactly favourable to the work. 
In this part of the work the author has not always succeeded in avoiding the impression of tedious repetition, and the limitation which he has imposed on the freedom of his exposition by placing the purely historical point of view at the forefront is particularly noticeable here. For the economy of the whole, it would have been better to assign part of the material dealt with here to the second or third part, while on the other hand many things would have been taken into consideration if the author had been less strictly bound to the year 1860 as the limit of historical development. The one-sided physical standpoint which the author adopts with full awareness means that he has only briefly touched on the part played by technology in the development of the concept of energy, and has not taken account of the philosophical circles of ideas at all. The Faculty would have wished for a more comprehensive presentation and a more detailed appreciation of these influences.

The Faculty has taken note of the second section of the treatise with great interest; here the methodical way of thinking, the thorough mathematical-physical education of the author and the prudence of his judgement come into their own. The love with which the author has immersed himself in the subject of his enquiry is matched by the care with which he is able to elucidate it in all directions. The Faculty states with lively satisfaction that, apart from a few factually insignificant inaccuracies, the question of the formulation and proof of the energy principle has found a beautiful and complete solution in this part of the treatise.

The Faculty's appreciation of the last part of the work is not equally unqualified. It regrets that the author's limited time prevented him from giving his account of the \dashuline{\,various types of energy} the desirable completeness and uniformity. As attractive as the author's observations are, and as much instruction can be drawn from them, the Faculty misses a more general examination of the question of \dashuline{\,how many types of energy} are to be distinguished and how each of them is to be defined. Instead, the author has confined himself to demonstrating in detail how the \dashuline{\,principle of energy} can be used as a reliable and uniform basis of representation in the various fields of physics. The skilful treatment which he gives to mechanics from this point of view would have gained greater significance if the author had subjected the scope of the \dashuline{\,principle of the superposition of different energies} introduced by him to more detailed criticism. Also, in the opinion of the Faculty, a somewhat more detailed consideration of the \dashuline{\,reflection and refraction of light} would by no means have exceeded the scope of the task.

Like the treatment of \dashuline{\,optics}, that of \dashuline{\,thermal and chemical energy} also appears somewhat brief. In particular, the author has omitted a critical discussion of those experimental investigations on which our knowledge of the \dashuline{\,numerical value of the mechanical heat equivalent} is based. The treatment of \dashuline{\,electrical and magnetic energy} sufficiently demonstrates the comprehensive and thorough knowledge which the author possesses in this field, but in some cases his observations lack the clarity, consistency and coherence which have made the study of his work such a pleasant task. Finally, the Faculties must withhold their approval from the remarks by which the author seeks to come to terms with \dashuline{\,Weber's law}. They would have considered a thorough examination of Weber's ideas necessary.

It will not be difficult for the author to complete his treatise in the points mentioned before it is published. In this hope and in full recognition of his achievements, the Faculty awards this treatise the second prize.\,}''

If I have nevertheless decided, in response to the Faculty's expressed wish, to hand over the work to the public essentially unchanged, this has not been done without serious misgivings and careful consideration, since it must be in my interest to give the work, once completed to a certain degree, the completeness and rounding off that is considered desirable by such high authorities. However, it was precisely this consideration that first gave me serious doubts as to whether I would even succeed in realising the hope expressed by the Faculty. 

This applies in particular to a possible revision of the third section, especially the chapter on \dashuline{\,electrical energy}, in line with the comments contained in the Faculty's judgement. I am, of course, far from wishing to detract from the great importance of \dashuline{\,Weber's ideas}, which they have gained over the last few decades, especially in Germany. The powerful support, the great expansion of the overview, which is due to the work of this brilliant researcher in the entire field of electricity and beyond, is obvious to every physicist.

But neither can I avoid, indeed I consider it my duty, to confess openly that through careful study and mature reflection I have come to the firm conviction that the more speculative, deductive direction which Weber teaches has already borne its most valuable fruit, and that further substantial progress can therefore no longer be expected from it to the same extent in the future. In my personal opinion, such progress can only be achieved in the near future by closely following the inductive method, and I have tried to express this point of view in the description of electrical phenomena by restricting myself essentially to the investigation of uniform closed currents, whose laws have been sufficiently established by experience and which can be deduced even without Weber's general law, whereas I have focussed on the area in which its actual significance begins: the effects of unclosed currents or moving electrical mass points, or at least I thought I should only touch on it.

Since, of course, every physicist, in order to increase the clarity of the phenomenal world, adapts a basic view to the observed laws of nature according to his personal taste, I will also gladly confess that I currently count myself among the supporters of that theory which abandons the assumption of the direct remote effect and thus also the idea of the primary existence of an electrical basic law in the manner of Weber's at all. The decision on this question must, of course, be left to the future, but for me there is one important circumstance to consider, namely that in the event of a possible attempt at a reworking in the sense indicated, I would in any case not set to work with the same enthusiasm as the first time, and this had to be an absolute prerequisite if I wanted to count on fulfilling the Faculty's expectations, which were so honourable for me.
If I add to this uncertainty in the prospect of a satisfactory outcome the time that would be necessary for such a reworking, and which would hit me doubly hard just now 
I have turned to other studies, as well as the associated delay in the publication of the work, the completion of which is now almost a year behind, I believe that I cannot entirely relinquish the hope that my decision to leave it at a few 
editorial changes,$\,$\footnote{$\:$Apart from a newly added note at the end of the third section.}
and otherwise prefer to publish the work entirely in the form in which it has found the judgement of the Faculty communicated in the foregoing, may also find the excuse from the high Faculty, which is so important to me

\vspace*{2mm}

Kiel, July 1887. 
The author {\it\color{red}(Max Planck)}\,. 

\newpage
\section{\underline{Historical development} (p.1-91)}
\label{Section-1}
\vspace*{-2mm}

\vspace*{2mm} 
\begin{center}
{\it\color{red}\bf
========
The principles (p.1)
========
}
\end{center}
\vspace*{-3mm}

There are two propositions which serve as the foundation of the present structure of the exact sciences: the \dashuline{\,principle of the conservation of matter} and the \dashuline{\,principle of the conservation of energy}. 
Above all other laws of physics, however comprehensive, they assert their undeniable precedence, because even Newton's great axioms (the laws of inertia, of the proportionality of force and acceleration, and of the equality of action and counteraction)  extend only to a special branch of physics: mechanics --for which, moreover, they can all be deduced from the \dashuline{\,principle of the conservation of energy}, under certain conditions to be explained later (see the Section~3) 
  In recent times, however, it has become more and more likely that all natural processes can be traced back to phenomena of motion, 
i.e. to the laws of mechanics, but this transfer is still far from being successful to any extent, which allows the direct application of these mechanical axioms to any natural phenomenon.

The \dashuline{\,principle of the conservation of energy}, on the other hand, documents its universal character precisely by the fact that, if a completely new natural phenomenon were to be discovered today, a measure and a law for the new phenomenon could easily be obtained from it, while there is no other axiom that could be extended with the same confidence to all processes in nature. This has been shown particularly clearly in the justification of the various theories of electricity, in that, apart from the experimentally established facts (effects of closed currents), the recognition of the \dashuline{\,principle of the conservation of energy} forms the only common starting point for all theories that claim to be admissible.

The two principles mentioned at the beginning are to a certain extent opposed to each other in a coordinated manner, in that one expresses the \dashuline{\,indestructibility of matter} (better: of mass measured by weight), \dashuline{\,the other that of force} (in the corresponding sense of this word) --an analogy which can be carried out even more in detail and which has contributed a great deal to the clarification of the concept. 
As related as the two propositions appear in their content, however, they have undergone such a different history of development. While the \dashuline{\,constancy of matter} was already asserted by the ancient Greek natural philosophers, especially Democritus, and was maintained by all atomists, and finally brought to unqualified recognition by Black and Lavoisier by the proposition that the weight of a system of bodies is not altered by any chemical process, not even by combustion, the discovery of the \dashuline{\,principle of the conservation of energy} must be regarded as an achievement of more recent times, and in its most precise and general form, of the most recent times.

\vspace*{-2mm} 
\begin{center}
{\it\color{red}\bf
======== 
Perpetuum mobile  (p.2)
========
}
\end{center}
\vspace*{-2mm}

The first trace of the existence of such a principle appeared in the experience gained centuries ago, partly through laborious and costly experiments, i.e. inductively, that \dashuline{\,it was not possible to build a 
perpetual motion machine},$\,$\footnote{$\:$ In contrast to its literal meaning, the expression: \dashuline{\,Perpetuum mobile} is usually not used in the sense of constant movement, but in the sense of constant work performance.} 
i.e. to construct a (periodically acting) machine through which any amount of work or 
\dashuline{\,living force} {\it\color{red}(kinetic energy)}$\,$\footnote{\label{label_footnote_vis_viva}$\,${\it\color{red}Here, Max Planck retained the old terminology coming from the old Latin term ``\,Vis viva\,'' coined by Gottfried Wilhelm Leibniz and after the work of Christian Huygens, or equivalently  ``\,force vive\,'' in French, or ``\,living force\,'' in English, or ``\,lebendige Kraft\,'' in German.
Technically, the ``\,Vis viva\,'' $(m\:v^2)$ was twice what is nowadays called the ``\,kinetic energy\,'' $(m\:v^2/2)$ after the result of the work of Gaspard-Gustave Coriolis and Jean-Victor Poncelet.
The ``\,lebendige Kraft\,'' (living force) of Max Planck is therefore nothing else than the ``\,kinetic energy\,'' $(m\:v^2/2)$, in opposition with the other kind of ``\,dead force\,'' or ``\,potential energy\,'' in English. 
 (P. Marquet)}.} 
can be gained without a corresponding expenditure of any other agent, be it consumption of certain materials, be it loss of other work or living force {\it\color{red}(kinetic energy)}, in other words: without a certain other change connected with it, which --to use an expression that R. Clausius uses, albeit on a completely different occasion-- has the peculiarity that it cannot be reversed without on its part, be it indirectly or directly, to cause a consumption of work or living energy {\it\color{red}(kinetic energy)}. This change can be viewed as compensation, as an equivalent of the work done, and one can then say in short: {\it\dashuline{performance of work or production of living force {\it\color{red}(kinetic energy)} cannot take place without some form of compensation}}, or even more briefly: {\it\dashuline{\it it is impossible to obtain work from nothing}}.

\vspace*{-2mm} 
\begin{center}
{\it\color{red}\bf
---------
(p.3) 
---------
}
\end{center}
\vspace*{-2mm}

The extent to which this theorem can serve to prove the \dashuline{principle of the conservation of energy} in its generality will be shown in the next section of this paper, although it can already be seen here that there was still a long way to go from the realisation of this \dashuline{empirical theorem} to the exact mathematical formulation of the \dashuline{general principle}. After all, it took quite some time to arrive at the further, here highly essential, realisation that this theorem could also be reversed, i.e. that there is no device by which \dashuline{work} or \dashuline{living force} could be continuously consumed without some other change that could be regarded as \dashuline{compensation}. Because the more general recognition of this latter proposition, as we shall see clearly later, is of much more recent date, and is further striking proof that we are dealing here with \dashuline{a pure fact of experience}, since it has always been more important to men to gain (work-power)  ``\dashuline{\,capacity for work}'' than a lose of it.
The difficulties still to be overcome related mainly to the answer to the question in which processes the above-mentioned compensation was to be sought, in what connection the amount of compensation was to be found with the work performed (or consumed), and what was to be regarded as the measure of compensation, as the equivalent value of the work performed. 
We shall have occasion to remark frequently hereafter that most of the important differences of opinion and misunderstandings which have arisen in the course of time in the application of our principle have not referred either to the recognition or denial of the proposition itself -- its validity was generally admitted -- but rather to the \dashuline{measurement of compensation}, to the \dashuline{equivalent value of the work performed}. 
This can be traced right up to recent times.

\vspace*{-2mm} 
\begin{center}
{\it\color{red}\bf
========
S. Stevin (1608/Deutch-Latin--1634/French) (p.4)
========
}
\end{center}
\vspace*{-2mm}

Despite its imperfect form, the primitive proposition that \dashuline{work, like matter, cannot arise from nothing}, 
already show a certain fruitfulness in earlier times, and as it became more and more deeply imprinted on human thought, 
the best preparation was thereby gained for its later specification by the general principle.
We also find the theorem of the impossibility of perpetual motion used repeatedly to reach scientific conclusions in mechanics. 
The proof that 
\dashuline{S. Stevin}$\,$\footnote{$\;$See E. Mach: Die Mechanik in ihrer Entwickelung historischkritisch dargestellt (Mechanics in its development presented in a historically critical manner), Leipzig 1883, p.24. E. D\"uhring: Kritische Geschichte der allgemeinen Principien der Mechanik (Critical history of the general principles of mechanics), Berlin 1873, p.61..}
gives in his 
``\,{\it Hypomnemata mathematica\,}''$\,$\footnote{$\:$S. Stevinus: Hypomnemata mathematica (Mathematics memos), in French by Girard. Leiden 1634, p. 448.}
published in Leiden in 1608, for the \dashuline{{\it laws of equilibrium on an inclined plane}}, and uses it as the basis for his entire system of statics, is famous. 
If you imagine a heavy chain placed over the tip and the adjoining legs of a vertical triangle (with a horizontal base line), the ends of which are connected on both sides of the base line, then it is clear, according to Stevin, that the chain is in equilibrium.
Because if this were not the case, it would begin to slide to one side, and this movement would continue indefinitely, since the configuration of the system always remains the same.
You could therefore use this device to \dashuline{gain work ad infinitum without any corresponding compensation}. 
From the \dashuline{impossibility of such an apparatus}, Stevin concludes the existence of equilibrium, which remains undisturbed even if the two chain ends hanging symmetrically from the end points of the base line are cut off at the same time, so that the result is the sentence that is not at all obvious at first glance, that an unclosed chain placed over a vertical triangle is in equilibrium if its end points lie in the same horizontal plane.

\vspace*{9mm} 
\begin{center}
{\it\color{red}\bf
==========
Galileo Galilei (p.5)
==========
\\
=
(letter-to-Sarpi-1605,  Dialogo-1632, Discorsi-1638)
=
\\
===============================
}
\end{center}
\vspace*{-2mm}

\dashuline{Galileo Galilei} also seems to have started from a similar assumption as Stevin, when proving the theorem that the speed acquired by a heavy body by falling on any path depends only on the vertical distance between the initial and final positions, {\it\color{red}with} the supposition that if this proposition were incorrect, a means might at once be given of raising a body to a greater height solely by the action of its own gravity, by making it fall on a certain curve and rise again on another suitable one let.
This would of course result in \dashuline{\,perpetuum mobile}.

\vspace*{-2mm} 
\begin{center}
{\it\color{red}\bf
=====
Christian Huygens (1656-1669-1673) (p.5)
=====
\\
======
Johannes Bernoulli (1695-1717-1742)
======
}
\end{center}
\vspace*{-2mm}

Closely related to this is the insight, also gained by Galileo,
that if a greater load is (slowly) lifted by a falling weight, the products of the weights in the simultaneously travelled distances are equal 
--a theorem that was later extended, especially by 
\dashuline{Johannes Bernoulli} 
(1717)$\,$\footnote{$\:$Joh. Bernoulli: Opera, 1742, T. III.} 
to the \dashuline{principle of virtual displacements} 
(velocities).

More precisely, the statement that a body cannot rise through its own gravity, or more generally: {\it that a system of heavy points or bodies cannot move its center of gravity higher through the driving force of its own gravity}, 
was of the greatest importance for the development of mechanics by \dashuline{C. Huygens}.
Indeed, as is well known, the 
same {\it\color{red}\dashuline{C. Huygens}}$\,$\footnote{$\:$C. Huygens: Horologium oscillatorium, Paris, 1673.}
based his theory of the \dashuline{\,physical pendulum} on the theorem that a system of firmly connected mathematical pendulums (which every physical pendulum can be viewed as) cannot move its center of gravity higher during the ascending movement due to the resulting speed, as if the pendulums were all swinging independently of each other (with the same initial angular velocity). 
Huygens does not seem to have considered it necessary to provide a proof of this theorem, so the instinctive conviction of its correctness, i.e. the recognition of the \dashuline{\,impossibility of perpetuum mobile}, must already have been present in him.
If one admits the statement to be correct, then the theory of the center of oscillation follows immediately.

\vspace*{-2mm} 
\begin{center}
{\it\color{red}\bf
---------
(p.6) 
---------
}
\end{center}
\vspace*{-3mm}

\dashuline{Huygens' theorem} already contains the principle of 
\dashuline{\,living force}$\,$\footnote{\it\color{red}$\:$See the footnote~\ref{label_footnote_vis_viva} about the ``vis viva'' $(m\:v^2)$ and the ``kinetic energy'' $(m\:v^2/2)$. (P. Marquet)}
in its application to gravity.
Because it was already known through Galileo that \dashuline{the heigh} to which a body thrown upwards can rise  \dashuline{is proportional to the square of its velocity}, 
and therefore, after 
\dashuline{Leibnitz}$\,$\footnote{$\:$G. W. Leibnitz: Acta Erud. Lips. 1695.} 
in 1695, the name ``\dashuline{\,vis viva\,}'' was introduced for the quantity $m\:v^2$ (the factor $1/2$ is probably first found in 
Coriolis$\,$\footnote{$\:$Coriolis: Calcul de l'effet de machines, Paris 1829.
{\it\color{red}It is indeed p.17 of this book that Gustave Coriolis \dashuline{\,explicitly} suggested for the first time in 1829 to modify the ``force vive'' by introducing the factor $1/2$. 
Similarly, Jean-Victor Poncelet \dashuline{\,explicitly} explained in the same year 1829 that:
``{\color{blue}the quantity of work developed by the gravity (...) is equal to half of the product of the velocity squared by the mass of this body (...) the quantity of work is half the ``vis viva'' and therefore with the factor 1/2 included.}''
Moreover, this factor $1/2$ was previously \dashuline{\,partly/implicitly} introduced by \dashuline{Johannes Bernoulli} in 1724-1727 (where the quantity $\frac{1}{2}\:v\:v$ appeared); also by \dashuline{Daniele Bernoulli} in 1736-1738 (where the quantity $\frac{1}{2}\:m\:v\:v$ appeared); also by \dashuline{Lazare Carnot} (the father of Sadi Carnot) in 1786 (where the quantity $\frac{1}{2}\:M\:V^2$ appeared); and then by \dashuline{de La Grange (next Lagrange)} in 1788 (where the quantity $\iiint \frac{1}{2}\:v^2\:dm$ appeared. (P. Marquet)}}), 
uttering the proposition that the \dashuline{\,living force} of a body that moves under the influence of gravity, 
whether completely free or limited by fixed connections (axes of rotation, etc.), depends only on the height of the centre of gravity.

Unfortunately, due to a confusion with Newton's concept of force, \dashuline{\,Leibnitz's terminology}, which has been preserved to this day, has brought about a \dashuline{\,disastrous confusion of ideas} and a countless number of misunderstandings, which could not be avoided because Leibnitz was for the pressure of a heavy body at rest (i.e. the Newtonian force), and wanted to use the distinctive term ``\dashuline{\,vis mortua\,}.'' 
The two types of forces (\dashuline{\,living and dead forces}) were of completely different dimensions. 
We will return to this point later when discussing \dashuline{R. Mayer}'s work.

The importance of the concept of \dashuline{living force} for the \dashuline{laws of collision} was recognized earlier (1669) by Wren and 
Huygens$\,$\footnote{\it\color{red}$\:$In fact, C. Huygens  discovered as early as in 1656 --and thus long before Leibnitz, Wreng or Wallis-- 
the ``{\color{blue}Regles du} 
{\color{blue}mouvement dans la rencontre des corps   (Laws of motion in the encounter of bodies)}''
and first published in France in 1669 (in old French and in the Journal des Sçavans), then in 1669 in England (in Latin and in the Philosophical translations), and then in Holland (in Latin and in the ``Opuscula postuma'') in 1703, thus in his last posthumous paper ``\,De motu corporum ex percussione\,'' (On the movement of bodies by percussion), the interest of the quantity $m\:v^2$ when he discovered the laws of encounter of bodies and the rules of motion during collisions of bodies (with still modern applications to the game of billiards).
In particular, the sixth (new) rule of Huygens was: ``\,{\color{blue}the sum of the product of the mass multiplied by the velocity squared (namely $\sum_km_k\:v_k^2$) is the same before and after the encounter.}''
This clearly corresponds to the conservation of what will next be called the ``vis viva'' and the ``kinetic energy'' as well, although C. Huygens did not give a special name to this quantity $\sum_km_k\:v_k^2$ ... and here is likely the reason why C. Huygens is so often forgotten in the story of the discover of the conservation of the vis viva (or kinetic) energy. 
(P. Marquet)},
who in their theories of elastic collision unanimously came to the conclusion that no \dashuline{\,living force} is lost when two elastic bodies collide.
On the other hand, from the laws that \dashuline{Wallis} discovered at the same time in his investigations into the inelastic collision, there is a loss of \dashuline{\,living force} in this 
collision.$\,$\footnote{\it{\color{red}$\:$In fact, it is now well-known that both Wallis (1669) and Christopher (1669) \dashuline{stole the ideas of C. Huygens} when they published papers in the Philosophical Transactions, but without any credit of the ideas of Huygens, whereas they do learned the results of Huygens (1656) when Huygens \dashuline{visited and meet in London Drs. Moore, Brounker, Neil, Rook, Goddard, Wallis and Wren in 1661 (explaining to them, with the use of his new rules, their experimental facts that they were unable to explain)}... Lastly, the Secretary of the Royal Society of England, M. d'Oldembourg and others, have confirmed that ``Huygens was the first to know these new rules for collisions, and in particular the fifth and sixth of them'' ... even though they considered that: ``Wallis and Wren would have been right to use the ideas of Huygens and to publish them first on their side, and without even citing Huygens' contributions''... what a strange world! (P. Marquet)}}

\vspace*{-2mm} 
\begin{center}
{\it\color{red}\bf
==
Ren\'e Descarte - Gottfried Wilhelm Leibnitz  (1687-1695) (p.7)
==
}
\end{center}
\vspace*{-2mm}

However, the concept of \dashuline{\,living force} gained the most interest through the well-known controversy between \dashuline{Descartes} (\dashuline{Papin}) and \dashuline{Leibnitz}, which was continued with increasing violence by their supporters on both sides long after their deaths, \dashuline{about the true measure of the force of a body in motion}. 
Leibnitz relied on the experience that the same force (work) is required to lift a certain weight by $4$ feet as to lift $4$ times the weight by $1$ foot, since in both cases the entire effort can be broken down into $4$ individual powers, each consisting of the lifting of the simple weight by $1$ foot.

If you now imagine that the lifting of the weights is accomplished by giving them an upward speed that is just sufficient to bring them to the specific height, then according to the Galilean theorems you have to give the simple weight so that it has four times the height achieved, and not four times but twice the speed of that which has to be given to the four times the weight so that it reaches the single height. 
But \dashuline{since the same causes belong to the same effects}, Leibnitz conclude that, the force inherent in a single weight with twice the speed is also equal to the force inherent in the four-fold weight with one speed, from which the general expression $m\:v^2$ comes from as a measure of strength follows.

  Descartes and his disciples were different {\it\color{red}because for them}: 
a double force produces a double speed on the same body in the same time, consequently the quantity of movement ($m\:v$) forms the true measure of force$\,$\footnote{$\:$M. Zwerger: Die lebendige Kraft und ihr Maass. Ein Beitrag zur Geschichte der Physik. München {\it\color{red}(The living force and its measure. A contribution to the history of physics. Munich) 1885.}}.

\vspace*{-4mm} 
\begin{center}
{\it\color{red}\bf
---------
(p.8) 
---------
}
\end{center}
\vspace*{-3mm}

According to our current, more precise physical approach, which distinguishes precisely between \dashuline{\,force} and \dashuline{\,work}, we must of course initially explain this entire dispute as a purely verbal dispute.
Because one can only speak of a factual controversy when one has agreed on the (completely arbitrary from the outset) definition of the term in question. As long as there was no clear idea associated with the word \dashuline{\,strength}, a dispute about the measure of strength was completely pointless. 
However, there is no mistaking the fact that the dispute under discussion was based on a much deeper content, because the parties, even if this was expressed only occasionally and indistinctly, did in fact agree to a certain extent on what they wanted to mean by ``force.'' 
Both \dashuline{Descartes} and \dashuline{Leibnitz} certainly had 
{\it\color{red}a unique}, 
(although not entirely clearly specified) idea of the existence of a principle which expresses the \dashuline{\,immutability and indestructibility} of that from which all movement and effect in the world arises.
While \dashuline{Descartes} supported the validity of this principle through \dashuline{theological considerations} based on the \dashuline{\,eternity of the Creator}, \dashuline{Leibnitz} starts from the \dashuline{\,law of cause and effect}:
``{\it \dashuline{cause can only produce the effect that corresponds to it, no larger or smaller}}.'' 
So neither growth nor decrease can take place in the continuous chain of causes and effects from which the phenomena of the world are formed: ``{\it \dashuline{there is something there that remains constant}}.'' 
If we call this ``{\it something\,}'' \dashuline{\,force}, we have an idea, albeit a very imperfect one, of what \dashuline{the concept of force} formed the common starting point for the two different conceptions. 
For now a difference of opinion was very possible as to whether Descartes' \dashuline{quantity of motion} or whether Leibnitz's \dashuline{living force} was the true measure of that concept. 
If the dispute had been conducted in this somewhat more precise form, \dashuline{Leibnitz would have been right}.
We already have before us one of the cases mentioned above, where it is less about the recognition of the immutability of force than about the equivalent value of this quantity recognized by both parties as immutable, namely the \dashuline{measure of compensation}, which is expressed in the \dashuline{speed of one body} that occurs when its movement is used to produce a specific effect. The same thought will come back to us again and again.

\vspace*{-1mm} 
\begin{center}
{\it\color{red}\bf
=======
Isaac Newton (1687) (p.9)
=======
}
\end{center}
\vspace*{-2mm}

When, towards the end of the 17th century, mechanics, which at that time still constituted almost the only branch of physics, was brought to perfection by \dashuline{Isaac Newton}, which is still essentially unsurpassed 
today,\footnote{$\:$\it\color{red}This was written by Planck in 1887, and thus long before 1905-1915 and the publication of the papers of Einstein...} 
the \dashuline{concept of force}, it seems, also became important all times, finally established, \dashuline{in a sense that follows the measure of force used by Descartes}. 

\dashuline{Newton} (1687) understood the force directly as a pressure (as can be perceived through muscle feeling) and therefore \dashuline{measured the magnitude of a force by the quantity of movement that this pressure produces on the unit of mass in the unit of time}, from which the dimension of the force is derived \dashuline{as a product of mass and acceleration}. 
Of course, this size has nothing to do with the principle of conservation of \,``\,force,'' and this may also have been a reason why this principle lost some of its interest again for a while. 
\dashuline{\,Leibnitz's concept of force} now appears as the \dashuline{\,performance or work of Newton's force}.
The latter merely describes a necessary, but not yet sufficient, condition for achieving a performance.

\dashuline{Newton} himself seems to have never been particularly concerned with the concept of the \dashuline{\,power or work of a force}, although there are a few places in his works where he explains this concept in more detail. This includes the often quoted definition of 
\dashuline{\,actio agentis}$\,$\footnote{$\:$I. Newton, Philosophiae naturalis principia mathematica, Opera, ed. S. Horsley. Vol. II. Londini 1779. p. 28s.} 
(\,``{\it\color{red} action of the agent\:}'' or ``\:product of a force in the corresponding velocity component of its point of application\,''\,), which indicates the \dashuline{\,amount of work done by the force in a unit of time}. 
However, no further use is made of this definition {\color{red}by \dashuline{Newton}}. 
In general, it seems to me that attempts to derive the \dashuline{principle of conservation of energy} from this passage, which is taken from the commentary on the axiom of the \dashuline{\,equality of action and reaction}, do not promise success, if only because the content of the two sentences mentioned are in completely different areas heard. 
In any case, Newton accepted the fact that motion is lost through friction or through imperfect elasticity, without any hesitation or other remark
remark.\footnote{$\:$I. Newton: Opera, ed. S. Horsley. Vol. IV. Londini 1782. p. 258.}

\vspace*{2mm} 
\begin{center}
{\it\color{red}\bf
=
J. \& D. Bernoulli (1724-1727-1742) - Leonhard Euler (1745) (p.10)
=
\\
=====
(James Watt, 1782) -- (Jean-Victor Poncelet, 1826)
=====
}
\end{center}
\vspace*{-2mm}

We owe the further development of the \dashuline{concept of work and living force} to the physicists of Basel, especially {\it\color{red}(Jean)} \dashuline{Johannes Bernoulli}, who follows Leibnitz's view quite closely. He repeatedly speaks of the \dashuline{\,conservatio virium vivarum} {\it\color{red}(in Latin: conservation of living forces)} and emphasizes that when \dashuline{\,living strength} disappears, the \dashuline{\,ability to do work} 
(\dashuline{\,facultas agendi}$\,$\footnote{$\:$Joh. Bernoulli: Opera, 1742, T. III, p.239.})
is not lost, but only transformed and is 
converted$\,$\footnote{$\:$Joh. Bernoulli: Opera, Lausannae et Genevae 1742, T. III, p. 243.}
into a different form (e.g. compression). 

According to 
\dashuline{L. Euler}$\,$\footnote{$\:${\it\color{red}L. Euler: De la force de percussion et de sa v\'eritable mesure (On the percussion force and its true measurement), M\'emoires de l'Acad\'emie des Sciences de Berlin, p.21-53, 1745}.}, 
the \dashuline{\,living force} of a point that is attracted or repelled by a fixed center according to a power of distance is always the same as often as it returns to the same place in space, while otherwise its increase is due to work (``\dashuline{\,effort\,}'' {\it\color{red}in French}) of the force is measured (the expression ``\dashuline{\,travail\,}'' {\it\color{red}in French} comes from 
Poncelet$\,$\footnote{$\:$Poncelet: Cours de mécanique appliquée aux machines. Metz 1826.}).

\dashuline{Daniel Bernoulli} extended this sentence to \dashuline{several moving points} and also taught us the high fruitfulness of the principles for the laws of \dashuline{motion of fluids} developed by his 
father {\it\color{red}Johannes}$\,$\footnote{$\:$Dan. Bernoulli : Hydrodynamica, 1738. Ausserdem vgl. Remarques sur le principe de la conservation des forces vives pris dans un sens général. Histoire de l'Académie de Berlin, 1748, p.356.}.

The need for a closer study of the concept of work also became apparent \dashuline{in technology} and led to the introduction of the term ``\,\dashuline{horse power}\,'' (work of a horse per second) by \dashuline{J. Watt}.

\vspace*{-1mm} 
\begin{center}
{\it\color{red}\bf
=======
Thomas Young  (1807) (p.11) 
=======
}
\end{center}
\vspace*{-3mm}

It was \dashuline{Thomas Young} who first used the name \dashuline{\,energy} to describe the \dashuline{\,living force of a moving body}, and thus laid the foundation for the meaning of this expression today. The word 
\textgreek{>en'ergeia}
{\it\color{red}(and thus ``energeia'')} 
can already be found
in the physical sense 
in Aristotle and other physicists: Galileo and  
{\it\color{red}(Jean)} Johannes Bernoulli$\,$\footnote{$\:$Mr. Hagenbach highlights in a lecture on the merits of John and Dan. Bernoulli repeatedly points out (p.24 and p.28) about the theorem of \dashuline{\,conservation of energy} (Verh. d. naturf. Ges. zu Basel, Th. VII, 1884) that Joh. Bernoulli already gave the concept of work the name ``Energy.'' However, despite a careful review of all of Bernoulli's writings (Opera, 1742), I could not find this remark confirmed anywhere; the only time I noticed the word \dashuline{\,énergie} (T. III, p.45) is in one used in a completely different sense.
{\it\color{red}In fact, I have found in Johannes Bernoulli (1724-1727) in the book ``Discours sur les loix de la communication du mouvement'' (written in old French, with the ``s'' replaced by ``f'') twice the use of this word:  
``C'est cette force, en tant qu'elle est dans le corps
mis en mouvement par épuisement de la pression du ressort, qu'on doit apeller proprement la \dashuline{force vive}, en vertu de laquelle le corps fe tranfporte d'un lieu à l'autre, avec une certaine viteffe, plus ou moins grande felon l'\dashuline{énergie} du reffort'' {\color{blue}(It is this force, in so far as it is in the body set in motion by the exhaustion of the pressure of the spring, that must properly be called the \dashuline{living force}, by virtue of which the body is transported from one place to another  with a certain speed more or less great according to the \dashuline{energy} of the spring)} in the Chapter~V, 
p.34; and then: ``(...) puifque le moment ou l'\dashuline{\,énergie} des forces mortes (...)'' {\color{blue}(since the moment or \dashuline{energy} of the dead forces)} in the Chapter~VI, 
p.42
(P. Marquet)}}, 
who also use it occasionally, but without attaching a special meaning to it.
In his investigation of the laws of collision,  
\dashuline{Young}$\,$\footnote{$\:$Th. Young: A course of lectures on natural philosophy. London 1807. Vol. I. Lect. VIII. p.75. On collision.}
found, just like Wren, Wallis and Huygens before him, that in the central collision of two bodies the quantity of movement (i.e. the amount of movement of the center of gravity) is preserved under all circumstances, 
and, quite in the Descartesian sense, rejecting the opposing views expressed by Leibnitz and Smeaton, he describes this quantity as the true measure of the \dashuline{\,force inherent in the moving body}. 
However, he {\it\color{red}(Thomas Young)} considers the quantity that others call \dashuline{\,living force} to be important enough to give it a special name, that of the \dashuline{\,energy of the moving body}, especially since there are cases where the effect of the moving body is obviously measured by the \dashuline{\,square of its speed}.
For example, a ball with double velocity drills a hole four times as deep in a piece of soft clay or tallow than one with single velocity, and to double the velocity of a body, you have to drop it from four times the height. Young also emphasises that perfectly elastic balls retain their energy on impact. Nevertheless, he is still far removed from the \dashuline{\,general principle of the conservation of energy}, because the necessary extension of the \dashuline{\,concept of energy} was reserved for a later time.

\vspace*{-2mm} 
\begin{center}
{\it\color{red}\bf
---------
(p.12) 
---------
}
\end{center}
\vspace*{-2mm}

If we briefly survey the research carried out up to the end of the last {\it\color{red}(18th)} century and the beginning of the present {\it\color{red}(19th)} century in the field we have described, the mature fruit of this research is the realisation of the \dashuline{\,law of the conservation of living forces}. In a system of material points which are subject to central forces, the living force is only dependent on the momentary configuration of the system, namely on the value which the \dashuline{\,force function} (so named by R. Hamilton) has in this configuration. 
The change in the \dashuline{\,force function} thus measures the  \dashuline{\,work done by the forces}, by whatever means the change takes place, and on returning to the same configuration the living force is again the same. 
This theorem rules out the construction of a perpetuum mobile by purely mechanical effects. 
However, the validity of this theorem must be limited to a certain kind of forces, which are currently characterised as ``conservative.'' 
This theorem does not apply to friction, inelastic shock, etc., because here, 
on the contrary, loss of living force regularly occurs.

Only very few people at the time would have had any idea of the great generalisation that the \dashuline{\,law of living forces} was capable of. 
Nevertheless, it is a fact that by the end of the last century the \dashuline{\,impossibility of constructing the perpetual motion machine}, even by other than mechanical methods, had already been pretty much universally recognised, the best proof of which is provided by the fact that in 1775 the 
French Academy$\,$\footnote{$\:$Hist. de l'Acad. Roy. des Sciences. 1775, p. 61 and 65 {\it\color{red}(in French)}. {\it\color{red}Communicated in German by:} H. v. Helmholtz: Vorträge und Reden {\it\color{red}(Lectures and speeches)}. Brunswick 1884. I. p. 64.}
declared once and for all that \dashuline{\,it would no longer accept any alleged solutions to this problem}. 
To most contemporaries, this impossibility may have seemed a regrettable fact, a kind of necessary evil, without anyone thinking of capitalising on it for science, despite the successes that Stevin and Huygens had already achieved in this direction.

\vspace*{-1mm} 
\begin{center}
{\it\color{red}\bf
=======
Sadi Carnot (1824) (p.13)
=======
}
\end{center}
\vspace*{-3mm}

In 1824, 
\dashuline{Sadi Carnot}$\,$\footnote{$\:$S. Carnot: Réflexions sur la puissance motrice du feu, et sur les machines propres à développer cette puissance. Paris 1824 {\it\color{red}(the original book in French)}. {\it\color{red}Reprinted in} Ann. de l'école norm. (2) I, p.393, 1872.}
took the first decisive step towards demonstrating the applicability of this theorem to non-mechanical  
phenomena$\,$\footnote{\label{label_footnote_Exergie}\it{\color{red}$\:$Note that Planck did not mentioned the writings of the father of (Nicolas Léonard) \dashuline{Sadi Carnot}, namely \dashuline{Lazare} (Nicolas Marguerite) \dashuline{Carnot}, who published in 1786 (and then in 1803) a book in French entitled ``Essai sur les machines en général (An essay on the machine in general)'' where he clearly defined the concept of ``work done by the displacement of a force'' but called it as the ``moment of activity'' (``moment d’activité'').
Moreover, Lazare Carnot not only tried to establish the  \dashuline{impossibility of perpetuum mobile} based on the study of the ``living force'' (forces vives), but also with the introduction}
\color{red} of the concept of ``motive force'' and ``motive power'' (forces et puissances motrices).
This is the same term retained in the title of the book of his son Sadi Carnot in 1822: ``Réflexions sur la \dashuline{puissance Motrice} du feu et sur les machines propres à développer cette puisance'' (Reflections on the \dashuline{motive power} of fire and on the machines capable of developing this power).
In fact Sadi Carnot never considered, nor write, the term of ``living force'' (or energy $U$) but, instead, the one of ``motive power of heat,'' which will be next called ``Motivity' by William Thomson and ``available energy'' by Gibbs and Maxwell, and is nowadays called  the ``exergy'' of the system and depend on the quantity $U-T_0\:S$, where $U$ is the energy, $S$ the entropy and $T_0$ the absolute temperature of the thermostat or the (infinite) source of heat of the system (P. Marquet).}.
Since the invention of the steam engine had made the lack of a satisfactory \dashuline{theory of the mechanical effects of heat} most keenly felt, Carnot, starting from the idea of the \dashuline{impossibility of perpetuum mobile}, undertook to establish a \dashuline{new theory of heat}, which was later developed further by 
Clapeyron$\,$\footnote{$\:$Clapeyron: Mémoire sur la puissance motrice du feu. Journ. de l'école polytechnique, T. XIV, p.170, 1834 {\it\color{red}(in French)}. {\it\color{red}And then, in German:} Pogg. Ann. 59, p.446 and 566, 1843.}
in the same sense, but with somewhat more elegant, more easily comprehensible means of representation. 
Here, however, it again became apparent that the application of this principle requires the correct determination of the \dashuline{equivalent value of the work} performed as the most important condition. The question was: if work is produced by heat, what process must be regarded as compensation for the work done and how is it to be measured? 

\vspace*{2mm} 
\begin{center}
{\it\color{red}\bf
---------
(p.13-14) 
---------
}
\end{center}
\vspace*{-3mm}

Since in Carnot's time the \dashuline{theory of heat} was in full favour which regarded heat as an (indestructible) substance whose presence in greater or lesser quantity makes a body appear more or less warm, he had to come up with the idea that \dashuline{thermal matter produces living force} in a similar way to the gravity of ponderable matter.
The latter endeavours to fall from higher to lower levels.
The \dashuline{living force} generated in this process is measured by the product of gravity at the height through which it falls, and this product is therefore the equivalent of the living force generated. Product is therefore the equivalent of the \dashuline{living force} generated. 
Carnot concluded from this that the heat fluid has the tendency to pass from higher to lower temperatures, as can be recognised from the laws of heat conduction. 
However, this tendency can be utilised to generate a \dashuline{living force}, which is then measured by the product of the amount of heat transferred in the temperature interval passed through. 
Carnot therefore sought compensation for the production of work in the transition of heat from a higher to a lower temperature, and regarded the product of a quantity of heat into a temperature difference as the measure of this, i.e. as the \dashuline{equivalent of work}.

According to a {\it\color{red}(next)} calculation by Clapeyron, the transition of a calorie from 1° Celsius to 0° Celsius is just capable of lifting 1.41 kilograms 1 metre high, so the number 1.41 should be called \dashuline{Carnot's heat equivalent}. 
This number is nothing other than the Joule mechanical heat equivalent divided by the absolute temperature in Celsius degrees of the melting ice.
As can be seen, these considerations are quite similar in form to those made later by Mayer and Joule.
\dashuline{Carnot's mistake} lies only in the fact that he brought with him a false idea of the nature of the process in which the compensation for the work produced is to be sought, an idea which, however, was essentially conditioned by the \dashuline{then prevailing theory of heat}.

In mechanics, work can be produced in two ways: by the performance of other work or by 
\dashuline{expending living force}. 
Instead of looking for the analogue of the \dashuline{production of work by heat} in the second process, i.e. seeing compensation in the disappearance of heat and measuring the work done by the quantity of heat destroyed, {\it\color{red}(Sady)} Carnot compared the effectiveness of heat with the \dashuline{work done by the gravity of ponderable matter}, which is indestructible in and of itself, and is only capable of producing effects by changing its position. 
For Carnot, therefore, heat was nothing other than force, as it was for Newton: a necessary but not yet sufficient condition for producing effects

\vspace*{-3mm} 
\begin{center}
{\it\color{red}\bf
---------
(p.15) 
---------
}
\end{center}
\vspace*{-3mm}

But it can also be seen that, since the Carnot-Clapeyron theory is essentially based on the principle of the \dashuline{excluded perpetuum mobile}, a \dashuline{principle of the conservation of energy} could very well be built on it, only then the \dashuline{energy of heat} must be regarded not as a simple quantity of heat, but as the \dashuline{product of a quantity of heat into a temperature}, and it is therefore incorrect to assume that the principle of the conservation of energy in itself involves a contradiction to the material theory of heat: 
on the contrary, Carnot is completely in favour of this principle. 
In agreement with this view, 
Helmholtz$\,$\footnote{$\:$v. Helmholtz : Wiss. Abh. I. Leipzig 1882, p. 33.},
in his ``Erhaltung der Kraft'' (Conservation of Force), considers both theories, the material and the mechanical, to be equal from the outset, and rejects the former only for the reason that it has been proved by experiment that the quantity of heat can change.

Incidentally, for the assessment of Carnot's achievements, the fact is very important and here all the more remarkable because it may have remained completely unknown in wider circles that Carnot, as can be seen from a handwritten essay left behind, which was given to the French Academy by his surviving 
brother$\,$\footnote{$\:$Carnot : Lettre. Compt. Rend. 87, p.967, 1878.},
felt compelled some time after the publication of his main work to \dashuline{abandon the material theory of heat} that he had hitherto advocated and to declare heat to be motion. As much of the content of the essay mentioned in the Comptes rendus shows that Carnot was just as clearly aware of the consequences for the new view flowing from the \dashuline{principle of the conservation of energy} as J. R. Mayer and J. P. Joule were soon after him. 

It is said {\it\color{red}in Carnot (1878)}, among other things: \dashuline{wherever work disappears} (où il y a destruction de \dashuline{puissance motrice}), \dashuline{heat production} (production de chaleur) \dashuline{takes place, and vice versa, in proportional quantities}. 
According to an unspecified calculation, $1$ unit of work (the lifting of a cubic metre of water by $1$ metre) is equivalent to a heating of $2.70$ calories --a number that is the mechanical heat equivalent of $370$ kgr-m (cf. Mayer's number). 
If one considers that \dashuline{Carnot made this calculation at least $10$ years earlier than Mayer} (he died in 1832), \dashuline{he definitely deserves the credit for the first evaluation of the mechanical heat equivalent}. 
Unfortunately, due to the lack of a timely publication, this discovery could no longer be utilised for scientific purposes.

The \dashuline{Carnot-Clapeyron theory of heat} effects was further developed, especially in England.
\dashuline{In 1848, 
W. Thomson}$\,$\footnote{$\:$W. Thomson : On an \dashuline{absolute thermometric scale founded on Carnot's theory of the motive power of heat}, and calculated from Regnault's observations. Phil. Mag. (3) 33, p.313, 1848.} 
\dashuline{based his absolute temperature scale on it}, because it is clear that, if the product of heat and temperature is equivalent to work, a definition of temperature can be derived from this equation if the measure of heat is given. 
A certain temperature interval is then completely defined by the amount of work which a calorie ``descending'' through this interval is able to perform.

\vspace*{-2mm} 
\begin{center}
{\it\color{red}\bf
===
Sadi Carnot (1824) - \'Emile Clapeyron (1834-1843) (p.16)
===
}
\end{center}
\vspace*{-2mm}

A major weakness of this {\it\color{red}Carnot-Clapeyron} theory, however, lies in the assumption that work, although it cannot arise from nothing, can at least decay into nothing. 
Clapeyron$\,$\footnote{$\:$Clapeyron: Mémoire sur la puissance motrice du feu, Journ. de l'école polytechnique, T. XIV, p.170, 1834. Pogg. Ann. 59, p.446 and 566, 1843.}
states this clearly 
and says: {\it when heat is conducted directly from a warmer body into a colder one, an effective quantity (the \dashuline{ability to perform work}) is lost}. 
Thus, according to him, one can very well lose work without gaining any equivalent serving as compensation. 
He thought the same of friction: {\it it destroys living force without providing an equivalent}. 
Thomson, on the other hand, saw this point as a major difficulty in Carnot's theory, as he was evidently already convinced at the time that the \dashuline{law of perpetual motion} was also reversible. 
He expressed himself as 
follows$\,$\footnote{$\:$W. Thomson: An account of Carnot's theory of the motive power of heat. Transact. of the Roy. Soc. of Edinburgh, vol. XVI, p.541, 1849.}:
{\it if by direct conduction of heat from a higher to a lower temperature a thermal effect is produced, what becomes of the mechanical effect which could be obtained by this transition? 
\dashuline{Nothing can be lost in nature}, \dashuline{energy is indestructible}, so the question arises as to what effect it is that takes the place of the transferred heat}. 
He considers this question to be very perplexing and believes that a perfect theory of heat must provide a satisfactory answer. 
Nevertheless, in the above-mentioned treatise he still adheres to Carnot's theory, estimating the difficulties arising from its abandonment to be incomparably greater.

\vspace*{-2mm} 
\begin{center}
{\it\color{red}\bf
---------
(p.17) 
---------
}
\end{center}
\vspace*{-4mm}

And yet these difficulties were overcome so surprisingly quickly. The experiences that forced us to \dashuline{abandon the theorem of the indestructibility of heat} became more and more frequent, until finally the consequences of the brilliant \dashuline{discovery of the mechanical equivalent of heat} put a rapid end to the material theory of heat. If \dashuline{heat is regarded as motion}, it is self-evident that the compensation of the work done by heat is to be sought in the disappearance of heat, in that the lost living force of heat motion must then be obtained as the equivalent of the work done.
It is the confirmation of the consequences of this proposition by experience which has helped the \dashuline{mechanical theory of heat} to gain the decisive preponderance. 
The difficulty of explaining the heat produced by friction had long led some physicists to \dashuline{the opinion that heat} could not be invariable in quantity, i.e. that it \dashuline{could not be a substance}. 
According to the material theory, the frictional heat must either be supplied from outside or the rubbed bodies must have reduced their heat capacity to such an extent that the same heat causes a much higher temperature in them.

\vspace*{-1mm} 
\begin{center}
{\it\color{red}\bf
---------
(p.18) 
---------
}
\end{center}
\vspace*{-3mm}

\dashuline{Rumford}$\,$\footnote{$\:$Rumford: An inquiry concerning the source of the heat which is excited by friction. Trans. of the Roy. Soc. London 1798, Jan. 25.}
showed in a striking manner that both assumptions are unfounded, by setting a blunt drill, which was pressed against the bottom of a cannon barrel, in rotation by horse power, and even brought a considerable quantity of water to boiling by the frictional heat.
The heat capacity of the metal was not altered at all. 
As the heat produced in this way could be increased to any extent by continuing the process, Rumford related it to the force applied, without, however, engaging in a numerical comparison of the work performed and the heat generated. 
\dashuline{Davy}$\,$\footnote{$\:$Davy: An essay on heat, light and the combinations of light, in Beddoe's Contributions to physical and medical knowledge, Bristol 1799. Works vol.II, London 1836, p.11.}
proved the same thing almost simultaneously by the friction of two pieces of metal by means of a self-acting clockwork mechanism under the air pump, and even more strikingly by the friction of two pieces of ice completely isolated from external influences, which were brought to the point of melting. 
The fact that the heat capacity of water is almost twice that of ice is particularly significant. 
Since then, many other experiments have been carried out which clearly indicate that \dashuline{heat can be generated}, in particular \dashuline{through the absorption of light or heat rays}, the identity of which has been proven since Melloni's experiments, \dashuline{and also through the mediation of electricity}, whether this is \dashuline{generated chemically} or \dashuline{through 
the use of mechanical work}.

Even if every single one of these facts speaks convincingly against the material conception of heat, the \dashuline{advocates of the mechanical theory of heat}, among whom, in addition to those mentioned, \dashuline{Th. Young, Ampère and Fresnel} should be mentioned in particular, were nevertheless in a decided \dashuline{minority until the middle of this century}, and until then \dashuline{no serious attempt had ever been made to develop the principle of the impossibility of perpetual motion in a similar way to Carnot's material theory for the mechanical theory}.

\vspace*{-4mm} 
\begin{center}
{\it\color{red}\bf
---------
(p.19) 
---------
}
\end{center}
\vspace*{-3mm}

Occasional traces of such endeavours can, however, be found: in the book ``{\it Etude sur l'influence des chemins de fer\,}'' (1839) by 
\dashuline{Séguin aîné}$\,$\footnote{$\:$Séguin aîné: \'Etude sur l'influence des chemins de fer. Paris. 1839, p.378. Vgl. Compt. Rend. XXV, p.420, 1847.}, 
the following remark is made: 
``{\it Steam is only the means of generating power. The \dashuline{driving cause is heat}, which, \dashuline{like living force}, is capable of \dashuline{producing power}.}'' 
Séguin attributes the authorship of this idea to his uncle, the famous J. M. Montgolfier (1740-1810).

But these considerations extended not only to the field of heat, but also to other natural phenomena, whereby we again find confirmed the fact, already often emphasised, that the validity of the principle itself was not doubted by anyone, but that only the conception of the consequences gave rise to differences of opinion.

\dashuline{Roget}$\,$\footnote{$\:$Roget: Treatise on galvanism, 1829, p.113 (Library of useful knowledge).}, 
for example, believed he could use the principle as evidence against the electric contact theory, arguing as follows: 
``{\it All forces and sources of motion, with whose cause we are acquainted, are, when they exert their peculiar effect, expended in the same proportion as these effects are produced, and from this arises the \dashuline{impossibility of producing by them} a perpetual effect, or in other words, \dashuline{a perpetual motion}.}'' 
Roget therefore declares it impossible to produce a perpetual current without the corresponding expenditure of another agent (here chemical affinity) and thus opposes the contact theory.
He would also be completely right if the contact theory allowed such a process.

\vspace*{-2mm} 
\begin{center}
{\it\color{red}\bf
---------
(p.20) 
---------
}
\end{center}
\vspace*{-3mm}

\dashuline{Faraday}$\,$\footnote{$\:$M. Faraday: Exp. Researches. Phil. Trans. London pț. I. p.93, 1840. Pogg. Ann. 53, p.548, 1841.}
expresses himself in a similar way: 
``{\it The contact theory assumes that a force capable of overcoming powerful resistances can arise from nothing. This would be a creation of force that takes place nowhere else without a corresponding exhaustion of something that gives it nourishment. If the contact theory were correct, the equality of cause and effect would have to be denied. Then the perpetuum mobile would also be possible and it would be easy to achieve incessant mechanical effects on the first case of an electric current generated by contact.}'' 
There is no need to explain here that these objections to the contact theory are based on a misunderstanding. The whole controversy does not refer at all to the mode of maintaining the electric current, but to the cause of the initiation of a current, because the fact that a current cannot be maintained without a constant consumption of energy is now as self-evident according to the contact theory as it is according to the chemical theory.

We also find applications of the principle in chemistry. The idea that the total amount of heat produced by a series of successive chemical reactions is independent of the way or order in which the individual reactions are carried out, if only the initial state and the final state of the system remain the same, has gradually and silently become naturalised in theoretical chemistry. It is perhaps first mentioned explicitly by 
\dashuline{Hess}$\,$\footnote{$\:$H. Hess : Thermochemische Untersuchungen . Pogg. Ann. 50, p. 392, 1840.}
with the words: 
``{\it When a reaction takes place, the amount of heat evolved is constant, whether the reaction is direct or indirect.}'' 
There is no doubt that the convincing truth of this proposition arises from the idea that \dashuline{heat cannot be produced from nothing}. This idea is of course supported by the idea of the indestructibility of the heat substance, but it is even more general and independent of this idea.

\vspace*{-4mm} 
\begin{center}
{\it\color{red}\bf
---------
(p.21) 
---------
}
\end{center}
\vspace*{-2mm}

However, the following passage from a treatise by 
\dashuline{K. Fr. Mohr}$\,$\footnote{$\:$K. Fr. Mohr: Über die Natur der Wärme. Zeitschr. f. Physik v. Baumgärtner, V, p.419, 1837. Ann. d. Pharmacie 24, p.141, 1837.}
on the nature of heat, in which the author, mainly inspired by the experiments of \dashuline{Melloni and Rumford}, vividly advocates the \dashuline{dynamic theory of heat}, shows how far individual physicists had already come in recognising the \dashuline{unity and mutual transformability of the various forces of nature}: 
``{\it Apart from the 54 known chemical elements, there is only one \dashuline{agent in the nature of things}, and this is \dashuline{called force}; under the right conditions it can emerge as \dashuline{motion, chemical affinity, cohesion, electricity, light, heat and magnetism}, and from each of these manifestations all the others can be produced. The same force, when it lifts the hammer, can, when applied differently, produce any of the other phenomena.}'' 

As you can see, it is only one step to the question of the \dashuline{common measure of all these natural forces} recognised as similar.

\vspace*{-2mm} 
\begin{center}
{\it\color{red}\bf
====
Julius Robert Mayer (1842, 1845, 1867, 1874) (p.21)
====
}
\end{center}
\vspace*{-2mm}

This step was taken almost simultaneously from different sides and in different ways. If we follow the chronological order of the individual publications, we must first turn our attention to the work of the physician from Heilbronn 
\dashuline{Dr Julius Robert Mayer}$\,$\footnote{$\:$J. R. Mayer: Die Mechanik der Wärme {\it\color{red}(The mechanics of heat)}, Stuttgart 1867, 2nd presumably ed. Stuttgart 1874.}.
In accordance with \dashuline{Mayer's whole school of thought}, who preferred to generalise philosophically rather than build up empirically piece by piece, the form of his reasoning was deductive. In his first short essay, published in 
May 1842$\,$\footnote{$\:$J. R. Mayer: Lieb. Ann. 42, p.233, 1842. Bemerkungen über die Kräfte der unbelebten Natur {\it\color{red}(Remarks on the forces of inanimate nature)}. Phil. Mag. (3) 24, p.371, 1844.}, 
he expresses himself as follows: 
``{\it An effect can never arise without a cause, or conversely a cause can never remain without an effect: 
Ex nihilo nihil fit {\it\color{red}(From nothing comes nothing)}, 
and vice versa: 
Nil fit ad nihilum {\it\color{red}(Nothing comes to nothing)}. 
On the contrary, every cause has a very definite effect corresponding to it, neither a greater nor a lesser one. The cause therefore contains everything that determines the effect, and is completely reflected in the effect, albeit in a different form. Cause and effect are thus in a certain sense equal to each other: Causa aequat effectum}.

\vspace*{7mm} 
\begin{center}
{\it\color{red}\bf
---------
(p.22) 
---------
}
\end{center}
\vspace*{-3mm}

Mayer therefore characterises causes as (quantitatively) indestructible and (qualitatively) mutable objects. Now he divides all causes into two parts: the one he classifies as \dashuline{matter}, the other as \dashuline{force}; each of these two kinds is indestructible, and there is no transition between them, i.e. matter cannot be transformed into force, nor vice versa, but \dashuline{matter as well as force can be transformed in many different ways}. 
But while there are still many different kinds of matter (the elements that cannot be transformed into one another), we only know one kind of force, because \dashuline{all forces can be transformed into one another}, all forces are different manifestations of one and the same object, one and the same cause.

It goes without saying that the word \dashuline{force} is taken here in Leibnitz's sense, which seems all the less striking as this meaning of the word was still quite common at the time. 
In any case, Mayer cannot therefore be accused of ambiguity  because, as the following shows, he knew very well how to distinguish between these terms. 
He calls \dashuline{Newtonian force} a property. 
Going into the nature of the various forces in more detail, he identifies three forms: \dashuline{heat, falling force and motion}. 
They are different in themselves, but can be transformed into each other according to certain numerical ratios, and are therefore also subject to a common measure. 
Falling force and motion are measured by the same measure anyway, so it remains to compare the \dashuline{unit of heat} with this measure. 
From experiments on the ``compression of air'' Mayer calculates that one calorie is equivalent to lifting one kilogramme by 365 metres at average values of acceleration of gravity
{\it\color{red}(to be compared with the equivalent value of $\:370$~kgr-m previously derived by Carnot before 1832 from the value of $2.70$~calorie, but published later in 1878)}.

\vspace*{-2mm} 
\begin{center}
{\it\color{red}\bf
---------
(p.23) 
---------
}
\end{center}
\vspace*{-3mm}

This calculation {\it\color{red}made by Mayer}, the details of which will only be shared in a subsequent 
paper$\,$\footnote{$\:$Mayer: Die organische Bewegung in ihrem Zusammenhang mit dem Stoffwechsel {\it\color{red}(Organic movement in its connection with metabolism)}, 1845 (Mechanik d. Wärme, {\it\color{red}Mechanics of Heat}, Stuttg. 1874.)},
was determined through many experiments and is based on the idea that the \dashuline{difference in the amounts of heat} that must be supplied from outside to \dashuline{a quantum of air} to achieve a certain temperature increase when the heating once at constant pressure, the other time at constant volume, \dashuline{is equivalent to the work done} in the first case \dashuline{by the expansion of the air}. 
However, it is tacitly assumed here that \dashuline{the excess of the heat capacity at constant pressure over that at constant volume only benefits the external work performance}, an assumption that is not at all admissible (since most gases and vapors show a noticeable reduction in temperature when they have expanded without any external work). 
For the so-called perfect gases, however, the assumption that subsequently plays a role in the further development of the heat theory under the name \dashuline{Mayer's hypothesis} has proven to be correct.

Once the \dashuline{mechanical equivalent of heat} has been determined, you can now \dashuline{measure heat} with the same measure \dashuline{as mechanical force} by always setting {\bf\dashuline{one calorie equal to $365$~g force (work) units}} (based on kilograms, meters and seconds). 
What is remarkable is the fact that \dashuline{Mayer did not start from the view that heat is movement}, but rather he cautiously \dashuline{left the question of the nature of heat completely out of the picture}. 
He expressly says: 
``{\it Heat, movement and falling force can be converted into each other according to certain numerical ratios.
Just as one cannot draw the conclusion from this that falling force and motion are identical, just as one cannot conclude that heat exists in motion.}'' 
In fact, \dashuline{the entire heat theory}, as it was later built up by \dashuline{R. Clausius} on his two main theorems, \dashuline{\,can also be derived without the idea of the mechanical nature of heat if} one only sticks to the assumption that \dashuline{heat under certain conditions can be converted into movement}. 
Only the gas theory, which was developed later, gave a more definite form to the ideas we have about the nature of heat.

\vspace*{-2mm} 
\begin{center}
{\it\color{red}\bf
---------
(p.24) 
---------
}
\end{center}
\vspace*{-3mm}

\dashuline{Mayer} does not stop at the statements just described. In a second, somewhat more detailed 
treatise$\,$\footnote{$\:$Mayer: Die organische Bewegung in ihrem Zusammenhang mit dem Stoffwechsel {\it\color{red}(Organic movement in its connection with metabolism)}, 1845 (Mechanik d. Wärme\,/{\it\color{red}Mechanics of heat\,}, Stuttg. 1874.)}, 
he expands his theory to other branches of natural science. He describes \dashuline{chemistry} as the study of the transformations of matter, \dashuline{physics} as the study of the transformations of force, and then summarizes his views on the \dashuline{equivalence of heat and motion}. 
Then electricity is also drawn into the circle of considerations, but again using terminology that unfortunately contradicted the prevailing usage. 
He calls \dashuline{electricity} a force like heat, and of course he means what we call \dashuline{electrical potential}. 
He explains the effect of the electrophorus in terms of the mechanical work involved. The ``chemical difference'' between two bodies is also introduced as a force, because heat can be produced through it, the amount of which then provides the measure of the force used. Finally, {\it\color{red}\dashuline{Mayer}} \dashuline{counts six different forces} that can be converted into one another according to certain equivalents as they work in inorganic nature on: \dashuline{falling force}, \dashuline{movement}, \dashuline{heat}, \dashuline{magnetism}, \dashuline{electricity}, \dashuline{chemical difference}. 
The consequences are also extended to organic nature, in particular Mayer develops here the importance of the \dashuline{assimilation process in plants} for the \dashuline{preservation of all animal life}.

Another special 
work$\,$\footnote{$\:$Mayer: Beiträge zur Dynamik des Himmels {\it\color{red}(Contributions to the dynamics of the sky)}, 1848 (Mech. d. W./{\it\color{red}Mechanics of heat\,}, Stuttg. 1874).}
{\it\color{red}\dashuline{Mayer}\,} made was the application of his theories to \dashuline{cosmic phenomena}. 
He gave the first rational explanation for the \dashuline{source of solar heat}, first showing that no chemical process (combustion) could replace the tremendous heat output produced by solar radiation, and then expressing the view that {\it\color{red}the way} that heat was supplied is caused by the \dashuline{living power} of the masses of meteorites constantly falling into the solar body (cf. the views of Helmholtz and Thomson below).
He explained the glow of meteors by the loss of \dashuline{living power} that they experience through friction in the atmosphere, draw attention to the fact that the phenomenon of tides must necessarily have an inhibiting effect on the speed of rotation of the earth due to the friction of the tide, and that all the work that can be achieved through the movement of the ebb and flow is gained at the expense of the \dashuline{living force} of the earth's rotation.

\vspace*{-5mm} 
\begin{center}
{\it\color{red}\bf
---------
(p.25) 
---------
}
\end{center}
\vspace*{-3mm}

Each of these various considerations shows that, even if \dashuline{Mayer} sometimes used an \dashuline{unusual} nomenclature, he was nevertheless well aware of the importance of the views he represented. He also explains this in detail in another 
treatise$\,$\footnote{$\:$Mayer: Bemerkungen über das mechanische Äquivalent der Wärme {\it\color{red}(Remarks on the mechanical equivalent of the Heat)}, Heilbronn 1850 (Mech. d. W./{\it\color{red}Mechanics of heat\,}, 1867, p.237.)}
and discusses very clearly the various concepts that Leibnitz and Newton assigned to the name ``force'' as well as the inexpediency of distinguishing between ``dead'' (Newtonian) and ``living'' (Leibnitz') force. 
Taken in this sense, \dashuline{the word force} would have to \dashuline{denote a common generic concept} in which the two more specific concepts are contained as special types. 
But this seems absurd because the two concepts have quantities of completely different dimensions. 
So they are not comparable at all.
One is therefore forced to give up one of the two terms. 
\dashuline{Mayer} decides to \dashuline{use the word force in the Leibnitzian sense} because he sees this term as the more fundamental one. 
This also gives the term a correct one, in contrast to the word matter, and today we still need the comparison: \dashuline{force and matter}, where we actually mean: \dashuline{energy and matter}. 
Both are indestructible.

\vspace*{-2mm} 
\begin{center}
{\it\color{red}\bf
---------
(p.26) 
---------
}
\end{center}
\vspace*{-3mm}

\dashuline{Mayer} may be right in these considerations: given the current state of natural science, Leibnitz's concept has actually become the more important one. 
But he did not take into account the power of the historical development of science. 
Physics was based on mechanics, and Newton's concept had already become too common in mechanics to be easily replaced by another name. 
\dashuline{So Mayer's suggestion did not prevail}, even though Leibnitz's term remained in some expressions (\dashuline{living force}, \dashuline{preservation of force}). 
Only consideration of history can explain this \dashuline{inconsistency}, which today at least does not carry the risk of misunderstanding as it once did. 

We have discussed \dashuline{Mayer's various related works} in context, although there are several years between their publication, in order to facilitate an overview of the \dashuline{new ideas that he introduced into natural science}. 
That he lacked a strictly scientific school, that he could perhaps have expressed himself more clearly and succinctly for professional physicists in some points, especially in the first treatises, but that the entire foundation of his teaching, which almost bordered on metaphysical, was on extremely weak foundations, one will probably have to admit to \dashuline{those who do not want to fully recognize its importance}. 
But it is irrefutably certain that he was the first to not only publicly express the idea that is characteristic of our contemporary view of nature, but also, what is most important, to evaluate it in measure and number and to all natural phenomena accessible to him applied in detail.

\vspace*{-3mm} 
\begin{center}
{\it\color{red}\bf
---------
(p.26-27) 
---------
}
\end{center}
\vspace*{-4mm}

And as far as the justification of the theorem is concerned, we must not forget that, as we will explain in more detail in the next section, it is not capable of any strictly deductive proof at all precisely because of its generality, i.e. that \dashuline{what Mayer failed to achieve in his proof procedure has not been achieved by any other physicist either}.
The most direct proof, which is as free from all presuppositions as possible, is provided by the investigation of the individual consequences, and Mayer, although he did not experiment himself, made a significant contribution to this through direct suggestion. 
%
But if we try to visualise the principle clearly, i.e. to relate it to other ideas and propositions with which we are familiar, \dashuline{Mayer's explanations, which are based on the idea that no effect is lost in nature, are still among the best of their kind}.
%
The importance of these must not be underestimated because, if we are not mistaken, the comparatively surprising rapidity and ease with which a proposition of such enormous consequence as the \dashuline{conservation of energy}, after overcoming the first difficulties, made itself at home in the minds, is to be attributed not only to the many individual inductive proofs, but also, to a great extent, to the conception of its connection with the \dashuline{law of cause and effect}.

Although we certainly cannot attribute any physical 
proof 
to \dashuline{Mayer's philosophical considerations}, they nevertheless have an eminent \dashuline{practical importance}, in that they facilitate an overview of the entire content of the principle, and thus \dashuline{indicate the guiding ideas} according to which the question is addressed and must take place in nature. 

People so often love to contrast \dashuline{Mayer} (who philosophizes somewhat vaguely) with his partner \dashuline{Joule} as the sober, exact empiricist 
 who adheres only to the individual facts).
But how could it be conceivable that \dashuline{Joule} would have carried out his famous experiments with this restless zeal and this tenacious perseverance (and would have devoted part of his life to answering one 
question), if he had not been enthusiastic about the new 
idea 
from the start and during his first experiments (which certainly did not in themselves give rise to such a great generalization idea and would have immediately grasped it in its generality). 

It is well known that \dashuline{Mayer's achievements} have been fully appreciated in recent times (in England they were first put into the right light by 
J. Tyndall$\,$\footnote{$\:$J. Tyndall: On force. Proc. of Roy. Inst. June 6, 1862. Phil. Mag. (4) 24, p.57, 1862.}, 
and have been recognized in a completely satisfactory manner by our first 
academics {\it\color{red}by von Helmholtz, I. Braunschw and R. Clausius}$\,$\footnote{$\:$von Helmholtz: Robert Mayer's Priorität. Vorträge und Reden {\it\color{red}(Robert Mayer's priority. Lectures and speeches)}, I. Braunschw. 1884, p. 60. R. Clausius : Über das Bekanntwerden der Schriften Robert Mayer's. Wied {\it\color{red}(On how Robert Mayer's writings became known. Again)}. Ann. 8, Anhang, 1879.}).

\vspace*{-3mm} 
\begin{center}
{\it\color{red}\bf
---------
(p.28) 
---------
}
\end{center}
\vspace*{-4mm}

However, it remains a fact that can no longer be changed that \dashuline{Mayer}, at least in the first period of his public appearance, \dashuline{had almost no influence at all} on the spread and development of the new principle: the same would probably have happened just as quickly without him, especially since almost simultaneously and completely independently, both from him and from each other, the same ideas appeared in different forms and with different reasons.

\vspace*{-2mm} 
\begin{center}
{\it\color{red}\bf
====
James Prescott  Joule (1843, 1846) (p.28-29)
====
}
\end{center}
\vspace*{-3mm}

On January 24, 1843, \dashuline{James Prescott Joule} (brewer in Salford) presented at the Manchester Philosophical Society 
a paper$\,$\footnote{$\:$J. P. Joule: On the heat evolved during the electrolysis of water. Mem. of the liter. and phil. soc. of Manchester. (2) vol.VII, 1846, p.87 and p.96.} 
on the \dashuline{connection between the thermal and chemical effects} of galvanic current. 
He had already been led to the conviction by the results of two 
previous works$\,$\footnote{$\:$J. P. Joule: On the heat evolved by metallic conductors of electricity and in the cells of a battery during electrolysis. Phil. Mag. (3) 19, p.260, 1841. J. P. Joule: On the electric origin of the heat of combustion. Phil. Mag. (3) 20, p. 98, 1842.}
that the amount of heat generated by a current in the closing circuit is identical to that which can be obtained by direct oxidation of the metals active in the chain (including hydrogen), and from this he formed the view that chemical heat in general was essentially of electrical origin. 

\vspace*{-3mm} 
\begin{center}
{\it\color{red}\bf
---------
(p.29) 
---------
}
\end{center}
\vspace*{-4mm}

It is of great interest to observe how \dashuline{Joule's initially somewhat vague ideas gradually} and carefully \dashuline{developed into} the full clarity of his awareness of the \dashuline{validity of the general principle}. 
It was only in the above-mentioned treatise, in which the statement that the thermal effects of a current are equivalent to the chemical ones was further developed and confirmed (cf. also the experiments by 
Becquerel$\,$\footnote{$\:$E. Becquerel : Des lois du dégagement de la chaleur pendant le passage des courants électriques à travers les corps solides et liquides. Compt. Rend. t.16, p.724, 1843.}), 
that \dashuline{the more general remark is found for the first time}, {\it\color{red}namely} that {\it\dashuline{in nature the destruction (annihilation) of work power (power) does not take place without a corresponding effect (effect)}}. 
At the same time, the assumption is expressed that if one were to let a current do work by switching on an electromagnetic machine, it would be in proportion to the chemical effect heat produced would decrease, proportionally to the 
work done$\,$\footnote{$\:$E. Becquerel (1oc. cit.) p. 96 u. 104.}.

This thought prompted \dashuline{Joule} to carry out a special investigation, the result of 
which$\,$\footnote{$\:$Joule: On the calorific effects of magneto-electricity and on the mechanical value of heat. Phil. Mag. (3) 23, p.263, 347, 435, 1843.} 
he communicated to the mathematical-physical section of the British Association at the  meeting in Cork on August 2nd of the same year {\it\color{red}(1843)\,}: 
``{\it On the \dashuline{calorific effects} of magneto-electricity and on the \dashuline{mechanical value of heat}.}'' 
In this work, \dashuline{Joule} first sets out his views on the \dashuline{nature of heat} and the processes in the galvanic chain. He explains heat as a \dashuline{type of movement} that consists of \dashuline{vibrations} and expresses the conviction that in the galvanic chain there is no generation but only a \dashuline{distribution (arrangement) of heat}. 
So the heat developed in the hydroelectric current arises from the combustion in the element, that in \dashuline{the magnetoelectric current} (generated by the movement of magnets) from the mechanical work expended, and \dashuline{is always just as great as if the combustion or the work had directly supplied heat}.

\vspace*{-2mm} 
\begin{center}
{\it\color{red}\bf
---------
(p.30) 
---------
}
\end{center}
\vspace*{-3mm}


These claims are {\it\color{red}to be} justified by the experiment. 
For this purpose, \dashuline{Joule considered 
an induction spiral}, which was located in a horizontal glass tube filled with water and served as a calorimeter, to rotate around a vertical axis between two strong magnetic poles using falling weights \dashuline{and compared the heat generated} by the induction currents in the spiral \dashuline{with that of mechanical work} done by the weights. It turned out that heating any quantity of water by 1°Fahrenheit would raise 838 times that quantity by 1 degree.feet, or the simple quantity corresponds to $838$ feet. \dashuline{This means that for 1 Celsius calorie, one kilogram is raised by $460$ meters}. -- On the other hand, \dashuline{Joule also measures the direct conversion of mechanical work into heat}, by means of the friction that occurs when water is forced through narrow tubes. This time \dashuline{the mechanical heat equivalent} (based on F.grade) was $770$ feet (or \dashuline{for C.grade $423$ M.}).

In view of the many sources of error, \dashuline{Joule considered the two numbers he found} {\it\color{red}(i.e. $838$ and $770$, or   $460$ and $423$)} \dashuline{to be sufficiently similar to be able to base the same claim as Mayer} on them: The basic forces of nature are indestructible, and wherever force is used, a quantity of heat corresponding to the effort is created. From this point of view he also explains \dashuline{latent heat} and the \dashuline{heat generated by chemical processes}. 
Latent heat also represents a force like gravity, and for a given case it can be converted into real heat, just as a wound clockwork is capable of performing mechanical work at any moment.


\vspace*{0mm} 
\begin{center}
{\it\color{red}\bf
=
A. Colding (1843, 1864) -- C. Holtzmann (1845, 1848, 1851) (p.30-31)
=
}
\end{center}
\vspace*{-4mm}

\dashuline{Joule did not achieve any particular success with his first works}, on the contrary: as is not surprising with such an innovation, \dashuline{most physicists were essentially hostile to the views presented here}. 
But in the same year in which Joule's first works appeared, on November 1, 1843, the Danish engineer \dashuline{Ludwig A. Colding} communicated to the Academy of Copenhagen, under the title 
``{\it Thesis on force,}'' 
experiments$\,$\footnote{$\:$A. Colding: Det kongel. danske vidensk. selsk. naturv. og math. afh. (5) II, 1843, p.121, 167. On the history of the principle of the conservation of energy, Phil. Mag. (4) 27, p.56, 1864.}
according to which the heat developed by the friction of solid bodies is in a constant relationship with the amount of work consumed, and at the same time expressed the view that the \dashuline{law of conservation of force is a generally valid one}. 

\vspace*{-5mm} 
\begin{center}
{\it\color{red}\bf
---------
(p.31) 
---------
}
\end{center}
\vspace*{-4mm}

He, like Mayer, came to this theorem through deductive considerations, which, however, extend far into the field of metaphysics, in that he started from the view that the forces of nature are spiritual and immaterial beings, and as such could not possibly be subject to transience, and therefore he describes the force as immortal. -- From a considerable series of friction tests with various solid bodies, he found the mechanical heat equivalent, based on Cels.Degrees and Danish feet, to be $1185.4$, or approx. $370$~m
{\it\color{red}(to be compared with $460$~m and $423$~m for Joule)}.

%
A fourth calculation of the \dashuline{mechanical heat equivalent}, which dates back to the same time, comes from 
\dashuline{C. Holtzmann}$\,$\footnote{$\:$C. Holtzmann: Über die Wärme und Elasticität der Gase u. Dämpfe {\it\color{red}(On the heat and elasticity of gases and vapors)}. Mannheim 1845. Auszug in Pogg. Ann. Erg. II, p.183, 1848.}. 
He found, essentially in the same way as Mayer, that the heat which increases the temperature of a kilogram of water by 1ºC is able to lift $374$ kg by $1$ meter high.
However, it must be noted here that \dashuline{Holtzmann} in no way represented the viewpoint of the \dashuline{mechanical theory of heat} in the sense that he assumed that heat would disappear. Rather, he essentially stuck to the \dashuline{material theory}, since he still defended the \dashuline{theorem of the immutability of the thermal 
fluid}$\,$\footnote{$\:$C. Holtzmann : Über die bewegende Kraft der Wärme {\it\color{red}(On the moving power of heat)}. Pogg. Ann. 82, p.445, 1851.}.

\vspace*{-2mm} 
\begin{center}
{\it\color{red}\bf
==
William Scoresby (1846) -- James Prescott Joule (1845) (p.31-32)
==
}
\end{center}
\vspace*{-4mm}

Meanwhile, the restlessly active \dashuline{Joule} continued his experiments on the heat equivalent using completely different methods. 
First he 
compared$\,$\footnote{$\:$Joule: On the changes of temperature produced by the rarefaction and condensation of air. Phil. Mag. (3) 26, p.369, 1845.}
the \dashuline{mechanical work} involved in compressing air with the resulting \dashuline{increase in temperature} and again found the assumed proportionality. 
He showed in a special experiment that this increase in temperature is caused by external work and not by a change in the heat capacity of the air by allowing up to $22$ atmospheres of compressed air to flow out into an evacuated room. 
In this case (in agreement with a result previously obtained by Gay Lussac) there was no decrease in temperature after equilibrium had occurred, corresponding to the fact that the air had to do no external work when flowing out. 
This experiment is so important because it proves what Mayer had tacitly assumed (p.23): that no internal work is done when the volume of air changes. 
If, on the other hand, he allowed compressed air to flow out into the free atmosphere, there was a decrease in temperature, proportional to the work done in overcoming the resistance. 
From this, after various series of tests, Joule calculated the mechanical heat equivalent to $823$, then to $795$ 
(or $452$, then $436$~m.Cels).

Shortly thereafter, other observations appeared 
again$\,$\footnote{$\:$Joule: On the existence of an equivalent relation between heat and the ordinary forms of mechanical power. Phil. Mag. (3) 27, p.205, 1845.}.
This time the \dashuline{mechanical work was converted into heat through friction}. A paddle wheel was set in rotation by descending weights in a water bath and the friction caused the temperature of the water to increase. From this the heat equivalent was $890$ ($488$m.Cels.), while the number found when forcing water through narrow tubes was $774$ ($425$m.Cels.).

The relatively good agreement of his results prompted Joule to publish a short summary of his views, confirmed by all these different experiments, on the laws of force generation through mechanical, chemical, galvanic, electromagnetic, thermal effects, as well as through animal 
work$\,$\footnote{$\:$Scoresby and Joule: Experiments and observations on the mechanical powers of electro-magnetism steam and horses. Phil. Mag. (3) 28, p.448, 1846.}.

\vspace*{-1mm} 
\begin{center}
{\it\color{red}\bf
===
Franz Ernst Neumann (1845, 1846) -- J. Liebig (1845) (p.33)
===
}
\end{center}
\vspace*{-3mm}

While the number of works which promoted the application of the \dashuline{principle of conservation of force} on the basis of the \dashuline{new conception of the nature of heat} increased, and while Joule's continually repeated and varied experiments in particular gradually attracted the attention of his colleagues, the concept of energy was also introduced from other sides into other branches of natural science.

The investigations by 
\dashuline{F. Neumann\,}$\,$\footnote{$\:$Franz Ernst Neumann: Allgemeine Gesetze der inducierten Ströme {\it\color{red}(General laws of induced currents)}. Abh. d. kgl. Akad. d. Wiss. Berlin 1845. Pogg. Ann. 67, p.31, 1846.} 
(1845 and 1847) into the laws of induced currents can be attributed to this, which led to the result that the galvanic induction in a conductor only depends on the change in the electrodynamic potential of the inducing current system the conductor, regardless of whether this change arises from a relative movement of the ponderable parts of the conductor and the current system, or from a change in the intensities of the currents. However, in Neumann's two writings which deal with this subject, the close connection between his theorem and our principle does not yet emerge directly.

The new ideas were also transferred to the field of \dashuline{organic nature}. 
\dashuline{J. Liebig}, who was based on the view that a \dashuline{steam engine} cannot produce more heat than it originally received from the boiler, and that a galvanic current in the closing circuit does not produce more heat than through the usual chemical reaction of the substances reacting in the element is obtained, vigorously defended the proposition that the \dashuline{heat produced by the animal body} is supplied entirely by the combustion of foodstuffs in a 
direct manner$\,$\footnote{$\:$J. Liebig: Über die thierische Wärme (On animal warmth). Lieb. Ann. 53, p.63, 1845 Planck, Energie.}. 

\vspace*{-5mm} 
\begin{center}
{\it\color{red}\bf
---------
(p.34) 
---------
}
\end{center}
\vspace*{-3mm}

However, the difficulty for him was that the calorimetric experiments carried out by \dashuline{Dulong} and \dashuline{Despretz} on the heat given off by the animal body produced \dashuline{numerical values} that appeared to be \dashuline{considerably too large}, since the direct heat of combustion of the corresponding amounts of hydrogen and carbon was only $70$ to $90$\% the amount actually given by the animal. The correct explanation of this circumstance was given by \dashuline{Helmholtz}, 
who emphasized$\,$\footnote{\label{label_footnote_Helmholtz_a}$\:$von Helmholtz: Fortschr. d. Phys. v. J. 1845, p.346, Berlin 1847. Wiss. Abh. I p.8.} 
that instead of the heat obtained by burning food, one should not simply substitute the heat of combustion of the elements contained in it.

\vspace*{-1mm} 
\begin{center}
{\it\color{red}\bf
=======
Hermann von Helmholtz (1847)  (p.34)
=======
}
\end{center}
\vspace*{-3mm}

In the above-mentioned essay by \dashuline{Helmholtz} {\it\color{red}(1847, see the footnote~\ref{label_footnote_Helmholtz_a})} we also find a very brief overview of the various consequences that the general implementation of the ``{\it principle of the constancy of the force equivalent in the excitation of one natural force by another\,}'' in various areas of physics given the state of the art at that time Research. With regard to the transformation of mechanical force into heat, \dashuline{Helmholtz} does not yet get to the discussion of the \dashuline{mechanical equivalent of heat}, although for obvious reasons (already mentioned above) he decides against the previous material theory and in favor of a \dashuline{theory of motion}. 
On the other hand, \dashuline{he mentions the law of the constancy of chemical heat production}, regardless of the way in which the connection is made. With regard to the processes taking place in constant hydroelectric currents, \dashuline{Ohm}'s law in conjunction with \dashuline{Lenz}'s (\dashuline{Joule}'s) leads Law on heat generation in the closed circuit and \dashuline{Faraday}'s electrolytic law to the proposition that the total heat generated in the closed circuit is equivalent to the amount of electrochemical conversion in the chain, regardless of the other arrangement. For heat generation through static electricity, it follows from \dashuline{Riess}'s theorems about the discharge heat that this is equal to the product of the discharged quantity of electricity in its density (now better: voltage).

\vspace*{-2mm} 
\begin{center}
{\it\color{red}\bf
---------
(p.35) 
---------
}
\end{center}
\vspace*{-3mm}

These considerations are further developed and given a more systematic form in \dashuline{Helmholtz}'s paper, which followed on the heels of this essay, and in which for the first time \dashuline{the universal significance of the principle of the conservation of force for all natural phenomena was developed from the height of the development of physics} at that time \dashuline{\,in a precise treatment with a concise overview}.

On July 23, 1847, \dashuline{Hermann Helmholtz} gave a lecture on the \dashuline{principle of conservation of force} at the meeting of the Physical Society in 
Berlin$\,$\footnote{$\:$von Helmholtz: Über die Erhaltung der Kraft {\it\color{red}(On the Conservation of Force)}. Berlin, Reimer 1847. Wiss. Abh, I p.12.}. 
In this, as in the other expressions he used ``\,living force\,'' and ``\,tension force,'' \dashuline{Helmholtz} (like \dashuline{Mayer}) follows \dashuline{Leibnitz}'s concept of force, although he otherwise retains \dashuline{Newton}'s usual scientific terminology.
Characteristic of the way the Principle is introduced is the fact that \dashuline{Helmholtz}, based entirely on the standpoint of the mechanical view of nature, understands the Principle as a direct generalisation of the \dashuline{mechanical theorem of the conservation of living force} (p.6f).
The introduction to the treatise forms a series of deductive considerations, from which the aim of the physical sciences is the task of \dashuline{tracing natural phenomena back to movements of individual material points that act on one another with attractive or repulsive forces that depend in a certain way on their distances}. 
The fact that the \dashuline{principle of conservation of living force} (in the more specific and more general sense) can be derived from this assumption with the help of \dashuline{Newton's axioms} is taught in mechanics. 
\dashuline{Helmholtz} also shows, conversely, that instead of using this starting point as completely equivalent, one can also use the one that \dashuline{Carnot} and \dashuline{Clapeyron} already made the basis of their theories, namely the \dashuline{impossibility of perpetual motion}.

\vspace*{-2mm} 
\begin{center}
{\it\color{red}\bf
---------
(p.36) 
---------
}
\end{center}
\vspace*{-4mm}

When applied to the mechanical processes of nature, \dashuline{Helmholtz} expresses this sentence as follows: 
``\,{\it Let us imagine a system of natural bodies which stand in certain spatial relationships to one another and begin to move under the influence of their mutual forces until they have reached certain other positions, we can consider the velocity they gain as a certain mechanical work, and transform it into such work. If we now want to let the same forces take effect a second time in order to obtain the same work again
(and thus obtain a periodically working machine), 
we must bring the bodies into the initial conditions in some way by applying different ones put us back at the command of forces at our command. 
We will therefore use up a certain amount of the latter's work for this purpose. 
In this case, the working quantity, which is gained when the bodies of the system move from the initial position to the second, and is lost when they pass from the second to the first, is always the same, regardless of the type, path, position or speed of this transition may be.\,}''

Put into mathematical form, this sentence presents itself as the \dashuline{principle of living force}. This principle, in conjunction with the assumption that all forces can be broken down into those that only act from point to point, then leads with the help of Newton's axioms to the conclusion that \dashuline{the elementary forces are central forces, i.e. attractive or repelling with an intensity that only depends on the distance}, and this assumption is precisely what was assumed above, in the first case.

The transformation which \dashuline{Helmholtz} made of the \dashuline{principle of living force}, in order to make it appear as the \dashuline{principle of the conservation of force}, consists in the fact that in the equation $T-A = const.$ that expresses the \dashuline{invariability of the difference} of the \dashuline{living forces} $T$ and the \dashuline{work done by the acting forces} $A$, he introduces, instead of the concept of work $A$, that of the quantity of tensioning forces $U$, which quantity is equal and opposite to the work $A$. 

\vspace*{5mm} 
\begin{center}
{\it\color{red}\bf
---------
(p.37) 
---------
}
\end{center}
\vspace*{-4mm}

The tensioning force, like work, depends only on the instantaneous state of the system. Thus the tension force, like the work, depends only on the instantaneous state of the system, and the above equation now expresses itself as follows: 
``\,{\it the sum of the quantities of the living and tension forces is invariable with time: $T + U = const.$\,}'' 
If we designate \dashuline{this sum briefly as the force inherent in the system}, we have the \dashuline{theorem of the conservation of force}.

As slight as this reinterpretation may seem at first glance, the perspective that it opens up in all physical areas is still incalculably wide, because the generalization to any natural phenomenon is now easy to see. 
The main reason for this circumstance may be that the \dashuline{principle of conservation of force} now \dashuline{comes into parallel with} the \dashuline{principle of conservation of matter}, which has long been familiar to us and has, so to speak, become instinctive. 
Just as the quantity of matter present in a body system (measured by weight) cannot be reduced or increased by any means (although the most diverse physical and chemical transformations can be carried out with it) so too the quantity of force present in the system is an independent one, completely unchangeable size. 
Like matter, force can also be brought into a variety of forms. 
At first it appears in two main forms: as \dashuline{living force} or as \dashuline{tension}, but both can also appear to us in a variety of ways: 
{\it\underline{living force}} as \dashuline{visible movement}, as \dashuline{Light}, as \dashuline{heat}; 
the {\it\underline{tension}} as the \dashuline{lifting of a weight}, as \dashuline{elastic}, as \dashuline{electrical tension}, as \dashuline{chemical difference}, etc. 
But the sum of all these reserves of energy (stored up, so to speak, in different magazines) remains unchangeably the same, and \dashuline{all processes in nature consist only in converting the individual ones into one another}.

\dashuline{Helmholtz}'s view differs significantly from \dashuline{Mayer}'s in that the latter assumes a number of \dashuline{qualitatively different forms of force}, such as movement, gravity, heat, electricity, etc. s. w., while here, according to the \dashuline{mechanical view}, all different manifestations are subsumed under \dashuline{the two concepts of living force and tension} (a further step in the simplification of all natural phenomena).

\vspace*{-2mm} 
\begin{center}
{\it\color{red}\bf
---------
(p.38) 
---------
}
\end{center}
\vspace*{-3mm}

In order to apply the principle to any process that takes place in a system of bodies, one only needs to bring together at some point in time \dashuline{all the different types} of \dashuline{living force} and \dashuline{tension} and combine them into one sum. 
This sum then represents the \dashuline{total power inherent in the system and is unchangeable over time}, of course only as long as any external influence is excluded. 
It is understood that all individual members of this sum, i.e. all individual quantities of force, must be measured according to the same unit. 
But since we now have different types of force in physics, such as the \dashuline{living force of visible movement}, \dashuline{heat}, etc., {\it\color{red}if they are} measured according to different, conventional measures, then it will be necessary to reduce each type of force to the common (mechanical) measure before summing it up, i.e. to determine their mechanical equivalent, and here lies a certain difficulty which stands in the way of the application of the principle from the outset. 
There is \dashuline{no general rule} according \dashuline{to which the equivalent value can be calculated in advance} in every case, regardless of the principle. 
In fact, we have already seen repeatedly that the application of the principle by using false equivalence values as a basis has led to wrong conclusions. 
It is therefore \dashuline{necessary to examine the value of the corresponding equivalent for each type of force} in particular, and this is best done by applying the principle to a particularly simple, easily overlooked case. 
From this point of view it is interesting to follow the line of thought that \dashuline{Helmholtz} takes when discussing the various physical phenomena in their connection with the \dashuline{principle of conservation of force}.

\vspace*{-5mm} 
\begin{center}
{\it\color{red}\bf
---------
(p.38-39) 
---------
}
\end{center}
\vspace*{-3mm}

First, the field of mechanics is considered in the narrower sense, for which, as already mentioned, the \dashuline{general principle is the well-known theorem of living forces}. 
The movements that take place under the influence of the general gravitational force obey this law, the movements of incompressible solid and liquid bodies (as long as the living force of visible movement is not lost through friction or inelastic impact), and finally the movements of completely elastic solid and liquid bodies. 
In addition, the phenomena of sound and light, as well as radiant heat, are included here insofar as the destruction of movement does not occur through absorption. 
In all excepted cases, according to the \dashuline{law of conservation of force}, the lost living force must appear in some other form of force.
In fact, it shows itself in the absorption of heat rays as heat, in the absorption of light rays, its identity with the warming ones and chemically acting rays \dashuline{Helmholtz} already recognized, either as light (phosphorescence) or as heat or as a chemical effect. 
The \dashuline{principle of conservation of force} also requires \dashuline{compensation in some other form} for the loss of force caused by the impact of inelastic bodies and by friction, and \dashuline{Helmholtz} finds this replacement in a change in the molecular constitution that is accompanied by an increase in the quantity of internal tension forces of rubbing or pushing bodies, then in acoustic and electrical, but especially in thermal effects.
\dashuline{Helmholtz} concludes that in all cases of friction, where molecular changes, development of electricity, etc. are avoided, according to the \dashuline{principle of conservation of force}, \dashuline{for every loss of mechanical force a certain quantity of heat must arise}, which is equivalent to the work expended.
\dashuline{Joule}'s first experiments are cited here, although the measurement methods do not appear to be precise enough to deserve complete trust (\dashuline{R. Mayer}'s work was still virtually unknown at the time.) 
\dashuline{Carnot-Clapeyron}'s view that heat as such is imperishable and can therefore only produce mechanical work through its transition from higher to lower temperatures 
 is also dealt with and described as untenable in a detailed discussion.

\vspace*{-2mm} 
\begin{center}
{\it\color{red}\bf
---------
(p.40) 
---------
}
\end{center}
\vspace*{-3mm}

The experiments of \dashuline{Davy} (p.18) already speak in favour of the assumption of an absolute increase in the quantity of heat by friction, but then the generation of heat by electrical motion, namely by means of the charging of a bottle by the electrophorus, or by means of the excitation of a current by a magnet, forces to the same conclusion.
It follows from this that heat does not consist in the presence but in changes or movements of a substance, so that:  
``\,{\it the quantity of heat contained in a body must be understood as the sum of the living force of thermal movement (free heat) and the Quantity of those tension forces in the atoms {\it\color{red}(in German: Atomen)} which can produce such a thermal movement when their arrangement is changed (latent heat, internal work).\,}''
As far as the \dashuline{chemical generation of heat} is concerned, \dashuline{Helmholtz} cites the statement expressed by \dashuline{\,Hess} (p.20) that the same amount of heat is always generated in a chemical compound, regardless of the order and intermediate stages in which the compound takes place. Although this theorem originally emerged from the idea of the \dashuline{immutability of the heat substance}, it also turns out to be \dashuline{a consequence of the principle of conservation of force}.

What now follows is a consideration of the effects of heat, of which the generation of mechanical force is primarily examined. \dashuline{Joule}'s experiments (p.31f) are mentioned here, in which compressed air flows out into the atmosphere at one time and into an airless vessel at another time. In the first case, a decrease in the temperature of the air occurs in accordance with the work done in overcoming the air pressure, while in the latter case no overall change in temperature can be perceived. Finally, the theoretical studies by \dashuline{Clapeyron} and \dashuline{Holtzmann} are discussed.
A significant part of the treatise is made up of the applications of the principle to electricity and magnetism, most of which appear to be completely new. 
First, static electricity is dealt with, the effects of which are differentiated as mechanical (movement of electricity with the conductors) and thermal (movement in the conductors).
Here \dashuline{Helmholtz} uses the mechanical, now so-called electrostatic measuring system.

\vspace*{-5mm} 
\begin{center}
{\it\color{red}\bf
---------
(p.41) 
---------
}
\end{center}
\vspace*{-3mm}

The value of the quantity of the electrical tension forces is provided by the sum of the potentials of the various electrically charged bodies present in the system under consideration on each other and on themselves (Helmholtz here uses a definition of the potential that is somewhat different from the one that is now usual, in that he uses it firstly takes the opposite sign and also calculates the potential of a charge on itself to be twice as large as what happens now). 
If \dashuline{living force} (visible movement or heat) is generated by the effect of electricity, its size is measured by the decrease in the electrical tension. 
If only heat is generated by discharging electricity, then it is equal to the increase (now: decrease) in the total electrical potential, from which the law follows for batteries whose external occupancy is derived that the heat of discharge is proportional to the square of the discharged quantity of electricity and the reciprocal discharge quantity (capacity) of the battery, regardless of the shape of the closing wire, as is essentially confirmed by \dashuline{Riess}' experiments. 
In addition (because of the use of the mechanical measurement system), the \dashuline{mechanical equivalent of heat} can be found as a factor in the denominator in the expression for the discharge heat.

Moving on to \dashuline{galvanism}, \dashuline{Helmholtz} first discusses the two opposing hypotheses regarding the cause of the galvanic current: the \dashuline{contact theory} and the \dashuline{chemical theory}.
The first seeks the location of the excitation of the current at the contact surface of the metals, the second in the chemical processes of the chain. 
Helmholtz finds the \dashuline{equivalent of the work done} by the galvanic current in the chemical decomposition of the second-class conductors caused by it, from which it follows that \dashuline{the contact theory would then come into contradiction with the principle of conservation of force} if there were a single second-class conductor (i.e. that does not follow the voltage law) that would not be decomposed by the current.

\vspace*{-5mm} 
\begin{center}
{\it\color{red}\bf
---------
(p.42) 
---------
}
\end{center}
\vspace*{-3mm}

We have already mentioned the attacks against contact theory derived from this assumption above (p.19f). 
But if one considers every second-class conductor as an electrolyte from the outset, \dashuline{the assumption of the contact force not only does not involve any contradiction}, but also gives a simple and convenient view of the nature of electrical voltage by imagining that the different metals have different attractive forces affect the electricity. 
In a state of equilibrium, the electrical voltage must then be equal to the difference in the living forces that a unit of electricity would gain due to these attractive forces when passing into the interior of every metal, i.e. regardless of the size and shape of the contact surface. 
Then the validity of the voltage law also arises directly for a series of metals connected one after the other, in that the electrical voltage between the first and last metal then becomes independent of the metals in between.
Of the galvanic chains, \dashuline{Helmholtz} first considers those that only produce \dashuline{chemical decomposition} but no polarization. 
Here it follows from the equality of the electrical and chemical heat, based on \dashuline{Ohm}'s law in relation to the current strength and \dashuline{Lenz}'s law in relation to the heat development in the closure, that the electromotive force of an element (\dashuline{Daniell}, \dashuline{Grove}) is equal to the difference in the heat tones that occur when the equivalents of the two metals are oxidized and the oxide is dissolved in the acid. 
It also follows that all chains in which the same chemical processes take place also have the same electromotive forces, for which \dashuline{Poggendorff}'s experiments are cited. 
Chains with polarization are also examined, initially those in which only polarization but no noticeable chemical decomposition takes place. Here you get inconstant, usually soon disappearing currents, which essentially only serve to create the electrical equilibrium between liquid and metals.

\vspace*{-2mm} 
\begin{center}
{\it\color{red}\bf
---------
(p.43) 
---------
}
\end{center}
\vspace*{-3mm}

If the polarization of originally identical plates (in composite chains) is caused by external electromotive forces, the lost force of the original current can be regained as a secondary (depolarization) current by switching off the exciting elements. 
If the phenomena of polarization and chemical decomposition take place simultaneously, the resulting current can be divided into two parts, the polarization current and the decomposition current, and these two parts can be considered individually in the same way as before. 
Again and again, \dashuline{the heat generated throughout the entire circuit}, be it the development of it everywhere proportional to the square of the current intensity or, as Helmholtz also assumes, in certain places according to a different law, \dashuline{is identical with the heat that generates the chemical heat} that takes place in the elements processes would cause if they proceeded in the usual way, without the development of electricity.

While chemical processes are to be considered as the source of hydroelectric currents, \dashuline{Helmholtz} finds the effects of such currents discovered by \dashuline{Peltier} on the solder joints of two metals as an equivalent of the force generated by thermoelectric currents, and here the \dashuline{principle of conservation of force} requires that the the heat developed inside the conductor is equal to the heat absorbed at the solder joints as a whole. The consequence of this assumption is, among other things, the theorem that the \dashuline{Peltier} effect at a soldering point is proportional to the current strength, and that the electromotive force of the thermal chain grows in the same ratio as the heat absorbed by the current unit at both soldering points together.

\dashuline{Helmholtz} then deals with the \dashuline{effects of magnetism} in exactly the same way as those of static electricity. The magnetic tension is measured by the analogously defined magnetic potential (of the magnets on each other and on themselves), and the increase in this quantity results in the growth of living force. A distinction is made here between permanent magnetism and magnetism that can be changed by induction. In the first case, with permanent steel magnets, the potential of the magnet on itself is constant and can therefore be omitted entirely. 
With induced magnetism, however, this potential is changeable.

\vspace*{-5mm} 
\begin{center}
{\it\color{red}\bf
---------
(p.44) 
---------
}
\end{center}
\vspace*{-3mm}

Helmholtz limits himself here to considering bodies (made of soft iron) in which the magnetism is induced to the point of perfect binding, i.e. so that the magnetic surface coating, which can always be substituted for the internal distribution, is formed according to exactly the same law as the electrical surface distribution in an electrically induced conductor that is not charged to begin with. As is well known, this assumption is contained as a special case in Poisson's theory of magnetic induction.

Moving on to the phenomena of electromagnetism, the principle of conservation of force is applied to closed currents based on the laws of electrodynamic effects developed by \dashuline{Ampère} and \dashuline{F. Neumann}. 
If a permanent magnet initially moves under the influence of a hydroelectric stationary current $J$, the following force equivalents must be taken into account: 
\\ \hspace*{5mm}
1) the living force of the movement of the magnet, 
\\ \hspace*{5mm}
2) the heat developed by the current in its conduction: and finally
\\ \hspace*{5mm}
3) the chemical work produced in the elements.  
\\
The algebraic sum of these equivalents must have a value that does not change over time, so their change over time is $= 0$.
Now the living force of the magnet obtained in the time element $dt$ results with the help of the potential of the current on the magnet: $J\,.\,V$ (the current $J$ is thought to be replaced by a magnetic double layer according to Ampere) to $J\,.\;dV/ dt$, furthermore the heat developed in the circuit at the same time to $J^2\:.\:W\:.\:dt$ (with $W$ the resistance), and finally the chemical work generated to $-\,A\ dt $ (with $A$ the electromotive force of the elements), so we get the condition: 
$$ 
       J\,.\;\frac{dV}{dT}\:.\:dt 
 \;+\; J^2\:.\:W\:.\:dt 
 \;-\; A\:.\:J\:.\:dt 
 \;=\; 0 \; ,
$$ 
or 
\vspace*{-3mm}
$$
  J \;=\; \frac{\displaystyle \: A \:-\: \frac{dV}{dt}\:}{W} \; .
$$
\vspace*{-6mm} 
\begin{center}
{\it\color{red}\bf
---------
(p.45) 
---------
}
\end{center}
\vspace*{-3mm}

\noindent By comparing with Ohm's formula, the law of magneto-electric induction follows in such a way that every change in position of the magnet in a closed conductor induces an electromotive force: 
$-\,dV/dt$, which is measured by the speed of the change in potential of the magnet onto the conductor, the latter thought to be traversed by the current unit.
This law essentially agrees with the principle derived by F. \dashuline{Neumann} for electrical induction and differs from it in form only in that \dashuline{Neumann} still has to multiply the change in potential by an undetermined constant $\varepsilon$ in order to obtain the value of the induced electromotive force, whereas in Helmholtz's derivation this constant is given a specific value, which depends only on the units of measurement chosen (it is $= 1$ in the magnetic measurement system now in use, and for those used at the time by Helmholtz is $=$  ``\,the reciprocal value of the mechanical equivalent of heat,\,'' since there he measures the resistance by the heat generated in it by the unit of current in the unit of time).

 The described way of applying the \dashuline{principle of the conservation of force} leads us to a fundamentally important consideration. 
 The question could be raised as to whether it is justified from the outset to apply the various equivalents of force in the same way as was done above: the \dashuline{living force} of the magnet, \dashuline{heat development} in the circuit, \dashuline{chemical work}. 
 By analogy with earlier theorems, it would be reasonable to assume that the potential $J\,.\,V$ of the current on the magnet could also be regarded as a certain type of force and included as a member in the sum of the various force equivalents, whereby the equation of the conservation of force would then have an additional member: $d(J\,.\,V)/dt\,.\, dt$.

\vspace*{-2mm} 
\begin{center}
{\it\color{red}\bf
---------
(p.45-46) 
---------
}
\end{center}
\vspace*{-3mm}

Just as the magnetic tension force is counted as a specific force equivalent in the movement of two magnets (and measured by the value of the magnetic potential), \dashuline{the idea arises that an electromagnetic tension force should also be cited as a special equivalent} (alongside the other types of force in the case of electromagnetic forces, which are so closely related to magnetic forces). 
However, \dashuline{the conservation of force would then lead to a conclusion that deviates from the law of induction} derived above and from experience.

The correct answer to the question can only be that there is in fact no means of deciding from the outset, without the aid of experience, whether the electromagnetic potential is to be regarded as a special kind of force or not (See p.38). Only the fact that the assumption of the stated conjecture, applying the law of conservation of force, leads to a contradiction with experience, entitles us to conclude that \dashuline{there is in fact no electromagnetic potential force}, just as there is a magnetic one, \dashuline{at least if we retain the ideas of magnetism used so far}.

Therefore, the frequently stated assertion that \dashuline{magnetoelectric induction} is a \dashuline{direct consequence} of the \dashuline{principle of conservation of force} is \dashuline{not entirely correct}. 
For example, one could just as well assume from the outset that a current that interacts with a magnet also behaves completely like a permanent magnet. Then, when a movement occurs, the current strength would remain constant, the chemical work would be completely converted into current heat, and the principle of conservation of force would be satisfied in exactly the same way as with the movement of a constant magnetic double layer interacting with a magnet. A closed metallic conductor would then behave just as indifferently towards a magnet as towards any non-magnetic body, whereas a stationary constant magnet would exert certain induction effects on a stationary variable current.

The phenomena of induction cannot therefore be deduced from the \dashuline{principle of the conservation of force} alone, but only simultaneously from experience.
They are not a consequence of the principle in and of itself, but a consequence of the further assumption that there are no other kinds of force than those considered above. 
However, from this assumption they emerge as being completely numerically determined.

\vspace*{-5mm} 
\begin{center}
{\it\color{red}\bf
---------
(p.47) 
---------
}
\end{center}
\vspace*{-3mm}

 The practical importance of these considerations becomes apparent when the principle is applied to the \dashuline{interactions between two currents}, 
   where the equation given by Helmholtz is incomplete.
It contains as force equivalents only the \dashuline{living force} of the movement of the current conductors, the heat generated in the conductors and the chemical work consumed, while it later turned out that there is also an \dashuline{electrodynamic force equivalent}, the now so-called \dashuline{electrokinetic energy}, which is measured by the \dashuline{(negative) potential of the two currents on each other}.
 In order to form the complete equation for the conservation of force, this quantity must be introduced as a member of the sum of the individual types of force, 
 and only then does this equation become generally correct. 
 -- just like the potential of the currents on each other, the potential of a current on itself also provides a special type of force, which, strictly speaking, should also have been taken into account above in the electromagnetic effects (self-induction). 
 We will undertake a systematic discussion of these questions in the \dashuline{third section of this document}. 
 Here it should only be pointed out once again how important it is to introduce the \dashuline{correct force equivalent} for each individual phenomenon when applying the principle of conservation of force.

A reference to the processes in organic nature, insofar as they can be brought into connection with our principle, in particular to the accumulation of chemical tension in plants under the influence of the chemically active sun's rays, and to the heat production of the animal body, which we already have is discussed, as well as the rejection of some objections to the principle, form the \dashuline{conclusion of Helmholtz's treatise}. 
Laid out according to a magnificent plan, and containing in a small space a wealth of facts and ideas, some of which were only taken up and further developed individually by other researchers over the course of the years, this work will forever be one of the most remarkable and instructive monuments in the history of development of the \dashuline{principle of conservation of strength}.

\vspace*{-2mm} 
\begin{center}
{\it\color{red}\bf
---------
(p.48) 
---------
}
\end{center}
\vspace*{-3mm}

The impression the work {\it\color{red}(of Helmholtz)} made on his peers when it was published was not a significant one.
The new principle was downright unpopular at the time, {\it\color{red}(because)} it demanded such a radical {\it\color{red}change?/upheaval?/revolution?/transformation?} of all physical views that it was understandably generally viewed 
  with surprise and mostly rejection.
  As a result, this essay, which later became so famous, did not even reach wider circles at first (\dashuline{W. Thomson}, for example, did not see it until 1852, according to his own account).
  
A few other impulses had to be added before the change in general opinion took place.

Before we move on to this, we would like to mention \dashuline{Clausius}' critical comments on \dashuline{Helmholtz's treatise}, the first part of 
which$\,$\footnote{$\:$R. Clausius: Über einige Stellen in der Schrift von Helmholtz über die Erhaltung der Kraft {\it\color{red}(About some passages in Helmholtz's work on the conservation of strength)}. Pogg. Ann. 89, p.568, 1853.} 
appeared in 1853 (when the victory of the principle was already decided). 
In these remarks, on the one hand, some applications that Helmholtz had made of the principle of conservation of force are critically discussed, in particular the way in which the potential of a conductor on itself is defined (p.41), and then the consistency of \dashuline{Riess}'s experiments about the thermal effects of an electrical discharge with the theory, along with the \dashuline{independence of the heat} generated from the nature of the closing wire, as well as \dashuline{Holtzmann}'s conception of the \dashuline{equivalence of heat and work}. 
On the other hand, \dashuline{an objection is raised against} the theorem put forward by Helmholtz (p.36), \dashuline{that the solvability of natural forces into central forces} (acting from point to point in the direction of the connecting line with an intensity that depends only on the distance) \dashuline{is a necessary consequence of the theorem of living power}.

\vspace*{-5mm} 
\begin{center}
{\it\color{red}\bf
---------
(p.49) 
---------
}
\end{center}
\vspace*{-3mm}

\dashuline{Helmholtz}$\,$\footnote{$\:$H. Helmholtz: Erwiderung auf die Bemerkungen von H. Clausius {\it\color{red}(Reply to the comments of H. Clausius)}. Pogg. Ann. 91, p.241, 1854. Wiss. Abh. I p.76.}
responds to these comments {\it\color{red}(of Clausius)} in detail and explains, particularly with regard to the last-mentioned sentence (always based on the mechanical view of nature), that its derivation depends only on the one premise that real effects have their complete basis in the relationships of real things have each other. 
The \dashuline{relative position of one point to another is determined only by the distance}, and it follows that the \dashuline{living force}, if it depends only on the position of the points (which today does not appear to be a necessary consequence of the \dashuline{principle of living force}, cf. the basic electrical law of W. \dashuline{Weber}) can also only depend on the distance.
This then leads to the assumption of central forces (see the third section for more information). 
It is different if a \dashuline{physically {\bf extended}, infinitely small element} is substituted \dashuline{instead of a point}. 
Here, in general, \dashuline{there are {\bf directions} of different values}, and it is quite conceivable that \dashuline{the living force of a point} that moves under the influence of an active element \dashuline{has different values depending on the {\bf direction}} of its distance from the element.

But \dashuline{Helmholtz} now proves that if the living force of the point is an arbitrarily given function of its coordinates, an arrangement of points can always be found (in an infinitely varied way) within or on the surface of the element, which in turn is simply based on central forces act and completely replace the effect of the element. 
In this way, this general case can also be traced back to the existence of central forces. 
As is well known, we make use of this sentence by e.g. imagine the long-distance effects of an elementary magnet as arising from the interaction of two poles acting with simple central forces. 
Finally, \dashuline{Helmholtz} completes his earlier applications of the principle to magnetism and electrodynamics based on \dashuline{Poisson}'s theory of magnetic induction, and his own investigations into current fluctuations of induced
currents$\,$\footnote{$\:$H. Helmholtz: Über die Dauer und den Verlauf der durch Stromesschwankungen inducierten elektrischen Ströme {\it\color{red}(On the duration and course of electrical currents induced by current fluctuations)}. Pogg. Ann. 83, p.505, 1851. Wiss. Abh. I p.429.}.

\vspace*{-5mm} 
\begin{center}
{\it\color{red}\bf
---------
(p.50) 
---------
}
\end{center}
\vspace*{-3mm}

It is {\it\color{red}(also)} found that a galvanic current, through its existence, in and of itself represents a force equivalent which is proportional to the square of its intensity (electrodynamic potential on itself). If the current is interrupted, this power supply is either transformed directly into heat (interrupted spark) or only indirectly, through the resulting extra current. 
F. \dashuline{Neumann}'s general law of induction by magnets or currents is also shown to be consistent with the \dashuline{principle of conservation of force}. 
This discussion came to a conclusion with a \dashuline{second reply from
Clausius}$\,$\footnote{$\:$R. Clausius : Über einige Stellen d. Schrift v. Helmh. üb. d. Erh. d. Kr. {\it\color{red}(About some places of writing  of von Helmholtz ex. d. Rec. d. Kr.)}. Zweite Notiz, Pogg. Ann. 91, p.601, 1854.}.

\begin{center}
{\it\color{red}\bf
=
James Prescott Joule (1848, 1857, 1850) 
-- Marc François Seguin (aîné) (1847)
=
\\
======
William Robert Grove (1848)
-- Julius Rober Mayer (1848)
======
\\
=======
Ernst Mach (1872)
-- Ludwig A. Colding (1843, 1864)
=======
\\
=====
William Thomson (next Lord Kelvin) (1848, 1849, 1850)
=====
\\
=========
Joseph John Thomson (1849) / (p.50)
=========
}
\end{center}
\vspace*{-3mm}

If we now turn back to the year 1847, we first find a 
new work$\,$\footnote{$\:$Joule: On the mechanical equivalent of heat as determined by the heat evolved by the friction of fluids. Phil. Mag. (3) 31, p.173, 1857. Pogg. Ann. 73, p.479, 1848.}
by Joule, who published a series of experiments to \dashuline{determine the mechanical equivalent of heat}, which related to the generation of heat through friction in liquids. 
A paddle wheel made of brass or iron immersed in a liquid (water, Wallrath oil, mercury) was made to rotate by falling weights, and the heat generated by the friction in the liquid was compared with the work expended. 
The ratio gave the \dashuline{average mechanical equivalent of heat} of $430$ kilogram meters.


Now the number of those who, following Joule's process, helped to develop the new theory in the wide area that was opening up gradually increased. 
The above-mentioned \dashuline{Séguin aîné}, who was strengthened in his views on the nature of heat, now also calculated the mechanical heat equivalent, namely through the cooling that water vapor experiences (from $180$° to $80$° Celsius) when it does work during 
expansion$\,$\footnote{$\:$Séguin aîné: Note à l'appui de l'opinion émise par M. Joule, sur l'identité du mouvement et du calorique {\it\color{red}(Note in support of the opinion expressed by Mr. Joule, on the identity of the movement and the caloric)}. Compt. Rend. 25, p.420, 1847.}. 

\vspace*{-5mm} 
\begin{center}
{\it\color{red}\bf
---------
(p.51) 
---------
}
\end{center}
\vspace*{-3mm}

The average result was $449$~kg. 
Also to be mentioned here are the works of 
W. \dashuline{Grove}$\,$\footnote{$\:$W. Grove: Résumé de quelques leçons sur les rapports des divers agents ou forces physiques {\it\color{red}(Summary of some lessons on the relationships between various agents or physical forces)}. L'Institut Nr. 750-753, 1848.}, 
who spent a long time studying the laws of the transformation of the various natural forces into one another, as they result from the mechanical view (and his work on this subject by the lectures given at the Royal institution in London) were combined into a more popular 
book$\,$\footnote{$\:$W. Grove: The correlation of physical forces. 3. Aufl. 1855.} 
on the relationship of natural forces, which was translated into French by \dashuline{Moigno} in 1856, and later 
repeatedly$\,$\footnote{$\:$W. Grove: Die Verwandtschaft der Naturkräfte deutsch von E. v. Russdorf {\it\color{red}(The relationship of the forces of nature German by E. v. Russdorf)}, Berlin 1863, von Schaper, Braunschweig 1871.} 
into German.

It is noteworthy that with the \dashuline{discovery of the mechanical equivalent of heat} and the development of the \dashuline{general principle of conservation of energy}, the development of the view that all natural phenomena are based on motion went so immediately hand in hand and was often even identified. 
Strictly speaking, the principle teaches nothing other than the \dashuline{transformability of the individual natural forces into one another} according to fixed conditions, but it gives absolutely no information about the way in which this transformation comes about. 
The necessity of the \dashuline{mechanical view of nature} cannot be deduced from the validity of the principle, while conversely the principle turns out to be a necessary consequence of this view, at least if one starts from central forces (p.35). 
This latter circumstance, in connection with the need to form a \dashuline{uniform idea of the operation of natural forces}, sufficiently explains the fact that the \dashuline{mechanical theory} was accepted so quickly and without contradiction, and which \dashuline{has in fact been brilliantly confirmed everywhere so far}.
At least at the moment, I don't think I should share the fears that are linked to the general feasibility of this theory, as an excessively \dashuline{narrow-minded view of natural 
phenomena}$\,$\footnote{$\:$E. Mach : Die Geschichte und die Wurzel des Satzes von der Erhaltung der Arbeit {\it\color{red}(The history and the root of the theorem of conservation of work)}. Prag 1872. Calve.}.

\vspace*{-5mm} 
\begin{center}
{\it\color{red}\bf
---------
(p.52) 
---------
}
\end{center}
\vspace*{-3mm}

While \dashuline{Joule's work gradually achieved deserved recognition, especially in Germany}, and even the priority disputes (between 
Mayer$\,$\footnote{$\:$R. Mayer: Sur la transformation de la force vive en chaleur et réciproquement {\it\color{red}(On the transformation of living force into heat and vice versa)}. Compt. Rend. 27, p.385, 1848 etc.}, 
Joule, 
Séguin$\,$\footnote{$\:$Séguin aîné: Note à l'appui de l'opinion émise par M. Joule, sur l'identité du mouvement et du calorique {\it\color{red}(Note in support of the opinion expressed by Mr. Joule, on the identity of the movement and the caloric)}. Compt. Rend. 25, p.420, 1847.}, 
Colding$\,$\footnote{$\:$A. Colding: Det kongel. danske vidensk. selsk, naturv. og math. afh. (5) II, 1843 p.121, 167. On the history of the principle of the conservation of energy, Phil. Mag. (4) 27, p.56, 1864.}) 
were already beginning, \dashuline{the more important physicists in England behaved reticent towards the new theory for a while}. 
However, in a 
communication$\,$\footnote{$\:$W. Thomson: Report of the 18. Meeting of the British Association for the adv. of sc. Notices and abstr. of communic. p.9, 1848. On the theory of electromagnetic induction.} 
to the British Association from \dashuline{1848} on the theory of electromagnetic induction, \dashuline{W. Thomson} draws attention to the close connection that exists between the work related to the movement of the inducing magnet and the intensity of the current caused by this movement, but he remains with the assumption that a certain amount of mechanical effect is lost due to the induced current, without further asking about a replacement for this loss. 
Nevertheless, he {\it\color{red}(W. Thomson)} recognized the importance and fruitfulness of the theorem that \dashuline{it is impossible to obtain work from nothing}, so clearly that he set about \dashuline{taking up Carnot's theory}, which is based on the same idea, anew and, 
using the latest observational data (namely from Regnault), made it suitable for application to the \dashuline{moving force of 
heat engines}$\,$\footnote{$\:$W. Thomson: An account of Carnot's theory of the motive power of heat. Transact. of the Roy. Soc. of Edinburgh, vol.XVI p.541, 1849.}. 

\vspace*{-5mm} 
\begin{center}
{\it\color{red}\bf
---------
(p.53) 
---------
}
\end{center}
\vspace*{-3mm}

We have already considered the difficulties that he {\it\color{red}(W. Thomson)} encountered above (p.17).
In any case, he did not consider it impossible to overcome them on the path he had chosen. 
Incidentally, since Carnot's old theory is partly based on a correct basis, some of its conclusions appeared to be in agreement with experience, for example: the conclusion drawn by 
J. Thomson$\,$\footnote{$\:$J. Thomson: Theoretical considerations on the effect of pressure in lowering the freezing-point of water. Trans. Roy. Soc. Edinburgh XVI, p.575, 1849.}, 
and subsequently confirmed experimentally by 
W. Thomson$\,$\footnote{$\:$W. Thomson: The effect of pressure in lowering the freezingpoint of water experimentally demonstrated. Phil. Mag. (3) 37, p.123. 1850. Pogg. Ann. 81, p.163, 1850.}, 
that external pressure lowers the freezing point of water.

Meanwhile, \dashuline{Joule} had continued his work with iron endurance and increasingly precise methods. While previously his primary concern was to prove the \dashuline{existence of the mechanical equivalent of heat}, i.e. the \dashuline{constancy of the ratio between heat and work}, in a wide variety of transformation processes, he now set out to do so as precisely as possible, based on his diverse experiences to \dashuline{determine the value of this equivalent}. 
Of all the methods he had previously used, he {\it\color{red}(\dashuline{Joule})} chose the generation of heat by moving a paddle wheel in water or mercury, as well as by rubbing cast iron discs against each other, as the most reliable, and thus,
taking into account as far as possible all conceivable sources of error,
\dashuline{determined from numerous experiments the mechanical equivalent of the unit of heat}, 
the latter (based on English pounds and F. degrees)
to be $772$~foot pounds, 
or (the heat unit based on Kgr. and C. degrees)
to be $423.55$~kilogram 
meters$\,$\footnote{$\:$J. P. Joule: On the mechanical equivalent of heat. Phil. Trans. London 1850, p.61.}, 
a number which from then on 
was regarded for some time as the most reliable value of this important constant
(it is probably a little too small, cf. section 3.)

\vspace*{-2mm} 
\begin{center}
{\it\color{red}\bf
---------
(p.54) 
---------
}
\end{center}
\vspace*{-3mm}

Although the \dashuline{theorem of the equivalence of work and heat} gained more and more reputation as \dashuline{Joule's work} progressed, no attempt had yet been made by any side to make this theorem the \dashuline{basis of a detailed theory like Carnot}'s was to do. 
It was up to \dashuline{Rudolph Clausius} to enrich science with such a theory.

\vspace*{-1mm} 
\begin{center}
{\it\color{red}\bf
====
Rudolph Clausius (1850, 1864, 1876)
====
}
\end{center}
\vspace*{-3mm}

After he had accepted the assumption in a recently published 
work$\,$\footnote{$\:$Clausius: Über die Veränderungen, welche in den bisher gebräuchlichen Formeln für das Gleichgewicht und die Bewegung elastischer fester Körper durch neuere Beobachtungen notwendig geworden sind {\it\color{red}(On the changes that have become necessary due to recent observations in the previously used formulas for the balance and movement of elastic solid bodies)}. Pogg. Ann. 76, p.46, 1849.}
that \dashuline{heat was an indestructible substance}, \dashuline{Clausius} published 
a treatise$\,$\footnote{$\:$Clausius: Über die bewegende Kraft der Wärme und die Gesetze, welche sich daraus für die Wärmelehre selbst ableiten lassen {\it\color{red}(About the moving force of heat and the laws that can be derived from it for the theory of heat itself)}. Pogg. Ann. 79, p.368, 500, 1850. Vgl. auch : R. Clausius. Abhandl. üb. d. mechanische Wärmetheorie {\it\color{red}(Treatise ex. d. mechanical heat theory)}. 1. Aufl. Braunschw. 1864, 2. umg. Aufl. Braunschw. 1876.} 
on the \dashuline{moving power of heat} in the Annals of Physics by Poggendorff in 1850, in who, in accordance with the ideas that can be found sporadically in the writings of \dashuline{Helmholtz} and \dashuline{Joule}, brought the basic idea of the \dashuline{equivalence of heat and work} to further development. 
He expresses this basic idea in the following sentence: 
``\,{\it in all cases where work is created through heat, an amount of heat proportional to the work produced is consumed, and conversely, when an equal amount of work is consumed, the same amount of heat is generated.\,}'' 
\dashuline{This is the opposite of Carnot's theory} and a number of Carnot's terms have been overturned. 
If a body goes through a circular process and finally returns to its old initial state (determined by temperature and density), then, according to \dashuline{Carnot}, the total amount of heat absorbed from outside 
  during the process must be equal to the amount of heat released, regardless of the external work performed by the body as a whole.
  From this it follows that in order for a body \dashuline{to pass from a certain {\bf(arbitrarily fixed) zero state} {\it\color{red}(\,``\,Nullzustand\,''\,)} to a certain other state}, it must absorb a certain amount of heat from the outside, regardless of the way in which the transfer takes place.
 

\vspace*{-5mm} 
\begin{center}
{\it\color{red}\bf
---------
(p.55) 
---------
}
\end{center}
\vspace*{-3mm}

\dashuline{Clausius} now showed that this concept of total heat was no longer permissible in the new theory, because the amount of heat that a body has to absorb from outside (through conduction or radiation) in order to move from one state to another depends essentially on the external work that it does during the transfer, i.e. on the path of the transfer. 
Unfortunately, \dashuline{Clausius}' designation ``\,total heat\,'' gave rise to a misunderstanding in that other physicists assigned this name to a different quantity, which in turn really only depends on the current state. 
As already mentioned above (p.40), Helmholtz called with the name: ``\,Quantity of heat contained in a body\,'' the sum of the \dashuline{living forces} and \dashuline{tension forces} contained in it, and this quantity is of course only dependent on the current state, no matter what more specific idea one has of the mutual action of the smallest particles of the body because it represents the reserve of strength contained in the body. 
For this latter quantity, 
\dashuline{Clausius} later$\,$\footnote{$\:$R. Clausius: Abh. üb. d. mech. W. 1. Aufl. I. p.281, 1864. 2. Aufl. I. p. 33, 1876.} 
adopted the expression ``\,internal energy of the body,\,'' which was used by \dashuline{W. Thomson} and is now generally used, while in the treatise currently under discussion there is still no special name for it. 
Rather, \dashuline{Clausius} breaks it down The entire supply of energy, just like \dashuline{Helmholtz}, is divided into two parts: the \dashuline{free heat} (\dashuline{sum of the living forces}) and the \dashuline{internal work} (\dashuline{sum of the tension forces}). 
Each of these quantities in itself is a specific function of the state. 
The \dashuline{old concept of latent heat} is of course canceled out by this: if heat does not produce an increase in temperature, it does not become latent, but disappears 
altogether, since it is transformed into work (labour).

\vspace*{-5mm} 
\begin{center}
{\it\color{red}\bf
---------
(p.56) 
---------
}
\end{center}
\vspace*{-3mm}

To establish the basic equations of the theory, \dashuline{Clausius} uses \dashuline{Clapeyron}'s method of circular processes and applies the same to permanent gases and saturated vapors. 
In doing so, he already introduces the assumption, which was later developed in great detail, that the regularity which is expressed in the behavior of all permanent gases against pressure and temperature changes through the laws of \dashuline{Mariotte and Gay Lussac} has its basis in a consistent constitution of these gases. 
In them, {\bf\dashuline{the bond between the molecules}} is so loosened that when the gas expands, no internal work is done (\dashuline{Mayer}'s hypothesis, p. 23) and so the entire internal supply of force, insofar as it is changeable, relies on the \dashuline{free heat} (\dashuline{living force}) reduced. 
Then all external work is only done at the expense of the heat of the gas. 
One can see that this idea leads directly to {\bf\dashuline{the modern gas theory}}. 
To these assumptions, \dashuline{Clausius} adds the further fact that the \dashuline{specific heat of a permanent gas is constant} (independent of the temperature) at a constant volume, and thus arrives at various theorems about the specific heats, some of which are completely new, and some of which have already been derived from experiments appear confirmed. 
The laws of expansion under certain external conditions (at constant temperature, at constant pressure, and when the supply of heat from outside is prevented) are also derived \dashuline{in essential agreement with experience}.
The second part of the treatise contains a significant expansion of the theory by incorporating \dashuline{Carnot}'s  \dashuline{principle of work performance} through the transfer of heat from higher to lower temperatures. 
\dashuline{Clausius finds that this principle}, even though it contradicts the principles of the mechanical heat theory in its original version, \dashuline{still contains a correct and very valuable idea} that just needs to be brought into the appropriate form, in order to play an important role in the new theory to play. 

\vspace*{-2mm} 
\begin{center}
{\it\color{red}\bf
---------
(p.57) 
---------
}
\end{center}
\vspace*{-3mm}

This idea essentially says that: \dashuline{when heat is converted into mechanical work through a cycle, the production of work is necessarily accompanied by a transfer of a certain (different) amount of heat from a higher to a lower temperature}. 
The heat tends to move from a higher temperature to a lower one, and this effort can be used to generate work (converting heat into work), although \dashuline{there is a certain maximum of the work to be gained} that only depends on the Temperatures between which heat transfers, but not on the nature of the bodies in question. 
In order to cause a change in the heat in the opposite direction (i.e. a transition from a lower to a higher temperature) a certain amount of work is always required (conversion of work into heat), which is at least as great as the maximum work that can be gained through the reverse process.
%

In this modification, the \dashuline{Carnot's principle}, whose further development \dashuline{later led Clausius to his second law of the mechanical theory of heat}, does not form a contradiction to, but rather \dashuline{a complement to, the principle of the equivalence of heat and work}, since it regulates the conditions for the mutual convertibility of equivalent types of force. 
From here on, in the further \dashuline{development of the heat theory}, these \dashuline{two principles}, which are in fact not in any logical connection with one another, separate sharply from one another, and from now on in this work we will only deal with the presentation of one of them have to deal with, which is directly linked to the \dashuline{general law of 
conservation of energy}.$\,$\footnote{$\:${\it\color{red}I have recalled in the footnote~\ref{label_footnote_Exergie} that the ``puissance motrice du feu'' (``motive power of fire'') of Sadi Carnot, next called} 
{\it\color{red}``Motivity' by William Thomson and ``available energy'' by Gibbs and Maxwell, was a mix of the first and second principles of thermodynamics. Technically, this ``available energy'' $A=E_x-E_{x0} = (U-U_0) - T_0\:(S-S_0)$ (also called ``exergy'' nowadays, for the same concept of ``maximum of work'' mentioned by Clausius) depends on the quantities $E_x = U-T_0\:S$ and $E_{x0} = U_0-T_0\:S_0$, where $U$ is the energy ($1$st law), $S$ the entropy ($2$nd law) and $T_0$ the absolute temperature ($0$th law) of the thermostat or the (infinite) source of heat of the system. It seems that Planck (like many others) missed these points when he described the vision of Carnot, with the efficiency given by $1-T_0/T$ (P. Marquet).}}

\dashuline{Clausius}' treatise concludes with a description of various methods for calculating the \dashuline{mechanical equivalent of heat}, in which Joule's experiments are also considered.
\dashuline{The most probable value is} finally assumed to be about 
$400$~Kgr\,.\,m -- It is probably not wrong to date the period in which the \dashuline{mechanical theory of heat} gained the upper hand from this epoch-making treatise.

\vspace*{-1mm} 
\begin{center}
{\it\color{red}\bf
===
William John Macquorn Rankine (1850, 1851, 1853) (p.58)
===
}
\end{center}
\vspace*{-3mm} 

Almost 
simultaneously$\,$\footnote{$\:$W. J. M. Rankine: Über die mechanische Theorie der Wärme {\it\color{red}(On the mechanical theory of heat)}, Pogg. Ann. 81, p.172, 1850. (Brief.) Phil. Mag. (4) 2, p.61, 1851. On the centrifugal theory of elasticity, as applied to gases and vapours. Phil. Mag. (4) 2, p.509, 1851.} 
with \dashuline{Clausius}, W. J. M. \dashuline{Rankine} began to work on the \dashuline{theory of heat}  from the \dashuline{new mechanical point of view}  (here we will ignore the attempts to establish a mechanical theory of heat insofar as they have not received any notable implementation, like those of 
\dashuline{Buys-Ballot}$\,$\footnote{$\:$2) Buys-Ballot: Schets eener physiologie van het onbewerktuigde ryk der natuur {\it\color{red}(It's a physiology of how nature works)}. Utrecht 1849.}, 
\dashuline{Wilhelmy}$\,$\footnote{$\:$L. Wilhelmy: Versuch einer mathematisch-physikalischen Wärmetheorie {\it\color{red}(Attempt at a mathematical-physical theory of heat)}. Heidelberg 1851.} 
and others). 
However, like \dashuline{Clausius}, 
\dashuline{Rankine}$\,$\footnote{$\:$M. Rankine: On the mechanical action of heat. Trans. Roy. Soc. Edinburgh (geles. Febr.4 1850) vol. XX p.147, 191, 195, 205, 425, 441, 565, 1853.} 
did not content himself with the simple presupposition of the mutual transformability of heat and work, but added a series of \dashuline{more specific ideas about the special nature of the movement that we perceive as heat}. 
According to him {\it\color{red}(\dashuline{Rankine})}, \dashuline{the heat movement} consists in \dashuline{a violent vortex movement of the atmosphere surrounding the physical atoms}{\it\color{red}(!)}, the \dashuline{living force} of which accounts for the \dashuline{quantity of heat} present. 

If heat is supplied to a body from outside, only part of it is used to increase this living force (true specific heat), and the rest is used to change the arrangement of the atoms. 
Here too, \dashuline{as with Clausius}, we have {\it\color{red}(\dashuline{with Rankine})} the distinction between \dashuline{free heat} and \dashuline{internal work}. 
But it can be immediately seen that \dashuline{Clausius}' investigations are based on a more reliable basis, because they contain only the strict consequences of the \dashuline{principle of the equivalence of heat and work}, and \dashuline{not the least arbitrary assumption about the nature of heat}. 

Since we want to keep our presentation \dashuline{as independent as possible from molecular hypotheses}{\it\color{red}(?)}, \dashuline{we have no reason to go into Rankine's ideas in more detail} here.

\vspace*{1mm} 
\begin{center}
{\it\color{red}\bf
===
William Thomson (next Lord Kelvin) (1851) (p.59)
===
}
\end{center}
\vspace*{-3mm} 

Around the same time, \dashuline{William Thomson} was also induced by investigations into galvanic processes to join the ranks of the \dashuline{pioneers of the mechanical theory}, and from then on he contributed to the \dashuline{development of the theory} in a \dashuline{considerable series of treatises} containing the most diverse applications of the new principle worked \dashuline{in a similar way as Joule} did experimentally \dashuline{to establish the existence and numerical value of the mechanical equivalent of heat}.

%

In his first paper on this 
subject$\,$\footnote{$\:$W. Thomson: On the mechanical theory of electrolysis. Phil. Mag. (4) 2, p.429, 1851.}
\dashuline{Thomson} starts from the principle that the current of a magneto-electric machine generates an amount of heat throughout the entire circuit that is equivalent to the work used to generate the current.
But if the same current produces electrolytic effects at the same time, the heat generated is smaller, namely by the amount of the heat that would arise from the reunification of the decomposed substances. 
\dashuline{Thomson} therefore calls the latter heat \dashuline{the heat equivalent of the chemical effect} that has taken place. 
This principle is applied to the currents that arise from the mechanical rotation of a circular metal disk under the inducing influence of earth's magnetism. 
By appropriately applying wires to the disk, one can obtain a current in the wires, which also causes chemical decomposition, and is then able to check the correctness of the above theorem. 
From this, \dashuline{Thomson} derives the conclusion, in a similar way to \dashuline{Helmholtz} (p. 42), that the electromotive force of a galvanic element is equal in absolute measure to the mechanical equivalent of the chemical effect, which in it varies from the unit of current to the unit of time is produced.

%

In a following 
paper$\,$\footnote{$\:$W. Thomson: Applications of the principle of mechanical effect to the measurement of electro-motive forces and of galvanic resistances in absolute units. Phil. Mag. (4) 2, p.551, 1851.} 
the heat developed by a current in a conductor is used to measure the resistance of this conductor in absolute magnetic terms.

\vspace*{-2mm} 
\begin{center}
{\it\color{red}\bf
---------
(p.60) 
---------
}
\end{center}
\vspace*{-3mm}

If you divide the amount of heat developed in the unit of time, measured in calories, with the mechanical heat equivalent, you get the current heat $J^2\,.\,W$ in the mechanical measure, and dividing this quantity by the square of the current intensity $J$ measured in the magnetic measure gives the expression for the resistance $W$ of the conductor in absolute measure. 
The unit of resistance is then designated by the fact that in it the current $1$ develops heat equal to the mechanical work $1$ in time $1$. 
The calculation carried out for the specific resistance of silver and mercury gave results which agree well with those found by W. \dashuline{Weber} in a completely different way. 
He based his definition of 
resistance$\,$\footnote{$\:$W. Weber: Elektrodynamische Maassbestimmungen, insbesondere Widerstandsmessungen {\it\color{red}(Electrodynamic measurements, especially resistance measurements)}. Abh. d. Leipz. Akad. I p.197. Pogg. Ann. 82, p.337, 1851.}
on setting \dashuline{Neumann}'s induction constant $\varepsilon$ (p.45) equal to $1$. 
Both definitions therefore lead to consistent resistance values, as must also be the case according to the theoretical derivation of Helmholtz's law of induction.

The same treatise also contains applications of the above-mentioned \dashuline{theorem of the equality of heat} generated by galvanic and ordinary chemical means. 
If the galvanic heat only occurs as Joule (Lenz) heat, i.e. if it is proportional everywhere to the square of the current intensity, then the electromotive force $E$ is, as already mentioned above, equal to the chemical heat $A$ in the element reduced to the current unit and time unit. 
Because then it follows from the equivalence of the thermal effect: $J^2\,.\,W$ and the chemical effect: $J\,.\,A$: 
\vspace*{-4mm}
$$ \hspace*{20mm}
\mbox{$\,A=J\,.\,W\,$  and according to Ohm's law $\,=\,E\,$\:.} $$

%

\dashuline{Thomson} finds this assumption confirmed for \dashuline{Daniell}'s element, using \dashuline{Andrews}' numbers of the heat tone of the processes taking place in this element. 
But it is also very conceivable --and remarkably, \dashuline{Thomson} explicitly draws attention to this case here-- that the heat corresponding to the chemical processes does not only appear as \dashuline{Joule} heat, i.e. is not completely converted into current work, but a part of it appears as local or secondary heat, especially at the boundary between two conductors.

\vspace*{-2mm} 
\begin{center}
{\it\color{red}\bf
---------
(p.61) 
---------
}
\end{center}
\vspace*{-3mm}

This local heat can follow a completely different law than \dashuline{Joule}'s.
In particular, it can be proportional to the simple current intensity and can therefore also become negative. 
Then the electromotive force $E$ of the element will be reduced by an amount corresponding to the local heat generated. 
If this latter heat is approximately equal to $J\,.\,C$, then if we again set the total heat equal to that to be generated by the chemical processes according to the energy principle, we have: 
$$ J^2\,.\,W \;+\; J\,.\,C \;=\; A\,.\,C \; , $$ 
\vspace*{-8mm} \\
from which follows: 
$$ J\,.\,W \:\;(\,= E\:) \;=\; A \:-\:C \; .  $$
\vspace*{-6mm}

\dashuline{Thomson} here cites the view expressed by 
\dashuline{Faraday}$\,$\footnote{$\:$M. Faraday: Exp. Res. Phil. Trans. London 1834 Apr., Ch.~919.}
with whom \dashuline{Joule} also agreed, that in \dashuline{Daniell}'s element only the processes of the oxidation of zinc and the reduction of copper oxide are electromotively effective, while on the other hand the heat resulting from the dissolution of the zinc oxide in the sulfuric acid and the precipitation of copper oxide from the copper vitriol, as a special local heat (positive at the anode, negative at the cathode) appears independently of the current excitation. 
The fact that such local heat generation generally exists also follows from the behavior of the \dashuline{Smee} column (platinized silver, sulfuric acid, zinc), whose electromotive force is smaller than that calculated from the chemical heat conversion. 
A local heat development must therefore occur here, which is equivalent to the excess of chemical work over \dashuline{Joule} heat.

\vspace*{-1mm} 
\begin{center}
{\it\color{red}\bf
=====
James Prescott Joule (1851, 1857) (p.62)
=====
}
\end{center}
\vspace*{-3mm} 

The year 1851 brought another publication by 
\dashuline{Joule}$\,$\footnote{$\:$J. P. Joule: Some remarks on heat and the constitution of elastic fluids. Mem, of the Phil. Soc. of Manchester (geles. Oct. 3, 1848) (2) vol. IX p.107, 1851. Phil. Mag. (4) 14, p.211, 1857.}
which initially received little attention, but later became the basis of a new branch of physics: {\bf\dashuline{the modern gas theory}} {\it\color{red}(\,in fact the Statistical Physics)}. 
Starting from the idea that {\bf\dashuline{the heat}} of a gas {\bf\dashuline{consists in the living force {\it\color{red}(\,kinetic energy)} of the movement of the smallest particles of the body}}, \dashuline{Joule} did not, like \dashuline{Davy} and later \dashuline{Rankine}, conceive of this movement as consisting in vibrations or rotations, but thought himself to be in accordance with ideas from Daniel 
\dashuline{Bernoulli}$\,$\footnote{$\:$Daniel Bernoulli's Ansicht über die Constitution der Gase {\it\color{red}(The view of Daniel Bernoulli on the constitution of gases. A contribution (written by the Annalen) from the ``Hydrodynamics'' of Daniel Bernoulli: Argentorat, Sectio decima: De affectionibus atque motibus fluidorum elasticorurn praecipue autem äeris, 1738)}. Pogg. Ann. 107, p.490, 1859.}
and 
\dashuline{Herapath}$\,$\footnote{$\:$Herapath: On the dynamical theory of airs. Athen. 1, p.722, 1860.},
\dashuline{the gas molecules fly around freely} and, through their constant \dashuline{collision with the enclosing vessel walls}, create the force that is perceived as the pressure of the gas. 
A single \dashuline{molecule moves in a straight direction at a constant speed until it either hits another molecule or the vessel wall} and here follows the \dashuline{laws of collision of completely elastic bodies}. 
This simple assumption also makes it possible to establish numerical values for the \dashuline{average speed of a molecule}. 
Joule made the calculation easier by imagining the gas in a hollow cube and each molecule moving with the same speed in one of the three edge directions. 
This resulted in the validity of \dashuline{Boyle-Mariotte's law}, as well as the proportionality of heat content (total living force), temperature (living force of a single molecule) and pressure. For the speed of a hydrogen molecule at 60ºF. and 30 inches of mercury resulted in the value of 6225 English. feet per second. 
For different gases it was theorized that equal volumes (at the same pressure) contain the same living force.

\dashuline{For Joule}, a difficulty that could not be overcome at the moment arose from the \dashuline{assumption that the molecules should be viewed as simple material points}, i.e. that the entire heat content, the entire living force of a gas comes solely from the progressive movement of the molecules. 
If the expression for the \dashuline{living force of the progressive motion of all the molecules} is formed on the basis of the velocity calculated above, this would give the heat contained in the gas in calories by division with the mechanical heat equivalent, and \dashuline{the specific heat of the gas at constant volume could be calculated from this}.
%

\vspace*{-2mm} 
\begin{center}
{\it\color{red}\bf
---------
(p.63) 
---------
}
\end{center}
\vspace*{-3mm}

\dashuline{Joule} now \dashuline{found the specific heat} calculated in this way to be \dashuline{considerably smaller than that actually observed}, and was unable to find a satisfactory \dashuline{explanation for this circumstance}. This, along with the corresponding modification of the theory, was \dashuline{only given later by Clausius}.

\dashuline{Joule's ideas about the nature of gases} are essentially repeated, and in some cases somewhat further developed, in a communication which 
J. J. \dashuline{Waterston\:}$\,$\footnote{$\:$ J. J. Waterston: On a general theory of gases. Rep. of the 21. Meeting of the Brit. Ass. 1851, Notices and abstracts, p.6.}
presented to the British Association in the same year. 
Initially completely unnoticed, these ideas were later brought to the fore again when the ground for them began to become more favorable and, supported by the views of the chemists that were favorable to them, \dashuline{achieved general recognition relatively quickly}. In any case, it is only from here on that the theory of heat deserves to be called a mechanical one, since the mere ability to transform it into work does not say anything about \dashuline{the nature of heat}.

\vspace*{-1mm} 
\begin{center}
{\it\color{red}\bf
=====
William Thomson (next Lord Kelvin):
=====
\\
=====
the ``\,zero level state\,'' (1852, 1855)
=====
}
\end{center}
\vspace*{-3mm}

Let us now return to considering the further \dashuline{development of the law of conservation of energy}. 
Once the general validity of this principle and its eminent fruitfulness had been established in a sufficient number of cases, the series of applications and extensions quickly accumulated one after the other, and interest in it spread to ever wider and wider circles. 
Every year now brought a considerable number of new achievements in this field. 

First, 
\dashuline{W. Thomson}$\,$\footnote{$\:$W. Thomson: On the dynamical theory of heat. Phil. Mag. (4) 4, 
p.8   {\it\color{red}(Part.I)}, 
p.105 {\it\color{red}(Part.II)}, 
p.168 {\it\color{red}(Part.III)}, 
p.424 {\it\color{red}(Part.IV)}, 
1852.}
worked out \dashuline{a dynamic theory of heat} in a very similar way to \dashuline{Clausius}, now finally breaking with \dashuline{Carnot}'s old conception. 

\vspace*{-2mm} 
\begin{center}
{\it\color{red}\bf
---------
(p.64) 
---------
}
\end{center}
\vspace*{-3mm}

\noindent He based it on the idea that heat is based on movement, i.e. that the work it does requires a corresponding expenditure of \dashuline{living power from the vibrating molecules} (and any internal work). 
At the same time, he {\it\color{red}(W. Thomson)} also introduced the \dashuline{modified Carnot principle} with its applications to reversible processes into his theory. 
In this treatise we find for the first time the definition of  \dashuline{the mechanical energy contained in a body in the general sense} in which we now use it.
As is well known, the heat that must be imparted to a body from outside so that it can move 
\dashuline{from a certain {\bf zero state} {\it\color{red}(\,``\,Nullzustand\,''\,)}} to another certain state depends essentially on the external mechanical work that the body performs during this transition (the larger this is, the more heat the body will have to absorb from outside) (see p.55).
%

But if you subtract the amount of work done from that of the heat imparted (measured mechanically), you always get the same quantity, whatever the route of transfer. 
\dashuline{Thomson calls this quantity the mechanical energy of the body} in the assumed state.
\dashuline{It is completely determined by the state itself, {\bf except for an additive constant} that depends on the {\bf chosen zero level} {\it\color{red}(\,``\,Nullstand\,''\,)}}. 
As you can see, this is exactly the same function that was described by \dashuline{Helmholtz} as the \dashuline{quantity of the body's total heat (sum of the internal living forces and tension forces)}. 
However, Thomson's form of definition has the advantage that, with its help, a \dashuline{numerical determination of the value of the function} can be immediately imagined.

%

\vspace*{-5mm} 
\begin{center}
{\it\color{red}\bf
---------
(p.65) 
---------
}
\end{center}
\vspace*{-4mm}

The properties and means of calculating the energy of a body are discussed in even more detail in another 
treatise$\,$\footnote{$\:$W. Thomson {\it\color{red}(On the dynamical theory of heat/Part.V)}: On the quantities of mechanical energy contained in a fluid mass, in different states, as to temperature and density. Phil. Trans. Edinburgh (read  Dec. 3, 1851) vol. XX p.475, 1853. Phil. Mag. (4) 3, p.529, 1852 ; with more details: Phil. Mag. (4) 9, p.523, 1855.}
by \dashuline{W. Thomson}, in which it is stated that the total amount of the mechanical effect (algebraic sum of heat and work that a body has at the transition from one state to another) is only dependent on these two states, but not on the type of transition. 
The \dashuline{mechanical energy} indicates the total effect that is obtained \dashuline{when the body changes from its state to the {\bf (arbitrarily chosen) 
zero state} {\it\color{red}(\,``\,Nullzustand\,''\,)}}.$\,$\footnote{$\:${\it\color{red}More precisely, it is in the part~V (1853 p.478 or 1855 p.526, before Eq.~8) that W. Thomson explicitly considered from ``\,The theory of the integration of functions of two independent variables'' (...) ``\,for determining the value of $e$ (the mechanical energy of the mass) for every value of $v$ (the volume)'' (...) and ``\,if the fluid mass consist of water and vapour of water at the (Celsius) temperature $t$''  (...) to:  ``\,\dashuline{regard the zero or ``\,standard'' state} of the mass as being \dashuline{liquid water at the temperature $0$°}'' (P. Marquet)}.} 
\dashuline{{\bf The absolute zero state} would be one from which the body could no longer produce any positive effect}, i.e. \dashuline{neither heat nor work} (but this is \dashuline{{\bf unattainable by our means}}).
As noted above, \dashuline{W. Thomson's term energy} was first adopted by \dashuline{Clausius} and then gradually came into common use.
%

\vspace*{-1mm} 
\begin{center}
{\it\color{red}\bf
=
W. Thomson (1852, 1853): Energy of radiant heat and light -- Ether (p.65)
=
}
\end{center}
\vspace*{-3mm}

These \dashuline{theoretical works by W. Thomson} on the \dashuline{mechanical performance of heat} were followed, as an application of the theory, by an investigation into the \dashuline{effects of radiant heat and light, 
 as well as the sources of power offered to mankind by 
nature}$\,$\footnote{$\:$W. Thomson: On the mechanical action of radiant heat or light; on the power of animated creatures over matter; on the sources available to man for the production of mechanical effect. Phil. Mag. (4) 4, p.256, 1852.}.
Light and radiant heat are here identified, the importance of sunlight for the assimilation of plants and thus for the respiration of animals (cf. R. Mayer p. 24) is appreciated, and on the earth as a whole \dashuline{three main sources of work} are distinguished: primarily the \dashuline{radiation of the sun,} then secondarily the \dashuline{relative motion of earth}, sun and moon (tides), finally to a lesser extent also \dashuline{terrestrial sources of power}.

The further question of the origin and constant replacement of the emitted solar heat was dealt with in an article by 
\dashuline{J. J. Waterston}$\,$\footnote{$\:$J. J. Waterston: On dynamical sequences in kosmos. Athen. 1853, p.1099.}, 
who made a calculation of the heat that a body generates when it falls on the earth or sun from an infinite distance through the living energy gained in the process produces strength. 

\vspace*{-5mm} 
\begin{center}
{\it\color{red}\bf
---------
(p.66) 
---------
}
\end{center}
\vspace*{-4mm}

Based on this calculation, \dashuline{Waterston} concluded, similar to \dashuline{Mayer} before him (see p.25), that the heat expended by the sun was covered at the expense of the work of \dashuline{Newton}'s gravitational forces, either through the falling of cosmic masses into the sun it is due to the constantly increasing condensation of the solar body itself. 
The former idea was later developed further by \dashuline{W. Thomson} and the latter by \dashuline{Helmholtz} (see p.73).

%

From this time there was also an attempt at a theory of the interaction between the vibrations of the ether and those of the ponderable molecules, based on the principle of conservation of living forces, in a paper by 
J. Power$\,$\footnote{$\:$J. Power : Theory of the reciprocal action between the solar rays and the different media by which they are reflected, refracted or absorbed. Phil. Mag. (4) 6, p.218, 1853.}.
%

\vspace*{-1mm} 
\begin{center}
{\it\color{red}\bf
=
W. Thomson (1852, 1853, 1856): Energy of electricity and magnetism (p.66)
=
}
\end{center}
\vspace*{-3mm} 

From the heat theory, \dashuline{W. Thomson} turned back to the theory of \dashuline{electricity and magnetism} in order to apply the new principles there too. First, a study on thermoelectric currents 
appeared$\,$\footnote{$\:$W. Thomson: On a mechanical theory of thermo-electric currents. Phil. Mag. (4) 3, p.529, 1852. Proc. of Edinb. Soc. III, p.91, 1852.}. 
Based on the reversal of thermoelectric currents discovered by 
Cumming$\,$\footnote{$\:$Cumming: Phil. Trans. Cambridge 1823, p.61.}
  in 1823, the latter delivered the new and surprising result, albeit with substantial use of the \dashuline{improved Carnot principle}, that a galvanic current, in addition to Joule heat, which is proportional to the square of its intensity, also produces another heat in an unequally heated conductor, the later so-called \dashuline{Thomson} heat, which is proportional to the simple intensity and therefore changes its sign when the current is reversed.
While \dashuline{Joule}'s heat is to be considered as coming from the resistance of the conductor, \dashuline{W. Thomson}'s heat comes from an electromotive force acting inside the conductor (3rd section). 
 \dashuline{W. Thomson} later found this purely theoretically derived theorem confirmed experimentally after much
effort$\,$\footnote{$\:$W. Thomson: On the dynamical theory of heat. Thermo-electric currents. Phil. Mag. (4) 11, p. 214, 281, 379, 433, 1856. Ferner: 8, p. 62, 1854.
W. Thomson: On the electrodynamic properties of metals. Phil.
Trans. London 1856, p.649.}.
%

\vspace*{-2mm} 
\begin{center}
{\it\color{red}\bf
---------
(p.67) 
---------
}
\end{center}
\vspace*{-3mm}

Another important application of the theory is contained in a 
paper$\,$\footnote{$\:$W. Thomson: On transient electric currents. Phil. Mag. (4) 5, p. 393, 1853.} 
by \dashuline{W. Thomson} on the \dashuline{discharge current of an electrified conductor} (e.g. a sphere) connected to the ground by a thin wire. 
While the current flows from the conductor to earth, the total energy of the system must remain constant. 
However, this is made up of the following three parts: 
\\ \hspace*{15mm}
1) Electrostatic potential, 
\\ \hspace*{15mm}
2) Heat generated by the discharge current, 
\\ \hspace*{15mm}
3) Electrodynamic energy of the current. 
\\
The fact that an energy of the last kind exists, i.e. that a galvanic current through its existence in and of itself represents a certain supply of energy (actual energy or mechanical value of the current), is evident from the fact that a current through its disappearance without any effort can produce heat from other energy (e.g. as extra electricity). 
This energy is essentially positive, i.e. proportional to the square of the current strength, but it should not be understood as the living force of the electrical particles moving in the current and endowed with inert mass, because experience has shown that its value depends essentially on the shape of the current conductor. 
The same is equal to $0$ if the (linear) conductor is shaped in such a way that an opposite current element runs immediately next to it (because then the extra current disappears 
completely$\,$\footnote{$\:$M. Faraday: Exp. Res. Phil. Trans. London 1835, p.50. Chapt. 1096.}).

%

\vspace*{-2mm} 
\begin{center}
{\it\color{red}\bf
---------
(p.67-68) 
---------
}
\end{center}
\vspace*{-3mm}

If you now form the expression of the \dashuline{total energy of the system} by summing the above three expressions and set the change in this quantity for each time element $= 0$, you get an equation for determining the dependence of the current on time. 
\dashuline{W. Thomson} found from this that the discharge can take place in two completely different ways, depending on the value of the various constants contained in the problem. 
\dashuline{Either:} the discharge current always flows in the same direction, with initially increasing intensity and then gradually decreasing again, \dashuline{or:} the current direction oscillates and thereby charges the conductor alternately positively and negatively. 
The absolute value of the current gradually decreases towards $O$. Both types of discharges have since been repeatedly demonstrated, both experimentally and theoretically.

\vspace*{-3mm} 
\begin{center}
{\it\color{red}\bf
======
William Rankine (1853, 1855) (p.68)
======
}
\end{center}
\vspace*{-3mm} 

Next to \dashuline{Thomson}, \dashuline{Rankine} distinguished himself most among the English physicists in developing new ideas. In addition to some more specific 
applications$\,$\footnote{$\:$W. Rankine: Mechanical theory of heat. Phil. Mag. (4) 5, p.437, 1853. W. Rankine: On the application of the law of the conservation of energy to the determination of the magnetic meridian on board ship. Phil. Mag. (4) 6, p.140, 1853.}, 
he also tried to give \dashuline{a general definition of the concept of energy}, the application of which should not only be limited to the area of heat, but should encompass all 
natural forces$\,$\footnote{$\:$W. Rankine: On the general law of the transformation of energy. Phil. Mag. (4) 5, p.106, 1853.}. 
\dashuline{He calls energy} ``{\it Every affection of substances which constitutes or is commensurable with a power of producing change in opposition to resistance,\:}'' which \dashuline{Helmholtz} 
translates$\,$\footnote{$\:$Fortschr. d. Phys. v. J. 1853, Berlin 1856, p.407.}: 
``{\it Every affection of a substance which consists in or is comparable to a force capable of bringing about change that requires overcoming resistance.\:}'' 
\dashuline{Rankine} distinguishes and defines \dashuline{two main types of energy}, the \dashuline{actual} (kinetic) and the \dashuline{potential} energy (energy of movement and energy of position), the meaning of which corresponds entirely to that of \dashuline{Helmholtz}'s terms: \dashuline{living force} and \dashuline{tension}. 
He {\it\color{red}(Rankine)} also carries out a general investigation into the transformation of the different types of energy into one 
another$\,$\footnote{$\:$See also W. Rankine: Outlines of the science of energetics. Edinb. Journ. (2) II p.120, 1855.}, 
in which, however, the achievement of unlimited generality is only realized at the expense of a noticeable loss of precision in the expression.

\vspace*{-2mm} 
\begin{center}
{\it\color{red}\bf
---------
(p.69) 
---------
}
\end{center}
\vspace*{-3mm}

In general, \dashuline{the above definition of energy} does not seem to have a particularly high value from a physical point of view, as it 
\dashuline{is far too vague}$\,$\footnote{$\:${\it\color{red}In particular, W. Rankine based his approach on a ``\,theory of circulating streams of elastic vortices whose volumes spontaneously adapted to their environment\,'' and thus based on ``\,molecular vortices.\,'' Moreover, W. Rankine defined a ``\,thermodynamic function\,'' which he later realised was identical to the entropy of Clausius (P. Marquet).}}. 
A definition can only be called physically useful if it is possible to use it to \dashuline{specify the numerical value of the defined quantity} for any given case --an achievement whose fulfillment the present definition (particularly in contrast to \dashuline{Thomson}'s is apparently far away, see p.64)

%

\vspace*{-1mm} 
\begin{center}
{\it\color{red}\bf
======
James Prescott Joule (1852) (p.69)
======
}
\end{center}
\vspace*{-3mm} 

Also worth mentioning here is a 
work$\,$\footnote{$\:$J. P. Joule: On the heat disengaged in chemical combinations. Phil. Mag. (4) 3, p.481, 1852.} 
by \dashuline{Joule} on the calculation of some \dashuline{chemical compound heats} (copper oxide, zinc oxide, water) by galvanic means. If a decomposition apparatus is switched into a current instead of a metallic conductor in such a way that the current strength is the same as before, then the energy supplied by the current to the decomposition apparatus is obviously the same as that previously supplied to the conductor, since the rest of the circuit is exactly the same carry out the same processes again. 
The \dashuline{energy} supplied to the conductor consists only of \dashuline{heat}, but that supplied to the decomposition apparatus consists of \dashuline{heat and chemical work}, so the heat value of the chemical work is found by the lesser amount of heat developed in the decomposition apparatus, and this is precisely the heat of connection of the decomposed substance.
%

\vspace*{-1mm} 
\begin{center}
{\it\color{red}\bf
=======
Rudolf Clausius (1852, 1853) (p.69)
=======
}
\end{center}
\vspace*{-3mm} 

In our last statements we followed the \dashuline{development of the energy principle by the English physicists} up to the year 1853. The decisive predominance that the new theory was already enjoying in the scientific world at this time is aptly documented by a speech with which W. Hopkins, as President, opened the 23rd meeting of the British Association in Hull, and in which the merits be highlighted by \dashuline{Rumford}, \dashuline{Joule}, \dashuline{Rankine} and 
\dashuline{W. Thomson}$\,$\footnote{$\:$W. Hopkins: Dynamical theory of heat. Rep. of Brit. Ass. 23. Meeting Hull 1853. p.XLV.}.

%

\vspace*{-5mm} 
\begin{center}
{\it\color{red}\bf
---------
(p.70) 
---------
}
\end{center}
\vspace*{-4mm}

While in England there was constant work on developing the theory, there was no idleness in Germany either. 
Here \dashuline{Clausius} earned the main merit in further developing the principles he had already stated in his first treatise on the \dashuline{moving power of heat}. 

First, a study on the effects of an electrical discharge 
appeared$\,$\footnote{$\:$R. Clausius : Über das mechanische Äquivalent einer elektrischen Entladung und die dabei stattfindende Erwärmung des Leitungsdrahtes {\it\color{red}(About the mechanical equivalent of an electrical discharge and the resulting heating of the conductor wire)}. Pogg . Ann. 86, p 337, 1852.}. 
These can be of various types, especially mechanical or thermal, but in all cases the \dashuline{total energy} produced by the discharge, i.e. the sum of the mechanical work done and the heat generated, is equal to the decrease in electrical energy, i.e. the electrostatic potential. 
\dashuline{Clausius} expresses this sentence in the form: 
``{\it The sum of all effects produced by an electrical discharge is equal to the increase in the potential of the entire electricity on itself\:}'' 
 --in that the potential here is still taken with the opposite sign. The realisation of this theorem is followed by arguments about the correspondence of the theory with individual experiments.

In another work, 
\dashuline{Clausius}$\,$\footnote{$\:$R. Clausius: Über die bei einem stationären Strom in dem Leiter gethane Arbeit und erzeugte Wärme {\it\color{red}(About the work done and heat generated in the conductor with a stationary current)}. Pogg. Ann. 87, p.415, 1852.}
moves on from the electrostatic effects to those caused by a stationary galvanic current inside a metallic conductor. If a constant current flows through a metallic conductor, which does not suffer any induction effects from outside, the current is driven solely by the free static electricity accumulated on the surfaces of the various conductors. 
\dashuline{Clausius} therefore comes to the statement: 
``{\it The work done by the force acting in the conductor during a certain movement of a quantity of electricity is equal to the increase (decrease) in the potential of this quantity of electricity and the free electricity on each other that occurred during the movement.\,}'' 

\vspace*{-2mm} 
\begin{center}
{\it\color{red}\bf
---------
(p.71) 
---------
}
\end{center}
\vspace*{-3mm}

Now, in a metallic, if neither mechanical nor chemical effects take place in the conductor, this work is completely transformed into heat, from which \dashuline{Joule}'s heating law emerges (while the flowing electricity inside the conductor drops from higher to lower values of potential and thereby continuously produces heat, on the other hand, there must be places where in its circuit the closed chain raises it back to the original height of the potential, and these places are found at the interfaces of two neighboring conductors, where the potential suffers a jump and thus the electricity flowing through it a finite amount of work is done or consumed in an infinitely small way).

Depending on the circumstances, this \dashuline{work} can turn into \dashuline{heat} or \dashuline{chemical energy}. 
The former is the case with thermoelectric chains, the effectiveness of which \dashuline{Clausius} makes the subject of further 
investigation$\,$\footnote{$\:$R. Clausius: Über die Anwendung der mechanischen Wärmetheorie auf die thermoelektrischen Erscheinungen {\it\color{red}(On the application of the mechanical heat theory to thermoelectric phenomena}. Pogg. Ann. 90, p.513, 1853.}. 
Here the work done by electricity when it passes through a soldering point between two thermoelectrically active metals is expressed in the heat generation or absorption discovered by \dashuline{Peltier}. Regarding the question of whether \dashuline{Peltier}'s heat is always equivalent to the work done by overcoming the electrical voltage at the soldering point, \dashuline{Clausius} brings up a fundamentally important point in detail that has not yet been fully clarified to this day. As is well known, the \dashuline{Peltier} effect at the soldering point of two metals is not at all proportional to the electroscopic (voltaic) voltage that the same metals show when they touch each other on the electrometer. We will only discuss this fact, which is important for the application of the \dashuline{principle of conservation of energy}, in detail in the third section. 

It should also be mentioned briefly that \dashuline{Clausius}, like \dashuline{Thomson} before him, also applies the \dashuline{second law of the mechanical heat theory} to thermoelectric phenomena and thereby arrives at the corresponding results.

%

\vspace*{-3mm} 
\begin{center}
{\it\color{red}\bf
====
French physicists -- (Henri) Victor Regnault (1853) (p.72)
====
}
\end{center}
\vspace*{-3mm} 

The \dashuline{reticence which the French physicists} maintained towards the rapidly advancing development of the newly discovered principle \dashuline{until the mid-1850s} must seem somewhat striking. 
Apparently one could not decide so quickly to \dashuline{give up the heat theory}, which had been developed mainly in France on the basis of the \dashuline{assumption of the heat material}, although the important experiences made in the areas of calorimetry and heat conduction could ultimately be carried over into \dashuline{the new view} almost completely unchanged. 
In 1854, in the 
Comptes rendus$\,$\footnote{$\:$Hermite: Théorie et description d'une machine à courants électriques {\it\color{red}(Theory and description of an electric current machine)}. Compt. Rend. 39. p.1200, 1854.}
(by an off-duty officer: \dashuline{Hermite}) one even found a description of a machine (working in the manner of an electrophorus) that was supposed to \dashuline{deliver electricity and work at the same time!}

%

The first of the important French physicists who undertook to break with the old views was \dashuline{V. Regnault} in his great experimental 
work$\,$\footnote{$\:$V. Regnault: Recherches sur les chaleurs spécifiques des fluides élastiques {\it\color{red}(Research on the specific heats of elastic fluids)}. Compt. Rend. 36, p.676, 1853.} 
on the specific heat of gases, in which he conclusively established through exact measurements that the specific heat of the permanent gases is essentially independent of the volume, so that the heat generated by the compression cannot, as was previously assumed, be attributed to a change in the heat capacity.
%

\vspace*{-1mm} 
\begin{center}
{\it\color{red}\bf
==
Applications to Solar Radiations -- Light ether (1854) (p.72)
==
}
\end{center}
\vspace*{-3mm} 

In the meantime, the movement that had arisen through the new teaching in science took on an ever broader flow, \dashuline{new points of view and new applications} were constantly being sought out. 
Either the conclusions were confirmed by experience, or they provided interesting, previously hidden insights into the \dashuline{economy of nature}. 

\dashuline{In February 1854 Helmholtz} gave a popular scientific 
lecture$\,$\footnote{$\:$H. Helmholtz: Über die Wechselwirkung der Naturkräfte {\it\color{red}(On the interaction of natural forces)}. Königsb. 1854. Vortr. und Reden I p.25.} 
in Königsberg on the \dashuline{interaction of natural forces}, the main content of which was the new principle, primarily in its application to the \dashuline{theory of heat}. 

\vspace*{-5mm} 
\begin{center}
{\it\color{red}\bf
---------
(p.73) 
---------
}
\end{center}
\vspace*{-3mm}

The question of how to replace the \dashuline{radiated heat from the sun}, the source of all life on earth, is solved by the assumption that the sun continually heats up through continued condensation. It turns out that a reduction in the diameter of the sun by $10,000$ parts of its current size produces an amount of heat that is capable of maintaining the sun's radiation at its current strength for $2,100$ (according to a later calculation, $2,289$) years.

%

\dashuline{W. Thomson}$\,$\footnote{$\:$W. Thomson: Mémoire sur l'énergie mécanique du système solaire {\it\color{red}(Memoir on the mechanical energy of the solar system)}. Compt. Rend. 39, p.682, 1854.} 
answers the same question in a slightly different way by assuming that it is \dashuline{alien cosmic masses} that cause the latter to heat up by falling into the solar body. 
\dashuline{W. Thomson} calculates that if the Sun's heat output were always just met in this way, the Sun's diameter would have to increase by $1/10$ of an arc second in $4000$ years.

%

Another application of the mechanical heat theory to the energy of solar radiation is made by 
\dashuline{W. Thomson}$\,$\footnote{$\:$W. Thomson: Note sur la densité possible du milieu lumineux et sur la puissance mécanique d'un mille cube de lumière solaire {\it\color{red}(Note on the possible density of the luminous medium and on the mechanical power of a cubic thousand of solar light)}. Compt. Rend. 39, p.529, 1854.}
in relation to the \dashuline{density of the light ether}. 
If one takes light and heat rays to be identical, then all the heat energy given off to the earth by the sun's radiation is supplied by the living force of the vibrating ether particles, and one can calculate the density of the light ether from this, based on \dashuline{Pouillet}'s measurements of radiation intensity. 
However, this also includes knowing the \dashuline{speed of the oscillations of an ether particle}, or, since we know the oscillation period, the amplitude of these oscillations. 
\dashuline{Thomson} assumes that the speed at which an ether particle passes through its equilibrium position is less than $50$. 
Part of the speed of propagation of light, which one can do with certainty since the ratio of the oscillation amplitudes to the wavelength is a very small one, and concludes from this that a (English) cubic foot of ether has a greater mass than the $156$ trillionth part of a pound.
%

\vspace*{1mm} 
\begin{center}
{\it\color{red}\bf
=
Mechanical equivalent of heat: Numerical values (1854, 1855, 1858) (p.74)
=
}
\end{center}
\vspace*{-3mm} 

All of these calculations naturally require \dashuline{knowledge of the mechanical equivalent of heat}, and it is obvious that the need to determine these important constants as accurately as possible soon became apparent.
Because even if the \dashuline{Joule} measurements could be considered relatively reliable, the most precise experimental results still differed in the third digit. 
Here the physicists were given a fertile field for experimental investigations, and so we gradually see a whole literature of determinations of the mechanical heat equivalent emerge, most of which, however, were carried out using already known methods and only a few of which are comparable in exactness to \dashuline{Joule}'s calculations could measure.
In the following we will only make special mention of those that are based on essentially new ideas. 

\dashuline{Next to Joule}, whose experimental figures were compiled by  
\dashuline{L. Soret}$\,$\footnote{$\:$L. Soret: Sur l'équivalence du travail mécanique et de la chaleur {\it\color{red}(On the equivalence of mechanical work and heat)}. Arch. d. scienc. phys. et nat. 26, p.33, 1854.}, 
\dashuline{G. A. Hirn} made important contributions to this field through a long series of experimental work. 
Initially prompted by a competition from the Berlin Physical Society, in 1855 he shared the results of several series of 
experiments$\,$\footnote{$\:$G. A. Hirn: Recherches expérimentales sur la valeur de l'équivalent mécanique de la chaleur {\it\color{red}(On the equivalence of mechanical work and heat)}. 1855. Fortschr. d. Phys. v. J. 1855. (Von Clausius reports).
G. A. Hirn: Recherches sur l'équivalent mécanique de la chaleur
présentées à la société de physique de Berlin {\it\color{red}(Research on the mechanical equivalent of heat presented at the Berlin Physical Society)}. Colmar 1858.},
which had been carried out using very different, more or less original methods.

However, \dashuline{the numbers for the mechanical equivalent of heat} obtained from the individual experiments differ not insignificantly from one another. 
The observation of heat generation by friction of a cast iron drum on a metal body gave the number $371.6$~Kgr-m, and by drilling a piece of metal the number $425$~Kgr-m. 

\vspace*{-5mm} 
\begin{center}
{\it\color{red}\bf
---------
(p.75) 
---------
}
\end{center}
\vspace*{-3mm}

The calculation of the \dashuline{equivalent of heat} by comparing the work done by a steam engine with the heat consumed in it was of fundamental importance, since here for the first time work performance, not work consumption, was made the basis of the measurement (apart from the few \dashuline{Joule} experiments on the Expansion of air when pressure is overcome). 
The result was (according to a calculation improved by 
\dashuline{Clausius}$\,$\footnote{$\:${\it\color{red}(R. Clausius)} Fortschr. d. Phys. v. J. 1855, p.XXIII.}) 
413 Kgr-m {\it\color{red}(\,i.e. in modern units: $\Delta U = M\:g\:\Delta z=$\,4051\,J for $g=$9.81\,m${}^{2}$\,s${}^{-1}$ and for a mass $M$ of 1~kg (Kgr) elevated by 413~m. This corresponds to about \dashuline{$4.05$~J for one calorie} and for a mass of 1\:g. In fact, the modern  \dashuline{calorie is defined as}: ``\,the energy needed for elevating by $\Delta T=\:$1°C\, a mass $m=\:$1\:g of liquid water\,'' and thus corresponding to $\Delta u = m\:C_p\:\Delta T \approx\:$\dashuline{4.184\:J~cal${}^{\:-1}$} \:for $C_p\approx\:$4184~J~K${}^{-1}$~kg${}^{-1}$, and thus to 4184 / 9.81  $\approx$ \dashuline{426.5~Kgr-m} / P. Marquet)}. 

Finally, \dashuline{Hirn} also examined the amount of heat given off to the outside by the human body in a state of rest and in a state of work. 
In the latter case, this heat is more considerable, due to the increased respiration of the working body, but according to the theory, the additional amount of heat produced must be less than would correspond to the increased oxygen consumption, because of the work done at the same time. 
\dashuline{Hirn} actually found this confirmed, but received very different quantitative results due to the many uncontrollable sources of error. 
This led him to the \dashuline{strange conclusion} that the \dashuline{equivalent of heat was not constant after all, and that the mechanical theory of heat was based on a false basis}.

%

But \dashuline{Hirn didn't give up} on the matter at this point: later experiments were carried out with the friction of solid bodies under strong 
pressure$\,$\footnote{$\:$G. A. Hirn: Recherches sur l'équivalent mécanique de la chaleur présentées à la société de physique de Berlin {\it\color{red}(Research on the mechanical equivalent of heat)}. Colmar 1858.}, 
which found the equivalent number $451$~Kgr-m and especially the results obtained through new experiments on steam engines finally led him to \dashuline{agree to the assumption of the constancy of the  
equivalent of heat}$\,$\footnote{$\:$G. A. Hirn: Equivalent mécanique de la chaleur {\it\color{red}(Mechanical equivalent of heat)}. Cosmos XVI, p.313, 1860.}
This idea is also carried out {\it\color{red}\dashuline{by Hirn}} in his extensive 
work$\,$\footnote{$\:$G. A. Hirn: Exposition analytique et expérimentale de la théorie mécanique de la chaleur {\it\color{red}(Analytical and experimental exposition of the mechanical theory of heat)}. Paris et Colmar 1862.}
on the mechanical theory of heat (1862), in which he does not want to understand heat as movement, but, in a similar way to \dashuline{Mayer}, as a ``\,special principle.\,''

\vspace*{-5mm} 
\begin{center}
{\it\color{red}\bf
---------
(p.76) 
---------
}
\end{center}
\vspace*{-3mm}

In this work he {\it\color{red}\dashuline{(Hirn)}} publishes a number of new experiments to \dashuline{determine the equivalent of heat}. These are based firstly on the friction of a liquid enclosed between a solid and a hollow cylinder (result: $432$~Kgr-m) and then on the outflow of water under high pressure (result : $433$~kg), also on the impact of solid bodies. An iron cylinder, which is suspended horizontally on two pairs of ropes, falls sideways against a sandstone block that serves as an anvil and is also suspended, while between them there is a lead cylinder which is heated by the impact (result: $425$~Kgr-m, {\it\color{red}\,i.e. 425 * 9.81 / 1000 $\,=\,$ 4.17\:J~cal${}^{\:-1}$\,}). Still other experiments are carried out with the expansion of gases (result: $441.6$~Kgr-m, {\it\color{red}\,i.e. 4.33\:J~cal${}^{\:-1}$}). 

\dashuline{As the average} of all his experiments, \dashuline{Hirn} ultimately decides on the value $432$~Kgr-m {\it\color{red}(\,i.e. 4.24\:J~cal${}^{\:-1}$)} of the mechanical equivalent of heat, \dashuline{which is the joule}. This value is exceeded by around $2$\% 
{\it\color{red}(\,in fact, this value is exceeded by around $1.3$\% only with respect to the modern value of 
\:426.5~Kgr-m\,  
or 
\:4.184\:J~cal${}^{\:-1}$\, 
/ P. Marquet)}. 

In contrast, 
\dashuline{Favre}$\,$\footnote{$\:$P. A. Favre: Recherches sur l'équivalent mécanique de la chaleur {\it\color{red}(Research on the mechanical equivalent of heat)}. Compt. Rend. 46, p.337, 1858.} 
found the number $413.2$~Kgr-m {\it\color{red}(\,i.e. 4.05\:J~cal${}^{\:-1}$)} in 1858 from observations about the frictional heat that arises when steel springs are pressed against a rotating disk
{\it\color{red}(\,this value is too small by about $3$\% only with respect to the modern value of 
\:426.5~Kgr-m\, 
or 
\:4.184\:J 
/ P. Marquet)}.
%

\vspace*{1mm} 
\begin{center}
{\it\color{red}\bf
====
Other developments by: G. Kirchhoff -- C. Person -- W. Thomson 
====
\\
===
A. Krönig -- J. P. Joule -- R. Clausius -- J. H. Koosen -- J. P. le Roux
===
\\
==
P. A. Favre -- G. v. Quintus Icilius -- E. Lenz  -- C. Holtzmann -- E. Lenz
==
\\
====
J. Bosscha -- Grotthuss -- Marié-Davy and Troost -- M. Faraday
====
\\
====
(1844, 1851, 1854, 1856, 1857, 1858, 1859)
====
}
\end{center}
\vspace*{-3mm} 

Before we move on to determining \dashuline{other determinations of the  equivalent of heat} obtained electrically, we would like to take a look at the further development of heat theory around that time.
 Once the general validity of the theorem of the equivalence of heat and work was thoroughly recognized, general interest began to turn more and more away from this theorem and towards the \dashuline{second law of the mechanical theory of heat}, the  \dashuline{modified Carnot's principle}, 
so that the presentation of the progress of this theory must be significantly limited here. 

 \dashuline{Most of the investigations}, both experimental and theoretical, that have been carried out in this area since that time, \dashuline{deal with} (or have as a prerequisite)  \dashuline{the validity of Carnot's principle}, so that one rarely comes across \dashuline{a new theorem} 
  which proves to be \dashuline{a pure consequence of the principle of the conservation of energy}.

\vspace*{-5mm} 
\begin{center}
{\it\color{red}\bf
---------
(p.77) 
---------
}
\end{center}
\vspace*{-3mm}

One such example can be found in a 
treatise$\,$\footnote{$\:$G. Kirchhoff: Über einen Satz der mechanischen Wärmetheorie und einige Anwendungen desselben {\it\color{red}(On a theorem of the mechanical heat theory and some applications of the same)}. Pogg. Ann. 103, p. 177 (203), 1858.} 
by G. \dashuline{Kirchhoff}.
It concerns the dependence of the heat that is released to the outside during the chemical connection of two bodies on the temperature at which the reaction takes place and on the specific heats the body and the connection, and is based only on the premise that  \dashuline{the internal energy} (according to \dashuline{Kirchhoff}: ``\,effect/action function\,'') of a body  \dashuline{is completely determined by its current state}  
(because it makes no difference to their value, and consequently to the whole effect given off to the outside world, whether the chemical process is carried out directly, or whether the two bodies are first brought separately to a different temperature and then the process is initiated, if only the final state is the same again).

%

\dashuline{Person}$\,$\footnote{$\:$C. Person: Recherches sur la chaleur latente de dissolution {\it\color{red}(Research on latent heat of dissolution)}. Ann. d. chim. et d. phys. (3) 33, p.448, 1851.}
had already developed a very similar theorem for the \dashuline{heat of solution of salts} at different temperatures in 1851, but he was still based on the old idea of a heat substance. In fact, both views lead to the same goal here (if one disregards the external work), as we already noticed earlier with \dashuline{Hess's theorem about chemical heat}, which rests on entirely the same basis.

%

\dashuline{W. Thomson} deals with a new type of energy in an 
investigation$\,$\footnote{$\:$W. Thomson: On the thermal effect of drawing out a film of liquid. Proc. Roy. Soc. London IX, 255, 1858.}
into the expansion of a liquid membrane, in which it is shown that surface tension is capable of delivering work. 
But the application of Carnot's principle plays too important a role here to allow us to go into the content in more detail.
%

\vspace*{1mm} 
\begin{center}
{\it\color{red}\bf
---------
(p.78) 
---------
}
\end{center}
\vspace*{-3mm}

In the meantime, \dashuline{the mechanical theory of heat}, in the narrower sense of the word, had also taken an important step forward. First, in 1856, 
a treatise$\,$\footnote{$\:$A. Krönig: Grundzüge einer Theorie der Gase {\it\color{red}(Fundamentals of a Theory of Gases)}. Berlin 1856. Pogg. Ann. 99, p.315, 1856.}
by \dashuline{Krönig}, the content of which is essentially the same as that of \dashuline{Joule}'s paper from 1851, but without being in the least dependent on it, drew general attention to the hypothesis according to which the individual gas particles fly around in rectilinear paths in space at a constant speed and, through their impact against a solid wall, make the pressure of the gas and, through their living force, the heat content of the same noticeable. 
\dashuline{Krönig} attaches particular importance to upholding the law of the equivalence of heat and work by demonstrating that as soon as the expansion force of the gas performs work, the living force of the particles is reduced to a corresponding degree. For the rest, he, like \dashuline{Joule}, still adheres to the assumption that \dashuline{the gas molecules are to be conceived as simple material points}, and thus had to encounter the same difficulty that we have already mentioned above, namely that \dashuline{the heat capacity calculated from the mechanical conception at constant volume is considerably smaller than that observed by calorimetric means}.
\dashuline{Only Clausius}$\,$\footnote{$\:$R. Clausius : Über die Art der Bewegung, welche wir Wärme nennen {\it\color{red}(On the nature of the movement which we call heat)}. Pogg. Ann. 100, p. 353, 1857.}
succeeded in settling this critical point, and in the happiest way, by assuming that the \dashuline{heat content of a gas} (i.e. the entire inner living force) was not to be sought solely in the \dashuline{travelling motion of the molecules}, but that apart from this there was still an \dashuline{oscillating motion within the molecules}, the living force of which must be added to the first according to \dashuline{a general mechanical theorem} in order \dashuline{to produce the total living force}. 

According to this idea, therefore, the individual molecule, even of a simple gas, still breaks down into smaller constituents (\dashuline{atoms}), and while the pressure of the gas is determined solely by the travelling motion of the molecules, the heat is composed of this and the oscillating motion.

\vspace*{-2mm} 
\begin{center}
{\it\color{red}\bf
---------
(p.79) 
---------
}
\end{center}
\vspace*{-3mm}

But since in perfect gases both the pressure and the heat are proportional to the absolute temperature, it follows that the living force of the travelling motion is in a constant proportion to the total living force contained in the gas, depending at most on the nature of the gas. 
\dashuline{Clausius} finds \dashuline{this ratio to be 
equal to $0.6315\,$}\footnote{$\:${\it\color{red}We now know that this ratio should be $3/2$ over $5/2$, and thus $0.6$, if all the impact of the $2$ degrees of freedom for rotation were fully activated, and with a larger ratio is they are not fully activated for the absolute temperature $T$ not infinite with respect to the Debyes's temperature $\Theta_D$ that was unknown at that time by Clausius and Planck (P. Marquet)}.}
\dashuline{for all diatomic gases} whereas in all gases in which \dashuline{more than two atoms} are combined to form a molecule, a proportionately larger fraction of the total living force is used for \dashuline{intramolecular 
vibrations}$\,$\footnote{$\:${\it\color{red}We recognize here the idea that, in Statistical Physics: 
only the translational degree of freedom is to be considered for (simple) mono-atomic gases;
with moreover the rotational degrees of freedom for diatomic gases; 
and with moreover the vibrational degrees of freedom for  gases with more than two atoms (P. Marquet)}.}.
But here we come to more specific questions that are further removed from \dashuline{our task of describing the development of the concept of energy}. 
It is sufficient to point out that the attempt to \dashuline{understand the heat of a gas as the living force of the individual particles} can actually be regarded as essentially successful and that as a result two types of energy, that of heat, were previously different in form and concept and those of the movement, have been merged into one. 
It is clear from the outset that \dashuline{the heat phenomena}, viewed from this new point of view, consistently \dashuline{obey the law of conservation of energy}, as long as one (as is generally the case in gas theory) considers the forces acting between molecules and atoms as central forces, or any such collision between two particles is assumed to be completely elastic.
Because, here we find ourselves in the \dashuline{area of pure mechanics}, for which \dashuline{Helmholtz} already emphasized the agreement of the \dashuline{law of living forces} with that of the \dashuline{conservation of energy}. 
However, what the latter could only describe with the general expressions ``\,inner tension\,'' and ``\,living force\,'' has been made more precise in detail by \dashuline{the new gas theory}, and although the theory has so far been extended to liquids and solid bodies because of the more complicated conditions has not made any significant progress, these difficulties are due only to the imperfection of the methods and not to the nature of the matter.

%

\vspace*{-5mm} 
\begin{center}
{\it\color{red}\bf
---------
(p.80) 
---------
}
\end{center}
\vspace*{-3mm}

It remains for us to remember the further development of the applications which our principle found in the field of electricity and magnetism in the middle and at the end of the fifties. Here we first come across a 
work$\,$\footnote{$\:$W. Thomson: On the mechanical values of distributions of electricity, magnetism and galvanism. Phil. Mag. (4) 7, p.192, 1854.} 
by \dashuline{W. Thomson}, which gives a kind of \dashuline{overview of the different types of energy} that come into play in electrical and magnetic processes. 
If work is produced by such a process, which of course always happens at the expense of some energy supply, then, according to \dashuline{Thomson}, this work comes from one of three different types of energy: 
\\ \hspace*{15mm}
1) electrostatic energy; 
\\ \hspace*{15mm}
2) magnetic energy; 
\\ \hspace*{15mm}
3) electrokinetic (electrodynamic) energy;
\\
contained in galvanic currents.
If the electrical particles had inertia, the living force of their movement in the current would also contribute to this latter type of energy. 
By summing these three individual types of energy, one obtains the \dashuline{total electrical-magnetic energy contained in a system of bodies}.

The \dashuline{type of conversion of energy} that occurs in a closed galvanic circuit has been described in detail by 
\dashuline{Koosen}$\,$\footnote{$\:$J. H. Koosen: Über die Gesetze der Entwickelung von Wärme und mechanische Kraft durch den Schliessungsdraht der galvanischen Kette {\it\color{red}(On the laws of the development of heat and mechanical force through the closing wire of the galvanic chain)}. Pogg. Ann. 91, p.427, 1854.}. 
The \dashuline{generation of heat} by the current suggested the idea of using this process for the \dashuline{numerical calculation of the mechanical equivalent of heat}, and a number of determinations were also carried out using this method. 

Since \dashuline{Joule}'s experiments previously mentioned, which led to the value of $460$~Kgr-m\,{\it\color{red}(\,i.e. 4.51\:J~cal${}^{\:-1}$\,)}, \dashuline{le Roux} was the first to take up this question 
again$\,$\footnote{$\:$J. P. le Roux: Mémoire sur les machines magnéto-électriques {\it\color{red}(Memoir on magneto-electric machines)}. Ann. d. chim. (3) 50, p.463, 1857.}. 
He had a current generated by a magneto-electric machine and compared the calorific effect of this with the mechanical work required. 

\vspace*{-2mm} 
\begin{center}
{\it\color{red}\bf
---------
(p.81) 
---------
}
\end{center}
\vspace*{-3mm}

The result was $458$~Kgr-m\,{\it\color{red}(\,i.e. 4.49\:J~cal${}^{\:-1}$\,)}, for the heat unit. 
Favre$\,$\footnote{$\:$P. A. Favre: Recherches sur les courants hydro-électriques {\it\color{red}(Research on hydroelectric currents)} Compt. Rend. 45, p.56, 1857.} 
proceeded somewhat differently: he did not generate the current magnetically through induction, but through a hydrochain, but then switched on an electric motor and measured the heat developed by the current, first with and then without the work output of the motor. 
In the latter case, the heat generated was of course more significant in relation to the chemical action. The comparison with the work done in the first case resulted in numbers that differed from 426 to 464~Kgr-m as the equivalent value\,{\it\color{red}(\,i.e.\:4.18 to 4.55\:J~cal${}^{\:-1}$\,)}.

\dashuline{Quintus Icilius}$\,$\footnote{$\:$G. v. Quintus Icilius: Über den numerischen Wert der Constanten in der Formel für die elektro-dynamische Erwärmung in Metalldrähten {\it\color{red}(On the numerical value of the constants in the formula for electro-dynamic heating in metal wires)}. Pogg. Ann. 101, p.69, 1857.} 
obtained a not insignificantly different value ($399.7$~Kgr-m)\,{\it\color{red}(\,i.e. 3.92\:J~cal${}^{\:-1}$\,)} by direct measurement of the \dashuline{Joule heat}.
He was prompted to carry out this investigation for the fragrant  contradiction that arose when comparing the values for the current heat derived from 
Lenz's experiments$\,$\footnote{$\:$E. Lenz: Über die Gesetze der Wärmeentwickelung durch den galvanischen Strom {\it\color{red}(On the laws of heat development through galvanic current)}. Pogg. Ann. 61, p.18, 1844.} 
on galvanic heat generation with the values required by theory. 
This contradiction related to the size of the constant coefficient $c$ in the formula for the electricity heat generated in a unit of time: $Q = c \:.\: J^2 \:.\: W$. 
If $Q$ is measured in mechanical measure, then according to the theory this constant is $= 1$ (assuming that the current $J$ and the resistance $W$ are expressed in absolute measure.
But if the heat is given in calories, so $c$ is equal to the reciprocal value of the mechanical heat equivalent. 
On the other hand, 
\dashuline{Holtzmann}$\,$\footnote{$\:$C. Holtzmann: Die mechanische Arbeit, welche zur Erhaltung eines elektrischen Stromes erforderlich ist {\it\color{red}(The mechanical work required to maintain an electric current)}. Pogg. Ann. 91, p.260, 1854.} 
had calculated a value for $c$ that was over 4 times as large from Lenz's 
experiments$\,$\footnote{$\:$E. Lenz: Über die Gesetze der Wärmeentwickelung durch den galvanischen Strom {\it\color{red}(On the laws of heat development through galvanic current)}. Pogg. Ann. 61, p.18, 1844.}. 

\vspace*{-5mm} 
\begin{center}
{\it\color{red}\bf
---------
(p.82) 
---------
}
\end{center}
\vspace*{-3mm}

\dashuline{Quintus Icilius} now undertook to resolve the conflict between theory and observation, although the decision could not have been doubtful from the outset, since \dashuline{Thomson} had previously carried out very similar investigations and found sufficiently good agreement.

In fact, the experiments of \dashuline{Quintus Icilius} resulted in a value for that constant which was significantly smaller than that of \dashuline{Lenz} and was much closer to that required by the theory. 
If one really viewed it, as the theory would have it, as \dashuline{the reciprocal value of the mechanical equivalent oh heat}, the above-mentioned number $399.7$~Kgr-m\,{\it\color{red}(\,i.e. 3.92\:J~cal${}^{\:-1}$\,)} resulted. 
Since the question of the cause of the divergence of the results of \dashuline{Lenz} and \dashuline{Quintus Icilius} still remained unanswered, 
\dashuline{J. Bosscha}$\,$\footnote{$\:$J. Bosscha: Über das mechanische Äquivalent der Wärme, berechnet aus galvanischen Messungen {\it\color{red}(On the mechanical equivalent of heat, calculated from galvanic measurements)}. Pogg. Ann. 108, p. 162, 1859.} 
later felt compelled to return to the matter and then stated that the presumed reason was the \dashuline{incorrectness of the results} from the Lenz's experiments accounted for the \dashuline{considerable difference in the types of copper} that \dashuline{Lenz} had used in his experiments and \dashuline{Holtzmann} had subsequently used in his calculations.

%

\dashuline{Bosscha} takes another way to gain knowledge of the mechanical 
\dashuline{heat equivalent}$\,$\footnote{$\:${\it\color{red}(J. Bosscha, 1859)} p.168 l. c.} 
by expressing the electromotive force of a galvanic element (Daniell) first in heat units and then in absolute magnetic measures (in the units of time and current related) and divide these two numbers by each other.
The quotient provides the mechanical equivalent of heat. 
So, using some information from joules for the equivalent of heat, he finds the number $421.1$~Kgr-m\,{\it\color{red}(\,i.e. 4.131\:J~cal${}^{\:-1}$\,)}.
%

In 1858, in response to a general need, \dashuline{Bosscha} also 
produced$\,$\footnote{$\:$J. Bosscha: Het behond van arbeitsvermogen in den galvanischen stroom {\it\color{red}(The ability to work in the galvanic current)}. Leiden 1858, completed by von E. Jochmann: Fortschr. d. Phys. v. J. 1858, p.351.} 
\dashuline{the first compilation of all the determinations of the mechanical equivalent of heat} that had been made up to that point by various physicists, whether by mere calculation or by direct experiment 
--a table which, however, still has quite varied rows of numbers.  

\vspace*{-5mm} 
\begin{center}
{\it\color{red}\bf
---------
(p.83) 
---------
}
\end{center}
\vspace*{-3mm}

However, the most reliable experiments, which still primarily include those on joules and brains, essentially only fluctuate between the limits of $420$ and $430$~Kgr-m\,{\it\color{red}(\,i.e. between 4.12 and 4.22\:J~cal${}^{\:-1}$, which is indeed an interval including the modern value of 4.184\:J~cal${}^{\:-1}$ and $426.5$~Kgr-m\,)}. 
We will talk about the later determinations of the heat equivalent in the third section of this work.

Since that time, particular interest has taken place in the application of the new mechanical principles to the theory of the galvanic chain and the sometimes very complicated processes of electrolysis that take place therein. 
\dashuline{Grotthuss} had already attempted to base a theory of galvanic decomposition on purely mechanical ideas.
\dashuline{Clausius}$\,$\footnote{$\:$R. Clausius : Über die Elektricitätsleitung in Elektrolyten {\it\color{red}(On electricity conduction in electrolytes)}. Pogg. Ann. 101, p.338, 1857.} 
then developed these ideas further, whereby, however, in order to achieve a better agreement with Ohm's law, he introduced some essential modifications to the older view, above all the assumption that the molecules of an electrolyte in their natural state do not have a certain equilibrium position around which they oscillate, but move quite irregularly through each other, whereby it can easily happen, through random combinations, that a molecule splits into its two constituent parts, the electropositive and the electronegative (the partial molecules), or vice versa, so that two individual partial molecules unite to form a whole molecule when they meet by chance.

This includes, first of all, the investigation of the work which the current performs inside an electrolyte and which, as with first-class conductors, is generally completely converted into heat. 
\dashuline{Clausius} finds their amount to be exactly the same as in metallic 
conductors$\,$\footnote{$\:${\it\color{red}(Clausius, 1857)} 
p.340. l. c.}.

\vspace*{-2mm} 
\begin{center}
{\it\color{red}\bf
---------
(p.84) 
---------
}
\end{center}
\vspace*{-3mm}

\dashuline{Joule}'s heating law also applies here, as the experiment had long since confirmed. 
It is a different matter if a porous diaphragm, an animal membrane or the like is inserted into the electrolyte. 
Then the phenomenon of electrical endosmosis generally occurs, i.e. a certain quantity of fluid is forced through the septum by the current. 
In this case, the current has to do other work in addition to the work already stated, which is caused by overcoming the hydrostatic pressure counteracting the movement of the liquid, and also by the friction of the liquid in the pores of the 
wall$\,$\footnote{$\:${\it\color{red}(Clausius, 1857)} 
p.357. l. c.}.
Even more important for the theory is the determination of the general connection between the chemical work of the current and its thermal output --a question that we have often had occasion to touch on and that continues to concern physicists until recently.
\dashuline{Bosscha} first devoted himself more specifically to it. The basic idea that guides his and all subsequent related investigations is the \dashuline{direct consequence of the law of conservation of energy}, namely that in a stationary galvanic current the sum of the heat produced by the current is equivalent to the total expenditure of chemical energy. 
However, \dashuline{Bosscha} initially started from the \dashuline{unfounded assumption} that the thermal effects of the current are limited solely to the \dashuline{Joule} heat, from which, according to the considerations already made earlier by \dashuline{Helmholtz} and \dashuline{Thomson}, it would follow that the electromotive force is directly measured by the chemical work consumed in the unit of electricity in the unit of time. 
However, this theorem is confirmed specifically for Daniell's element. 
By first determining the electromotive force of such an element electromagnetically, and then comparing its magnitude with the thermochemical experiments of \dashuline{Favre} and \dashuline{Silbermann}, \dashuline{Andrews} and \dashuline{Joule}, \dashuline{Bosscha} found a completely satisfactory 
agreement.\footnote{$\:$J. Bosscha: Über die mechanische Theorie der Elektrolyse {\it\color{red}(On the mechanical theory of electrolysis)}. Pogg. Ann. 101, p.517, 1857.}

\vspace*{-2mm} 
\begin{center}
{\it\color{red}\bf
---------
(p.85) 
---------
}
\end{center}
\vspace*{-3mm}

This belief in the equivalence of the electromotive force of a chain with its corresponding chemical heat seems to have been fairly widespread at the time.
So \dashuline{Marié-Davy and Troost}$\,$\footnote{$\:$Marié-Davy et Troost: Mém. sur l'emploi de la pile comme moyen de mésure des quantités de chaleur développées dans l'acte des combinaisons chimiques {\it\color{red}(Memoir on the use of the battery as a means of measuring the quantities of heat developed in the act of chemical combinations)}. Ann. d. chim. (3) 53, p.423, 1858. Ferner Compt. Rend. 46, p. 936, 1858.}
wanted to avoid the thermochemical experiments altogether and instead just observe the magnetic needle, which ultimately turned out to be a bit premature. 
As soon as one became aware that, in addition to Joule's heat, other thermal effects also take place in the circuit (Peltier's heat was known for a long time anyway), this conclusion from the basic principle was bound to falter. 
We have already stated that every secondary developed or absorbed heat necessarily requires a modification of this rule.
Conversely, any deviation in the magnitude of the electromotive force from the above rule must necessarily indicate a heat effect in the chain that deviates from Joule's law.

\dashuline{Bosscha} also recognized this conclusion and tried to prove it in 
detail$\,$\footnote{$\:$J. Bosscha: Über die mechanische Theorie der Elektrolyse {\it\color{red}(On the mechanical theory of electrolysis)}. Pogg. Ann. 103, p.487, 1858; and also 105, p.396, 1858.}. 
\dashuline{Faraday}'s$\,$\footnote{$\:$M. Faraday: Exp. Res. Phil. Trans. London 1834 Apr., Section~ 919.} 
view, already stated above, that in Daniell's chain only the oxidation of zinc and copper is electromotively effective, rejected the idea that the dissolution (or precipitation) of the oxides only produced local heat, because the electromotive force calculated on the basis of this assumption from the experiments of Favre and Silbermann was too small.
On the other hand, he felt compelled to assume that the decomposition of water produced local heat. 

\vspace*{-5mm} 
\begin{center}
{\it\color{red}\bf
---------
(p.86) 
---------
}
\end{center}
\vspace*{-3mm}

If you calculate the electromotive force of polarization during the decomposition of water between platinum electrodes from the heat of combustion of the hydrogen and then from the weakening of the electric current caused by switching on the decomposition apparatus, you will find the second number is about $60$\:\% larger than the first. 
\dashuline{Bosscha} now thinks that this deviation arises from the fact that the electrolytically developed gases do not immediately appear in their usual form, but first go through the ``\,active\,'' state, an allotropic modification, which is characterized by the fact that each gas in them has a larger  potential energy (a more loose connection between the atoms), as in the natural state, which is why a greater heat value must occur when the active gases are combined with one another.
The electromotive force that is observed during polarization corresponds, according to \dashuline{Bosscha}, to this greater heat of connection in active state.
On the other hand, when the gases return to their natural state, a certain local amount of heat is released in each gas, which has no direct connection with the electrical processes. According to this view, the excess energy that the electromotive force of the polarization current delivers compared to the chemical heat of combustion of hydrogen would be equivalent to the local heat that is released in both gases during the transition from the active to the natural state. This local heat naturally causes an increased weakening of the current in the entire chain in which the water decomposition takes place, since the energy expenditure that it requires is withdrawn from the electromotive effect.

%

\vspace*{-5mm} 
\begin{center}
{\it\color{red}\bf
---------
(p.86-87) 
---------
}
\end{center}
\vspace*{-3mm}

Incidentally, since experience has shown that the electromotive force of polarisation is not constant, but is essentially dependent on the density of the current at the electrodes, \dashuline{Bosscha} must add to his assumptions that, in general, not all of the heat released during the transition of a gas from the active to the normal state occurs locally, but that, especially with a larger electrode surface, where a certain proportion of the gas already passes directly at the electrode into the normal state, the corresponding amount of energy still benefits the entire current and is thus converted into Joule heat.
This would then explain the dependence of the polarisation current on the current density, as well as on the nature of the electrode metal and the electrolyte.
%

It goes without saying that completely different views than \dashuline{Bosscha}'s can be put forward as to the immediate cause of local heat generation, namely the occurrence of a transition resistance. This is a question that the \dashuline{principle of conservation of energy} deals with from the outset no information is given: it is sufficient if the local heat appears at the corresponding point in the amount calculated in advance. 
For more information about this, see the 3rd section.

%

\dashuline{Bosscha} speaks most clearly about these deviations of galvanic heat generation from Joule's law, which are required by the energy principle, in a later 
work$\,$\footnote{$\:$J. Bosscha: Über das Gesetz der galvanischen Wärmeentwickelung in Elektrolyten {\it\color{red}(On the law of galvanic heat development in electrolytes)}. Pogg. Ann. 108, p.312, 1859.}, 
in which \dashuline{Smee}'s column and its local heat development are mainly examined.

\vspace*{-1mm} 
\begin{center}
{\it\color{red}\bf
=====
The principle of conservation of energy: a general 
=====
\\
=====
recognition in spite of residual misunderstandings 
=====
\\
=====
(1847, 1857, 1859, 1867, 1882) == (p.87)
=====
}
\end{center}
\vspace*{-3mm} 

\dashuline{In the short period of barely 18 years}, \dashuline{this theorem} had risen from a completely hidden or at least completely ignored existence to \dashuline{become a dominant force in the entire field of natural science}, as had previously only been experienced with Newton's great discovery, which in turn only related to a limited area of natural phenomena. Given the speed with which it spread, it is not surprising that there were still some people around this time, especially among the older physicists, who were not yet entirely comfortable with the new ideas.

\vspace*{-5mm} 
\begin{center}
{\it\color{red}\bf
---------
(p.88) 
---------
}
\end{center}
\vspace*{-3mm}

It is known from the genius 
\dashuline{Faraday}$\,$\footnote{$\:$M. Faraday: On the conservation of force. Phil. Mag. (4), 13, p.225, 1857; 17, p.166, 1859 etc. Vgl. M. Rankine :On the phrase ``\,potential energy\,'' and on the definitions of physical quantities. Phil. Mag. (4) 33, p.88, 1867.}
that he found many things to criticise about the concept of potential energy as \dashuline{Rankine} understood it. He was not content with the simple assumption that two bodies attracting each other at a greater distance have a greater potential energy, but sought a special physical substrate for this energy in a changed nature of the intermediate medium, although it should be borne in mind that this does not change the situation significantly, because this peculiar state of the intermediate medium would only consist in a changed position of its smallest particles, so that we would ultimately have to look for the potential energy again only in a changed static arrangement of matter in space.

The fact that \dashuline{misunderstandings} occasionally occurred \dashuline{in the application of the new principle} should not be seen as particularly noticeable given the unusual nature of the way of thinking and reasoning that it entails. In his ``\,Conservation of Force,\,'' 
\dashuline{Helmholtz}(2)$\,$\footnote{$\:$H. v. Helmholtz: Wiss. Abh. 1, p.66, 1882.} 
already discusses some of the objections that 
\dashuline{Matteucci}(3)$\,$\footnote{$\:$Matteucci: Bibl. univ. Genève Suppl. Nr.16, 1847, p.375.}
 had raised from his own experiments against the admissibility of the new view, which, however, had not yet been formulated into a specific principle at that time, and the reasons for them best shows how difficult it might have been for a trained physicist at the time to penetrate the spirit of the principle. Among other things, he derives an objection from the fact that zinc, when dissolved in sulfuric acid, generates just as much heat as when dissolved in ordinary chemical ways, as if you first form a galvanic (Smee's) chain using a platinum plate as a second electrode. 

\vspace*{-2mm} 
\begin{center}
{\it\color{red}\bf
---------
(p.89) 
---------
}
\end{center}
\vspace*{-3mm}

Indeed, in the first case only heat is produced by the process, but in the second case heat and electricity are produced at the same time.
It must be therefore (according to the theorem in question) the heat produced in the latter case will be smaller by the amount of the equivalent of the electricity produced. 
Matteucci did not take into account that this equivalent is $= 0$ at the end of the process, because the electricity produced has disappeared again and it only served as an intermediate link for the conversion of chemical potential energy into thermal energy.

Even more outrageous seems to us the assertion that, according to the principle, a current must generate less heat in the closing circuit if it receives a magnetic needle in deflection than if it does not, because this conclusion is based on \dashuline{a complete misunderstanding of the concept of work}.

But even in later times we sometimes come across \dashuline{erroneous conceptions}. In 1857, for example, 
Soret$\,$\footnote{$\:$L. Soret : Recherches sur la corrélation de l'électricité dynamique et des autres forces physiques {\it\color{red}(Research on the correlation of dynamic electricity and other physical forces)}. Arch. d. sc. phys. 36, p.38, 1857.} 
considered it a \dashuline{postulate of principle} that a galvanic current which performs mechanical work through the electromagnetic effects of one part of its circuit develops less Joule heat in this part than in another of the same galvanic resistance which has no such effects - \dashuline{an error which}, incidentally, has \dashuline{also happened to other physicists}.

If we \dashuline{disregard these very isolated cases of misunderstanding}, it can be asserted with all certainty that \dashuline{around the year 1860} the struggle for the \dashuline{recognition of the new theory had come to an end} and the decision had finally been made in its favour. 
The \dashuline{principle of the conservation of energy had proved to be perfectly valid} wherever the development of experimental methods had enabled a test to be carried out, and was now included in the number of axioms serving as the basis and starting point for further research. Very gradually, the word ``\,energy\,'' also became established on the continent from England, especially since it had been adopted by \dashuline{Clausius in thermodynamics}.

\vspace*{0mm} 
\begin{center}
{\it\color{red}\bf
---------
(p.90) 
---------
}
\end{center}
\vspace*{-3mm}

From this time onwards, \dashuline{a new epoch dates for the development of all exact natural sciences}. 
Hitherto, wherever it had not already been possible, as in mechanics and astronomy, to find the fundamental laws from which all individual phenomena emerge, one had to rely on the purely inductive method.
From now on, one was \dashuline{in possession of a principle which}, having been tested in all known fields by careful investigations, now also \dashuline{provided an excellent guide} for completely unknown and unexplored regions. 
Firstly, the whole line of enquiry, which is one of the most essential elements of any investigation that promises success, had already been set on the right track, and then, at all points along this path once trodden, one always had an infallible control at hand, the application of which never failed. 
Since then, \dashuline{the principle of energy} has formed \dashuline{the most solid starting point} for all scientific speculations and has in fact been \dashuline{used for this purpose many times}.

At the same time, however, we see what \dashuline{this change of role in the position of the Principle} in relation to the other laws of nature 
\dashuline{has as a result 
for our present investigation}. 
Whereas previously this position was a narrow one, and the literature relating to it appeared more or less sharply separated from the rest, now the \dashuline{applications are beginning to extend in all directions} and gradually to lose themselves in the special fields, and if we endeavour to follow the \dashuline{traces of the historical development of the Principle}, even if it be only in physics, we come into the individual and most singular questions, which for the most part still \dashuline{await final solution}.

\vspace*{-2mm} 
\begin{center}
{\it\color{red}\bf
---------
(p.90-91) 
---------
}
\end{center}
\vspace*{-3mm}

For the sake of clarity, it therefore seems advisable not to deal with these questions here, insofar as they still belong to the \dashuline{history of the principle} itself and do not merely concern applications of the principle to other unproven hypotheses, but only in connection with the presentation of the various individual types of energy in the \dashuline{third section} of this paper, while \dashuline{the second} will be devoted to the task of \dashuline{establishing the concept of energy in its generality} on the basis of the preceding historical development, separating and clearly organising \dashuline{the various ways of formulating the principle of the conservation of energy}, and finally taking \dashuline{a critical look at the evidence that can be adduced for the validity of the principle and its efficiency}.

\vspace*{10mm}
\section{\underline{Formulation and proof of the principle} (p.92-142)}
\label{Section-2}
\vspace*{-1mm}

\vspace*{0mm} 
\begin{center}
{\it\color{red}\bf
---------
(p.92) 
---------
}
\end{center}
\vspace*{-3mm}

Every physical definition that claims to be useful must ultimately reduce the concept to be defined to \dashuline{concepts that arise from immediate perception through the senses}, so that \dashuline{only direct observation} is required to \dashuline{determine the quantity in question} more or less exactly in numerical values to express.
 
Since we are made aware of the phenomena of nature through the most diverse senses, it would be of no use to us here, where we are initially concerned with \dashuline{establishing a general definition of the concept of energy}, if we wanted to take the standpoint of the mechanical view of nature from the outset.
Indeed, the mechanical measure with which we can measure any phenomenon is not immediately given to us, and on the contrary, it has to be sought in the first place. 
It will therefore initially be our task to \dashuline{base the definition of energy purely on measurable facts}, independently of any particular conception of nature.

%

\vspace*{-4mm} 
\begin{center}
--------------------------------------------------- 
\end{center}
\vspace*{-2mm}

From this point of view we can proceed in two ways. 

\dashuline{We can {\it\color{red}(first)} define the energy} of a material system as a function whose value depends in a certain way on the variables that determine the state of the system (i.e. on the positions, speeds, temperatures, etc., namely the material elements of the system). 
However, this definition already presupposes the \dashuline{general validity of the principle of conservation of energy}, because in order to know that such a function actually exists and how it is composed of those individual quantities, you have to already know and apply the principle.

%

\vspace*{-5mm} 
\begin{center}
{\it\color{red}\bf
---------
(p.93) 
---------
}
\end{center}
\vspace*{-3mm}

\dashuline{We therefore} initially \dashuline{give preference to  another {\it\color{red}(second)} definition}, essentially \dashuline{from W. Thomson} (see p.64), which, completely without regard to the validity or invalidity of the principle, allows the value of the \dashuline{energy of a material system} to be calculated solely from the external effects accessible to observation that the system produces, when its state changes to a certain extent. 
We then refer to \dashuline{the energy (ability to do work) of a material system} in a certain state as \dashuline{the amount, measured in mechanical work units, of all effects} that are caused outside the system when it is changed from its state in any way to an \dashuline{arbitrarily {\bf fixed zero state} {\it\color{red}(\,``\,Nullzustand\,''\,)} in any way, and that the work value of the external effects  transforms}.

%

\vspace*{-2mm} 
\begin{center}
--------------------------------------------------- 
\end{center}
\vspace*{-2mm}

First of all, some expressions contained in the wording of this sentence require special explanation. 
By the ``\,effects produced outside the system\,'' (or, in short, by the ``\,external effects\,'') we want to understand all the changes that occur in nature at the end of the process that are related to the position and nature of the surrounding bodies (i.e. not included in the system), including, for example, the change in the position of the system relative to the environment (because this depends --apart from the position of the system itself-- on the position of the surrounding bodies). 
  In order to obtain the external effects in their completeness, it is best to first think of the system as completely isolated in infinite space, and only then bring those bodies into the vicinity whose influence is suitable for bringing about the required transition.
  
If, for example, the system moves under the influence of gravity, the external effects also include the change in the position of the system relative to the earth.
The amount of this measured in units of work is the work done by gravity during the movement, which therefore forms the measure of the energy of the system (here its living force plus an additive constant, with more details on this below).

\vspace*{-2mm} 
\begin{center}
{\it\color{red}\bf
---------
(p.94) 
---------
}
\end{center}
\vspace*{-3mm}

It is, of course, something different if one includes the earth in the material system under consideration, because then the external effects disappear completely.
Furthermore, as regards the expression used in the definition: ``\,the amount \dashuline{measured in mechanical units of work}\,'' (in short: work value, mechanical equivalent) of the external effects, it has, of course, a definite meaning only on condition that \dashuline{either the external effects themselves are merely mechanical in nature} (i.e. consist in the production or consumption of living force or labour --in the strict sense), \dashuline{or that their mechanical equivalent is already known in some other way} (if they are of any other kind).

If, however, this condition is not fulfilled (if we assume, for example, that the external effects consist in the production of some peculiar change, such as a certain agent whose work value is unknown) the definition naturally fails at first, and one must try to help oneself by getting rid of the newly produced agent in some way (for example by using it to perform mechanical work or to produce such effects that are reducible to the measure of mechanical work).

If this attempt is successful, one can ultimately express all external effects in work equivalents and thus achieve the goal. Then the mechanical equivalent of an effect is represented as the amount of work into which this effect can be transformed (incidentally, it remains entirely open whether the amount of work turns out to be different if the transformation is carried out in different ways). 

\vspace*{-2mm} 
\begin{center}
--------------------------------------------------- 
\end{center}
\vspace*{-2mm}

\dashuline{But it is also very conceivable that} it is impossible to transform the new agent entirely into mechanical effects, and in this case the explanation given for the concept of work value, including \dashuline{the definition of energy, becomes invalid}.

\vspace*{-5mm} 
\begin{center}
{\it\color{red}\bf
---------
(p.94-95) 
---------
}
\end{center}
\vspace*{-3mm}

Let us assume, for example, that \dashuline{the mechanical equivalent of heat} is still unknown and that we have to \dashuline{calculate the energy of some body} at a certain temperature (under normal atmospheric pressure), \dashuline{with {\bf the zero state of the body} characterised by a certain lower temperature (about 0°\,Celsius)}.

It is now very easy to \dashuline{bring the body into {\bf the zero state}} by removing a certain quantum of heat (by dissipating heat), but the external effect produced in this way (the heating of the surroundings) cannot be completely converted into mechanical work by any means.
No matter what experiments one wanted to make for this purpose, there would always be a certain amount of change that would not be directly measured in units of work. 
One could, for example, convert that heat into work by expanding the medium of the same, but then one would again have a certain effect in this expansion, the mechanical equivalent of which is not known, and which cannot be traced back to mechanically measurable changes.
In short: \dashuline{one would never arrive at an expression for the value of the energy in this way}.


\vspace*{-2mm} 
\begin{center}
--------------------------------------------------- 
\end{center}
\vspace*{-2mm}

 From this it follows that in the case stated the explanation given for the concept of the ``\,work value\,'' of an effect requires a suitable supplement, and we can base this supplement on the fact that, if a certain effect cannot be completely transformed into mechanical labour, it can always be produced by the expenditure of a certain amount of work (see, however, the objection p.96 and around). its encounter).
Therefore, in all cases where the first explanation is not sufficient, we refer to the amount of work that is expended as the mechanical equivalent of an external effect in order to produce this effect, or in short: which can be transformed into this effect (in any way).
Then, under all circumstances we gain an expression for the work value of the external effects, and thus for the energy of the system under consideration. 

\vspace*{-5mm} 
\begin{center}
{\it\color{red}\bf
---------
(p.95-96) 
---------
}
\end{center}
\vspace*{-3mm}

This becomes apparent in fact directly based on the example described, where the external effect consists in the heating of a body. 
While it is impossible to transform this effect completely into work, there are various methods available to bring about this effect by purely mechanical means, i.e. to bring the body from the originally lower temperature to the higher temperature, such as: impact, friction or compression (when using the last method, care must be taken to ensure that the body expands again after compression without external work, so that the heat generated by the compression is not subsequently lost again when the body is brought back to its original pressure).
\dashuline{The mechanical equivalent of heating is therefore equal to the amount of work whose consumption causes the heating}.


\vspace*{-2mm} 
\begin{center}
{\it\color{red}\bf
---------
(p.96) 
---------
}
\end{center}
\vspace*{-3mm}

We see from this that, using the stated definitions, \dashuline{the definition of the energy of any material system} in all cases \dashuline{provides} (at least) one (positive or negative) \dashuline{numerical value in known units}, which, depending on the accuracy of the experimental methods available, can be larger or larger can be determined with lower accuracy. 
Of course, \dashuline{the definition is independent of any hypothetical idea} that one can form \dashuline{about the nature of the various agents active in nature}, especially also of the mechanical view, \dashuline{since it is based solely on the direct measurement of} mechanical working \dashuline{quantities}.

Furthermore, what is particularly remarkable, \dashuline{it is completely independent of the validity of the principle of conservation of energy}, 
because it leaves it completely undecided whether different values are obtained when using different methods of \dashuline{transferring the material system from the given state to {\bf the zero state the energy}} arrives or not, just as it is left completely open whether every external effect corresponds to a clearly defined mechanical equivalent, or not.

However, we must specifically address one objection that can be raised against the usefulness of the given definition. 
\dashuline{It could be that }the transition of the system from the given state to  \dashuline{the {\bf (arbitrarily} {\bf fixed) zero state} is not possible at all}. 
Let us assume, for example, that the material system consists of a certain quantity of carbon which, in the given state whose energy is to be determined, appears as amorphous coal, while in the zero state it forms the modification of the diamond.

\vspace*{-2mm} 
\begin{center}
{\it\color{red}\bf
---------
(p.97) 
---------
}
\end{center}
\vspace*{-3mm}

Here \dashuline{the transition to {\bf the zero state} cannot be accomplished by any experimental means} (although the reverse transition is possible) and \dashuline{the definition of energy fails from the outset}. 

\vspace*{-2mm} 
\begin{center}
--------------------------------------------------- 
\end{center}
\vspace*{-3mm}

Yes, we can go even further.

\dashuline{Cases} are certainly conceivable \dashuline{where the transition cannot be accomplished in any direction}, \dashuline{neither from the given state {\bf to the zero state, nor vice versa}}, while (which is of course a prerequisite) it is the same material system (i.e. the same chemical elements) that we have in front of us in both states. 

Let’s choose a specific example for this too.


Dextrose and levulose are two chemical individuals of exactly the same quantitative composition, so the same atoms can be thought of as being combined once to form dextrose, and once to form levulose.
However, the two compounds cannot currently be converted into one another, nor can either of them be represented synthetically from their elements, and therefore (given the current state of science) it cannot be made possible by any external means (not even by breaking them down into the elements) to bring the system from one state to the other. 
Therefore, \dashuline{if the energy of a quantum} of dextrose (at any temperature, etc.) \dashuline{were to be determined 
  in relation to
the same quantum of levulose {\bf as a zero state}} (a very important task under certain circumstances), 
\dashuline{the given definition of the energy 
would completely fail}.
There are of course many others that can be added to this example.


We can address the objection raised in two different ways. 

First of all, \dashuline{we could claim} with some justification \dashuline{that this is not a really feasible measurement of the value of energy}, \dashuline{which can never be carried out with absolute accuracy} anyway, and for which we will find other better methods below but rather to make the meaning of the concept of energy sufficiently clear, regardless of whether the path by which we arrive at this concept is only feasible for imagination or also for experiment. 

\vspace*{-5mm} 
\begin{center}
{\it\color{red}\bf
---------
(p.98) 
---------
}
\end{center}
\vspace*{-3mm}

For this purpose it would certainly be sufficient to prove that the described \dashuline{transition from the given state to {\bf the zero state}} is actually possible in nature, i.e. that the existing forces of nature, with appropriate cooperation, would be able to bring about the transition.

\vspace*{-3mm} 
\begin{center}
--------------------------------------------------- 
\end{center}
\vspace*{-3mm}

Now it should be borne in mind that experimentation consists only in the more or less arbitrary combination of certain natural forces, and that the area to which these combinations extend can in any case be called an extremely limited one in comparison to the variety of things that occur without our intervention (inorganic and organic world effects) that take place every day.


Even if we are not able to transform amorphous coal into diamond at will, there is nothing to stop us from assuming (and indeed there are many analogies) that diamond can be formed from a solution of an ordinary carbon compound, perhaps through a crystallization process lasting thousands of years can separate, and once this is admitted, one can certainly also talk about certain external effects and their work value. 
In any case, no fact is yet known which prevents us from believing that the forces of nature are capable of converting all substances, including organic and organized bodies, into all others, provided they are formed only from the same chemical elements.
We don't need to go any further.

However, the fundamental importance of the statements presented here requires that, in order not to leave even the slightest gap in the \dashuline{definition of the concept of energy}, we take into account the possibility of \dashuline{an assumption}, even if it is \dashuline{in itself unlikely}. 
In fact, in every case where, for whatever reason, our definition does not lead to the goal, we can still help ourselves in another way, namely by \dashuline{excluding the case in question} completely from consideration \dashuline{for the time being} and only making up the \dashuline{definition of the concept of energy} to be given for it \dashuline{on a later occasion} (p. 101), when we will be in possession of various theorems which \dashuline{allow the value of energy to be calculated under all circumstances}.

\vspace*{-1mm} 
\begin{center}
{\it\color{red}\bf
---------
(p.99) 
---------
}
\end{center}
\vspace*{-4mm}

We arrive at these propositions by establishing the \dashuline{principle of conservation of energy}, which we can now express as follows for all cases to which the definition of energy applies at all: {\it \dashuline{the energy of a material system in a certain state}, \dashuline{taken in relation to {\bf a certain 
 other state as a zero state}}, 
\dashuline{has a unique value}}. 

In other words, if we substitute the wording of the definition (p.93) here: 
{\it \dashuline{the amount measured in mechanical work units} (the mechanical equivalent, the work value) \dashuline{of all 
the effects} which a material system produces in its external environment when it passes \dashuline{from a certain state} in any way \dashuline{into an {\bf arbitrarily fixed zero state}}, \dashuline{has a definite value} (i.e. is independent of the nature of the transition)}.

\vspace*{-2mm} 
\begin{center}
--------------------------------------------------- 
\end{center}
\vspace*{-2mm}

  By referring the question of the \dashuline{provability of this proposition to another place}, we shall for the time being \dashuline{take it for granted} here, and in the following only prove that all other forms in which the principle is usually represented can be deduced from this one, and are therefore contained in it.

First of all, {\it\color{red}(let us consider)} \dashuline{the principle of the impossibility of perpetual motion}. 

  \dashuline{If we fix {\bf the zero state} of the material system} in such a way that it is \dashuline{identical to the given state} whose energy is to be determined, \dashuline{we have to {\bf set the value $0$} for the energy}, since it then obviously requires no external change at all in order to pass from the original state to the zero state.

 \dashuline{But {\bf this value $0$} is unique} for any kind of transition, so we have the theorem: {\it the mechanical equivalent of the effects that a material system produces in its external environment when, starting from a certain state, it is changed in any way and finally returns to its initial state (in short: when it undergoes a circular process) is $=0$}.

\vspace*{-5mm} 
\begin{center}
{\it\color{red}\bf
---------
(p.100) 
---------
}
\end{center}
\vspace*{-3mm}

 While this proposition, which \dashuline{excludes the possibility of the construction of a perpetuum mobile}, follows necessarily from the principle stated above, it does not in turn have that principle as a logical consequence, as we shall now note and later demonstrate in more detail.

\vspace*{-3mm} 
\begin{center}
--------------------------------------------------- 
\end{center}
\vspace*{-3mm}

Another consequence of the principle is as follows. 

We can imagine the entire process $A-N$, which leads the material system from the initial state $A$ (through certain intermediate states $B, C, ... M$) to the final state $N$, broken down into any number successive individual processes: $A-B$, $B-C$, ..., $M-N$, in such a way that the final state of each individual process (except for the last one) also forms the initial state of the following one. 

Then the work value of the external effects for the entire process $A-N$ is obviously equal to the sum of the respective amounts attributable to the individual processes, and from this follows the statement: 
{\it \dashuline{the energy of the system in state $A$, relative to the zero state $N$, is equal to the sum of the energies in the states $A, B, C, ..., M$, related to {\bf the respective zero states}: $B, C, D, ... , N$, or {\it\color{red}(equivalently)} in an easy-to-understand form:}} 
$$
[\,A\,N\,] \;=\; 
[\,A\,B\,] \;+\; [\,B\,C\,] \;+\; [\,C\,D\,] 
           \;+\; .\:.\:. \;+\; [\,M\,N\,] \; . 
$$
To this sentence we add a second, equally simple one. According to the equation just derived in connection with the \dashuline{principle of the impossibility of perpetual motion}: 
$$
[\,A\,N\,] \;+\; [\,N\,A\,] \;=\; [\,A\,A\,] \;=\; 0 \; , 
$$
$$
 \mbox{and from this:} \;\;\;  
 [\,A\,N\,] \;=\; -\:[\,N\,A\,] \; ,
 \hspace*{29mm}  
$$
i.e. {\it\dashuline{the energy of the system in state $A$, taken with respect to {\bf state $N$ as a zero state}, is equal and opposite to the energy of the system in state $N$, taken with respect to state $A$ {\bf as a zero state}}}. 

The same calculation laws obviously exist for  
{\it\color{red}(any other sets of)/(the applied)}\: 
symbols.

\vspace*{-2mm} 
\begin{center}
{\it\color{red}\bf
---------
(p.101) 
---------
}
\end{center}
\vspace*{-3mm}

We can now use the sentences derived here to complete the \dashuline{general definition of the concept of energy} by extending it to those cases that previously had to be excluded from consideration (p.98). 

If the transition from state A to state N cannot be carried out, but the transition from N to A can (as in one of the examples given above), then we define the sought energy $[\,A\,N\,]$ as the opposite of the energy of $[\,N\,A\,]$, and if the relationship is even more complicated, we \dashuline{introduce arbitrary intermediate states} $B, C, .\:.\:.$ and thereby break down the entire transition from $A$ to $N$ into \dashuline{a series of individual transitions}, which are chosen in such a way that each individual {\it\color{red}(processes)} can be carried out \dashuline{either in direct or in reverse direction}. 

 In this way \dashuline{we must always arrive at an expression of energy} by the application of the propositions stated, because if this were not the case: if we could not convey a transition from $A$ to $N$ successively by inserting corresponding intermediate states, we would not have the same material system before us in these two states at all, then the question of the value of energy would be absurd from the beginning. 


The fact that \dashuline{this way of expanding the definition of energy} does not entail an artificial complication of the term, but \dashuline{is based on the nature of the matter}, can best be seen from the fact that in \dashuline{every practical calculation of the energy of a material system} (e.g. in thermochemistry) actually depends on the process as prescribed by our definition, because there is no means or method of measurement that allows us to circumvent \dashuline{the way in which we arrived at the determination of the value of energy}.

\dashuline{If we relate the energy of a material system in a certain state $A$} at once {\it\color{red}(\,``\,einmal\,''/instantly)} \dashuline{to the state $N$}, then \dashuline{to another state $N'$ {\bf as a zero state}}, it follows from the relation:
$$
[\,A\,N\,] \;-\; [\,A\,N'\,] \;=\; 
[\,A\,N\,] \;+\; [\,N'\,A\,] \;=\; [\,N'\,N\,] 
\; 
$$
that \dashuline{the difference in the values of the energy} of $A$ caused by the \dashuline{{\bf different choice of the zero states}} is given by a quantity which \dashuline{does not depend at all on the properties of the state $A$}, but \dashuline{{\bf solely on the}} \dashuline{{\bf nature of the two zero states}}.

\vspace*{-4mm} 
\begin{center}
{\it\color{red}\bf
---------
(p.102) 
---------
}
\end{center}
\vspace*{-3mm}

If we therefore \dashuline{leave the {\bf choice of the zero state} completely open when determining the energy}, only \dashuline{a certain {\bf additive constant is left undetermined} in the expression of the energy}.

\vspace*{-3mm} 
\begin{center}
--------------------------------------------------- 
\end{center}
\vspace*{-3mm}

We now want to \dashuline{give the Principe another version}, which will be of the greatest importance for future considerations. 

If we imagine a material system being transferred from a certain state $A$ to another state $B$ through some process, then the work value of the external effects that occur is, in the notation we use, equal to 
$[\,A\,B\,] = [\,A\,N\,] - [\,B\,N\,]$, 
where $N$ represents \dashuline{a {\bf completely arbitrarily chosen state} of the same system}.
Therefore: through the process carried out, \dashuline{the energy of the system} \dashuline{relative to {\bf an arbitrarily} {\bf fixed zero state} $N$} has been reduced by the work value of the external effects produced, or, what is the same thing, \dashuline{the energy of the system has been increased (changed) by} the Work value of the effects that have disappeared (used, expended) outside the system in order to bring about \dashuline{{\bf the change in state}}.


{\it\color{red}More} specifically, if the process takes place in such a way that no effects take place in the external environment, then $[\,A\,B\,] = 0$, so $[\,A\,N\,] = [\,B\,N\,]$: the energy in the state $A$ equal to that in state $B$.

\vspace*{-3mm} 
\begin{center}
--------------------------------------------------- 
\end{center}
\vspace*{-3mm}

\dashuline{{\it The energy of a material system does not change}} if no external change occurs when some process is carried out, or in other words: \dashuline{{\it if only internal effects take place in the system}}. 

\dashuline{In this form the principle presents} itself as that of the conservation of energy, and it is this form which, through \dashuline{a slightly different conception of the concept of energy}, proves to be so incredibly convenient for direct observation and fruitful for further treatment. 

So far we have always considered \dashuline{the energy of a system as a quantity}, the concept of which \dashuline{is essentially linked to that of the external effects} that the system produces in the event of any change, because according to the definition, \dashuline{the amount of energy is only measured by these external effects}, and if one therefore wants to assign some material substrate to the energy in thought, one has to look for it in the environment of the system.

\vspace*{-3mm} 
\begin{center}
{\it\color{red}\bf
---------
(p.103) 
---------
}
\end{center}
\vspace*{-3mm}

\dashuline{Here alone energy finds its explanation and consequently also its conceptual existence}. 

As long as one completely abstracts from the external effects of a material system, one cannot talk about its energy, since it is then not defined. 

\dashuline{Kirchhoff'`s designation of energy as an ``\,effect function\,'' is in accordance with this view}.

\vspace*{-3mm} 
\begin{center}
--------------------------------------------------- 
\end{center}
\vspace*{-3mm}

Now, on the other hand, we see from the most recently derived form of the principle that \dashuline{the energy of a system remains constant if a process carried out with it does not produce any external effects}, no matter how extensive and varied the internal effects may be. 
This sentence leads us to understand \dashuline{the energy contained in a system as a quantity that is conceptually independent of external effects}. 

\dashuline{The system has a certain quantum of energy, which is completely determined by the current state {\bf (if the zero state is fixed)} and could be calculated at any time {\bf (by transferring it to the zero}} \dashuline{{\bf state)}}.

This quantum {\it\color{red}(of energy)} remains constant (is preserved) as long as the system does not emit or receive any external effects, and the internal effects only change its form, not its size. 

Therefore, 
\dashuline{we have to imagine the energy as being in the system itself}, as \dashuline{a kind of stock} (or according to C. Neumann: \dashuline{a ``\,capital\,''}), which is \dashuline{indestructible by internal effects}, and this conception is extremely convenient for immediate 
observation due to 
its analogy with the behaviour of matter, which can also be transformed into different forms, but is unchangeable according to its quantity (mass).


\dashuline{Just as the total mass of a body} is the sum of the masses of the individual chemical substances contained in it, \dashuline{the energy of a system is made up of the summation of the individual types of energy}, and one can also see the changes and transformations of these different types can be followed in the smallest detail, such as the changes in matter, for which we will find numerous examples in the following. 

\vspace*{-2mm} 
\begin{center}
{\it\color{red}\bf
---------
(p.104) 
---------
}
\end{center}
\vspace*{-3mm}

Undoubtedly, the relatively surprising ease and victorious clarity with which \dashuline{the principle of} \dashuline{conserva\-tion of energy gained general recognition within a few years and} became firmly established in the conviction of everyone \dashuline{rests largely on this analogy} {\it\color{red}(i.e. between the \dashuline{conservation of mass} and the \dashuline{conservation of energy} / P. Marquet)}.


\vspace*{-3mm} 
\begin{center}
--------------------------------------------------- 
\end{center}
\vspace*{-3mm}

 One could raise the question here as to whether it is really useful for the healthy further development of the principle to deviate in this way from the primary definition of the concept, and to give it a specific physical interpretation, which is ultimately based only on an analogy, and therefore does not in itself justify any conclusions.

In fact, it must be admitted that this question is not at all inadmissible from the outset.
Indeed, it can even be shown that it is precisely through this changed-{\it\color{red}(modified)} conception that the concept of energy (not its value, which is given once and for all by the general definition) acquires something indeterminate.
Think for example to the different interpretations that can be given to the concept of the electrostatic energy of a system of charged conductors in a state of equilibrium.  
Some seek the energy in a forced state of the dielectrics surrounding the conductor system, spatially extending across all dielectrics, others in a distant effect of the electrical charges of the conductors, spread out on the surfaces of the conductors. 

As long as the \dashuline{conflict between the two theories is left undecided} (i.e. if the question is limited to the consideration of natural processes that are equally satisfactorily explained by both) this question remains completely open.
\dashuline{The indeterminacy then lies in the concept of energy}, and one does not know the place to which one should assign it, and also has no means of finding it. 

If, on the other hand, \dashuline{one had adhere 
to the original definition}, one would have understood \dashuline{energy only as a certain number, as a certain amount of work}, whereby any indeterminacy of the concept is of course excluded.

\vspace*{-3mm} 
\begin{center}
--------------------------------------------------- 
\end{center}
\vspace*{-3mm}

However, it is precisely in the example mentioned, which others will follow later, that it is unmistakable that the substantial \dashuline{interpretation of the concept of energy} in question here is associated not only with an increase in clarity, but also with a direct advance in knowledge. 

\vspace*{-2mm} 
\begin{center}
{\it\color{red}\bf
---------
(p.105) 
---------
}
\end{center}
\vspace*{-3mm}

This progress is based on the stimulus for further physical research. One will no longer be content with knowing the numerical value of the energy of the system, but one will try to prove in detail the existence of the different types of energy in the different elements of the system and the transition into other forms and to other elements as well as the movement of a quantum of matter in space. 

But as soon as one addresses this question, the indeterminacy that previously lay in the concept itself takes the form of a physical problem capable of solution, and in fact it is to be expected that in this way, through research into the functioning of everything in nature active agents down to the individual, the \dashuline{physical meaning of the energy} will also become very specific, so that we can then view \dashuline{the entire energy of a material system} as an \dashuline{aggregate of individual elements}, each of which has its particular, 
special place 
in the matter.

It must certainly be admitted that this (so to speak material) \dashuline{conception of energy as a reservoir of effects}, the amount of which is \dashuline{determined by the current state of the material system}, may later have served its purpose and give way to another, more general and higher conception: at present it is in any case the task of physical research to develop this view, as the most clear and fruitful, down to the last detail and to test its consequences on the basis of experience.

As we will see later, many new points of view can be found in this direction.

\vspace*{-3mm} 
\begin{center}
--------------------------------------------------- 
\end{center}
\vspace*{-3mm}

As we now set about the task of carrying out this implementation systematically, whereby we will also become acquainted with the most convenient forms of the principle for its application, we start from the consideration of any material system in nature ongoing processes. 
Such a process always consists of a series of changes that the system undergoes, and we can always distinguish between \dashuline{two cases}. 

\vspace*{-2mm} 
\begin{center}
{\it\color{red}\bf
---------
(p.106) 
---------
}
\end{center}
\vspace*{-3mm}

\dashuline{Either:} the changes taking place in the system are completely independent of the external environment in which the system is located, so that the process would proceed in exactly the same way if one imagines that all matter that does not belong to the system is removed from space.
In this case we are only talking about \dashuline{internal} effects. 

\dashuline{Or:} the course of the process is essentially influenced by the presence of external bodies, and then in addition to any internal effects we also have to take \dashuline{external} ones into account. 

It is clear that this \dashuline{difference between internal and external effects is not an absolute one, but depends essentially on the choice of the material system.}
Indeed, we can make every external effect an internal one by including in the system the bodies in which (or between which) it takes place, and will therefore always be able (for any given process, by a proper extension of the system) to make all changes appear as internal effects.

Strictly speaking, \dashuline{there is no process at all that consists only of internal effects}, since \dashuline{all the bodies in the universe are in constant interaction with one another}, so that no matter how far we extend our material system, there will still be matter outside of it, which has an effect on it. 

However, \dashuline{wherever {\it\color{red}(only)} numbers are important}, it is sufficient to only take into account those \dashuline{quantities which lie above a certain small limit}, so that we can actually always achieve this in every natural phenomenon, despite the exclusion of an infinite number of bodies, only having to examine internal effects. 
We want to always think of this choice of system as having been made for the following considerations, so that initially \dashuline{we only have to talk about internal effects}.
 
\vspace*{-3mm} 
\begin{center}
--------------------------------------------------- 
\end{center}
\vspace*{-3mm}

In this case, \dashuline{the principle of conservation of energy} (according to page 102) is expressed in the form that \dashuline{the energy of the system is a constant quantity that does not change over time}. 

\vspace*{-2mm} 
\begin{center}
{\it\color{red}\bf
---------
(p.107) 
---------
}
\end{center}
\vspace*{-3mm}

So if we designate the state of the body system at the time when the changes begin as {the initial state} and another (in a finite or infinitely small time) as {the final state} of the process, 
then: \dashuline{the energy of the system in the initial state is equal to that in the final state}, 
or: \dashuline{the difference of the energies in the initial and final states is equal to $0$}. 

In order to be able to use this theorem with advantage, \dashuline{it is necessary to know the general expression of the energy of the system}. 
However, as we know, the \dashuline{energy is completely determined by the current state of the system {\bf (except for an additive constant)}}, so it must be able to be represented as a clear function of the quantities that determine this state. 

The main question now is: \dashuline{what are the variables that determine the state of a material system}? and this question leads to a closer \dashuline{discussion of the concept of state} in general.
 
\vspace*{-3mm} 
\begin{center}
--------------------------------------------------- 
\end{center}
\vspace*{-3mm}

If we restrict ourselves to the consideration of phenomena of motion, the state of a system of points can be described 
by the 
positions and velocities of all the points of the system. 
The characteristics 
of the state are therefore the spatial coordinates of the points {\it\color{red}(say: $q$)} and their first differential quotients with respect to time {\it\color{red}(say: $\dot{q}=dq/dt$)}.

The energy of the system depends on these quantities alone, and if they are given, the entire course of the motion, and hence all the variables of the system, is determined as functions of time. 

For any physical phenomena, however, this definition of the state is not sufficient, and we shall therefore define it more generally as follows: 
``\,{\it \dashuline{the state of a material system} at a given time is  
represented by 
all \dashuline{those quantities} whose instantaneous value determines the entire temporal progression of the material in the system\,}'' 
(where external effects are excluded).

The \dashuline{energy of the system} then appears as \dashuline{a certain function of these quantities}.

\vspace*{-3mm} 
\begin{center}
--------------------------------------------------- 
\end{center}
\vspace*{-3mm}

These ``\,determinants of the state\,'' (in short: ``\,state variables\,'') include, in addition to the already mentioned variables relating to mechanics, the temperature, the electrical and magnetic density, the galvanic current intensity, etc.

\vspace*{-5mm} 
\begin{center}
{\it\color{red}\bf
---------
(p.108) 
---------
}
\end{center}
\vspace*{-3mm}

Excluded, however, are quantities such as acceleration flow, speed of the conducted heat, etc., because these variables are always determined by the previous ones, so their knowledge is no longer necessary for determining the time course of a process.

This also applies if, as with Weber's basic electrical law, the force is set to depend not only on the position and the speed, but also on the acceleration. 
Because, on the other hand, if the force is assumed to be proportional to the 
acceleration$\,$\footnote{$\:$W. Weber: Elektrodynamische Maassbestimmungen, insbesondere über das Princip der Erhaltung der Energie {\it\color{red}(Electrodynamic measurements, especially on the principle of conservation of energy)}. Abh. d. k. sächs. Ges. d. Wiss. X. Nr.1, p.1, 1871. Vgl. auch IX, p.573, 1864.}, 
you always end up with a relation that once and for all traces the acceleration back to position and speed, so that it can never be given arbitrarily --a point that has sometimes been overlooked.

In general, the state variables will all be independent of each other, so in order to fully specify the state, you need to know the value of each of these variables.
However, it also often happens that fixed conditions that are given in advance prescribe a number of relations between these different quantities, which make some of them appear to be dependent on the others. 
We have such a case, for example, in mechanics, when certain equations exist between the coordinates of the moving points, i.e. when two points are connected to each other by a straight line of constant length. 
Then positions and speeds are obviously not independent of each other, and the state is determined by fewer variables than if the points are completely free.

We often encounter similar cases in other processes. 
For example, the Ohm's law, when applied to the stationary current of a galvanic battery, represents such a relationship between state variables. 
For the electromotive force of the chain (the sum of the electrical voltages of two conductors in contact), the resistance and the current intensity are all state variables, and in general the values of all three quantities must be given independently of one another if the current state of the entire body system that conducts the current is to be determined.

\vspace*{-2mm} 
\begin{center}
{\it\color{red}\bf
---------
(p.109) 
---------
}
\end{center}
\vspace*{-3mm}

Only the condition set from the outset that the state is stationary causes these quantities to depend on each other, so that one of them appears to be determined by the other two. 
But if you drop this condition, nothing prevents you from assuming that, at a moment, there is a current intensity that does not correspond to the value required by Ohm's law for the stationary state, and yes, as long as it is not particularly given, you can do not accept the situation as definite.

The current intensity will then generally not remain constant, but will change in a certain way in order to eventually change to a stationary state. 
This is shown, for example, in the phenomenon of a gradual increase in a current from the moment the line is closed.
The intensity then grows from $0$ to its constant level in a shorter or longer time
(taken in its generality, Ohm's law does not at all express any relationship between state variables, since the general expression of the electromotive force contains a term that comes from the induction of one's own or another's current, and this term in turn contains the differential quotient the current intensity over time, which is not one of the state variables).

Let's consider another example: temporary magnetization.

According to the common theory founded by Poisson, if one sets in a magnetically induced body (e.g. soft iron) the magnetic moment is proportional to the magnetizing force, so you get a relationship between all the state variables, because the magnetic force can also be expressed directly through the magnetic state of the body and the environment. 
But on the other hand, it is also known that this relation only corresponds to a certain state of equilibrium that occurs after a finite time, and that if this time is taken into account, one can very well assume a state in which the magnetic moment has not yet reached the value that it would assume in a state of equilibrium according to a certain magnetizing force. 

\vspace*{-5mm} 
\begin{center}
{\it\color{red}\bf
---------
(p.110) 
---------
}
\end{center}
\vspace*{-3mm}

In general, it is clear that every stationary and every equilibrium state involves a special equation of condition between state variables.

\vspace*{-3mm} 
\begin{center}
--------------------------------------------------- 
\end{center}
\vspace*{-3mm}

What makes this dependence of the state variables on each other, which occurs in certain cases, so important for our considerations is the fact that in every such case the expression of energy can be brought into different forms and this leads to the indeterminacy of the concept already mentioned on page 104 gives reason. 

Let us now see how one arrives at the \dashuline{expression of energy} in a specific case presented, for a given material system in a specific state. 

First of all, it should be noted that the numerical value of this quantity, as it emerges from the general definition, is always clearly determined.
You find it from the rules of the definition by \dashuline{first arbitrarily fixing {\bf a zero state}}, and then measuring in the specified way the work value of the external effects that are caused by the \dashuline{transition to {\bf the zero state}}. 

Once the numerical value has been found for a specific case, one does not yet have the general expression of the energy of the system in its dependence on the state variables, but must now further investigate how the numerical value found changes when the determining elements of the state are varied. 
This investigation falls under the general task of all experimental research, which is to determine by experiment the quantitative dependence of one phenomenon on another. 
Now, assuming that the law of this dependence has been found, the energy of the system can be expressed as a function of the state variables, and the problem has thus been solved.

\vspace*{-3mm} 
\begin{center}
--------------------------------------------------- 
\end{center}
\vspace*{-3mm}

However, two cases must be distinguished here. 

If the state variables were really varied in all possible ways, the value of the energy can only be represented in a single way as a function of these variables, since then the variables are all independent of one another. 

\vspace*{-5mm} 
\begin{center}
{\it\color{red}\bf
---------
(p.111) 
---------
}
\end{center}
\vspace*{-3mm}

In the following we will \dashuline{call this function the ``\,primary\,'' expression of energy\:}. 
It has general and unconditional validity. 

 However, it will often happen that not all possible combinations of values of the state variables are assumed, but only equilibrium or stationary states, or even states whose multiplicity (as described above) is restricted by one or more conditional equations between the state variables.

In each such case, \dashuline{the expression of energy} as a function of the state variables \dashuline{is not clearly determined}, but can be brought into different forms using these conditional equations by replacing any variable with any other. Then the decision about \dashuline{the form of the primary expression of the energy is impossible}, and remains so as long as one limits oneself to the states under consideration. 

 We see from this, among other things, that the calculation of energy from equilibrium (or even just stationary) states \dashuline{can never lead to the primary expression of the same}, as can already be seen from the example of an electrostatic equilibrium state given on page 104.

In fact, \dashuline{it has not yet been decided with certainty which of the two specified forms represents the primary expression of energy}, and until then one can arbitrarily regard each of them as the primary one (and the same applies to the other cases listed on page 108 and next). 
Let's take for example a stationary galvanic current of intensity $i$ with the electromotive force $e$ and the total resistance $w$, then the Joule heat generated in the resistance $w$ in the unit of time can be expressed in different ways: 
by $i^2\:w$, or by $e\:.\:i$, or by $e^2/w$. 
Which of these values should be viewed as the primary expression of heat production can only be decided when one moves from stationary flows to non-stationary ones, i.e. gives up the condition which here links the state variables to one another. One then finds that \dashuline{only $i^2\:w$ represents the primary expression sought}.

\vspace*{-3mm} 
\begin{center}
{\it\color{red}\bf
---------
(p.112) 
---------
}
\end{center}
\vspace*{-3mm}

We would like to bring up another case that also belongs here, which concerns the determination of the \dashuline{energy of an elastic body}.

If a completely elastic (solid or liquid) body performs movements (oscillations) thanks to the forces inherent in it, without experiencing any external influence that is associated with work, then, according to our principle, its \dashuline{energy is independent of time}. 

If we now further assume that the movement is such that any temperature differences that arise in its course due to the deformations are not equalised by heat conduction (as can generally be assumed for sound vibrations, for example), then the instantaneous state of the body is always already determined by the position (deformation) and velocity of all its particles.
In particular, the temperature of a particle depends only on its deformation, and the value of \dashuline{the energy can therefore also be represented as a function of position and velocity alone}.

Since the 
velocity 
is always contained in the form of the living force {\it\color{red}(kinetic energy)} in the expression of energy, we draw the conclusion from the \dashuline{immutability of the total energy} that: {\it \dashuline{the sum of the living force {\it\color{red}(kinetic energy)} and a certain function of deformation is constant throughout the entire movement}}. 

As is well known, this function of deformation is called the \dashuline{force function} or \dashuline{potential of elastic forces}, and it represents the kind of energy that is determined by the totality of the deformations. 
However, it should be borne in mind that this form of the value of energy is \dashuline{not derived from the most general state of the body}, but only from those states which emerge from each other by \dashuline{excluding heat conduction}. 

So, here again, we have the case of a condition between the state variables, and in fact it can be easily shown that \dashuline{the force function does not form the primary expression of energy}. 
It therefore \dashuline{loses its meaning} if one removes the restriction made here and moves on to consider more general states.

\vspace*{-3mm} 
\begin{center}
{\it\color{red}\bf
---------
(p.113) 
---------
}
\end{center}
\vspace*{-3mm}

This is most clearly evident in \dashuline{the movements of so-called perfect gases}, because the \dashuline{primary expression of energy} for them \dashuline{is generally known}. 

Let us consider a gas that oscillates, for example with a constant total volume, so that external effects are excluded, and let us first calculate its \dashuline{force function} under the assumption made above that there is no internal heat conduction.

Then the pressure $p$ of a mass particle is completely  determined by its volume $v$, 
namely:{\color{red}$\,$\footnote{$\:${\it\color{red}This relationship corresponds the adiabatic (isentropic) formula: $p\:v^{\,c_p/c_v} = p_0\:(v_0)^{\,c_p/c_v} = C = cste.$ / P. Marquet)}.}} 
$$ p \;=\; \frac{C}{v^{\,c_p/c_v}} \; ,$$ 
where $C$ is a constant dependent on the nature of the gas and $c_p$ and $c_v$ are the two mean specific heats.
Then {\it\color{red}(after an integration of the function $-\,v^{\,-\,c_p/c_v}$ of $v$)} the force function takes the value: 
$$ -\bigintssss p\;dv \;=\; 
\frac{1}{\left({c_p}/{c_v} \:-\: 1\right)} \:.\:
\frac{C}{v^{\,(c_p/c_v\,-\,1)}}
\;+\; const.
$$
According to \dashuline{the principle of conservation of energy}, the sum of the total living force of vibrations and the force function extending over all mass parts is independent of time.

However, the \dashuline{force function} does not form the \dashuline{primary expression of the energy}, 
because (for a unit mass) it is rather 
the quantity 
$c_v \:.\: \vartheta + const.$, 
which is completely independent of the volume, 
where $\vartheta$ is the absolute temperature and $c_v$ is related to mechanical work quantity. 
In any case, we can also express the same sentence if we use the expression 
$c_v \:.\: \vartheta + const.$ 
instead of the force function (related to the mass unit).

In fact, one is immediately convinced that under the conditions assumed here these two expressions are equivalent, since 
$$ 
\frac{1}{\left({c_p}/{c_v} \:-\: 1\right)} \:.\:
\frac{C}{v^{\,(c_p/c_v\,-\,1)}}
\;=\; c_v \:.\: \vartheta
\; ,
$$ 
because using the above value of $p$ we get: 
$$ 
\frac{1}{c_p \:-\: c_v} \:.\; p \: v \;=\; \vartheta \; ,
$$ 
an equation that applies generally to every state of a perfect 
gas.$\,${\color{red}\footnote{$\:${\it\color{red}This is merely the (molar) state equation: $p\:v = R \: T$, with $R=c_p-c_v$ the gas constant. / P. Marquet)}.}} 

\vspace*{-3mm} 
\begin{center}
{\it\color{red}\bf
---------
(p.114) 
---------
}
\end{center}
\vspace*{-3mm}

As long as one sticks to the type of movements considered here, it is completely irrelevant which of the two forms of energy one takes as the basis for the calculation and intuition.
The first-mentioned even has the advantage that changes in temperature do not need to be taken into account at all.
This is why they are mostly used in the theory of elasticity. 
However, as soon as the imposed limitation is broken, it is necessary to resort to the primary expression of energy.

\vspace*{-3mm} 
\begin{center}
--------------------------------------------------- 
\end{center}
\vspace*{-3mm}

Since, as we have seen, the acquisition of the primary form of energy is linked to the abolition of every limiting condition between the state quantities, one can never claim that we are really in possession of this primary form, because we are by no means sure whether the conditions with which we operate are in fact the most general. 
This is how we maintain (for example) that ``\,the product of the masses divided by their distance\,'' should be viewed as the primary expression of the energy of two masses gravitating towards each other (a view that arises from the idea of an immediate effect at a distance). 

But it would be very conceivable (and in view of the direction that the development of physical theories has recently taken, not even unlikely) that \dashuline{one would momentarily abandon this idea of a sudden distant attraction}, and replace it with one \dashuline{effect which propagates from particle to particle in a measurable time}, through the intermediate medium and by means of a peculiar deformation of the medium. 

\vspace*{3mm} 
\begin{center}
{\it\color{red}\bf
---------
(p.115) 
---------
}
\end{center}
\vspace*{-3mm}

Should this view really take hold, we could no longer regard the previously used expression of energy as the primary one, because its validity is linked to the condition that the effect emanating from one mass has already reached the other, and that a stationary state has formed in the intermediate medium.
But this state is then no longer the most general, and in fact the primary expression of energy would then change its form, and present itself as an integral that can be extended over the entire intermediate medium. 

Of course, this question remains open and each of the two expressions of energy is equally valid as long as one refrains from investigating such a more general state. We have already pointed out above (p.105) how stimulating the question of the primary expression of energy is for research into new phenomena.

\vspace*{-3mm} 
\begin{center}
--------------------------------------------------- 
\end{center}
\vspace*{-3mm}

For the following we want to make the assumption that the primary expression of energy is known to us, as far as the generality of the states we are considering extends. 
Wherever this is not the case with certainty, we prefer to limit that generality for the time being.

If, for example, in electrostatics we content ourselves with the consideration of states of equilibrium (of electricity), we may leave it entirely undecided whether the electrostatic energy is to be sought with Faraday-Maxwell in the interior of the dielectric, or with Coulomb-Weber on the surface of the conductor, and regard either of the corresponding expressions as the primary one at our discretion.

Furthermore, we want to imagine the underlying material system in the most diverse circumstances possible: there may be moving and stationary bodies in it, warmer and colder, luminous and dark, conductors and non-conductors, electrical bodies, through which currents flow, and magnetic bodies: in short all conceivable physical phenomena that may be represented in the system. 

Then the strange fact first becomes apparent that the (primary) expression of energy occurs in the form of a sum, the individual members of which are composed of certain state variables corresponding to the individual, special manifestations. 
Thus, the value of \dashuline{the entire energy breaks down into a number of individual energies that are independent of one another}, and each of which emerges in a special way from a single property of the state under consideration. 
This leads us to \dashuline{distinguish between different types of energy} in the system, such as \dashuline{mechanical, thermal, chemical, electrical and magnetic} energy. By summing them we get the total energy of the system.

\vspace*{-3mm} 
\begin{center}
{\it\color{red}\bf
---------
(p.116) 
---------
}
\end{center}
\vspace*{-3mm}

This fact, which we can call the {\bf\dashuline{principle of superposition of energies}}, is essentially related to the fact that \dashuline{many phenomena occurring in nature are completely independent of one another}: the heating of a body does not change its gravity, an electrostatic charge remains without influence on magnetism, etc., whether one regards this circumstance as the cause or as the consequence of that fact. 

It will be an extremely valuable aid for the development of further conclusions in the further treatment of our task to the conservation principle.
\dashuline{Let us simply accept this principle of the superposition of energies}, which expresses the generalisation of a whole series of propositions well known in physics, as given by experience.

In the further treatment of our task \dashuline{it will provide us with an extremely valuable aid for} the development of further \dashuline{conclusions from the conservation principle}.

\vspace*{-3mm} 
\begin{center}
--------------------------------------------------- 
\end{center}
\vspace*{-3mm}

 In order to facilitate the overview of the individual elements of this energy sum, it has been categorised according to various points of view, for example (in addition to the categorisation already mentioned by us, which is based on the diversity of the individual natural phenomena and can probably be regarded as the first) the categorisation into two summands: the \dashuline{actual and potential energy} (i.e. the \dashuline{energy of motion and position}, p.68).
This division is based on the assumption that \dashuline{all changes in nature are of a mechanical nature}: it counts all members that depend only on \dashuline{velocity as actual energy}; and all those that depend only on \dashuline{positions as potential energy}. 

However, since it was discovered that there are also \dashuline{types of energy} that are \dashuline{determined by position and velocity at the same time}, the way in which this classification is applied has become somewhat doubtful. 
This is for example the case with the so-called \dashuline{electrokinetic energy of a galvanic current}, which depends not only on the intensity of the current, but also on the relative position of the individual current elements.
It is not clear whether this energy has to be addressed as actual or as potential, but usually the former happens (see Section 3). 

\vspace*{-2mm} 
\begin{center}
{\it\color{red}\bf
---------
(p.117) 
---------
}
\end{center}
\vspace*{-3mm}

Basically, of course, nothing matters, since the value of the total energy remains unaffected by this difference in opinion. 

Another type of classification is the \dashuline{distinction between external and internal energies}, whereby \dashuline{external energy means essentially} the same thing as \dashuline{mechanical energy} in the narrower sense (energy of molar movement), but \dashuline{internal} energy means \dashuline{the rest} of the total energy. 

  From a still another point of view, namely the possibility of direct transformation into mechanical work, \dashuline{Helmholtz} more recently \dashuline{divided energy into free and bound {\it\color{red}(tied/stored/dead)} energies}.

\vspace*{-3mm} 
\begin{center}
--------------------------------------------------- 
\end{center}
\vspace*{-3mm}

By transferring the discussion of the individual types of energy to the last section of this work, we only want to point out the convenience that arises from the \dashuline{validity of the principle of superposition of energies} for the view of the concept, and for the \dashuline{calculation of the value of the total energy}. 

From this we can imagine the \dashuline{total energy of the system} as a \dashuline{reservoir created by} simply lining up the \dashuline{individual energies}, just as the total weight of a body results from the accumulation of the individual chemical elements it contains. 

The size of each individual type of energy can be calculated on its own, completely independent of other properties of the system under consideration, if one only knows the special state variables that correspond to it. 
In this way we come to \dashuline{mentally assign each type of energy its special place in matter}, and thereby achieve the practical advantage of a simplified overview of the individual types of energy, which saves us from the mistake of ignoring one of them when calculating the total energy. 

In general, \dashuline{every force active in the system}, or indeed every special property of the system, \dashuline{corresponds to a special type of energy}, which one can imagine to be located at the point at which that property appears.

\vspace*{-5mm} 
\begin{center}
{\it\color{red}\bf
---------
(p.118) 
---------
}
\end{center}
\vspace*{-3mm}

\dashuline{If} only those \dashuline{forces} are active in the system that \dashuline{only act at immeasurably small distances}, then the effect on any material particle will only depend on the state of this particle itself, or its immediate surroundings, and the energy of the system is then obtained simply by summing the energies of all its material particles. 

However, it is essentially different \dashuline{when forces occur that act directly at a distance}, since the energy caused by such a force will generally depend on the same quantities as the force itself, namely on the distance of the two elements acting on one another. 
\dashuline{In this case}, the concept of energy is based on the simultaneous position of the two elements, so its 
seat {\it\color{red}(place/origin)} 
is not at a single point in space, and \dashuline{one can no longer set the total energy of the system equal to the sum of the energies of the individual material elements} (and rather one has to add to this sum \dashuline{those types of energy} that are \dashuline{caused by the long-distance effects} of two elements). 

Assuming now that we have found the expression of the total energy as the sum of the individual types of energy, we have to set its value for every change in the system independent of time, while the individual types of energy can change their size at mutual expense. 
So every process that takes place in nature can be understood as a \dashuline{conversion of individual types of energy into one another}, \dashuline{while their sum}, \dashuline{the entire supply of energy present in the system}, \dashuline{can neither be increased nor decreased}.

\vspace*{-3mm} 
\begin{center}
--------------------------------------------------- 
\end{center}
\vspace*{-3mm}

We now want to go one step further. 

The previous considerations only related to changes in the system that only arise from internal effects, while the matter that does not belong to the system had no influence on it at all. 
If the principle were to be applied only to this case, relatively little benefit could be derived from it, because it then only provides a single equation, the one that expresses the constancy of the energy. 

\vspace*{-2mm} 
\begin{center}
{\it\color{red}\bf
---------
(p.119) 
---------
}
\end{center}
\vspace*{-3mm}

In addition, for a given process, in order to exclude all external effects, one would generally always have to include a considerable number of bodies in the system, which would make the number of variables determining the state a significant one, and one equation of the principle of conservation of energy would achieve little here. 

But we now want to show that for any process taking place in a material system we can generally derive not just one, but an infinite number of equations from the principle, so that it can often serve us to determine the entire temporal course of the process to be clearly determined.

\vspace*{-3mm} 
\begin{center}
--------------------------------------------------- 
\end{center}
\vspace*{-3mm}

Here the analogy of our principle with that of the preservation of matter in all its fertility becomes apparent. 

The sum of the ponderable masses contained in nature is unchangeable, but they change their position in space.
If we look at a defined volume of space, the mass contained in it is generally not constant, but the change (increase) in this mass over a certain period of time is equal to the mass that entered the volume from outside during this time. 

We derive a very similar theorem for the energy of a material system. 
Just as matter changes its position in space while its sum remains constant, so energy changes its position and form in matter, so that we can make the following consideration. 

In a material system that is not exposed to external effects, the energy remains constant. 

But if we select any complex of material elements from the system and consider it as a special system, then it will also have its special energy, which can be formed according to the pattern of the expression of the energy of the entire system. 
In general, this energy will not remain constant.
It would only be so {\it\color{red}(energy remains constant)} if the system under consideration did not suffer any external effects in the course of the process, which will generally not be the case.


Therefore {\it\color{red}(in general)} the energy changes, precisely in accordance with the external effects. 
Through these external effects, (positive or negative) energy is created (transferred) from outside into the system, in an amount that is given by the sentence developed on page 102: 

\vspace*{-4mm} 
\begin{center}
{\it\color{red}\bf
---------
(p.120) 
---------
}
\end{center}
\vspace*{-4mm}

\noindent
``\,{\it\dashuline{the change (increase) of energy corresponding to a certain change of state of a material system is equal to the work value of the effects which have been expended outside the system in order to bring about the change of state}}\,''.
Of course, the previous sentence is included in this sentence as a special case, since if no external effects take place, no energy can be transferred into the system.

\vspace*{-3mm} 
\begin{center}
--------------------------------------------------- 
\end{center}
\vspace*{-3mm}

\dashuline{The analogy} drawn with the change in the matter filling a certain volume of space \dashuline{only extends up to a certain limit}, which is that the total mass in a space is equal to the sum of the masses contained in the individual parts of the space, while a similar theorem for the total energy contained in a material system does not exist, at least not if the action in distances is permitted (p. 118). 

Rather, the \dashuline{energy of a system contains} \underline{other types of energy} \dashuline{in addition to the sum of the energies of the individual material parts}, and this makes its behavior a little more complicated.

\vspace*{-3mm} 
\begin{center}
--------------------------------------------------- 
\end{center}
\vspace*{-3mm}

If a material system undergoes a certain change in state as a result of a given process, the calculation of the work value of the external effects involved can often be made much easier by considering that: \\
{\it\color{red}1)} this work value is completely independent of the way in which the change in state is brought about; 
and \\ 
{\it\color{red}2)} instead of the given process and instead of the given external effects one can substitute any other 
(as long as they only cause the same change in the state of the system, because then the work value sought is the same again). 

Let us consider an example from mechanics. 

\dashuline{The energy of a body} in motion (\dashuline{of constant internal constitution}) is its living force {\it\color{red}(kinetic energy)}, which \dashuline{remains constant as long as no external effects take place}. 
But when mechanical forces from outside act on the body, they transfer energy to the body.

\vspace*{-3mm} 
\begin{center}
{\it\color{red}\bf
---------
(p.121) 
---------
}
\end{center}
\vspace*{-3mm}

The external effects can be of a very diverse nature, depending on the nature of the assumed forces (shock, friction, distant forces), so that external mechanical (or internal molecular, or thermal, or electrical) changes can have taken place in the environment: the work value of these external effects is always equal to the work that the assumed forces do on the body, regardless of where they come from. 
This is the \dashuline{amount of energy transferred into the body}, i.e. the \dashuline{increase of his living {\it\color{red}(kinetic)} force {\it\color{red}(power/strength)}}.

\vspace*{-3mm} 
\begin{center}
--------------------------------------------------- 
\end{center}
\vspace*{-3mm}

It is now easy to see that the number of applications of the principle in its final form to a given process is virtually infinite, both in terms of time and matter. 

Indeed we can, on the one hand, base our consideration on a time of any size, especially an infinitely small time (whereby we obtain the application to elementary processes), and on the other hand on a complex of material elements of any size (and especially an infinitely small one, i.e. an elementary body).
For each such complex the principle provides a special equation. 

Each time {\it\color{red}(i.e. for either elementary or complex systems)} you must choose the combination for which the calculation is most convenient, and of course you must always be careful to \dashuline{define the underlying material system} 
to be fixed exactly 
(we will call it the ``\,\dashuline{basic system}\,''  {\it\color{red}(Grundsystem)} for short).

\vspace*{-3mm} 
\begin{center}
--------------------------------------------------- 
\end{center}
\vspace*{-3mm}

Let's take for example any gas quantum whose state is changed by compression and by the supply of heat from outside. 

As long as we consider the gas alone as the basic system, the external effects involved consist in the change in position of the compressing body and in the heat release of the heat reservoir used, so the energy of the gas increases by the work value of the compression and the heat supplied. 

It is completely irrelevant whether the compression is caused by a heavy stamp that does work as it sinks, or by another gas that expands and thereby loses heat, etc. 
The only thing that matters here is the mechanical work, and it doesn't matter how this compression can be brought about, or in what way (p.121). 

\vspace*{-2mm} 
\begin{center}
{\it\color{red}\bf
---------
(p.122) 
---------
}
\end{center}
\vspace*{-3mm}

If we now assume more specifically that the compression is brought about by a stamp loaded with a weight, the weight of which maintains equilibrium with the pressure of the gas, and if we include this weight in the basic system, the compression work disappears as an external effect and instead occurs to this new addition is the work of gravity, which comes from the earth's attraction and acts on the weight, which is the same in magnitude as the previous one. 

If we also include the earth in the basic system, the work discussed as an external effect disappears completely, but in the expression of the energy of the basic system a new member appears, namely the energy of the gravity of the weight, as a function of height, on which it is located. 

As simple and self-evident as these considerations appear for the case presented, they become even more important as soon as one moves from these simple to somewhat more complicated states, in which for example the compressing weight has a certain speed, and the pressure of the gas is no longer equal to the weight of the weight.

\vspace*{-3mm} 
\begin{center}
--------------------------------------------------- 
\end{center}
\vspace*{-3mm}

Perhaps it would not seem inappropriate to take this opportunity to mention a usage which, if incorrectly understood, can easily give rise to misunderstandings. 

The energy of a heavy body is sometimes spoken of as the product of its weight and the height of its center of gravity. 
This term is inappropriate if you think of the body as a basic system, because the energy of a body always depends only on its own state, never at the same time on the storage of external masses. 
In fact, this expression is not found among other central forces. 

In order to be able to talk about the energy of gravity, one must always (even if tacitly) think of the earth as being included in the basic system. 
Otherwise, the work of gravity should not be taken into account as a type of energy, but rather as an external effect (of the earth that happens to be nearby, see page 93).

\vspace*{-5mm} 
\begin{center}
{\it\color{red}\bf
---------
(p.123) 
---------
}
\end{center}
\vspace*{-3mm}

If we consider more closely the meaning of the generalisation which we have made in the original version of our principle, it is essentially based on the fact that we have broken down the 
unique 
equation which expresses the conservation of energy (for a system deprived of all external effects) into a number of equations which regulate the changes of energy 
(via 
the absorption and release of this energy in the individual parts of the system, according to the measure of the corresponding external effects).

However, we would like to draw particular attention to one point. If we divide the entire system (which is not subject to any external effects and whose energy is therefore constant) into two parts, which we consider one after the other as a basic system, then it would be wrong to assume that the energy absorbed from one part in a certain time is equal to that released by the other part in the same time. 

This theorem would only apply if the energy of the entire system were equal to the sum of the energies of the two partial systems, which, as we have already emphasized repeatedly, is not generally the case. 

Let's take two material points as an example, which act on each other with a central force. 
The energy of the system is then the sum of the living forces {\it\color{red}(kinetic energies)} and the potential of the central force {\it\color{red}(gravitational potential energy)}.
It {\it\color{red}(this sum)} is unchanging over time. 

The energy of a single point is its living force, and its change is measured by the external impact that the point suffers (i.e. through the work that the force does on him). 

It can obviously happen that, over a certain period of time, positive energy from outside is transferred to each of the two points, and as a result the living forces of both grow at the same time. 

Only when the long-distance effects are completely eliminated (as in phenomena such as elastic wave movement and heat conduction) can one say that the energy transferred to one material complex is simultaneously withdrawn from another.

\vspace*{-3mm} 
\begin{center}
--------------------------------------------------- 
\end{center}
\vspace*{-3mm}

  In general, \dashuline{the ideas on which the conception of the effectiveness of natural forces is based} play an even more important role \underline{here} (where we speak of the energy of an arbitrarily selected material system) than above (\underline{there}, where we considered only those systems that are not subject to external effects). 

\vspace*{-2mm} 
\begin{center}
{\it\color{red}\bf
---------
(p.124) 
---------
}
\end{center}
\vspace*{-3mm}

\underline{There} (p.110) it was only a question of the primary form of energy, the magnitude of which was determined for each state of the system by the general definition.

It would also be so \underline{here}, if one were always able to realise the measurement as prescribed by the definition. 

But as this is not always the case (owing to imperfect means of observation), it may happen that not only different forms but quite different numerical values are obtained for the energy of a given material system, depending on the assumptions made about the nature of the acting forces, and that one is unable to settle the resulting difference by experimental means. 

An example of this is provided by the nature of the electric field, which we have discussed repeatedly. 

According to Faraday's conception, the energy of a randomly selected part of a dielectric is different from zero, and one can obtain work power from the transfer of this part from the forced state to the neutral state, while according to Weber's conception, the insulator is always in the same state (apart from any secondary changes), whether free electricity is present in the conductors or not. 

As long as no final decision has been reached on this question through more specific research, it will therefore be necessary  (before proceeding to the establishment of the equation which expresses the principle of energy) to first fix exactly the point of view which one wishes to adopt in the conception of the processes to be investigated.

\vspace*{-3mm} 
\begin{center}
--------------------------------------------------- 
\end{center}
\vspace*{-3mm}

The decomposition of the equation of the conservation of energy that we have carried out is based on the consideration of a material complex, selected arbitrarily from the original system, as the basic system, and the energy entering or leaving it.

Instead of decomposing the system into its material parts, \dashuline{we can}, with the same right, and sometimes with considerable advantage, \dashuline{carry out another decomposition}, namely \dashuline{into volume parts}. 

\vspace*{-2mm} 
\begin{center}
{\it\color{red}\bf
---------
(p.125) 
---------
}
\end{center}
\vspace*{-3mm}

A given volume of space always contains a certain material system at a certain time, and insofar as this system has a certain energy at that same time, one can speak of \dashuline{the energy of the volume}. 

\dashuline{The energy of a fixed volume of space} will not change over time if neither matter enters or leaves the volume, nor external effects on the matter contained in it, so the change in energy always comes from one of these two causes, so that we 
can express the proposition:
``\,{\it the energy transferred into a volume of space is determined on the one hand by the external effects on the matter contained in it, and on the other hand by the entry of new matter.\,}''
 
It will now depend on whether the formulation of the expressions for the energy thus transferred into the volume is made easier by the specific circumstances of the case under consideration.

\vspace*{-3mm} 
\begin{center}
--------------------------------------------------- 
\end{center}
\vspace*{-3mm}

In fact, there are many applications in which this theorem provides convenient services. 

Clausius, for example, essentially uses it to calculate the (Joule) heat generated in a conductor by a stationary galvanic current, whether, as with metallic 
conductors(1)$\,$\footnote{$\:$R. Clausius: Über die bei einem stationären Strom in dem Leiter gethane Arbeit und erzeugte Wärme {\it\color{red}(About the work done and heat generated in the conductor with a stationary current)}. Pogg. Ann. 87, p.415, 1852.}, 
the matter of the conductor is at rest or, as with electrolytic 
conductors(2)$\,$\footnote{$\:$R. Clausius: Über die Elektricitätsleitung in Elektrolyten {\it\color{red}(On electricity conduction in electrolytes)} Pogg. Ann. 101, p.338, 1857. p.340.}, 
the matter is travelling at the same time as the electricity.

If we imagine any fixed volume through which the current flows, the energy contained therein can be increased: \\ 
\hspace*{19mm} 1) by external effects; or \\
\hspace*{19mm} 2) by new matter entering it. \\ 
When calculating this magnification, however, it depends essentially on which idea about the nature of the galvanic current is based on. 

Let us first assume that the electrical particles behave like material atoms of infinitesimal inertia, which are driven forward on their path by the attractive (repulsive) force of the total free electricity present in the conductor system.

\vspace*{-2mm} 
\begin{center}
{\it\color{red}\bf
---------
(p.126) 
---------
}
\end{center}
\vspace*{-3mm}

\noindent 
{\it\color{red}At the same time}, for the sake of simplicity, we can assume only one type of electricity to be mobile, and the energy of a material particle (electricity, ion) 
moving in a constant current is independent of its position (not variable with the potential function, because this originates from external masses).

Since in a certain time just as much matter enters the ``\,basic volume\,'' as leaves it, no change in energy is caused by this circumstance, and only the external effects remain to be considered. 

These provide an increase in energy equal to the work performed by the forces of the overall system on the matter flowing through the entire volume. 
The energy of the volume is therefore increased by the amount of this work, and since the electrical energy is constant due to the stationary state, the increase must benefit the thermal energy.

\vspace*{-3mm} 
\begin{center}
--------------------------------------------------- 
\end{center}
\vspace*{-3mm}

The train of thought becomes somewhat different if you imagine electricity as a fine, incompressible fluid that is pressed through the conductor by a force (a kind of pressure) that only acts at immeasurably small distances. 
In this case too, the energy of a flowing particle is independent of location (cf. the energy of incompressible liquids, Section 3), so here, as above, the entry of matter into the volume under consideration does not cause an energy change in it. 

But as far as the external effects are concerned, these are reduced to the work of the forces acting on the surface of the volume (everything else is internal effects). 
This work is proportional to the value of the potential function at every location (just as it is to the pressure in liquids), so it is greater at the entry points of the flow than at the exit points, and thus we get a positive expression for the total work done by the external effects, whose value corresponds to that obtained from the above consideration.

\vspace*{-3mm} 
\begin{center}
--------------------------------------------------- 
\end{center}
\vspace*{-3mm}

If one finally assumes, as some would like, that the electrical current does not consist in a translation of matter, but, in the manner of heat conduction, in a propagation of special forms of motion, then in order to explain Joule heat it is necessary to assume that the living force {\it\color{red}(kinetic energy)} of these movements entering a volume at a certain time interval is greater by a certain amount than that emerging, so that the resistance of the conductor causes a kind of absorption of the vibrations that constitute the current.

\vspace*{-2mm} 
\begin{center}
{\it\color{red}\bf
---------
(p.127) 
---------
}
\end{center}
\vspace*{-3mm}

The derived theorems obviously enable us to derive a special equation for each material element (and each volume element) of a body from the principle of energy, and thus, as already noted above, we do indeed have an infinite number of equations available for any given process, which have a determining influence on its course. 
But if it is a question of clearly determining the temporal course of the process, we must not stop at the results obtained.
Because we are not yet able to solve this further task with the tools we have acquired so far. 
We would only be so if the changes in a single material element depended on a single variable (corresponding to the one equation we can set up for the element), which is not generally the case.

\vspace*{-5mm} 
\begin{center}
--------------------------------------------------- 
\end{center}
\vspace*{-3mm}

However, if we go a step further, in many cases we can obtain the means that are just sufficient to solve the problem in question.
This step consists in the inclusion of \dashuline{the principle of superposition of energies}, which was occasionally used above. 

Experience has shown that the energy of a material system is the sum of the individual types of energy, which are completely independent of each other, and therefore the total supply of energy breaks down conceptually into a series of individual energies, each of which can be determined independently. 

If effects are exerted on the system from outside, through which energy is transferred into it, then these effects can generally be broken down into different types. 
Each of these types of effects brings about a specific change in the corresponding type of energy in the system, so that the equation that expresses the connection between the change in the total energy and the external effects breaks down into a series of individual equations, each of which determines the change in a specific type of energy in its dependence governed by a special external effect. 

\vspace*{-2mm} 
\begin{center}
{\it\color{red}\bf
---------
(p.128) 
---------
}
\end{center}
\vspace*{-3mm}

So here we have a further decomposition of the energy equation, which differs from the previous one in that above we have divided the material system into individual material (or volume) parts, but here we have divided the energy into individual types of energy.

\vspace*{-3mm} 
\begin{center}
--------------------------------------------------- 
\end{center}
\vspace*{-3mm}

For example, let's imagine a body that moves freely in space. Its energy is divided into two parts: the living force {\it\color{red}(kinetic energy)} of its visible movement, and its internal (e.g. thermal) energy. 

As long as no external effects take place, the total energy remains constant. 
But not only this, but also each of the two types of energy remains constant: the body moves with constant velocity and remains at constant temperature, under no circumstances will a change of these quantities take place without a corresponding external effect (although, according to the general principle, a transformation of a single type of energy into another would be quite permissible). 

If we further imagine that certain effects are exerted on the body from outside, which may consist in a mechanical force coming from a distance (such as gravity), but also in a supply of heat (such as radiation), then the energy of the body will increase by the sum of the work done by the force and the quantity of heat supplied.

But we can say even more: the external effects are divided into two different types, each of which only influences the type of energy corresponding to it: the work of the force only changes the velocity, but not the temperature of the body, and the heat supplied only increases the temperature, but not the velocity. 
The various effects with their corresponding energies belong to completely separate areas, and therefore each provide a special equation.

\vspace*{-3mm} 
\begin{center}
--------------------------------------------------- 
\end{center}
\vspace*{-3mm}

  However, it should be borne in mind that the independence of the types of energy (as well as the external types of action from each other) can never be established a priori in this case (as in all similar cases), but must always first be established experimentally.

\vspace*{4mm} 
\begin{center}
{\it\color{red}\bf
---------
(p.129) 
---------
}
\end{center}
\vspace*{-3mm}

So one can easily imagine that the \dashuline{heat rays existing in ether vibrations} {\it\color{red}(old-fashioned concept)} exert a direct mechanical effect on the body, even if such an effect has not yet been proven with certainty, and on the other hand it is obvious that a mechanical force, if not from a distance (but e.g. acts as friction or shock on the surface of the body), can at least largely be converted directly into heat. 

It is also very conceivable that two types of energy, which have been considered independent of each other for a while, and can still be viewed as independent when calculating certain phenomena, will one day become dependent on one another upon closer understanding of the forces of nature.

\vspace*{-3mm} 
\begin{center}
--------------------------------------------------- 
\end{center}
\vspace*{-3mm}

Here we actually come across a limit to the applicability (not the validity) of \dashuline{the energy principle}, because if the individual types of energy no longer change independently of each other (each according to the external effects corresponding to it), then we have to stop with the decomposition of the equation of energy.

Then \dashuline{the energy principle for a material element} provides fewer equations than are required to calculate its change of state. 
This includes, among other things, all the cases in which the processes taking place inside an element cannot be brought into any direct connection with the external effects (such as: all explosive phenomena in which minimal external effects can cause the largest and most diverse transformations of individual types of energy into one another). 

The question now arises: 
{\it according to which law and in what sense does the turnover of energy take place in such cases?} 

This question goes beyond the scope of our current investigations, and its answer can no longer be approached from the standpoint of the simple law of conservation of energy, but must be based on \dashuline{completely new principles} independent of that law.

\vspace*{-2mm} 
\begin{center}
{\it\color{red}\bf
---------
(p.130) 
---------
}
\end{center}
\vspace*{-3mm}

We already have such a principle in \dashuline{the second law of the mechanical heat theory}, founded by \dashuline{Carnot and Clausius}, which gives us \dashuline{information about the direction in which the conversion of the different types of energy into one another takes place}.

\vspace*{-3mm} 
\begin{center}
--------------------------------------------------- 
\end{center}
\vspace*{-3mm}

But regardless of the limitations presented, the high value inherent in separating the external effects according to their influence on the individual types of energy remains undiminished. 
It is essentially based on gaining a fixed point of view from which one can understand the conditions for the transformation of the kinds of energy among each other, and thus the diversity of natural forces in general. 

Therefore you are always led to the new question: 
which conversions of energy take place independently of one another? 

The answer to this question offers the first means of bringing order to the apparently complicated processes that take place within the framework of the smallest process, and of making them individually accessible to experimental investigation. 

Without the principle of superposition of energies, one could not separate mechanics from heat, electricity from magnetism, and the division of the entire physics into its various areas would be inadmissible from the start. 

Whether in later times we will be able to distinguish between fewer types of energy than now, whether all of them can perhaps be reduced to a single one or two, is no more answerable than the question of whether ponderable matter is essentially different from light ether, or not. 

We will endeavor to show most clearly in the last section of this work 
that
we are actually able (based solely on the point of view we have now gained) \dashuline{to derive the basic laws of mechanics} as well as the other parts of physics in the same form as is usually done \dashuline{with the help of} the developed formulations of \dashuline{the principle of conservation of energy}. 

At the same time, this task will provide us with a wealth of remarkable applications of the statements presented here to the individual types of energy.

\vspace*{-3mm} 
\begin{center}
{\it\color{red}\bf
---------
(p.131) 
---------
}
\end{center}
\vspace*{-3mm}

In concluding herewith the investigations which have led us to the establishment of the principle of the conservation of energy in a form as convenient as possible for its application, we shall use the conclusion of this section to examine the number and importance of the proofs which can be adduced in favour of the correctness of the principle.

In recent times, however, the assertion has been made that the principle is neither capable of nor in need of proof, because it is a priori (i.e. a necessary form of our faculty of perception and thought given to us by nature).
Here, as with so many other truths, the knowledge of which has been fought for through centuries of labour, the same truths are afterwards, when the power of habit comes into its own, presented as self-evident and innate.

Therefore, all we need to justify ourselves (if we reject such an assertion out of hand) is a reference to the historical development of the principle.

\vspace*{-3mm} 
\begin{center}
--------------------------------------------------- 
\end{center}
\vspace*{-3mm}

Like every proof of a scientific theorem, our principle can also be described as having a double method: the deductive and the inductive method. 

While the former {\it\color{red}(\dashuline{deductive method})} makes the proposition appear in its entirety as a logical result of the combination of a series of other propositions that are generally recognized as correct (whether from experience or elsewhere), the {\it\dashuline{inductive method}}, on the other hand, proceeds on the basis of experience to examine the individual consequences that  proceed 
from the proposition to be proven (if one temporarily accepts it as valid, and then combines it with other sufficiently justified propositions). 

If a single conclusion then comes to light that does not agree with experience, the proposition must be decisively rejected.
If this is not the case, it can remain in place.
However, only a certain degree of probability of the truth of what is to be proven can be achieved by inductive means, which increases as the experiments are varied. 

\vspace*{-5mm} 
\begin{center}
{\it\color{red}\bf
---------
(p.132) 
---------
}
\end{center}
\vspace*{-3mm}

Nevertheless, \dashuline{a particularly high value will always be attached to the inductive method of proof}, because since the truth of \dashuline{our entire natural science is ultimately based on experience}, the belief in the correctness of a proposition will be all the more firmly rooted in our conviction the closer the proposition is connected to a \dashuline{fact that can be verified directly through experience}. 

That is why, whenever it comes to establishing a new principle, we will \dashuline{try to approach} it \dashuline{from all possible angles through experiment and observation}, and no physicist will be content with the pure deduction of a scientific law of some importance (he probably still will consult the highest authority, i.e. the experience).

\vspace*{-3mm} 
\begin{center}
--------------------------------------------------- 
\end{center}
\vspace*{-3mm}

If we look at the principle we are dealing with from this point of view, \dashuline{a glance at the applications} presented in the previous and the following section (not a single one of which contradicts experience) is sufficient to \dashuline{give us the totality of the inductive proofs}, 
which \dashuline{extend over all natural phenomena known to us}, appear to us as an almost imposing power, and in the most definite manner \dashuline{advocates the unlimited correctness of the principle}.

It would be necessary to repeat the entire history of its development if we wanted to try (at this point) to give an overview of the various empirical evidence that has been compiled over time.

And \dashuline{almost every new application} also \dashuline{brought new evidence}, \dashuline{from the heat of friction} (which, regardless of the material of the rubbing bodies, their speeds, temperatures, etc., is determined solely by the mechanical work expended)  \dashuline{to the processes of galvanic induction} (which causes the movement of a magnet, regardless of the nature of the conductor in which it is produced).

\vspace*{-3mm} 
\begin{center}
--------------------------------------------------- 
\end{center}
\vspace*{-3mm}

And yet: no matter how overwhelming the number and importance of these inductive proofs appear to us, no one should be such an inveterate empiricist that he would not \dashuline{feel the need for another proof} which, \dashuline{built on a deductive basis}, would demonstrate the principle in its entire, comprehensive meaning as a single, closed whole \dashuline{that can emerge from certain even more general truths}. 

\vspace*{-2mm} 
\begin{center}
{\it\color{red}\bf
---------
(p.133) 
---------
}
\end{center}
\vspace*{-3mm}

Indeed, even if \dashuline{the multitude of individual experiences} that have been made \dashuline{seem to necessarily urge us to accept this law}, no one can guarantee that an isolated class of facts, hitherto overlooked for some reason, cannot yet again be found which does not comply with the demands of the principle. 

Nor can it be disputed that we cannot obtain the full reassuring certainty (which the conviction of the truth of a proposition gives us by induction alone), but only at the same time by considering the proposition as a complete unity from a higher point of view. 

How else would it be conceivable that \dashuline{at a time when} (apart from the investigations by Mayer and Colding, which received little attention in wider circles) \dashuline{there were only a small number of Joule's experiments available} (and when one could hardly speak of an inductive proof) \dashuline{the idea of conservation of energy} would \dashuline{have taken root so surprisingly quickly in many places at the same time} (and stimulated new investigations on the most diverse sides) if \dashuline{the knowledge of that immediate unity} (which can only be achieved through deduction) had not simply and clearly \dashuline{forced itself on everyone's minds}. 

 The question now arises for us, however, \dashuline{in what way one can arrive at the statement of our principle by deduction}, or \dashuline{whether there is a deductive proof at all} which may lay claim to strictly scientific significance, \dashuline{as is demanded in natural science} today.

Let's take a closer look at this question.

\vspace*{-3mm} 
\begin{center}
--------------------------------------------------- 
\end{center}
\vspace*{-3mm}


Since every logical deduction presupposes \dashuline{a general principle that is generally recognized as true}, the scope of which must not be smaller than that of the theorem to be proven.

\vspace*{-2mm} 
\begin{center}
{\it\color{red}\bf
---------
(p.134) 
---------
}
\end{center}
\vspace*{-3mm}

The main difficulty in our case will be to \dashuline{find such a general principle that}, on the one hand, \dashuline{enjoys such universal recognition} that it can \dashuline{serve as a sure guarantee for the correctness of our principle}, but on the other hand is also comprehensive enough to include the entire principle with its enormous scope. 

It can be seen at first sight that the choice among the propositions which fulfil both requirements will not be a large one. Nevertheless, several of them can be named which in the course of time have claimed the place of an upper proposition in deductive reasoning.

\vspace*{-3mm} 
\begin{center}
--------------------------------------------------- 
\end{center}
\vspace*{-3mm}

The oldest deduction reaches no less far up than to the person of \dashuline{the Creator himself}, who in his \dashuline{eternity and immutability} also communicates these qualities of his to \dashuline{the nature he has created and its forces}, from which it then follows that the entire ``\,quantity of motion\,'' contained in the world has an indestructible, constant value for all time.

Since this consideration, which comes from Descartes (see p.8), obviously aims to \dashuline{establish a general law of nature} that regulates the sum of \dashuline{the forces active in nature} as well as \dashuline{the amount of matter} present, it can certainly be \dashuline{associated with the principle of conservation of energy}.
In any case, these ideas could also be easily transferred to the current form of our principle.

\vspace*{-3mm} 
\begin{center}
--------------------------------------------------- 
\end{center}
\vspace*{-3mm}

Essentially in the same line of thought, albeit from a somewhat more modest standpoint, is the evidence that \dashuline{Colding} (see p.30) sought to provide for the \dashuline{conservation of force}. Although he no longer appeals to the highest authority, he does see the reason for the immutability of the forces of nature in the fact that these forces, precisely because they dominate nature so completely, must themselves be supersensible, spiritual beings, which as such are impossible may be subject to natural death or transience. 

After all, he {\it\color{red}(\dashuline{Colding})} considers it \dashuline{appropriate to test this statement through experience} and carries out his experiments with this in mind.

And since then, according to today's views, every proof of \dashuline{a scientific proposition whose meaning is rooted in metaphysical ground has lost its power from the start}, and we can briefly ignore these and similar deductions.

\vspace*{-5mm} 
\begin{center}
{\it\color{red}\bf
---------
(p.135) 
---------
}
\end{center}
\vspace*{-3mm}

The way of thinking that Mayer (p.21) placed at the forefront of his discussions deserves more attention. 
Although it is still on a somewhat shaky basis, it can certainly no longer be described as metaphysical. 

He {\it\color{red}(Mayer)} explains his main principle: ``\,Causa aequat effectum\,'' i.e. ``\,\dashuline{the cause is equal to the effect}\,'', saying that in nature every cause has an effect specific to it, and vice versa: that \dashuline{nothing is contained in the effect that is not already in the cause} (under some form).

Therefore, all changes that take place in nature do not consist in the production, but only in the \dashuline{transformation of forces according to certain constant proportions}.

The various \dashuline{forces are therefore equivalent to one another in certain specific ratios}, they can therefore all be \dashuline{measured in a common {\it\color{red}unit}}, and \dashuline{the sum of all the forces present in the world}, expressed in this common {\it\color{red}unit}, \dashuline{remains constant over time}.

One may well admit that this derivation has something confusing, because \dashuline{the law of cause and effect} forms the postulate of our entire knowledge of nature.

But, on the other hand, it should be borne in mind that the appeal that Mayer's deduction exerts on us would lose significantly in strength if \dashuline{we had} not already \dashuline{recognized the truth of the sentence for other reasons and}, through years of practice, \dashuline{had become accustomed to the idea}, which he utters, would have been accustomed to.

It would hardly make much of an impression on someone who was completely new to the matter. 
If one can therefore accept the described train of ideas as an excellent explanation of the \dashuline{principle of conservation} a posteriori, then the status of a binding proof in the physical sense must definitely be denied. 
\dashuline{The meaning of ``\,aequat\,'' (equal) is far too vague for that}: if the cause were really the same as the effect, there would be no change at all in nature.

\vspace*{-3mm} 
\begin{center}
{\it\color{red}\bf
---------
(p.136) 
---------
}
\end{center}
\vspace*{-3mm}

The first really \dashuline{physical deduction} through which the \dashuline{energy principle} was proven in its entirety is that which \dashuline{Helmholtz gave in his treatise on the conservation of force}: it is based on the \dashuline{mechanical conception of nature}, more specifically \dashuline{on the assumption that everything in nature effective forces can be resolved into point forces}, for which Newton's axioms apply. 

This prerequisite is associated either with the \dashuline{assumption that all elementary forces are central forces}, or with the \dashuline{assumption that the construction of a perpetual motion machine is impossible}. 
We have already provided the most important information about the implementation of these ideas in the previous section (p.35). 

According to this, the \dashuline{principle of the conservation of energy} would essentially be reduced to the \dashuline{mechanical proposition of the conservation of living forces{\it\color{red}-(kinetic energy)})}, and we would have to understand \dashuline{the entire energy of the world as consisting of only two kinds}: the actual (living force{\it\color{red}-(kinetic energy)}) and the potential (tension force).

\vspace*{-3mm} 
\begin{center}
--------------------------------------------------- 
\end{center}
\vspace*{-3mm}

  If one considers that \dashuline{the mechanical view of nature has played an important role in natural philosophy} from time immemorial (long before the principle of energy became known, probably mainly because it fulfils our need for causality, which strives for the greatest possible unity of the forces underlying phenomena, so admirably), \dashuline{and furthermore if} we overlook the fact that \dashuline{from the mechanical point of view the definition of the concept of energy, the formulation and finally the proof of the principle can be given so clearly}, it is quite understandable that \dashuline{it is precisely this proof that has been favoured among the deductive methods} (and is probably the one most frequently used today).

After \dashuline{Helmholtz}'s process, he was adopted by other physicists (\dashuline{Mayer}, as is well known, \dashuline{did not share the mechanical view of nature}), while at the same time the mechanical theory was spread and recognized from England by \dashuline{Joule}, who was joined by \dashuline{Rankine} and \dashuline{Thomson}.

\vspace*{-2mm} 
\begin{center}
--------------------------------------------------- 
\end{center}
\vspace*{-4mm}

Nevertheless, it would seem to me that \dashuline{one would be} more \dashuline{justified in making the principle of the conservation of energy the support of the mechanical view of nature} than, conversely, in making the latter the basis of the deduction of the energy principle, since this principle is far more securely founded than the assumption, however plausible, that every change in nature can be traced back to motion.

\vspace*{-2mm} 
\begin{center}
{\it\color{red}\bf
---------
(p.137) 
---------
}
\end{center}
\vspace*{-3mm}

 In innumerable decisive cases \dashuline{the energy principle has proved to be correct}, \dashuline{while the reasons that can be adduced in favour of the mechanical theory} (at least in so far as they are based on direct experience) \dashuline{are for the most part} (not exclusively, cf. the theory of gases) \dashuline{based on the conservation of energy}, from which, by the way, they by no means necessarily follow (cf. p.51).

Efforts have so far been made in vain to trace the entirety of electrical and magnetic phenomena back to simple movements, and when applied to the organic world (to which we also want and must extend the proof of the principle of conservation of energy) it is possible there is not even a trace of a beginning.

\vspace*{-3mm} 
\begin{center}
--------------------------------------------------- 
\end{center}
\vspace*{-3mm}


However, we must resolutely oppose the view (which is now sometimes expressed) that the mechanical theory must be accepted as an a priori postulate of physical research.
It cannot free us from the obligation to justify that theory by legal means. 

\dashuline{Natural science} only knows one \dashuline{postulate}: the \dashuline{principle of causality}, because this is its condition of existence. 

We do not need to examine here whether this principle itself is drawn from experience or whether it forms a necessary form of our thinking.

\vspace*{-3mm} 
\begin{center}
--------------------------------------------------- 
\end{center}
\vspace*{-3mm}

  It therefore seems to me that it is more in keeping with the \dashuline{empirical character of our modern natural science} (so brilliantly proved hitherto) to \dashuline{regard the mechanical conception of nature as the possible and probable goal of research} than to anticipate prematurely a result not yet at all certain (in order to make it the starting-point of the proof of a proposition whose generality appears to be as certain as that of few others in the whole of natural science).

The \dashuline{great importance of the mechanical view of nature} remains completely undiminished by this consideration: it shows us the \dashuline{direction in which research must move}, because the question of the admissibility of this theory can only be decided through experience. 

\vspace*{-2mm} 
\begin{center}
{\it\color{red}\bf
---------
(p.138) 
---------
}
\end{center}
\vspace*{-3mm}

We will therefore \dashuline{use all possible means to carry out the mechanical conception in all areas of physics, chemistry, etc.}, right down to the last consequences (and in this sense the efforts aimed at this have their fundamental value, all the more so because they have already been brilliant so far brought results to light). 

But there is still a big difference whether one can \dashuline{regard a hypothesis as probable} or whether one \dashuline{places it at the head of such a deduction} as the one we are dealing with here.
By exercising caution, we also protect ourselves from unpleasant disappointments. 

Because if one really has \dashuline{the strange experience that our view of space and time is not general enough to describe the wealth of phenomena that nature presents 
to us}$\,$\footnote{$\:$E. Mach: Die Geschichte und die Wurzel des Satzes von der Erhaltung der Arbeit {\it\color{red}(The history and the root of the theorem of conservation of work)}. Prag, 1872, Calve. Übrigens kann ich mich nicht mit allen hier dargelegten Ansichten einverstanden erklären {\it\color{red}(By the way, I cannot agree with all of the views expressed here / M. Planck)}.},
we shall not, as has already happened in similar cases, drop other well-founded propositions at the same time, but will easily be able to separate the proven essential from the unproven 
non-essential.$\,${\color{red}\footnote{$\:${\it\color{red}Note, however, that these contributions from E. Mach will be studied by Albert Einstein at his ``\,Olympia Academy\,'' meetings (with Maurice Solovine, Conrad Habich and Mileva Marić-Einstein), and will prove to be interesting in that they have encouraged (and somehow  inspired) Albert Einstein to follow the path toward the special and general relativity theories (P. Marquet)}.}}

\vspace*{-3mm} 
\begin{center}
--------------------------------------------------- 
\end{center}
\vspace*{-3mm}

Since, after what has been said, \dashuline{we cannot decide to give the mechanical proof of the principle of conservation of energy the importance that it usually enjoys}, we assume all the more the \dashuline{obligation to look for another theorem that has a more solid foundation and is better suited to serve as a starting point for the deduction}. 

Now there is actually another sentence that seems to have the necessary properties in a sufficient way: it is \dashuline{the empirical} sentence which states the \dashuline{impossibility of perpetuum mobile and its reversal}, quite independently of any particular conception of nature. 

\vspace*{-2mm} 
\begin{center}
{\it\color{red}\bf
---------
(p.139) 
---------
}
\end{center}
\vspace*{-3mm}

Following our earlier terminology (p. 99), we can formulate it as follows: ``\,{\it {it is impossible to carry out a circular process with a material system (which returns the system exactly to its initial state) in such a way that the external effects have a value other than $0$} (i.e. either positive or negative) {work value}.}\,'' (on the concept of the labor value of an external effect, see p. 94f.) or more briefly: 
``\,{\it {positive work value can neither arise from nothing nor perish into nothing}}\,'' (and the reversal is an essential prerequisite).

\vspace*{-3mm} 
\begin{center}
--------------------------------------------------- 
\end{center}
\vspace*{-3mm}


As far as the justification of this sentence is concerned, it should be considered that centuries have been working on it.

There were people who were not afraid to risk their life and property to refute the claims of the sentence, by \dashuline{creating work value out of nothing}.{\it\color{red}(!)} 

Therefore, if one wants to accept an indirect proof gained through experience at all, then one has to do so in this case, and then one will not find the price at which the truth (so valuable for all of humanity) was bought, all that precious. 

In any case, the fact is that nowadays \dashuline{we are not in a position to simply declare anyone who tries to construct a perpetual motion machine to be a fool}.{\it\color{red}(!)}

\vspace*{2mm} 
\begin{center}
--------------------------------------------------- 
\end{center}
\vspace*{-3mm}


However, the proof of the converse theorem that work value cannot disappear into nothingness is somewhat weaker. 

There has hardly ever been a person who has dealt practically with \dashuline{the problem of destroying work}, any more than with the problem of \dashuline{turning gold into lead}.{\it\color{red}(!)} 

\dashuline{We cannot therefore speak of an empirical proof of the impossibility of} solving this problem in the fully important sense as that of the first sentence, but must limit ourselves to the statement of the fact that no process has ever been observed through which nothing further occurs is seen as \dashuline{the destruction of work value}. 

We must content ourselves with this fact instead of proof, because there can be no question of a deduction of the reverse proposition from the direct one, because not every natural process can be reversed. 

Logically speaking, there would be no contradiction in the assumption that \dashuline{work cannot arise from nothing, but} under certain circumstances it \dashuline{can perish into nothing} (Clapeyron's view p. 16).

\vspace*{-1mm} 
\begin{center}
{\it\color{red}\bf
---------
(p.140) 
---------
}
\end{center}
\vspace*{-4mm}


In general, it must be admitted that the empirical proof of the direct proposition (\dashuline{the impossibility of producing work from nothing}) has only been carried out on a relatively very limited part of the total range of natural forces. 

Today we already know and have access to a much more diverse range of phenomena than back then, when the aim was to achieve the practical development of perpetuum mobile. 

The extent to which we are now justified in extending the experience (previously gained in a narrower field) to all effects in nature is not easy to judge at present, since we are already too accustomed to the generality of this truth (through our familiarity with the energy principle) to be able to abstract from it completely for the time being.


\vspace*{-3mm} 
\begin{center}
{\bf\color{red}
=======
The ``\,zero state\,'' / ``\,Nullzustand\,'' (p.140)
=======
}
\end{center}
\vspace*{-3mm}

Be that as it may, we place at the top of the following statements the \dashuline{theorem of the impossibility of perpetual motion} and its reversal in the circle of the entire inorganic and organic nature, and want to investigate, completely independently of the mechanical view of nature, whether and under what conditions this theorem can be used to prove the \dashuline{principle of conservation of energy}. 

Let us first recall the remark (made on page 99) that all the different forms of the principle are contained in the one sentence: ``\,{\it\dashuline{the energy of a material system in a certain state, {\bf relative to a certain} {\bf zero state}, has a unique value.}}\,'' 

The only issue here is to \dashuline{deduce this theorem from the impossibility of perpetuum mobile}, \dashuline{based on the definition} that we have put forward (p. 93) \dashuline{for the concept of energy}. We choose the indirect method of proof by showing that in every single case where the definition would result in two different values of energy, the construction of a perpetual motion machine would be possible.

So let us assume that the material system has been brought \dashuline{from the given state $A$} \dashuline{to the {\bf zero state} {\it\color{red}(\,``\,Nullzustand\,''\,)} $N$} in any way, and that the work value of the external effects has been found to be equal to a.
But another way of transfer is also possible, and this would provide the work value $a'$ for the external effects, which is different from $a$. 

\vspace*{-2mm} 
\begin{center}
{\it\color{red}\bf
---------
(p.141) 
---------
}
\end{center}
\vspace*{-3mm}

Then it will always be possible to create a perpetual motion machine (although not in the way that is sometimes described) whereby the system is brought to the {\color{red}\bf (zero)} state $N$ in one way, and brings back to state $A$ in the other way, because the process in question need not be reversible. 
Rather, after the system has reached the {\color{red}\bf (zero)} state $N$ in one or the other specified way, we will transfer the system back to state $A$ in some arbitrary way and thereby close the cycle. 

If we designate the work value of the external effects occurring when {\color{red}\bf (the zero state)} $N$ returns to $A$ by $b$, then we have two circular processes at our disposal, which have the respective work values $(a + b)$ and $(a' + b)$ produce. Since, according to the assumptions made, these two quantities are unequal, at least one of them must be different from $0$, and this would give the possibility of perpetual motion (or its reversal).

\vspace*{-4mm} 
\begin{center}
--------------------------------------------------- 
\end{center}
\vspace*{-3mm}

 As an essential condition of the usefulness of this proof, however, we must recognise the general premise that the transformation of a material system from one given state into another is always possible in some way; without it the whole deduction becomes illusory.

 In fact, let us consider the external effects produced by the transformation of the diamond into amorphous carbon when it is carried out by chemical or physical (galvanic) means.

If these effects did not have the same mechanical work value, no one would still be able to use this circumstance to construct a perpetual motion machine, since we are not able to convert the coal back into diamond and thus close the cycle. 

However, I believe that the objection which one might derive from this circumstance against the general admissibility of the given deduction is unfounded. 
Indeed, what matters is not whether man's art is capable of making the transition from one state to another at will, but whether this transition really occurs in nature, or only occurs when natural forces work together appropriately could.

\vspace*{-2mm} 
\begin{center}
{\it\color{red}\bf
---------
(p.142) 
---------
}
\end{center}
\vspace*{-3mm}

If one did not want to recognize this conclusion, it would be tantamount to saying that the statement about the impossibility of perpetual motion does not arise from a natural law, but from man's lack of skill, which is certainly contrary to the essence of the statement. 

According to all our experiences, the condition stated can always be regarded as fulfilled, because nature continually creates everything it produces from the simplest elements, and it prepares inorganic substances and the most complicated organisms with equal ease (albeit in a way that is sometimes completely unknown to us), and then dissolves them back into their component parts.

We have already discussed the same question in the same spirit on an earlier occasion (p. 98).

\vspace*{-3mm} 
\begin{center}
--------------------------------------------------- 
\end{center}
\vspace*{-3mm}

We therefore believe we are not mistaken if we assume (in contrast to the limited transformation of the kinds of energy) that \dashuline{the transformability of matter from all possible states into all possible others} (if the chemical elements are preserved) \dashuline{is unlimited}, and thus \dashuline{the deduction of the law of the conservation of energy with all its consequences from the law of perpetuum mobile is assured}. 

We are not, indeed, at liberty to assign to this proof the foremost place among deductive methods, and it is not at all impossible that, when natural science has reached a higher stage of development, another theorem of experience (such as the mechanical view of nature) will be taken as the basis of deduction with better rights.

\vspace*{-3mm} 
\begin{center}
--------------------------------------------------- 
\end{center}
\vspace*{-3mm}

%

%

\newpage
\section{\underline{Different types of energy} (p.143-247)}
\label{Section-3}
\vspace*{1mm}

\vspace*{-2mm} 
\begin{center}
{\it\color{red}\bf
---------
(p.143) 
---------
}
\end{center}
\vspace*{-3mm}

Let us start with the task described in the previous section.

In order to evaluate the terms and sentences developed in this section individually by applying them accordingly to the various types of energy, we first want to take an orientating look at the area that lies before us, and at the same time fix the method that will guide us in the following investigations. 

 Whereas above it was merely a question of establishing the principles (and the particular cases discussed merely served to illustrate the general propositions) here we have to seek the purpose of our presentation in the systematic working through of those propositions through all the parts of physics, by which, however, the principles themselves are then again placed in a brighter light.

 What, however, distinguishes the applications to be made here from the principal arguments in the previous section is the fact that they do not (like previously) always retain their meaning unalterably (although they can themselves be modified in a certain way with the progressive development of our physical views).

It is therefore all the more important to emphasize this point in particular, so that the previously achieved results, which always continue to exist in the same way, do not appear to be at risk. 

  The definition of the concept and the \dashuline{principle of the conservation of energy} is invariably valid for all times, but \dashuline{the form of its application to a concrete natural phenomenon is subject to change}, namely because the concept of the types of energy (not their value expressed in numbers) is entirely dependent on the character of the respective view of nature.

\vspace*{-3mm} 
\begin{center}
{\it\color{red}\bf
---------
(p.144) 
---------
}
\end{center}
\vspace*{-3mm}

The historical development of physics affords more than one example of this, i.e. \dashuline{how often have the conceptions of the nature} {\it\color{red}(``\,das Wesens\,'')} \dashuline{of the agents at work in nature} {\it\color{red}(``\,in der Natur\,'')} \dashuline{changed} (and in our exposition we have had occasion to emphasise this fact, see p.125).

It is difficult to say to what ultimate end this \dashuline{constant change in the conception of the nature}  {\it\color{red}(``\,Wesens\,'')} \dashuline{of the forces of nature} {\it\color{red}(``\,Naturkräfte\,'')}  is tending: in the present state of development of physics, the chief momentum lies in \dashuline{the endeavour to reduce all natural phenomena to mechanical changes}. 

At the same time another tendency is also beginning to assert itself, namely, that of \dashuline{replacing all direct action at a distance by forces} which have appreciable magnitude only \dashuline{at infinitely small distances}.

In the following, we will repeatedly (especially when discussing electrical and magnetic energy) take the opportunity to return to this \dashuline{constant fluctuation of the basic concepts of our conception of nature}  
 (a fluctuation that, incidentally, does not appear as a wavering back and forth, but as a constant progression in 
   a certain direction, 
because \dashuline{the fact that a change in perception proves to be necessary} is always accompanied by \dashuline{an increase in the accuracy of the natural phenomena described}, and thus \dashuline{an increase in knowledge}).

\vspace*{-3mm} 
\begin{center}
--------------------------------------------------- 
\end{center}
\vspace*{-3mm}

For our present purposes, let us note one thing from what has just been said, that before proceeding to the \dashuline{application of the principle of the conservation of energy to a particular natural phenomenon}, it is above all \dashuline{necessary to obtain} from the outset (by experience) \dashuline{certain ideas about the nature of the phenomena to be investigated}, and to 
consistently record them in the following considerations. 

Only in this way can one keep oneself completely free of errors and, in particular, avoid the danger of either completely overlooking a certain type of energy or, as can also happen, of accidentally billing it twice.

\vspace*{-3mm} 
\begin{center}
{\it\color{red}\bf
---------
(p.145) 
---------
}
\end{center}
\vspace*{-3mm}

\dashuline{Depending on the accuracy of the results} claimed, the idea from which one starts will be made \dashuline{simpler or more complicated}. 

For example, if you operate with a \dashuline{liquid drop}, it is sufficient for certain purposes to think of it as completely incompressible and to apply the theorems that can be derived from the \dashuline{principle of energy for incompressible liquids}. 

In the interest of greater accuracy, however, it is important to think of the forces acting inside the liquid as being caused by changes in density. 
But even with this conception one will not be able to get by in some cases, but will be compelled to \dashuline{add certain other forces} to the assumed pressure forces, namely those \dashuline{which arise from the so-called viscosity of the liquid}, and which we \dashuline{characterise by the name of ``\,friction.\,''} 

This is not the end of the series of increases in the accuracy of the result.

{\it\color{red}For instance,} 
if the \dashuline{liquid} has so far been \dashuline{assumed to be a continuum}, a closer examination shows that it \dashuline{shows discontinuous properties in the smallest particles}, and the consideration of these requires \dashuline{a closer look at the forces that come into effect}, which then appear as \dashuline{molecular forces}.{\color{red}$\,$\footnote{$\:${\it\color{red}Note that these atomic-molecular considerations were discussed by Max Planck in 1887 long before this atomic-molecular status of matter will be proven and accepted by the study of the Brownian motion by Albert Einstein (1905) and Jean Perrin in 1907-1909 (P. Marquet)}.}} 

To \dashuline{each of these various conceptions} mentioned corresponds a \dashuline{particular form of the kinds of energy}, and therefore a \dashuline{different application of the principle of the conservation of energy}.
The more accurate the results are obtained, admittedly at the expense of the simplicity of the calculation, the higher the position in the listed order that the underlying view occupies.

\vspace*{-3mm} 
\begin{center}
--------------------------------------------------- 
\end{center}
\vspace*{-3mm}


  In order to properly characterise the point of view from which the calculation is to be based, it is necessary  
  to be precise about the \dashuline{number 
  and type of independent variables} on which the states of the material system under consideration are to be made to depend: the smaller this number, the simpler the view and the calculation.

\dashuline{The energy of the system} will then always present itself as \dashuline{a specific function of these independent variables}, although it remains completely irrelevant whether this function actually corresponds to the “\,primary\,” (p.111) form of the energy or not (see also p.114).

\vspace*{-3mm} 
\begin{center}
{\it\color{red}\bf
---------
(p.146) 
---------
}
\end{center}
\vspace*{-3mm}

For the rest of the calculations we will use the theorems derived in the previous section, the most important of which we would like to put together again here:
\begin{enumerate}[leftmargin=10mm,parsep=0mm,itemsep=1mm,topsep=-1mm,rightmargin=2mm]
\item 
\!\!\!) The change in energy corresponding to a certain change in the state of a material system is equal to the work value of the effects that must be expended outside the system in order to bring about the change in state (in some way) (p. 120). If no external effects take place, the energy of the system remains unchanged.
\item 
\!\!\!) The change in the energy of a certain volume is caused, on the one hand, by the external effects on the matter contained in the volume, and on the other hand by the entry of new matter into the volume (p. 125).
\item 
\!\!\!) The energy of a material system is the sum of the individual, independent types of energy present in the system, and every external effect only changes the type of energy that currently corresponds to it (principle of superposition, p. 127).
\end{enumerate}

\newpage
\subsection{\underline{Mechanical energy} (p.146-188)}
\label{Subsection-3-1}
\vspace*{-1mm}

\vspace*{2mm} 
\begin{center}
{\it\color{red}\bf
---------
(p.146) 
---------
}
\end{center}
\vspace*{-3mm}

The simplest material system is a material point whose internal nature is determined solely by its (unchangeable) mass $m$. Its energy is the living force {\it\color{red}(kinetic energy)}: 
$$
 \frac{m}{2} \:
 \left\{
  \left(\frac{dx}{dt}\right)^2
  \:+\:
  \left(\frac{dy}{dt}\right)^2
  \:+\:
  \left(\frac{dz}{dt}\right)^2
 \right\} \; , 
$$
plus an arbitrary constant, which we set equal to $0$, as usual.

\vspace*{-3mm} 
\begin{center}
--------------------------------------------------- 
\end{center}
\vspace*{-3mm}

According to the \dashuline{principle of conservation of energy}, the living force {\it\color{red}(kinetic energy)} remains constant as long as there are no external effects on the point. 
But if an \dashuline{external effect} occurs (i.e. if forces from other material points are exerted on the point under consideration) they \dashuline{cause a change in the energy} of the point over a certain period of time, the magnitude of which is given by \dashuline{the work value of these effects} (i.e. through the work that the forces do at the point under consideration in the assumed time), and this applies quite generally (from whatever source the forces mentioned may come from, see p.120). 

\vspace*{-2mm} 
\begin{center}
{\it\color{red}\bf
---------
(p.147) 
---------
}
\end{center}
\vspace*{-3mm}

If we restrict the application to a time element $dt$, \dashuline{the sum of the corresponding work of all forces} acting on the point has the form: 
     $$ X\:dx \;+\; Y\:dy \;+\; Z\:dz \; ,  $$ 
where $X, Y, Z$ are the components of the resulting force, taken according to the directions of the 3 coordinate axes. 

From the use of the stated theorem, it follows that the growth of energy in the infinitesimal time $dt$ (i.e. the differential of the above expression of the living force{\it\color{red}-kinetic energy}) is equal to the working quantity given above. 

Therefore, by applying the \dashuline{principle of conservation of energy}, we would have found an equation that obeys the motion of the point. 
However, this equation is not sufficient to determine the dependence of each of the three variables $x, y, z$ on time $t$.

\vspace*{-3mm} 
\begin{center}
--------------------------------------------------- 
\end{center}
\vspace*{-3mm}

However, we can obtain the required number of equations by applying the theorem that we reproduced on the previous page under 3). Let us note that the expression of the energy of the point under consideration is represented as a sum of 3 symmetrically constructed terms, each of which refers to a specific coordinate direction and depends solely on the variable in question. 

The total energy is therefore made up of 3 independent types of energy. But we also notice exactly the same property in the expression of the work of the forces acting from outside. This quantity also breaks down into 3 sums, each of which corresponds to a specific coordinate axis and has a value that is independent of the other two.

Each of the individual external effects is therefore assigned to a specific individual type of energy. 
It now makes sense to assume that not only the change in the total energy is measured by the total work of the external forces, but even more specifically that each of the individual energies mentioned is only influenced by the individual effect that corresponds to it, completely independent of the other two.

\vspace*{8mm} 
\begin{center}
{\it\color{red}\bf
---------
(p.148) 
---------
}
\end{center}
\vspace*{-3mm}

If we assume this idea to be correct, the equation listed above breaks down into 3 individual ones, each of which refers to a specific coordinate direction: 
\begin{align}
 d\:\left\{
 \frac{m}{2} \:
  \left(\frac{dx}{dt}\right)^2
 \right\} 
 \;=\; X \: dx
\; , \nonumber \\
 d\:\left\{
 \frac{m}{2} \:
  \left(\frac{dy}{dt}\right)^2
 \right\} 
 \;=\; Y \: dy
\; , \nonumber \\
 d\:\left\{
 \frac{m}{2} \:
  \left(\frac{dz}{dt}\right)^2
 \right\} 
 \;=\; Z \: dz
\; , \nonumber
\end{align}
and by carrying out the differentiation, Newton's equations of motion result: 
\begin{align}
 m \;\: \frac{d^2x}{dt^2} \;=\; X 
\; , \nonumber \\
 m \;\: \frac{d^2y}{dt^2} \;=\; Y 
\; , \nonumber \\
 m \;\: \frac{d^2z}{dt^2} \;=\; Z 
\; , \nonumber
\end{align}
which are sufficient to represent the entire movement.

\vspace*{-3mm} 
\begin{center}
--------------------------------------------------- 
\end{center}
\vspace*{-3mm}

Of course, this derivation cannot claim to be a proof of Newton's first two axioms, since the used theorem of the superposition of energies cannot be applied a priori. 
Rather, as we have already emphasized in more detail on page 130,  and as emphasized in the last presentation, its meaning is based essentially on its heuristic value. 

  Just as here the three coordinate directions provide the basis for the division of both the energy and the external effects into the corresponding parts, so at other times we have other points of view (for example of a thermal or electrical nature), which necessitate a division of the effects into various individual elements that are completely independent of each other (which are then simply added together to form the total effect).

\vspace*{-4mm} 
\begin{center}
{\it\color{red}\bf
---------
(p.149) 
---------
}
\end{center}
\vspace*{-3mm}

  But where this decomposition is really practicable and leads to correct conclusions, \dashuline{only experience can teach} (because, for example, the fact that the effects according to the 3 coordinate directions take place independently of each other is also a theorem of experience 
that we cannot go beyond
under any circumstances, with any kind of representation).

However, once one admits it, \dashuline{the above derivation from the principle of conservation of energy becomes completely strict}.

\vspace*{-3mm} 
\begin{center}
--------------------------------------------------- 
\end{center}
\vspace*{-3mm}

After we have seen that Newton's first two axioms (the third follows below on page 157 ff.) can be developed from the propositions provided by the energy principle, the question arises as to whether it is not in the interest of an even more rational view of mechanics is that when presenting it, a starting point similar to the one just used should definitely be used instead of the one now in use. 

At present, \dashuline{mechanics} are almost universally introduced with the \dashuline{principle of proportionality of force and acceleration}, even if, with Newton and 
\dashuline{W. Thomson}$\,$\footnote{$\:$W. Thomson und P. G. Tait: Handbuch der theoretischen Physik {\it\color{red}(Handbook of theoretical physics)}. Deutsch von H. Helmholtz und G. Wertheim. Braunschweig 1871, I. Chapter 207.}, 
the concept of force is ultimately based on that of \dashuline{pressure}, since for us it is transmitted directly through the \dashuline{muscular sense} (sense of touch, feeling), or that, along with 
\dashuline{Kirchhoff}$\,$\footnote{$\:$G. Kirchhoff: Vorlesungen über mathematische Physik. Mechanik {\it\color{red}(Lectures on mathematical physics. Mechanics)}, Leipzig 1877, p.5, 23.}, 
one \dashuline{identifies force and acceleration right from the start} through the definition (although the concept of force then loses meaning, since the sensations of the muscular sense are not affected is taken into consideration). 

\dashuline{The work, the energy}, etc. \dashuline{are then derived from the force}. 

In contrast to this is the other view, first cultivated 
by \dashuline{Huygens}$\,$\footnote{$\:${\it\color{red}See} E. Mach: Zur Geschichte des Arbeitbegriffes {\it\color{red}(On the history of the concept of work)}. Wien. Ber. (2) 68, p.479, 1873.}, 
which places \dashuline{the concept of energy (work, living force {\it\color{red} - or twice the kinetic energy}) at the 
head of mechanics}$\,${\color{red}\footnote{$\:${\it\color{red}Note that this vision of Ch. Huygens will be in agreement with (and at the founding of) the mass-energy equivalence of Einstein and the energy-tensor $T_{\mu\nu}$ as a source in the r-h-s of the Einstein equation: 
$R_{\mu\nu} +[\;\Lambda - R/2\;]\;g_{\mu\nu} = (8\:\pi\:G/c^4)\;T_{\mu\nu}$, and thus following certain of the visions of Ernst Mach (P. Marquet)}.}}
and relegates the other basic concepts, namely that of \dashuline{force, to a secondary  position}.

\vspace*{-1mm} 
\begin{center}
{\it\color{red}\bf
---------
(p.150) 
---------
}
\end{center}
\vspace*{-3mm}

Obviously, the latter point of view has the advantage that \dashuline{the concept of energy} (which characterises it) is a quantity defined for all the different branches of physics, so that not only mechanics but also the theory of heat and electricity can be based on the same concept, which undoubtedly \dashuline{gives rise to a more uniform, higher conception of physical phenomena}.
I also believe that, sooner or later, this conception will become widespread as soon as we have become more accustomed to this relatively underused {\color{red}(insufficiently used)} term through repeated 
practice.$\,${\color{red}\footnote{$\:${\it\color{red}An indeed, Energy will be one of the fundamental basic quantity for the next relativistic (momentum–energy quadri-vector and mass-energy tensor) and quantum (Hamiltonian) mechanics (P. Marquet)}.}}

\dashuline{On the other hand, however}, it should be noted that \dashuline{the concept of force} (on which mechanics has been exclusively based since Newton) \dashuline{has an advantage that is lacking in that of energy}: it is the fact that \dashuline{we possess a sense} (the muscular sense) \dashuline{through which we can directly feel a pressure} (although not exactly measure it, but certainly feel it), 
\dashuline{whereas we have no sense of energy at all}$\,${\color{red}\footnote{$\:${\it\color{red}Note that the same remarks has been made about my recent studies (since 2011) of the moist-air entropy and energy: most of people consider that ``\,only differences in energy and in entropy have a physical meaning,\,'' (with therefore all reference values undetermined and at our disposal independently for all substances) and that ``\,only \dashuline{temperature} is an observable quantity\,'' (it can be directly felt by our sense) and \dashuline{not the energy nor the entropy}... (P. Marquet)}}}
(cf. p.153 below with regard to heat).
This circumstance has also undoubtedly been one of the reasons why, in the course of the historical development of mechanics, the concept of force has been able to push that of work into the background and, for its part, to gain the decisive preponderance.

{\it\color{red}Accordingly}, \dashuline{the force appears to us} (at least in Newton's conception) \dashuline{as the primary thing}, as the cause, \dashuline{but the movement} (or the work performance, etc) \dashuline{as the effect} (even though force and acceleration coincide in time), and that for no other reason (as because when we move a body by muscular action) the physiological process within us actually precedes the movement that occurs. 

{\it\color{red}Similarly}, \dashuline{if a body starts moving independently of our muscular activity} (for example through the attraction of another) \dashuline{we can always imagine that} (after removing the attractive body) \dashuline{we bring about the same acceleration through our own effort}, and in this respect also in this body (in a very specific sense) \dashuline{we can speak of a force that causes this movement}. 
The fact that we can only obtain a quantitative measure of this force by observing the movement that has occurred is only due to the \dashuline{imperfection of our muscular sense and does not change the concept 
of force}.{\color{red}\footnote{$\:${\it\color{red}Note, however,  that this concept of force will disappear and become meaningless with the General Relativity and for the impact of gravitation, to arrive at the concept of ``\,free motion of bodies within a curved space-time.\,'' (P. Marquet)}}}

\vspace*{7mm} 
\begin{center}
{\it\color{red}\bf
---------
(p.151) 
---------
}
\end{center}
\vspace*{-3mm}

Since, after what has been said and in accordance with historical development, we see the essential meaning of \dashuline{the concept of force} in \dashuline{its connection with the sensations conveyed to us by the muscular sense}, \dashuline{we cannot decide to agree with Kirchhoff on the concept of force} (by abolishing this connection and \dashuline{to 
impress 
a purely kinematic one)}. 

{\it\color{red}Nevertheless,} it must certainly be admitted that \dashuline{a large part of mechanics can be constructed from the concept of acceleration alone}, especially all of \dashuline{astronomy}, and in general \dashuline{all those movement processes that are only perceived by the eye} {\it\color{red}(i.e. by opposition with the ``\,muscles\,'' and without the concept of force / P. Marquet)}.

However, 
\dashuline{physics has to do with the description of all phenomena} 
(and not only those which are conveyed to us by the sense of motion, or 
by the sense of muscle, the sense of temperature, the sense of colour, etc.), \dashuline{and accordingly the fundamental physical concepts are to be derived directly from specific sensations}.

\dashuline{We cannot precisely measure a temperature using the} ``\,\dashuline{temperature sense}\,'' 
({\it\color{red}i.e.} in the same way as 
we can measure a force using the muscle sense, or a color nuance using the color sense), {\it\color{red}simply} because the sharpness of \dashuline{our sensory perceptions is not sufficient for this}.
In order to achieve this purpose we have to \dashuline{look for other phenomena} that experience has shown to be \dashuline{in a necessary connection with the aforementioned sensations}, and offer the advantage of \dashuline{a quantitative measurement}.
These are usually phenomena of movement (in temperature, expansion, force, acceleration, color, wavelength, etc).
However, that does not mean we will feel compelled to explain temperature, force, color, etc. in terms of kinematics. 

Just as \dashuline{we do not primarily associate the word ``\,blue\,'' with the mechanical idea of a certain number of oscillations} (or a certain wavelength of the ether, which alone gives us the exact physical measure of the color), we should do the same when we talk about the ``\,\dashuline{attractive force}.\,'' 
{\it\color{red}Similarly}, when talking about \dashuline{a magnet on a piece of iron}, first of all don't think about the acceleration (multiplied by the mass) that the magnet gives to the iron, but rather about \dashuline{the pressure sensation that we feel in the muscles}, although not suitable for exact measurement (after removing the magnet, we give the iron the same acceleration 
through our own action).{\color{red}\footnote{$\:${\it\color{red}We see here how scientists could hope to explain physical observations with physiological analogies, at this time preceding the revolutions in physics that were the conceptions of Special and General Relativities (and the abandonment of the concepts of luminiferous ether and force of gravitation) and Quantum Mechanics (and the introduction of quanta and discrete values for energy exchanges) / P. Marquet.}}}

\vspace*{-2mm} 
\begin{center}
{\it\color{red}\bf
---------
(p.152) 
---------
}
\end{center}
\vspace*{-3mm}

In his admirable presentation of mechanics, \dashuline{Kirchhoff} \dashuline{reduced the concept of causal connection to} what it really means, namely \dashuline{the necessity of temporal succession}. Unfortunately, he also believed that he had \dashuline{to remove the sensations conveyed by the sense of muscle from the foundation of mechanics}, although they help us to arrive at physical concepts with exactly the same right as the sensations of the sense of the visual sense, 
which is, however, relatively sharper.

For the present state of mechanics, however, it may essentially amount to the same thing \dashuline{whether the concept of force} is connected from the first with \dashuline{the sensations of the muscular sense}, or whether this connection is \dashuline{only introduced afterwards} (for that it must be introduced at all is self-evident, if only in consideration of the theory of the first and oldest machines : those which are moved by muscular power).

But mechanics, like any other branch of physics, is not a closed science, even if it stands at a relatively high level: while the facts once observed remain, and are continually supplemented and completed by new ones, the views can change in often unexpected ways.

It is recognized that the only fixed and 
invulnerable 
starting point for us lies in \dashuline{the phenomena 
  supplied by the senses},
and it would therefore be highly rational \dashuline{to keep the use of all our senses at our disposal at all times}, and not to rely on the use of one of them from the outset which, as the name suggests, 
has rendered and will presumably continue to render the most important services to science for the \dashuline{understanding and development of the concept of force}.

\vspace*{0mm} 
\begin{center}
{\it\color{red}\bf
---------
(p.153) 
---------
}
\end{center}
\vspace*{-3mm}

If we now return to the question raised above, whether it is more expedient for the presentation of mechanics to \dashuline{derive the law of proportionality of force and acceleration from the principle of conservation of energy}, \dashuline{or vice versa}, we would like to repeat what we have already said above for the former opinion. 

On the other hand, in view of the immediacy of the \dashuline{concept of force} just discussed, it seems to us essential to \dashuline{first base the concept of work on that of force} (completely independent of acceleration), \dashuline{and then}, after formulating this concept, to \dashuline{apply the law of conservation of energy}, 
which then results in the \dashuline{proportionality of force and acceleration}. 

 We will not attempt to explain in detail how this should be done, especially as there is a certain amount of room for individual taste here. 
We would only like to emphasize one point on which we place particular emphasis, namely that the view presented here follows very closely the method which is already used with great success in other parts of physics, particularly in the theory of heat. 

We \dashuline{derive the effects of heat} from the \dashuline{principle of conservation of energy}, or the \dashuline{theorem of the equivalence of heat and work}, and yet in the theory of heat the concept of the amount of heat is not what is originally given, since we have no specific meaning for it (just like for work), but we only get to it  \dashuline{through the concept of temperature}, which is delivered to us directly through the sense of temperature (just like the concept of force through the muscle sense, or, as we can also say: force sense). 

In this respect, therefore, we have (curiously enough) already arrived at somewhat more mature views in the theory of heat than in mechanics, because in time we shall have to arrive at the conclusion that in mechanics, too, \dashuline{we must understand energy as what is primary existent}, \dashuline{and of force as an expression of this energy} (either potential or actual), just \dashuline{as we now already regard temperature as an expression of heat}.

Where there is no energy, neither a force, nor a temperature, nor any other sensation can arise. 

\vspace*{-3mm} 
\begin{center}
{\it\color{red}\bf
---------
(p.154) 
---------
}
\end{center}
\vspace*{-3mm}


Before we depart from the treatment of the movement of a single free material point, we would like to cite one more theorem that will be fruitful for further application. 

If a material point passes from a state of rest to a state of movement through the action of some force, then in the first moment of time of this movement the total work of the forces at work is always positive. 

Because the beginning of the movement is always associated with a growth in the energy of the point (its living force {\it\color{red}- kinetic energy}), the external effects which cause this growth must also be positive (the same sentence naturally follows directly from taking into account the fact that the  displacement of the point caused by the acting forces falls in the direction of the resulting force). 

If it happens that the point returns to its starting point in the course of its movement, then in general the total work of the forces will not be equal to $0$, and as a result the point will not have regained its old velocity. 

But this is always the case when the effects are caused by central forces emanating from certain resting masses, where there exists a potential $V$ of the acting forces such that: 
$$
X \;=\; -\:\frac{\partial V}{\partial x} 
\, , \;\;\;\;\;\;
Y \;=\; -\:\frac{\partial V}{\partial y} 
\, , \;\;\;\;\;\;
Z \;=\; -\:\frac{\partial V}{\partial z} 
\, . 
$$
In this case, the amount of work done in any finite {\it\color{red}interval of\,} time is measured simply by the decrease in this potential, whatever path the point follows. 

The last derived theorem reads as follows: {\it if a point at rest begins to move under the influence of forces that have a potential that only depends on the position of the point, this always happens in such a way that the potential decreases}.

\vspace*{-3mm} 
\begin{center}
{\it\color{red}\bf
---------
(p.155) 
---------
}
\end{center}
\vspace*{-3mm}

If we now move on to consider a material point whose mobility is limited by certain external conditions that have been determined in advance, we can initially distinguish between two types of forces:
\begin{enumerate}[leftmargin=10mm,parsep=0mm,itemsep=1mm,topsep=-1mm,rightmargin=2mm]
\item 
\!\!\!) those that strive to move the point in a certain way (in the following we will name the driving forces), their size and direction are generally immediately known; and 
\item 
\!\!\!) those which are caused by the existence of the fixed conditions (we will call them resistance forces). 
\end{enumerate}
These forces are only characterized by the fact that their effectiveness always succeeds in maintaining the fixed conditions under all circumstances. 
Both types of forces together determine the movement of the point according to the general laws of motion that apply to a free point. 

The indeterminacy that is still contained in the values of the resistance forces can be eliminated by the following theorem, which is based on the decomposition of the collective work into that of the individual forces: {\it if the realization and maintenance of the fixed conditions (to be achieved by any mechanical devices) is neither associated with effort nor with the production of energy, the work of the resistance forces at the material point under consideration is always $0$ (because in this case no energy can be imparted to the point through the effect of the fixed conditions, otherwise this energy would arise from nothing)}. 

This always occurs when the conditions do not depend on time (e.g.   when the point is forced to remain on a surface or curve that is fixed in space). 
Therefore, a point on which no driving forces act will move on a fixed surface (or curve) at a constant velocity.

\vspace*{-3mm} 
\begin{center}
--------------------------------------------------- 
\end{center}
\vspace*{-3mm}

If we now assume, while maintaining the assumed case, that the point is initially at rest, but begins to move as a result of the action of certain driving forces, then according to the theorem derived above, the total work of all the forces acting on the point must be positive. 

But since, as we just saw, the work of the resistance forces is equal to $0$, it follows, because the total work of all forces is the sum of the work of the individual ones, that: ``\,with every movement that occurs, the work of the driving forces per see only is positive.\,''
In other words: the direction of the resultant of the driving forces forms an acute angle with the direction of the movement that the point takes. 

\vspace*{-2mm} 
\begin{center}
{\it\color{red}\bf
---------
(p.156) 
---------
}
\end{center}
\vspace*{-3mm}

For the special case that the driving forces have a potential, it follows that at the beginning the potential of the movement decreases. 


From this follows directly the theorem of virtual displacements: 
{\it if among all the displacements which the point can undergo in consequence of the fixed conditions, there is not a single one for which the work of the driving forces is positive, no motion at all can come about, then equilibrium must exist}
(because if this were not the case, motion would occur, then for the resulting displacement the work of the driving forces would be $0$ or negative, which is incompatible with the above theorem). 

If $X, Y, Z$ are again the components of the resultant of the driving forces, then equilibrium exists if the {\it\color{red}(following)} condition applies for each permissible virtual displacement $\delta x, \delta y, \delta z$: 
$$
X\;\delta x \;+\; Y\;\delta y \;+\; Z\;\delta z 
\;\leq\; 0 \; .
$$

\vspace*{-3mm} 
\begin{center}
--------------------------------------------------- 
\end{center}
\vspace*{-3mm}

For the usual case where the fixed conditions are all expressed by equations (not by inequalities) between the coordinates of the moving point, if any displacement $\delta x,\: \delta y,\: \delta z$ is compatible with the conditions, it will always be the opposite one: $-\delta x,\: -\delta y,\: -\delta z$, must be permissible, so that the condition required for equilibrium is only fulfilled if for all permissible displacements: $X\:\delta x + Y\:\delta y +Z\:\delta z = 0$. 

If the driving forces have a potential, the equation is: $\delta V = 0$. 

This condition is always fulfilled when the value of the potential for the relevant point in space is a maximum or a minimum (it is immediately obvious that in the first case the equilibrium is unstable, and in the second case it is stable). 

If you move the material point to a place a little away from its equilibrium position, it will no longer be in equilibrium, but will start to move in such a way that the potential decreases. 

So if the potential is a minimum in the equilibrium position, it must return there (in the opposite case this is impossible). 

In between, there are cases in which the equilibrium is unstable for certain shifts but stable for others (then the value of the potential reaches neither a maximum nor a minimum).

\vspace*{-3mm} 
\begin{center}
{\it\color{red}\bf
---------
(p.157) 
---------
}
\end{center}
\vspace*{-3mm}


  We have deduced all these conclusions with a certain amount of awkwardness, which could have been shortened somewhat by using certain simple theorems (e.g. that the drag force of a surface or curve always acts perpendicular to its direction), but in return we have gained the advantage of transferring the applied consideration essentially unchanged to systems of any number of material points (see below). 

As is well known, not only the equilibrium conditions but also the equations of motion of the material point can be derived from the principle of virtual displacements, provided that the quantities 
$$
  -\:m\:\frac{d^2x}{dt^2} \; , \;\;\;\;
  -\:m\:\frac{d^2y}{dt^2} \; , \;\;\;\;
  -\:m\:\frac{d^2z}{dt^2} \; , 
$$ 
are added to the components of the driving forces.

\vspace*{-3mm} 
\begin{center}
--------------------------------------------------- 
\end{center}
\vspace*{-3mm}


We would therefore like to move straight on to the treatment of a system of several moving points (namely two points), in order to derive the validity of the principle of action and counteraction for this system (where all effects from other masses are initially excluded). 

Let us denote the coordinates of a point by $x\,y\,z$, its ``\,live force\,'' by $T$, the components of the force acting on it from the other point by $X\:Y\:Z$, where the attached index 1 or 2 corresponds to the point on which the force is acted.
Then for each individual point the growth of its energy is equal to the work of the force acting on it, i.e.: 
\begin{align}
d\,T_1 \;=\; 
 X_1 \: dx_1 \;+\; Y_1 \: dy_1 \;+\; Z_1 \: dz_1 \; ,
\nonumber \\
d\,T_1 \;=\; 
 X_2 \: dx_2 \;+\; Y_2 \: dy_2 \;+\; Z_2 \: dz_2 \; .
\nonumber
\end{align}
On the other hand, if we assume both points together as a ``\,basic system\,'' (p.121), then the external effects are $0$, and the energy is therefore constant. 

However, this quantity is of course not generally determined solely from the living forces (the ``\,actual\,'' {\it\color{red}kinetic} energy) of the two points, but an additional term will be added which also depends on the position of the points in space, and will be placed next to the other as a new type of energy. 

\vspace*{6mm} 
\begin{center}
{\it\color{red}\bf
---------
(p.158) 
---------
}
\end{center}
\vspace*{-3mm}

If we designate this type of energy (the ``\,potential\,'' energy) with $U$, then we have $T_1 + T_2 + U = const.$, and therefore:  
\vspace*{-3mm}
\begin{align}
d\,T_1 \;+\; d\,T_2 \;+\; dU \;=\; 0 \; .
\nonumber 
\end{align}
Taken together, the two equations above give: 
\begin{align}
 X_1 \: dx_1 \;+\; Y_1 \: dy_1 \;+\; Z_1 \: dz_1
 \;+\;  
 X_2 \: dx_2 \;+\; Y_2 \: dy_2 \;+\; Z_2 \: dz_2 
\;=\; -\: dU \; .
\nonumber
\end{align}
The total work of the acting forces therefore forms the complete time differential of a function that only depends on the instantaneous state (position and velocity) of the two points, and this condition is suitable for deriving certain necessary properties of the forces.

\vspace*{-3mm} 
\begin{center}
--------------------------------------------------- 
\end{center}
\vspace*{-3mm}


If we first let $U$ depend not only on the position but also on the velocity of the two points, then $dU/dt$ would also contain the acceleration, which follows from the last equation that the force components $X\:Y\:Z$ also would have to depend on the acceleration. 

This assumption is actually carried out in the basic laws that 
W. Weber$\,$\footnote{$\:$W. Weber: Elektrodynamische Maassbestimmungen {\it\color{red}(Electrodynamic measurements)}, Abh. d. k. sächs. Ges. d. Wiss. X, p.1, 1871. Vgl. auch Pogg. Ann. Jubelband, p.212, 1874.}, 
B. Riemann$\,$\footnote{$\:$B. Riemann: Schwere, Elektricität und Magnetismus {\it\color{red}(Gravity, Electricity and Magnetism)}, bearb. v. Hattendorff, Hannover 1876, p.326.} 
and 
R. Clausius$\,$\footnote{$\:$R. Clausius: Über ein neues Grundgesetz der Elektrodynamik {\it\color{red}(On a new basic law of electrodynamic)}. Pogg. Ann. 156, p.657, 1875. Crelle J.82, p.85, 1876. Die mechanische Behandlung der Elektricität {\it\color{red}(The mechanical treatment of electricity)}. Braunschweig 1879, p.277.} 
established for the effect of two electrical points on one another. 

However, this makes both the idea of how the forces work and the calculation itself much more complicated, and since the assumption of these laws does not seem necessary at all (and there are other specific reasons against each of them) we do not want to elaborate on the assumed case.

\vspace*{-3mm} 
\begin{center}
{\it\color{red}\bf
---------
(p.159) 
---------
}
\end{center}
\vspace*{-3mm}


Then there is nothing left but to assume that $U$ depends only on the position of the two acting points and, as we can immediately add, on their distance $r$, since this is the only physical quantity which is determined by the Location of the two points is fully defined. 
So we have: 
\begin{align}
dU &\:=\; 
 \frac{\partial U}{\partial x_1} \: dx_1 
 \;+\; 
 \frac{\partial U}{\partial y_1} \: dy_1 
 \;+\; 
 \frac{\partial U}{\partial z_1} \: dz_1 
\nonumber \\
  & \;+\; 
 \frac{\partial U}{\partial x_2} \: dx_2 
 \;+\; 
 \frac{\partial U}{\partial y_2} \: dy_2 
 \;+\; 
 \frac{\partial U}{\partial z_2} \: dz_2 
 \; ,
\nonumber
\end{align}
where 
$$ U \;=\; f(r) \; . $$

\vspace*{-3mm} 
\begin{center}
--------------------------------------------------- 
\end{center}
\vspace*{-3mm}

Even if the members into which the differential $dU$ is divided appear to be assigned individually to the quantities that make up the above expression of labour ($=-\,dU$), one is not yet entitled to equate $2$ corresponding members with each other, i.e. to make 
\begin{align}
X_1 &\:=\; -\,\frac{\partial  U  }{\partial x_1} 
     \;=\; -\,\frac{\partial f(r)}{\partial x_1} 
 \; , \;\;\;
Y_1 \;=\; -\,\frac{\partial  U  }{\partial y_1} 
     \;=\; -\,\frac{\partial f(r)}{\partial y_1} 
 \; , 
 \;\;\;\mbox{(and so on . . .)} \; 
\nonumber
\end{align}

\vspace*{-3mm} 
\begin{center}
--------------------------------------------------- 
\end{center}
\vspace*{-3mm}

\noindent
This would only be necessary if: 
\begin{enumerate}[leftmargin=8mm,parsep=0mm,itemsep=1mm,topsep=-1mm,rightmargin=2mm]
\item 
\!\!\!) the differentials $dx_1, dy_1 \: . \: . \: .$ would be completely independent of each other; and at the same time 
\item 
\!\!\!) the quantities $X_1 \: Y_1 \: . \: . \: .$ would be independent of these differentials (i.e. of the velocities);
\end{enumerate}
because if one of these two conditions is not fulfilled, it is always possible to find quantities different from $0$ which can be added to the values of the components 
$X_1 \: Y_1 \: . \: . \: .$ 
without changing the value of the work 
$X_1 \: dx_1 + Y_1 \: dy_1 + \: . \: . \: .$, 
and also that of $dU$. 

These ``\,additional\,'' forces therefore have the property that the work they do is equal to $0$. There are actually different names of such forces, and they fall into two separate types according to the two conditions stated. 

The former, which arise from a dependence of the co-ordinates on each other, are due to the existence of fixed conditions between the two acting points.
They play an important part in mechanics on account of the convenience of their mathematical treatment, and we shall return to them later in more detail.

\vspace*{-2mm} 
\begin{center}
{\it\color{red}\bf
---------
(p.160) 
---------
}
\end{center}
\vspace*{-3mm}

However, 
here we wish to emphasise the special thing that these kinds of forces (we have called them forces of resistance above) have absolutely no primary existence in nature. 

Indeed, they can always ultimately be resolved into forces that are given independently by the state of the points, 
because every fixed condition can only be produced in nature by certain mechanical means (i.e. by a suitable grouping of suitable bodies) and by resolving these bodies into their individual points (we break down the forces of resistance into their elements, which are all represented by ``\,driving\,'' forces, see p.154).

Ultimately, each point must be viewed as freely moving.

\vspace*{-3mm} 
\begin{center}
--------------------------------------------------- 
\end{center}
\vspace*{-3mm}

Therefore, all that remains to be discussed is the other type of additional forces: forces that depend on the velocities of the two (free) points in such a way that their total work at both points is always $0$.
We call their components 
$X'_1, \:\:Y'_1, \:\:Z'_1, \:\:X'_2, \:\:Y'_2, \:\:Z'_2$. 
We can safely assume that the magnitude of these forces does not depend on the absolute coordinates and velocities, but only on the relative values, since the former have no physical meaning at all. If the work of these forces: 
\begin{align}
 X'_1 \: dx_1 \;+\; Y'_1 \: dy_1 \;+\; Z'_1 \: dz_1
 \;+\;  
 X'_2 \: dx_2 \;+\; Y'_2 \: dy_2 \;+\; Z'_2 \: dz_2 
\nonumber
\end{align}
is to disappear identically, then, as can be easily shown, we have: 
$$
  X'_1 \;=\; -\:X'_2  \; , \;\;\;\;
  Y'_1 \;=\; -\:Y'_2   \; , \;\;\;\;
  Z'_1 \;=\; -\:Z'_2   \; , 
$$ 
i.e. the forces at both points are the same in magnitude and opposite in direction (which is arbitrary). 
If we further use the abbreviation 
$$
  x_1 \:-\: x_2 \;=\; x  \; , \;\;\;\;
  y_1 \:-\: y_2 \;=\; y   \; , \;\;\;\;
  z_1 \:-\: z_2 \;=\; z   \; , 
$$ 
then the condition 
\begin{align}
 X'_1 \: dx \;+\; Y'_1 \: dy \;+\; Z'_1 \: dz \;=\;  0
\nonumber
\end{align}
still has to be satisfied.

\vspace*{8mm} 
\begin{center}
{\it\color{red}\bf
---------
(p.161) 
---------
}
\end{center}
\vspace*{-3mm}

The general solution to this equation is: 
\vspace*{1mm} 
\begin{equation}
\left.
\begin{aligned}
  X'_1 & \: = \;
  Q \:\: \frac{dz}{dt} 
  \; - \;
  R \:\: \frac{dy}{dt} 
  \; , \vspace*{2mm} 
\\
  Y'_1 & \: = \;
  R \:\: \frac{dx}{dt} 
  \; - \;
  P \:\: \frac{dz}{dt} 
  \; , \vspace*{2mm} 
\\
  Z'_1 & \: = \;
  P \:\: \frac{dy}{dt} 
  \; - \;
  Q \:\: \frac{dx}{dt} 
  \; , \vspace*{2mm} 
\end{aligned}
\;\;\;\;
\right\} 
\nonumber 
\end{equation}
where $P$, $Q$, $R$ represent arbitrary functions of the (relative) coordinates and velocities.
Lipschitz$\,$\footnote{$\:$H. v. Helmholtz: Wiss. Abh. I, p.70.}
has drawn attention to the \dashuline{compatibility of such forces with the principle of conservation of energy}. 
If we specialize the ideas a little more in order to gain a clearer idea of the nature of these forces, we would first have to satisfy the condition that, when the coordinate system is rotated around the starting point, there is no change in the dependence of the force components on the coordinates, and their differential quotient occurs (as we have already assumed for a parallel shift of the coordinate axes). 
This condition is satisfied if we set: 
$$
  P \;=\; \frac{\partial \rho}{\partial x} \; , \;\;\;\;
  Q \;=\; \frac{\partial \rho}{\partial y}    \; , \;\;\;\;
  R \;=\; \frac{\partial \rho}{\partial z}    \; , 
$$ 
where $\rho$ represents any function of $r$. 
If $\rho$ is taken to be inversely proportional to the distance $r$, the force defined by this changes to that which a current element at rest exerts on a north pole at rest according to Ampère, provided that the current components are assumed to be proportional to the quantities $dx/dt, dy/dt, dz/dt$.
Such a force is therefore quite conceivable, but as it acts according to a less simple law than the central forces, it would only be ascribed a physical existence if it should be shown that certain phenomena of motion observed in nature could not come about without it (this has not yet been the case).

\vspace*{-3mm} 
\begin{center}
{\it\color{red}\bf
---------
(p.162) 
---------
}
\end{center}
\vspace*{-3mm}

One should not cite the above-mentioned interaction between current elements and magnetic poles as proof of the existence of such forces, because those effects mean nothing more than a brief and convenient summary of the forces which closed currents and complete magnets exert on one another.

A completely isolated current element has no physical existence at all, as there is always an external force to maintain the current. 
Only then would the introduction of the forces in question be justified and necessary if, in the interaction of two or more points or bodies completely isolated from external influences, a phenomenon could be observed that would be peculiar to those forces. 

Such a phenomenon would easily reveal itself: although the forces in question satisfy the \dashuline{theorem of the conservation of the movement of the center of gravity} (because they each act on two points of the same size and in opposite directions), they contradict the \dashuline{theorem of the conservation of surfaces} (because their directions do not coincide with the line connecting the points, but are perpendicular to it), and therefore provide a rotational moment. 

So, while the living force {\it\color{red}(kinetic energy)} of movement remains constant, the sum of the moments of the movement quantities in relation to a fixed axis is constantly changing. 
Therefore, any observed deviation of the movement of any point system that is not subject to external effects from the law of surfaces would lead to the assumption of the forces discussed here.

\vspace*{-3mm} 
\begin{center}
--------------------------------------------------- 
\end{center}
\vspace*{-3mm}

Since the direction of these forces does not fall in that of the line connecting the two points between which they act, they also \dashuline{contradict the principle of action and counteraction}, and therefore become invalid as soon as this principle is assumed to be generally valid. In this way 
Helmholtz$\,$\footnote{$\:$H. v. Helmholtz: Wiss. Abh. I, p.70.}
got rid of them.

However, for the sake of uniformity of treatment, we do not want to present \dashuline{the principle of effect and counter-effect as} given here, but rather explain under what conditions it turns out to be \dashuline{a consequence of the energy principle} that we apply everywhere. 

\vspace*{-2mm} 
\begin{center}
{\it\color{red}\bf
---------
(p.163) 
---------
}
\end{center}
\vspace*{-3mm}

If we start from the above equation of conservation of energy: 
\begin{align}
 X_1 \: dx_1 \;+\; Y_1 \: dy_1 \;+\; Z_1 \: dz_1
 \;+\;  
 X_2 \: dx_2 \;+\; Y_2 \: dy_2 \;+\; Z_2 \: dz_2 
 \;=\; -\: dU \; ,
\nonumber
\end{align}
where 
$$ U \;=\; f(r) \; , $$
i.e. 
\vspace*{-3mm} 
\begin{align}
dU &\:=\; 
 \frac{\partial U}{\partial x_1} \: 
 d\left(x_1\:-\:x_2\right) 
 \;+\; 
 \frac{\partial U}{\partial y_1} \: 
 d\left(y_1\:-\:y_2\right) 
 \;+\; 
 \frac{\partial U}{\partial z_1} \: 
 d\left(z_1\:-\:z_2\right) 
 \; ,
\nonumber
\end{align}
we can immediately assume that the force components only depend on the relative coordinates $x_1-x_2$, $y_1-y_2$, $z_1-z_2$, because their size is not changed by a parallel shift of the coordinate axes. Then it becomes necessary that: 
\vspace*{-3mm} 
\begin{align}
  X_1 \:+\: X_2 \;=\; 0 \; , \;\;\;\;
  Y_1 \:+\: Y_2 \;=\; 0 \; , \;\;\;\;
  Z_1 \:+\: Z_2 \;=\; 0 \; , 
\nonumber
\end{align}
\vspace*{-3mm} 
and 
\vspace*{-3mm} 
\begin{align}
 X_1 \: d\left(x_1\:-\:x_2\right) 
 \;+\;  
 Y_1 \: d\left(y_1\:-\:y_2\right) 
 \;+\;  
 Z_1 \: d\left(z_1\:-\:z_2\right) 
 \;=\; -\: dU \; .
\nonumber
\end{align}

\vspace*{-5mm} 
\begin{center}
--------------------------------------------------- 
\end{center}
\vspace*{-3mm}

If we now equate not only the two whole expressions for $dU$, but also the two corresponding members of these expressions, we get the result: 
$$
 X_1 \;=\; -\: \frac{\partial U}{\partial x_1} \: 
     \;=\; -\: \frac{\partial f(r)}{\partial r}
           \:.\: \frac{x_1\,-\,x_2}{r}
 \; , \;\; \mbox{etc.} \;\; ,
$$ 
i.e. the forces emanating from both points are equal and opposite to each other, and their directions coincide with the connecting line. 

The \dashuline{principle of action and counteraction} (Newton's 3rd axiom) follows from the \dashuline{principle of conservation of energy} with the help of the assumption that, not only the total work of the forces acting between two points expresses the \dashuline{change in their potential energy}, but also that each individual one of the parts related to the 3 coordinate axes (of which the total work is composed) measures the \dashuline{increase in energy} corresponding to the axis in question. 

Once again, this is nothing other than an application of the \dashuline{principle of the superposition of energies}, which has the meaning of bringing together a series of externally different sentences \dashuline{under a common point of view} (see the comments on pp.147 and p.130).

\vspace*{-3mm} 
\begin{center}
{\it\color{red}\bf
---------
(p.164) 
---------
}
\end{center}
\vspace*{-3mm}

We have thus reduced the forces acting between two points to central forces, which have a potential $U$ that only depends on the distance, where $U$ also expresses the value of the potential energy of the two points. 

So far, not a single phenomenon has been discovered in nature that contradicts the assumption that all forces can ultimately be traced back to such central forces (which, in particular, are independent of the velocities). 

No objection can be derived from the dependence of the electrodynamic effects on the current intensity (and even on its differential quotient over time), just as there is no objection to this from the fact that the pressure of a gas depends on the temperature, and therefore also on the living force {\it\color{red}(kinetic energy)} of the internal gas movements. 

Just as we learned a few decades ago to explain the latter force through the laws of elastic collision (which in turn can certainly be reduced to the effects of central forces), there is also a well-founded prospect of resolving the electrodynamic phenomena in a similar way. One only has to imagine the electrodynamic forces not as being primarily determined by the mere presence of flowing electricity, but rather as arising from a peculiar (as yet unknown) arrangement of the active centers, which itself is only a consequence of the activity of the current.

\vspace*{-3mm} 
\begin{center}
--------------------------------------------------- 
\end{center}
\vspace*{-3mm}

If we now move on to consider a system of several points (in a finite number) that act on one another at finite distances, we imagine the forces occurring here broken down into those that only act between two specific points, and thereby list all the effects Central forces back. 

The fact that such a decomposition is possible at all, and in particular that the magnitude and direction of the forces acting between two points are not influenced at all by the effects emanating from other points, is by no means self-evident, but is again a case of the application of \dashuline{the principle of superimposition of the effects}, which has repeatedly provided us with the means to utilize \dashuline{the principle of conservation of energy}. 

\vspace*{-2mm} 
\begin{center}
{\it\color{red}\bf
---------
(p.165) 
---------
}
\end{center}
\vspace*{-3mm}


This also explains why \dashuline{the theorem of the equality of action and counteraction} also applies to forces which do not appear directly in the form of central forces, such as friction, inelastic impact, etc. 

Such forces must always be regarded as composed of a number of central forces, and since the latter individually satisfy the said theorem, the resultants must also do so.

\vspace*{-3mm} 
\begin{center}
--------------------------------------------------- 
\end{center}
\vspace*{-3mm}

If we now have such a system of freely moving points on which no external effects are exerted (and whose energy is therefore constant), the components of the resultant of all forces acting on a point $x\:y\:z$ can be put into the form: 
$$
 X \;=\; -\: \frac{\partial U}{\partial x} 
 \;\; , \;\;\;\;\;\;\;\;
 Y \;=\; -\: \frac{\partial U}{\partial y} 
 \;\; , \;\;\;\;\;\;\;\;
 Z \;=\; -\: \frac{\partial U}{\partial z} 
 \;\; .
$$ 
Here, $U$ is the potential of the acting forces, which is formed by simply adding the potentials of two points to each other. 
For a single point, the statement applies again that the change in its energy (live force {\it\color{red}- kinetic energy}) is equal to the work of the force that acts on it. 
And if you add up all the resulting equations, and use $T$ to denote the sum of all living forces {\it\color{red}(kinetic energy)}, you get: 
$$
d\,T \;=\; 
\sum \: \left( \: X\:dx \:+\: Y\:dy \:+\: Z\:dz \: \right)
\;=\; -\:dU \; ,
$$ 
or: 
$$
  T \;+\; U \;=\; \mbox{const.}
$$

\vspace*{-3mm} 
\begin{center}
--------------------------------------------------- 
\end{center}
\vspace*{-3mm}

The \dashuline{energy of the system} always consists of two parts: the \dashuline{actual (or kinetic) energy} and the \dashuline{potential energy}, the first of which is formed by the \dashuline{sum of the living forces} {\it\color{red}(kinetic energy)}, the second by the \dashuline{potential of the central forces}. 

{\it  If external effects are added to the forces under consideration, the last equation changes to the effect that the change in energy (i.e. the energy transferred to the system from outside) occurring in a certain time is equal to the sum of the work performed by the external forces at all points of the system in the same time.}

This theorem enables one to \dashuline{apply the equation of energy to a complex of points 
arbitrarily selected from the system}.

\vspace*{-5mm} 
\begin{center}
{\it\color{red}\bf
---------
(p.166) 
---------
}
\end{center}
\vspace*{-3mm}

For a system of free points  (as we have assumed so far) the establishment of the equations of motion is just as simple as for a single free point. 

Things become somewhat different when the movements of the points are restricted by fixed conditions, since in this case certain resistance forces are added to the driving forces, the size and direction of which are not directly determined by the position of the points. 

But you can reach your goal here by using a property that is characteristic of all resistance forces. 

  Whatever solid conditions independent of time we may observe in nature (solid surfaces and lines, rigid bodies, completely flexible and inextensible threads and membranes, incompressible liquids), the law holds for all of them that: \dashuline{the total work of the resistive forces arising from them is always equal to $0$} (for the reason that the maintenance of these conditions is associated neither with effort nor with the generation of energy).

Because since neither the work of an external force is necessary to maintain the efficacy of the said compounds (nor do the bodies constituting the compounds undergo any internal change in any movement), \dashuline{no work or living force} {\it\color{red}(kinetic energy)} \dashuline{can arise from the effects of the resistive forces corresponding to the conditions in their totality} (otherwise the latter \dashuline{would have arisen from nothing}).

  However, as soon as one of the two circumstances mentioned ceases to exist, the arguments we have put forward cease to be relevant.

\vspace*{-3mm} 
\begin{center}
--------------------------------------------------- 
\end{center}
\vspace*{-3mm}

For example, the maintenance of fixed conditions generally requires the work of an external force whenever they are dependent on time (e.g. when a point is forced to remain on a surface that moves in a certain given way).
In this case, the resisting force of the surface will perform work at the point that is different from $0$: exactly the amount of work required to maintain the surface in its motion.

\vspace*{-2mm} 
\begin{center}
{\it\color{red}\bf
---------
(p.167) 
---------
}
\end{center}
\vspace*{-3mm}

On the other hand, we have to note an internal change in the bodies forming the fixed condition, for example when a point moves on a frictional surface (here the force emanating from the surface also performs work, but the surface does not remain unchanged, but is heated, electrified, etc). 

If we exclude such cases (which require special treatment) from consideration, we can generally state the theorem: {\it the total work of all resistive forces is $=0$}.

Of course, \dashuline{the resistance forces} can do work at individual points (such as the tension of an inextensible thread), i.e.: they \dashuline{can transfer energy from one point to another without changing its total amount}.

\vspace*{-3mm} 
\begin{center}
--------------------------------------------------- 
\end{center}
\vspace*{-3mm}

From this theorem, for a system of points that is subject to any fixed conditions and on which no external forces otherwise act, the above equation of conservation of energy follows immediately: 
$$ T \;+\; U \; =\; \mbox{const.} $$ 
completely independent of the fixed conditions.

\vspace*{-3mm} 
\begin{center}
--------------------------------------------------- 
\end{center}
\vspace*{-3mm}

If the system is put in motion from rest by the action of the driving forces, then since $\delta \, T > 0$, then $\delta \, U < 0$, i.e. the movement then always occurs in the sense that the potential energy (the potential of the driving forces) decreases. 

This immediately results in the equilibrium conditions of the system, because obviously no movement can occur if, for all displacements $\delta x , \delta y , \delta z$ that the points can suffer due to the fixed conditions, $\delta\,U \geq 0$, and then the condition necessary for movement to occur is not met. 

If the fixed conditions are such that, for each displacement compatible with the conditions the opposite displacement is also permissible (which is not the case, for example, if a thread is inextensible but not incompressible), the last condition can only be fulfilled by the fact that for all permissible displacements: $ \delta \, U = 0$ or $\sum \: ( \, X \: \delta \, x + Y \: \delta \, y + Z \: \delta \, z \, ) = 0$.

Equilibrium is therefore present in all states of the system for which $U$ is a maximum or a minimum. 

The fact that the first case corresponds to the absolutely unstable equilibrium, and the second to the absolutely stable equilibrium, is immediately obvious and has already been discussed by us for a single point (p. 156).

\vspace*{-5mm} 
\begin{center}
{\it\color{red}\bf
---------
(p.168) 
---------
}
\end{center}
\vspace*{-3mm}


Finally, using d'Alembert's process, we can reduce any state of motion to a state of equilibrium by adding at each point the quantities 
$-\,m\,d^{\,2} x/d\,t^2$, 
$-\,m\,d^{\,2} y/d\,t^2$ and 
$-\,m\,d^{\,2} z/d\,t^2$ 
to the components of the driving force $X, \:Y, \:Z$, respectively. 

We then get from the above theorem of virtual displacements 
$$
 \sum \:\left\{\:
 \left( X \:-\: m\:\frac{d^{\,2}x}{d\,t^{2}} \right)\:\delta\,x
 \:\;+\;
 \left( Y \:-\: m\:\frac{d^{\,2}y}{d\,t^{2}} \right)\:\delta\,y
 \:\;+\;
 \left( Z \:-\: m\:\frac{d^{\,2}z}{d\,t^{2}} \right)\:\delta\,z
 \:\right\}
\;=\; 0
$$ 
and can derive the equations of motion for each individual point in different forms using known methods (preferably developed by Lagrange and Hamilton). 

Since these investigations have essentially mathematical interest, we do not have to discuss them here, but we would like to add a note of fundamental importance before moving on to further tasks.

\vspace*{-3mm} 
\begin{center}
--------------------------------------------------- 
\end{center}
\vspace*{-3mm}

The energy of a system of points presents itself to us as consisting of two types, one of which, $U$  (the potential of the driving forces) only depends on the position, the other, $T$, the sum of the living forces {\it\color{red}(i.e. the kinetic energy)},  only depends on the velocity of the points. 

In the form we use, the values of the two types of energy are given by their primary expressions (see p.111), and they therefore consistently retain their validity and meaning, no matter how differently the fixed conditions may be assumed. 

However, it is often advantageous (using the given conditions) to use other variables instead of the rectangular coordinates of the points to determine the states of the system, and especially those that are independent of one another (which will not generally be the case with the coordinates). 

\vspace*{-2mm} 
\begin{center}
{\it\color{red}\bf
---------
(p.169) 
---------
}
\end{center}
\vspace*{-3mm}

  If, for example, we have $n$ points, i.e. $3\,n$ coordinates, and $m$ fixed conditions, it is often more convenient to attribute the movements of the system to $(3\,n - m)$ independent variables, especially since it is usually the independent variables that appear most directly accessible to natural observation.

As a result of this transformation, the expressions of the types of energy then lose their primary form, and sometimes take on other properties. 
If we call the independent variables $p_1, \: p_2, \:.\:.\:.$ and assume that the values of all $p$ {\it\color{red}(variables)}  determine those of all $x\:y\:z$, then the right-angled coordinates can all be expressed by the $p$ {\it\color{red}(variables)}, but the velocities by the $p$ and $dp/dt$ {\it\color{red}(variables)}  at the same time.

By substitution you then get the values of $U$ and $T$ represented by the new variables. 
But while the potential energy still appears as a function of the variables themselves, the kinetic energy completely changes its character. 
Although it remains a whole square homogeneous function of the differential quotients of the variables over time, in general it no longer only contains the pure squares, but also the products of two such differential quotients, and moreover the coefficients of this function are no longer constant, but dependent of the variables $p$. 

In this form, the kinetic energy ceases to be independent of the position of the points of the system, a circumstance which is of fundamental importance for the application of mechanical principles to heat and electricity (as can be seen from the expression of the kinetic energy of a system of galvanic currents derived by Maxwell from the general mechanical equations, which is a whole quadratic homogeneous function of the velocities of the conductors and the current intensities, while its coefficients depend on the position of the conductors).

\vspace*{-3mm} 
\begin{center}
--------------------------------------------------- 
\end{center}
\vspace*{-3mm}

The discussions we have made so far about mechanical energy only apply directly to the case in which the number of variables on which the state of the material system depends is finite.
But they need to be supplemented as soon as that number increases to infinity, for example as soon as the points under consideration belong to a continuously expanding body, as is the case in most applications. 

\vspace*{-2mm} 
\begin{center}
{\it\color{red}\bf
---------
(p.170) 
---------
}
\end{center}
\vspace*{-3mm}

One could now also transfer the results obtained above directly to the case described by means of a suitable border crossing, but it is far simpler and more interesting for the application of the energy principle to subject the questions that arise here to a special treatment. 

So we now want to deal first with a body that is assumed to be continuous, regardless of whether it is in the solid, liquid or gaseous state of aggregation.
 We can also count the \dashuline{light ether} {\it\color{red}(??)} among the solid bodies.{\color{red}$\,$\footnote{$\:${\it\color{red}Talking about this (luminiferous) ether is a heritage from the late 19th century that Einstein (and the quantum mechanics) will soon make it possible to forget (P. Marquet)}.}}

\vspace*{-3mm} 
\begin{center}
--------------------------------------------------- 
\end{center}
\vspace*{-3mm}


Let us first consider an element of the body and set up the equation for it which expresses the principle of energy. The change in energy corresponding to a particular change in the element's state is equal to the amount of mechanical work (or an equivalent effect) that must be expended outside the element in order to bring about the change in state in some way (p.146). 

In order to apply this equation to the change that the element undergoes in the course of any movement of the body during the time particle $dt$, we first consider the external effects that are capable of bringing about the change in state in question. These can be broken down into different types whose influences are simply superimposed. 

Firstly, we have to take into account the work of the forces that act from outside on the entire mass of the element, according to the type of gravity, and whose magnitude we want to assume as proportional to the mass and as known in advance. 
So, if $\partial\,\tau$ denotes the volume and $\mu$ the density of the element, we have an expression for the work of these forces of the form:
\begin{align}
( \, X \: d\, x \;+\; Y \: d \, y \;+\; Z \: d \, z \, )
\:.\: \mu \:.\: \partial\,\tau \; ,
\label{Eq_Planck_1887_1}
\end{align}
where 
$(X\:Y\:Z)$ are the components of the force acting on the mass unit, $(dx, \:dy, \:dz)$ are the components of the displacement of a material point during the time element $dt$ and whose coordinates at time $t$ are $(x, y, z)$. 

\vspace*{2mm} 
\begin{center}
{\it\color{red}\bf
---------
(p.171) 
---------
}
\end{center}
\vspace*{-3mm}

For the sake of better distinction, we have used a designation for the differential quantities, which we will also use everywhere in the following, namely by using the symbol $d$ for a differential in which time $t$ and some 3 are assumed as independent variables Quantities that characterize a certain material point, while the sign $\partial$ is intended to refer to a differential that is taken with respect to time t and the 3 space coordinates (here $x, y, z$) as independent Variables. 

For instance, the differential quotient $d\mu/dt$ corresponds to the change in time that the density undergoes in a given (moving) material point, whereas the differential quotient $\partial\,\mu/\partial\,t$ corresponds to the change in time which the density suffers at a certain point in space.{\color{red}$\,$\footnote{$\:${\it\color{red}The modern terminology used (for instance)  in fluid mechanics is that the material (or total, or Lagrangian) derivative $d\mu/dt = \partial \mu/\partial t
+ \vec{v}\:.\:\vec{\nabla}(\mu)$ 
is the sum of the local (or Eulerian) derivative $\partial \mu/\partial t$ and of the advection/convective term $\vec{v}\:.\:\vec{\nabla}(\mu)$: we recognize here the same two writings of Planck (and the same meanings) for both $d\mu/dt$ and $\partial \mu/\partial t$, and the same following equation written by Planck with the velocity vector $\vec{v} = (u, \:v, \:w)$ written as $(dx/dt, \:dy/dt, \:dz/dt)$ (P. Marquet)}.}}

Both quantities are related by the equation: 
\begin{align}
 \frac{d\mu}{dt} &\: = \;
 \frac{\partial \mu}{\partial t}
 \;+\; \frac{\partial \mu}{\partial x}\:\;\frac{dx}{dt}
 \;+\; \frac{\partial \mu}{\partial y}\:\;\frac{dy}{dt}
 \;+\; \frac{\partial \mu}{\partial z}\:\;\frac{dz}{dt}
 \; .
\nonumber 
\end{align}

\vspace*{-3mm} 
\begin{center}
--------------------------------------------------- 
\end{center}
\vspace*{-3mm}

However, the work of \dashuline{the forces acting on the mass of the element from a distance} is generally not sufficient to bring about the change in state of the element.
For example, it can never cause it to rotate.

On the other hand, we can always bring about the change in question, at least as far as the mechanical state is concerned, by imagining \dashuline{certain forces that act on the surface of the element from all sides}. 
\dashuline{The amount of work they do} is easy to calculate. 

  If we think of the volume of the element at time $t$ as a right-angled parallelepiped whose edges parallel to the coordinate axes have lengths $\partial\:x, \:\partial\:y, \:\partial\:z$ the total amount of work done in the vicinity of the element by the effect of these compressive forces is obtained by adding up the individual work done on the $6$ lateral faces of the parallelepiped.

The work done on a side surface will be proportional to the size of this surface, so that (for instance) for the side surface which passes through the point $x, y, z$ and is parallel to the $Y\:Z$-plane, the work value is: 
\vspace*{-3mm}
\begin{align}
( \, X_x \: d\, x \;+\; Y_x \: d \, y \;+\; Z_x \: d \, z \, )
\:.\: \partial\,y \:.\: \partial\,z \; ,
\nonumber 
\end{align}

\vspace*{-4mm} 
\begin{center}
{\it\color{red}\bf
---------
(p.172) 
---------
}
\end{center}
\vspace*{-4mm}

\noindent
where $X_x \: Y_x \: Z_x$ denote the components of the force acting from the outside on the unit area of that surface (whose internal normal is represented by the direction designated by the index $x$). 
In this sense, for example, the pressure $X_x$ in a gas is always positive, whereas in a wire stretched 
in the direction of the $X$ axis, it is negative.

\vspace*{-3mm} 
\begin{center}
--------------------------------------------------- 
\end{center}
\vspace*{-3mm}

However, on the opposite side surface, which passes through the point $(x+\partial\:x, \:y,\: z)$, work is done in the same time that is opposite in sign to that above (and only differs in magnitude from it in that $x$ has changed into $x +\partial\:x$), while $y$ and $z$ remain constant.
Therefore, we can inventory the amount of all the work on the pair of surfaces under consideration  {\it\color{red}(\:i.e. $x$ and $x +\partial\:x$)} and on the other two pairs of side surfaces {\it\color{red}(\:i.e. $y$ and $y +\partial\:y$ and $z$ and 
$z +\partial\:z$)}{\it\color{red}, where 
$\partial\:\tau = 
\partial\:x\:.\:\partial\:y\:.\:\partial\:z$ 
is the volume element}\,: 
\vspace*{0mm} 
\begin{equation}
\left.
\begin{aligned}
-\:\frac{\partial}{\partial x}
( \, X_x \: d\, x \;+\; Y_x \: d \, y \;+\; Z_x \: d \, z \, )
 \:.\: \partial\:\tau
  \; , \vspace*{2mm} 
\\
-\:\frac{\partial}{\partial y}
( \, X_y \: d\, x \;+\; Y_y \: d \, y \;+\; Z_y \: d \, z \, )
 \:.\: \partial\:\tau
  \; , \vspace*{2mm} 
\\
-\:\frac{\partial}{\partial z}
( \, X_z \: d\, x \;+\; Y_z \: d \, y \;+\; Z_z \: d \, z \, )
 \:.\: \partial\:\tau
  \; . \vspace*{2mm} 
\end{aligned}
\;\;\;\;
\right\} 
\label{Eq_Planck_1887_2} 
\end{equation}

The sum of the last three expressions {\it\color{red}in (\ref{Eq_Planck_1887_2})}, together with the expression (\ref{Eq_Planck_1887_1}), therefore represents the \dashuline{total mechanical work that is expended outside the element during the time $dt$}, and therefore contributes to \dashuline{the increase in  the energy of the element}.

\vspace*{-3mm} 
\begin{center}
--------------------------------------------------- 
\end{center}
\vspace*{-3mm}

Experience shows, however, that in general this mechanical work is not the only effect that takes place in the environment of the element, but that there are other effects that do not arise at the expense of the work just calculated, and are therefore already included in it (such as the compression heat), but which also occur in the environment at the same time as this work. 

This includes the phenomena of heat conduction (and radiation) caused by differences in temperature (which we can completely ignore here, because they are independent of the mechanical effects), as well as the processes of friction and shock caused by differences in velocity (through which energy is not generated only in the form of external mechanical work, but also in the form of molecular work, and heat is directly transferred).

\vspace*{-5mm} 
\begin{center}
{\it\color{red}\bf
---------
(p.173) 
---------
}
\end{center}
\vspace*{-3mm}

However, if we ignore these phenomena completely for the time being, i.e. if we limit the investigation to completely elastic bodies (which are not subject to after-effects), then we can at least consider the expression of mechanical work obtained above as the total work of the effects that occur outside the element in the time $dt$. 
It measures the simultaneous change in energy, which we will now move on to calculating. 

The total energy of the element consists of two parts: 
the external (kinetic) and the internal (potential) energy. 
The former has the value: 
\begin{align}
\frac{\mu \:.\: \partial\,\tau}{2}
\left\{\:
\left(\frac{dx}{dt}\right)^2
\;+\;
\left(\frac{dy}{dt}\right)^2
\;+\;
\left(\frac{dz}{dt}\right)^2
\:\right\}
\; ,
\nonumber 
\end{align}
and the latter depends only on the internal state of the element, which is determined not only by the storage of the smallest particles but also by the temperature of the element. 

However, in the case of completely elastic bodies in which no heat conduction takes place, the temperature is solely dependent on the mechanical change (deformation) of the element, since this determines the entire change in state, including the external effects. 

 We can therefore regard the internal energy of the element as a function solely of those quantities which determine the instantaneous deformation of this element, and of which (as is well known) there are always $6$ (for finite as well as for infinitely small changes, 
because every change in an element can be viewed as 
a linear change$\,$\footnote{$\:$G. Kirchhoff: Mechanik, 1877, p.107.} 
and is therefore given by $12$ coefficients). 

However, $6$ of these {\it\color{red}(quantities)} correspond to a translation and a rotation of the element, and therefore have no influence on the deformation (dilation in $3$ mutually perpendicular directions), with the other $6$ {\it\color{red}(quantities)} left to determine. 

\vspace*{-5mm} 
\begin{center}
{\it\color{red}\bf
---------
(p.174) 
---------
}
\end{center}
\vspace*{-3mm}

The displacements can be calculated from {\dashuline{an arbitrary {\bf zero state}}}: for solid bodies it is most convenient to calculate them from the natural state, and for gases from 
any 
state of uniform pressure.

If we also set the internal energy proportional to the mass, we get an expression of the form: $U\:.\:\mu\:.\:\partial\,\tau$. 
Of course, $U$ here is not the primary (p.111) expression of the internal energy of the mass unit, but only applies to the case under consideration, and of course it does not refer to processes that take place at a constant temperature, but to \dashuline{processes that in thermodynamics are called adiabatic}. 

We will postpone a little the closer determination of the form of the function $U$, so that we do not have to introduce the distinction between \dashuline{infinitely small} and \dashuline{finite movements} here.

\vspace*{0mm} 
\begin{center}
--------------------------------------------------- 
\end{center}
\vspace*{-3mm}

The growth of the energy of the element in time dt is therefore: 
\begin{align}
d\left[\;
\frac{\mu \:.\: \partial\,\tau}{2}
\left\{\:
\left(\frac{dx}{dt}\right)^2
\;+\;
\left(\frac{dy}{dt}\right)^2
\;+\;
\left(\frac{dz}{dt}\right)^2
\:\right\}
\;\right] 
\;+\; d\left[\: \mu \:.\: \partial\,\tau\:.\:U \:\right]
\; ,
\nonumber 
\end{align}
and, since the time change $d(\mu \:.\: d\,\tau)=0$: 
\begin{align}
\mu \:.\: \partial\,\tau
\left(\:
\frac{d^{\,2}x}{dt^2}\:\:dx
\;+\;
\frac{d^{\,2}y}{dt^2}\:\:dy
\;+\;
\frac{d^{\,2}z}{dt^2}\:\:dz
\:\right)
\;+\; \mu \:.\: \partial\,\tau\:.\:d\,U
\; .
\nonumber 
\end{align}
Then, equating this quantity to the sum of the expressions (\ref{Eq_Planck_1887_1}) and (\ref{Eq_Planck_1887_2}) results in the equation of \dashuline{the principle of conservation of energy for the element of a perfectly elastic body}.

\vspace*{-3mm} 
\begin{center}
--------------------------------------------------- 
\end{center}
\vspace*{-3mm}

  The principle itself does not allow any further conclusions to be drawn.
  
Nevertheless, by using the principle of superposition of effects, we can obtain just as many equations as are necessary to determine the motion by decomposing the above equation according to certain simple, immediately obvious premises (from which, incidentally, one can under no circumstances make oneself independent).  

  First of all, it can be seen from the form of expression (\ref{Eq_Planck_1887_2}) that the value of the work done outside the element is determined by two different circumstances, namely by the type of spatial variability: 
\begin{enumerate}[leftmargin=8mm,parsep=0mm,itemsep=1mm,topsep=-1mm,rightmargin=2mm]
\item 
\!\!\!) of the pressure components $X_x \;, Y_x \:.\:.\:. $\,;
\item 
\!\!\!) of the velocity components $(dx/dt, \; dy/dt, \; dz/dt)$\,; 
\end{enumerate}
and if both quantities are constant throughout space, all the work disappears.

\vspace*{-2mm} 
\begin{center}
{\it\color{red}\bf
---------
(p.175) 
---------
}
\end{center}
\vspace*{-3mm}

Now let us make the assumption that each of these two circumstances also affects the \dashuline{change of energy} in a special way: the variability of pressure in space is supposed to influence only the value of kinetic energy, {\it\color{red}whereas} the variability of velocity only the value of potential energy.

A change of the living force can therefore only occur if the pressure in space is unevenly distributed, and-{\color{red}whereas} a deformation can only occur if the velocity varies from place to place (conclusions whose justification is obvious).

It also goes without saying that the work performed by the forces acting from a distance on the entire mass of the element only benefits the kinetic energy.

\vspace*{-3mm} 
\begin{center}
--------------------------------------------------- 
\end{center}
\vspace*{-3mm}


So if we now break down the equation of the energy into the parts relating to the two different types of energy, we get, by omitting the factor $\partial\,\tau$ once for the kinetic energy: 
\vspace*{1mm} 
\begin{equation}
\left.
\begin{aligned}
\mu \:.
\left(\:
\frac{d^{\,2}x}{dt^2}\:\:dx
\;+\;
\frac{d^{\,2}y}{dt^2}\:\:dy
\;+\;
\frac{d^{\,2}z}{dt^2}\:\:dz
\:\right)
& \: = \; 
  \left(\:
      X\:\:dx \;+\; Y\:\:dy \;+\; Z\:\:dz
  \:\right) .\:\mu
  \; \vspace*{2mm} 
\\
& \; -\:
\left(\:
  \frac{\partial X_x}{\partial x} \; dx
   \;+\; 
  \frac{\partial Y_x}{\partial x} \; dy
   \;+\; 
  \frac{\partial Z_x}{\partial x} \; dz
\:\right)
  \; \vspace*{2mm} 
\\
& \; -\:
\left(\:
  \frac{\partial X_y}{\partial y} \; dx
   \;+\; 
  \frac{\partial Y_y}{\partial y} \; dy
   \;+\; 
  \frac{\partial Z_y}{\partial y} \; dz
\:\right)
  \; \vspace*{2mm} 
\\
& \; -\:
\left(\:
  \frac{\partial X_z}{\partial z} \; dx
   \;+\; 
  \frac{\partial Y_z}{\partial z} \; dy
   \;+\; 
  \frac{\partial Z_z}{\partial z} \; dz
\:\right)
  \;  \vspace*{2mm} 
\end{aligned}
\;\;\;\;
\right\} \: ,
\label{Eq_Planck_1887_3} 
\end{equation}
and from this with further decomposition the 3 coordinate axes by setting the coefficients of $dx, dy, dz$ on both sides equal to each other, the well-known Poisson's equations: 
\vspace*{1mm} 
\begin{equation}
\left.
\begin{aligned}
\mu \:\: \frac{d^{\,2}x}{dt^2}
& \:=\; \mu \:\: X 
  \:-\: \frac{\partial X_x}{\partial x} 
  \:-\: \frac{\partial X_y}{\partial y} 
  \:-\: \frac{\partial X_z}{\partial z} 
  \; , \vspace*{2mm} 
\\
\mu \:\: \frac{d^{\,2}y}{dt^2}
& \:=\; \mu \:\: Y 
  \;-\: \frac{\partial Y_x}{\partial x} 
\:\;-\: \frac{\partial Y_y}{\partial y} 
\:\;-\: \frac{\partial Y_z}{\partial z} 
  \; , \vspace*{2mm} 
\\
\mu \:\: \frac{d^{\,2}z}{dt^2}
& \:=\; \mu \:\: Z 
  \:-\: \frac{\partial Z_x}{\partial x} 
\:\:-\: \frac{\partial Z_y}{\partial y} 
\:\:-\: \frac{\partial Z_z}{\partial z} 
  \; . \vspace*{2mm} 
\end{aligned}
\;\;\;\;
\right\} 
\label{Eq_Planck_1887_4} 
\end{equation}

\vspace*{-5mm} 
\begin{center}
{\it\color{red}\bf
---------
(p.176) 
---------
}
\end{center}
\vspace*{-3mm}


On the other hand, the equation for the internal energy remains: 
\vspace*{1mm} 
\begin{equation} \tag{4a}
\left.
\begin{aligned}
\mu \:.\: d\,U
& \: = \; -\:
\left(\:
   X_x \:.\: \frac{\partial \,dx}{\partial x}
   \;+\; 
   Y_x \:.\: \frac{\partial \,dy}{\partial x} 
   \;+\; 
   Z_x \:.\: \frac{\partial \,dz}{\partial x} 
\:\right)
\\
& \quad\;\; - \:
\left(\:
   X_y \:.\: \frac{\partial \,dx}{\partial y}
   \;+\; 
   Y_y \:.\: \frac{\partial \,dy}{\partial y} 
   \;+\; 
   Z_y \:.\: \frac{\partial \,dz}{\partial y} 
\:\right)
\\
& \quad\;\; - \:
\left(\:
   X_z \:.\: \frac{\partial \,dx}{\partial z}
   \;+\; 
   Y_z \:.\: \frac{\partial \,dy}{\partial z} 
   \;+\; 
   Z_z \:.\: \frac{\partial \,dz}{\partial z} 
\:\right)
\end{aligned}
\;\;\;\;
\right\} \: ,
\label{Eq_Planck_1887_4a} 
\end{equation}
where $U$, the internal energy of the mass unit, is (according to the above explanations) a function of the $6$ quantities that determine the deformation of the element (and, in particular, $U$ is independent from the rotation suffered by the element as a whole).

Now, as is well known, the expressions: 
$$    \frac{\partial \,dz}{\partial y} 
\:-\: \frac{\partial \,dy}{\partial z} 
\; , \;\;\;\;\;\;\;\;\;\;\;\;
      \frac{\partial \,dx}{\partial z} 
\:-\: \frac{\partial \,dz}{\partial x} 
\; , \;\;\;\;\;\;\;\;\;\;\;\;
      \frac{\partial \,dy}{\partial x} 
\:-\: \frac{\partial \,dx}{\partial y} 
\; , 
$$ 
represent the double components of the infinitesimal rotation that the element undergoes in time $dt$, consequently you cannot depend on these differences but only on the corresponding sums, from which  follows immediately: 
\vspace*{1mm} 
\begin{equation}
\left.
\begin{aligned}
 Z_y & \: = \; Y_z \; , \;\;\;\;\;\;\;\;\;\;
 X_z   \; = \; Z_x \; , \;\;\;\;\;\;\;\;\;\;
 Y_x   \; = \; X_y \; ,
 \\
\mbox{and} \;\;\;
-\: \mu \:.\: d\,U
& \: = \; 
   X_x \:.\: \frac{\partial \,dx}{\partial x}
   \;+\; 
   Y_y \:.\: \frac{\partial \,dy}{\partial y} 
   \;+\; 
   Z_z \:.\: \frac{\partial \,dz}{\partial z} 
\\
& \quad\;\; + \:
   Z_y \:.\: 
   \left(
   \frac{\partial \,dz}{\partial y}
   \:+\:
   \frac{\partial \,dy}{\partial z}
   \right)
   \;+\; 
   X_z \:.\: 
   \left(
   \frac{\partial \,dx}{\partial z}
   \:+\:
   \frac{\partial \,dz}{\partial x}
   \right)
\\
& \quad\;\; + \:
   Y_x \:.\: 
   \left(
   \frac{\partial \,dy}{\partial x}
   \:+\:
   \frac{\partial \,dx}{\partial y}
   \right)
   \; .
\end{aligned}
\;\;\;\;
\right\} \: 
\label{Eq_Planck_1887_5} 
\end{equation}

\vspace*{-3mm} 
\begin{center}
--------------------------------------------------- 
\end{center}
\vspace*{-3mm}

Just as for an infinitely small element, these considerations naturally apply to \dashuline{any finite part of the body}, since the \dashuline{growth of energy} is always equal to the \dashuline{external work expended}. 

The same result can be reached directly by \dashuline{integrating the equation of energy} that applies to a mass element over a finite mass. If the whole body is left to its own devices, its total energy remains constant, whereas, if it passes from the state of rest to that of movement, the kinetic energy increases and the potential energy decreases, from which (like p.156) the theorem follows that: {\it the minimum of the function $U$ corresponds to the stable equilibrium state}.

\vspace*{3mm} 
\begin{center}
{\it\color{red}\bf
---------
(p.177) 
---------
}
\end{center}
\vspace*{-3mm}

The general finite expression of $U$ is only known for perfect gases and incompressible liquids, so we will initially limit ourselves to considering very small movements, such as those that can take place in solid bodies, liquids and gases, including sound and light movements are to be expected. In the case of solid bodies, this limitation is generally required anyway by the need to adhere to the elasticity limit.

\vspace*{-3mm} 
\begin{center}
--------------------------------------------------- 
\end{center}
\vspace*{-3mm}

For convenience, we introduce new names for the variables. 

We call $x, y, z$ the coordinates that a material point has in \dashuline{the {\bf(zero) state}} and from which the displacements are calculated, and $u, v, u$ these (small) displacements themselves. 

Then a certain material point is defined by $(x, y, z)$, which occupies the position $(x + u, y + v, z + w)$ at time $t$. In the equations so far we have set ourselves to think $(x + u, y + v, z + w)$ instead of $(x, y, z)$. 

 If $(u, v, w)$ do not have multiple maxima and minima in finite intervals of the variables, the differential quotients of these quantities according to time and place are also very small, and can be neglected against those of $(x, y, z)$ if the body is finitely extended. 

We then obtain from (\ref{Eq_Planck_1887_5}), since a permutation of the order of differentiation is now permissible:
\vspace*{1mm} 
\begin{equation}
\left.
\begin{aligned}
-\: \mu \:.\: d\,U
& \: = \; 
   X_x \:.\: d\:\frac{\partial u}{\partial x}
   \;+\; 
   Y_y \:.\: d\:\frac{\partial v}{\partial y} 
   \;+\; 
   Z_z \:.\: d\:\frac{\partial w}{\partial z} 
\\
& \quad\;\; + \:
   Z_y \:.\: d\,
   \left(
   \frac{\partial w}{\partial y}
   \:+\:
   \frac{\partial v}{\partial z}
   \right)
   \;+\; 
   X_z \:.\: d\,
   \left(
   \frac{\partial u}{\partial z}
   \:+\:
   \frac{\partial w}{\partial x}
   \right)
\\
& \quad\;\; + \:
   Y_x \:.\: d\,
   \left(
   \frac{\partial v}{\partial x}
   \:+\:
   \frac{\partial u}{\partial y}
   \right)
   \; .
\end{aligned}
\;\;\;\;
\right\} \: 
\nonumber 
\end{equation}
 We then use the abbreviations:
\vspace*{1mm} 
\begin{equation}
\left.
\begin{aligned}
\frac{\partial u}{\partial x} & \: = \; x_x \; , 
\;\;\;\;\;\;\;\;\;\;\;
\frac{\partial v}{\partial y}   \; = \; y_y \; , 
\;\;\;\;\;\;\;\;\;\;\;
\frac{\partial w}{\partial z}   \; = \; z_z \; , 
\\
\frac{\partial w}{\partial y} \:+\:
\frac{\partial v}{\partial z} 
& \: = \; z_y \;=\; y_z \; , 
\;\;\;\;
\frac{\partial u}{\partial z} \:+\:
\frac{\partial w}{\partial x}
  \; = \; x_z \;=\; z_x \; , 
\\
& \frac{\partial v}{\partial x} \:+\:
  \frac{\partial u}{\partial y} 
\; = \; y_x \;=\; x_y \; .
\end{aligned}
\;\;\;\;
\right. \: 
\nonumber 
\end{equation}

\vspace*{-6mm} 
\begin{center}
{\it\color{red}\bf
---------
(p.178) 
---------
}
\end{center}
\vspace*{-3mm}


These 6 quantities, independently of each other, determine the deformation of the element and consequently also the value of $U$. By comparing the two sides of the equation we get: 
\vspace*{1mm} 
\begin{equation}
\left.
\begin{aligned}
X_x & \: = \; 
 - \:\mu \:\: \frac{\partial\,U}{\partial\,x_x} \; , 
\;\;\;\;\;\;\;\;\;\;\;
Y_y   \; = \; 
 - \:\mu \:\: \frac{\partial\,U}{\partial\,y_y} \; , 
\;\;\;\;\;\;\;\;\;\;\;
Z_z   \; = \; 
 - \:\mu \:\: \frac{\partial\,U}{\partial\,z_z} \; , 
\\
Z_y & \: = \; 
 - \:\mu \:\: \frac{\partial\,U}{\partial\,z_y} \; , 
\;\;\;\;\;\;\;\;\;\;\;
X_z   \; = \; 
 - \:\mu \:\: \frac{\partial\,U}{\partial\,x_z} \; , 
\;\;\;\;\;\;\;\;\;\;\;
Y_x   \; = \; 
 - \:\mu \:\: \frac{\partial\,U}{\partial\,y_x} \; .
\end{aligned}
\;\;\;\;
\right. \: 
\nonumber 
\end{equation}
  This is still the expression of $U$.
Since the variables on which it depends are very small, it can be developed according to their powers, 
and if we remain with the quadratic members, the pressure components become linear functions of the variables.

Whether $U$ occurs in homogeneous form, or not, depends on the \dashuline{choice of the {\bf zero state}} 
$(x_x = 0 = x_y = . . . )$.
First, \dashuline{$U$ contains an {\bf arbitrary additive constant}} that we want to \dashuline{{\bf set equal to $0$}}, so that \dashuline{for the {\bf zero state $U = 0$}}. 
If we further make the assumption that \dashuline{the {\bf zero state}} represents \dashuline{{\bf an equilibrium state}} in which, \dashuline{{\bf with a uniform density $\mu_0$}}, there is \dashuline{{\bf a uniform pressure $p_0$}} (e.g. atmospheric pressure) acting perpendicularly on every surface element everywhere, then for {\bf this {\bf\color{red}(zero)} state}: 
$$
  X_x \;=\; Y_y \;=\; Z_z \;=\; p_0 \; ,
       \;\;\;\;\;\;\;\;\;\;\;
  Z_y \;=\; X_z \;=\; Y_x \;=\; 0 \; .
$$ 
Then \dashuline{the linear part of $U$} reduces to the expression: 
$-\,(p_0/\mu_0)\:(x_x + y_y + z_z)$. 
For solid bodies and liquids drops (for gases only in limited cases) $p_0$ can also be assumed to be $0$.

\vspace*{-3mm} 
\begin{center}
--------------------------------------------------- 
\end{center}
\vspace*{-3mm}


Finally, the quadratic part of $U$ generally contains $21$ constant coefficients, the number of which, however, decreases in the presence of symmetries in the structure of the body, and is reduced in a known manner to $2$ for isotropic solid bodies, but to $1$ for liquid and gaseous bodies. 

If now $U$ is known, the values of the pressure components, and thus the equations of motion in the whole body, result directly from this.

\vspace*{-4mm} 
\begin{center}
{\it\color{red}\bf
---------
(p.179) 
---------
}
\end{center}
\vspace*{-3mm}

The factor $\mu$, which appears in front of the differential quotient of $U$ in the expressions of the pressure components, and which is usually missing in the usual representation of the theory of elasticity, can be used for the small movements considered here (and are treated as the constant $=\mu_0$ at least when $p_0 = 0$), whereas in the case of finite movements (equation (\ref{Eq_Planck_1887_5})) its variability must be taken into account.

\vspace*{-3mm} 
\begin{center}
--------------------------------------------------- 
\end{center}
\vspace*{-3mm}

These equations are of particular importance for the derivation of the laws of motion of periodic oscillations, traveling or standing waves, in elastic media. 
The energy of every such oscillation consists of two parts, \dashuline{the kinetic and the potential energy}, the sum of which remains constant as long as no external effects take place. 
The fact that two different waves can weaken and even destroy each other through appropriate interference does \dashuline{not involve a contradiction to the principle of conservation of energy}. 

{\it\color{red}For example,} if you imagine two plane wave trains with the same period of oscillation, the same direction of propagation and the same amplitude superimposed with a path difference of half a wavelength, the resulting wave will disappear, but it should be noted that this interference phenomenon is not an independent process, but only one side of a much more comprehensive natural process. 

{\it\color{red}Indeed,} the waves that emanate from two different 
origins 
(light, sound sources) can never meet with the same phase everywhere, but they will always strengthen in some places, while weakening in others. 

{\it\color{red}Therefore,} the idea of a plane wave is actually just an abstraction. 
For example, if the two assumed plane waves are pieces of two spherical waves that emanate from infinitely distant centers, the path difference will of course be constant in the finite, but the ratio will be different at corresponding infinitely distant points, so that overall, no energy is lost, as follows from our equations.

\vspace*{-3mm} 
\begin{center}
--------------------------------------------------- 
\end{center}
\vspace*{-3mm}


The performance of our specific sensory organs is not sufficient to measure the energy of a light or sound source.
At most, these enable us to judge with greater or lesser accuracy the equality or the gradual difference in the energies of waves with the same period of oscillation. 

\vspace*{-2mm} 
\begin{center}
{\it\color{red}\bf
---------
(p.180) 
---------
}
\end{center}
\vspace*{-3mm}

\dashuline{An {\bf absolute measurement} of the energy} of a wave is only possible by \dashuline{transforming this energy} into \dashuline{another form of work}. 
It can be seen from this that it is accessible to a more precise measurement method. 
This includes, above all, the \dashuline{transformation into heat} through absorption.

\vspace*{-3mm} 
\begin{center}
--------------------------------------------------- 
\end{center}
\vspace*{-3mm}


A more detailed discussion of \dashuline{the theory of wave motion} would go too far here, since this theory, insofar as it contains new aspects independent of those discussed so far, is still somewhat \dashuline{too permeated by hypothetical ideas} to be presented as a consequence of the \dashuline{principle of the conservation of energy}.

The latter applies particularly to \dashuline{optics}, even though the beginning has recently been made to \dashuline{make the energy principle fruitful} for this part of physics (which has usually been treated quite separately from it).

\vspace*{-3mm} 
\begin{center}
--------------------------------------------------- 
\end{center}
\vspace*{-3mm}

Let us now turn back to the consideration of finite movements (in liquid and gaseous media), i.e. to equations (\ref{Eq_Planck_1887_4}) and (\ref{Eq_Planck_1887_5}), with the reintroduction of the notation there. 
The characteristic property of liquid and gaseous bodies is:
$$
  X_x \;=\; Y_y \;=\; Z_z \;=\; p \; ,
       \;\;\;\;\;\;\;\;\;\;\;
  X_y \;=\; Y_z \;=\; Z_x \;=\; 0 \; .
$$ 
where {\it\color{red}(the pressure)} $p$ is a certain function of {\it\color{red}(the density)} $\mu$ that depends on the nature of the medium.

Thus (\ref{Eq_Planck_1887_4}) {\it\color{red}given by}  
\vspace*{1mm} 
\begin{equation}
\left.
\begin{aligned}
\mu \:\: \frac{d^{\,2}x}{dt^2}
& \:=\; \mu \:\: X 
  \:-\: \frac{\partial p}{\partial x} 
  \; , \vspace*{2mm} 
\\
\mu \:\: \frac{d^{\,2}y}{dt^2}
& \:=\; \mu \:\: Y 
\:\;-\: \frac{\partial p}{\partial y} 
  \; , \vspace*{2mm} 
\\
\mu \:\: \frac{d^{\,2}z}{dt^2}
& \:=\; \mu \:\: Z 
\:\:-\: \frac{\partial p}{\partial z} 
  \; , \vspace*{2mm} 
\end{aligned}
\;\;\;\;
\right.
\nonumber 
\end{equation}
becomes the general hydrodynamic equations of motion, in the use of which either the differential signs $\partial$ (Eulerian form) or the signs $d$ (Lagrangian form) can be used at will.

\vspace*{-2mm} 
\begin{center}
{\it\color{red}\bf
---------
(p.181) 
---------
}
\end{center}
\vspace*{-3mm}

Furthermore, from (\ref{Eq_Planck_1887_5}) one obtains for the determination of the internal energy $U$ of the mass unit: 
\vspace*{1mm} 
\begin{equation}
\left.
\begin{aligned}
 -\: \mu \:.\: d\,U
& \: = \; 
   p \:.\: 
   \left(
   \frac{\partial \,dx}{\partial x}
   \;+\; 
   \frac{\partial \,dy}{\partial y}
   \;+\; 
   \frac{\partial \,dz}{\partial z}
   \right)
   \; .
\end{aligned}
\;\;\;\;
\right. \: 
\nonumber 
\end{equation}
Then, a simple kinematic consideration {\it\color{red}(\,i.e. the ``\,continuity\,'' equation 
$d\mu/dt + \mu \: \vec{\nabla} \:.\: \vec{v}$ with 
$\vec{v}=(dx/dt,\,dy/dt,\,dz/dt)$ the velocity and with $\vec{\nabla}=(\partial/\partial x,\,\partial/\partial y,\,\partial/\partial z)$ the ``\,Nabla\,'' gradient operator\,)} 
shows that:
\vspace*{1mm} 
\begin{equation}
\left.
\begin{aligned}
 d\,\mu
 \;+\;
 \mu \:.\: 
   \left(
   \frac{\partial \,dx}{\partial x}
   \;+\; 
   \frac{\partial \,dy}{\partial y}
   \;+\; 
   \frac{\partial \,dz}{\partial z}
   \right)
& \: = \; 0
   \; ,
\end{aligned}
\;\;\;\;
\right. \: 
\nonumber 
\end{equation}
and consequently {\it\color{red}(\,after an integration, and then an integration by parts\,)\,}:
\vspace*{1mm} 
\begin{equation}
\left.
\begin{aligned}
 d\,U  & \: = \; \frac{p}{\mu^2} \:.\: d\,\mu
   \; , \\
 U  \;=\; \bigintssss \frac{p}{\mu^2} \:.\: d\,\mu
   & \: = \; -\:\frac{p}{\mu} 
    \;+\; \bigintssss \frac{dp}{\mu}
   \; , 
\end{aligned}
\;\;\;\;
\right. \: 
\nonumber 
\end{equation}
or, if one sets :
\begin{align}
P &\:=\; \bigintssss \frac{dp}{\mu}
\nonumber \; , \\
U &\:=\; P \;-\; \frac{p}{\mu}
\label{Eq_Planck_1887_6}  \; .
\end{align}
\dashuline{An {\bf additive {\it\color{red}(integration)} constant remains arbitrary} in $P$ and $U$}
{\it\color{red}(\,and namely (\ref{Eq_Planck_1887_6}) should be written $U - U_0 = P - P_0 \:- p/\mu\,$)\,}. 
From this, the internal energy of the mass unit can be calculated as soon as $p$ is known as a function of $\mu$.

\vspace*{-3mm} 
\begin{center}
--------------------------------------------------- 
\end{center}
\vspace*{-3mm}

For instance, for a perfect gas you have: 
$$ p \;=\; C \; \mu^k   \;\;\;\;\;\;\;\;
   \mbox{(where the heat conduction is excluded)} \; , $$ 
where $C$ is a constant and $k$ is the ratio {\it\color{red}$(c_p/c_v)$} of the two specific heats. It follows {\it\color{red}(\,from $U - U_0 = P - P_0 \:- p/\mu\,$ and 
$\;p/\mu = C\:\mu^{k-1} \,$)\,}: 
$$
 P \: {\color{red}\:-\:\: P_0} 
 \;=\; \frac{C\:k}{k\,-\,1} \:\:.\:\: \mu^{k-1} \; , 
 \;\;\;\;\;
 U \: {\color{red}\:-\:\: U_0} 
 {\color{red}\;\:=\; C \,
 \left(
   \frac{k}{k\,-\,1} \:-\: 1
 \right) \,.\:\: \mu^{k-1} 
 }
 \;=\; \frac{C}{k\,-\,1} \:\:.\:\: \mu^{k-1} 
 \; .
$$
This value of internal energy is the same as that derived from the  temperature (see p.113 above).

\vspace*{-3mm} 
\begin{center}
--------------------------------------------------- 
\end{center}
\vspace*{-3mm}

For an incompressible fluid $\mu=const.$ (i.e. a fixed condition in the sense of p. 166), hence $P{\color{red}\:-P_0\:}=p/\mu$ and ${\color{red}d\:}U=0$ {\it\color{red}(\,i.e. $U=U_0=const.$\,)}. 
In fact, the internal state of an incompressible fluid remains constant, so no work can be done by changing it.

\vspace*{-3mm} 
\begin{center}
--------------------------------------------------- 
\end{center}
\vspace*{-3mm}

When applied to a finite part of the mass of the liquid, we also get the proposition that in any given state the total external and internal energy of the liquid is equal to the total amount of work that has been expended in the environment from a certain point in time onwards to bring about this state.

\vspace*{-2mm} 
\begin{center}
{\it\color{red}\bf
---------
(p.182) 
---------
}
\end{center}
\vspace*{-3mm}

This work comes first from the \dashuline{forces acting} on the mass \dashuline{from a distance}, and then from the \dashuline{pressure forces acting on the surface}.

\vspace*{-3mm} 
\begin{center}
--------------------------------------------------- 
\end{center}
\vspace*{-3mm}

In the case of stationary motion, the state at a certain point in space $x\:y\:z$ is independent of time, so that all differential quotients taken after $\partial\,t$ disappear. 
In this case, the equation of the energy, related to a single element, allows a general integration with respect to time $dt$, which gives: 
\begin{align}
\frac{1}{2}
\left\{\:
\left(\frac{dx}{dt}\right)^2
\;+\;
\left(\frac{dy}{dt}\right)^2
\;+\;
\left(\frac{dz}{dt}\right)^2
\:\right\}
\;+\; U
\;=\; const. \;-\; V \;-\; \frac{p}{\mu}
\; ,
\nonumber 
\end{align}
i.e. the {\it\color{red}(change in the)} total (external$\;+\;$internal) energy of the mass unit is equal to the decrease of the potential function of the mass forces $V$ (which is assumed to exist) and of the pressure divided by the density. 

This equation is somewhat simpler according to (6): 
\begin{align}
\frac{1}{2}
\left\{\:
\left(\frac{dx}{dt}\right)^2
\;+\;
\left(\frac{dy}{dt}\right)^2
\;+\;
\left(\frac{dz}{dt}\right)^2
\:\right\}
\;+\; P
\;+\; V
\;=\; const. 
\; 
\nonumber 
\end{align}
{\it\color{red}(\,this corresponds to what is called a Bernoulli's equation determining ``\,constant values\,'' during the motion of the fluid, with conversions occurring between kinetic, pressure and potential energies\,)}

\vspace*{-3mm} 
\begin{center}
--------------------------------------------------- 
\end{center}
\vspace*{-3mm}

 In concluding with these considerations about the motions of perfectly elastic bodies, 
we want to include a more complicated phenomena which can only be explained by a deviation from the simple assumptions we have made (and which are at present still somewhat more difficult to treat by the method we have adopted), namely: the frictional processes in the motion of an incompressible fluid.

The internal friction in a liquid can be brought into perspective with the phenomena of friction and inelastic collision of solid bodies. 
Both processes are caused by the occurrence of a force that only depends on the relative movement of masses in contact, and whose effectiveness is always aimed at equalizing the velocities. 

\vspace*{-2mm} 
\begin{center}
{\it\color{red}\bf
---------
(p.183) 
---------
}
\end{center}
\vspace*{-3mm}

In the process, the living force {\it\color{red}(kinetic energy)} of the molar movement is always lost, which is converted either into molecular movement (heat) or into molecular work. This idea is sufficient to derive the equations of motion of a frictional fluid using our principle in the path already taken above.

\vspace*{-3mm} 
\begin{center}
--------------------------------------------------- 
\end{center}
\vspace*{-3mm}

Let us first calculate the expression of the work (or the effects equivalent to it) that must be expended outside a liquid element in the time $dt$, or is actually expended, in order to bring about the change in state of the element that occurs in the same time. 

First of all, this includes the work of the force (gravity) acting from a distance on the mass of the element, the expression of which is given from (\ref{Eq_Planck_1887_1}) above: 
\begin{align}
( \, X \: d\, x \;+\; Y \: d \, y \;+\; Z \: d \, z \, )
\:.\: \mu \:.\: \partial\,\tau \; .
\label{Eq_Planck_1887_7}
\end{align}

\vspace*{-3mm} 
\begin{center}
--------------------------------------------------- 
\end{center}
\vspace*{-3mm}


Furthermore, all that remains for us to consider are the effects which are propagated through the surface of the element into the surroundings by the pressure of the liquid, in conjunction with the peculiar forces expressed by friction. 

These effects are generally of two natures, following the analogy of friction and shock between solid bodies. 

Firstly, they exert a force on the adjacent parts in the sense of changing the speeds (molar effect), secondly, they cause a change in the internal energy in the surrounding area (molecular effect), which can generally be associated with a certain deformation (although in our case it will only be documented as a change in temperature). 

According to this, we first have to set the amount of mechanical work for the external effects, which has exactly the same form as the general expression (\ref{Eq_Planck_1887_2}): 
\vspace*{1mm} 
\begin{equation}
\left.
\begin{aligned}
-\:\frac{\partial}{\partial x}
( \, X_x \: d\, x \;+\; Y_x \: d \, y \;+\; Z_x \: d \, z \, )
 \:.\: \partial\:\tau
  \; , \vspace*{2mm} 
\\
-\:\frac{\partial}{\partial y}
( \, X_y \: d\, x \;+\; Y_y \: d \, y \;+\; Z_y \: d \, z \, )
 \:.\: \partial\:\tau
  \; , \vspace*{2mm} 
\\
-\:\frac{\partial}{\partial z}
( \, X_z \: d\, x \;+\; Y_z \: d \, y \;+\; Z_z \: d \, z \, )
 \:.\: \partial\:\tau
  \; . \vspace*{2mm} 
\end{aligned}
\;\;\;\;
\right\} 
\label{Eq_Planck_1887_8} 
\end{equation}

\vspace*{-3mm} 
\begin{center}
{\it\color{red}\bf
---------
(p.184) 
---------
}
\end{center}
\vspace*{-3mm}


As far as the second part of the external effects is concerned (i.e. the generation of heat, which is caused outside the element by the work of the frictional forces acting on the surface), its amount is in any case proportional to the lateral surfaces of the element.

However, it is easy to see that this is infinitely small compared to the other effects that come into consideration, for the reason that the heat generated at the boundary layer disappears in any case compared to that which is produced throughout the interior of the element. 

There is an essential difference here compared to the phenomena that accompany the friction of solid bodies, since in the latter, molecular work is developed only on a single friction surface, which as a result has a value of the same order of magnitude as the molar work transferred by the friction.

\vspace*{-3mm} 
\begin{center}
--------------------------------------------------- 
\end{center}
\vspace*{-3mm}

Thus we would have found an expression for the external effects of exactly the same form as in the movement of elastic bodies. But things are different with the energy of the element: it consists of the living force {\it\color{red}(kinetic energy)}, the increase of which in the time element $dt$ is given by: 
\begin{align}
\left(\,  
      \frac{d^{\,2}x}{dt^2} \: d \, x 
\;+\; \frac{d^{\,2}y}{dt^2} \: d \, y 
\;+\; \frac{d^{\,2}z}{dt^2} \: d \, z 
\,\right) .\: \mu \:.\: \partial\,\tau \; ,
\label{Eq_Planck_1887_9}
\end{align}
and, since the fluid is incompressible, only of that through the internal friction generated heat. 

\dashuline{The heating that occurs in time $dt$} will generally \dashuline{depend on the velocity state of the element}, but \dashuline{not on the velocity components $(dx/dt,\:dy/dt,\:dz/dt)$, in short: $u,\:v,\:w$) themselves} (because at uniform velocity there is no friction at all) \dashuline{but from its local changes} (i.e. from the $9$ differential quotients of the quantities $u,\:v,\:w$ according to the coordinates $x,\:y,\:z$). 

But since the quantities: 
\vspace*{-1mm} 
\begin{align}
\frac{\partial w}{\partial y} 
\:-\; 
\frac{\partial v}{\partial z}  \; ,
               \;\;\;\;\;\;\;\;\;\;\;
\frac{\partial u}{\partial z} 
\:-\; 
\frac{\partial w}{\partial x}  \; ,
               \;\;\;\;\;\;\;\;\;\;\;
\frac{\partial v}{\partial x} 
\:-\; 
\frac{\partial u}{\partial y} \; , 
\nonumber 
\end{align}
only indicate a rotation of the element as a whole, not a deformation, the heating will only depend on the $6$ quantities: 
\vspace*{-5mm} 
\begin{center}
{\it\color{red}\bf
---------
(p.185) 
---------
}
\end{center}
\vspace*{-5mm}
\begin{align}
\frac{\partial u}{\partial x}   \; ,
 \;\;\;\;\;\;\;\;\;\;\;\;\;\;
&
\frac{\partial v}{\partial y}   \; ,
 \;\;\;\;\;\;\;\;\;\;\;\;\;\;
\frac{\partial w}{\partial z}   \; ,
\nonumber \\
\frac{\partial w}{\partial y} 
\:+\; 
\frac{\partial v}{\partial z}  \; ,
               \;\;\;\;\;\;\;\;\;\;\;
\frac{\partial u}{\partial z} 
&
\:+\; 
\frac{\partial w}{\partial x}  \; ,
               \;\;\;\;\;\;\;\;\;\;\;
\frac{\partial v}{\partial x} 
\:+\; 
\frac{\partial u}{\partial y} \; . 
\nonumber 
\end{align}


 If these are sufficiently small, we can stop at the second {\it\color{red}(term)} when developing the expression of the heating by powers of the variables.

The absolute term is equal to $0$ (because if the variables are $=0$, the friction disappears and hence the heating), and since the latter is essentially positive, the linear terms also disappear.
Furthermore, the form of the expression (i.e. the size of the coefficients) is independent of the choice of coordinate system. 

This condition, just as in the determination of the internal energy of the element of an isotropic elastic solid body (p.178), leads to an expression that depends only on two coefficients. 

If you finally add the condition of incompressibility 
\vspace*{0mm} 
\begin{equation}
   \frac{\partial \,dx}{\partial x}
   \;+\; 
   \frac{\partial \,dy}{\partial y}
   \;+\; 
   \frac{\partial \,dz}{\partial z}
   \; = \; 0 \; ,
\nonumber 
\end{equation}
the amount of heating of the fluid element caused by friction in the time $dt$ reduces to the mechanically measured value: 
\vspace*{1mm} 
\begin{equation}
\left.
\begin{aligned}
 2 \:k\:. & \left\{\:
 \left(\frac{\partial u}{\partial x}\right)^2
 \;+\;
 \left(\frac{\partial v}{\partial y}\right)^2
 \;+\;
 \left(\frac{\partial w}{\partial z}\right)^2
 \;+\;
 \frac{1}{2}\:
 \left(
   \frac{\partial w}{\partial y}
   \;+\;
   \frac{\partial v}{\partial z}
 \right)^2
 \:\right.
 \; \\
 & \left.\quad
 \;+\;
 \frac{1}{2}\:
 \left(
   \frac{\partial u}{\partial z}
   \;+\;
   \frac{\partial w}{\partial x}
 \right)^2
 \;+\;
 \frac{1}{2}\:
 \left(
   \frac{\partial v}{\partial x}
   \;+\;
   \frac{\partial u}{\partial y}
 \right)^2
 \:\right\} \,.\: dt \:.\: d\,\tau
 \; 
\end{aligned}
\;\;\;\;
\right.
\label{Eq_Planck_1887_10} 
\end{equation}
This is the only expression that satisfies the conditions set, and it contains only the single undetermined coefficient $k$ (the friction coefficient of the fluid),
this expression (\ref{Eq_Planck_1887_10}) being positive.

\vspace*{-3mm} 
\begin{center}
--------------------------------------------------- 
\end{center}
\vspace*{-3mm}


From this we now get the \dashuline{equation of the principle of conservation of energy} if we set the sum of expressions (\ref{Eq_Planck_1887_9}) and (\ref{Eq_Planck_1887_10}) and  \dashuline{the increase in energy of the element}, equal to the sum of expressions (\ref{Eq_Planck_1887_7}) and (\ref{Eq_Planck_1887_8}), with the amount of the external effects applied.

\vspace*{-3mm} 
\begin{center}
{\it\color{red}\bf
---------
(p.186) 
---------
}
\end{center}
\vspace*{-4mm}

Furthermore, by decomposing this equation into one {\it\color{red}{equation}} that relates to the external energy (living force {\it\color{red}- kinetic energy}) and another  {\it\color{red}{equation}} that relates to the internal energy (heat), \dashuline{the general equations of motion (\ref{Eq_Planck_1887_4}) result again}, but also \dashuline{an equation, which corresponds to equation (\ref{Eq_Planck_1887_4a})}: 
\vspace*{1mm} 
\begin{equation}
\left.
\begin{aligned}
 2 \:k\:. & \left\{\:
 \left(\frac{\partial u}{\partial x}\right)^2
 \;+\;
 \left(\frac{\partial v}{\partial y}\right)^2
 \;+\;
 \left(\frac{\partial w}{\partial z}\right)^2
 \;+\;
 \frac{1}{2}\:
 \left(
   \frac{\partial w}{\partial y}
   \;+\;
   \frac{\partial v}{\partial z}
 \right)^2
 \:\right.
 \; \\
 & \left.\quad
 \;+\;
 \frac{1}{2}\:
 \left(
   \frac{\partial u}{\partial z}
   \;+\;
   \frac{\partial w}{\partial x}
 \right)^2
 \;+\;
 \frac{1}{2}\:
 \left(
   \frac{\partial v}{\partial x}
   \;+\;
   \frac{\partial u}{\partial y}
 \right)^2
 \:\right\} 
 \; \\
 & = \: - \:
\left(\:
   X_x \:.\: \frac{\partial \,dx}{\partial x}
   \;+\; 
   Y_x \:.\: \frac{\partial \,dy}{\partial x} 
   \;+\; 
   Z_x \:.\: \frac{\partial \,dz}{\partial x} 
\:\right)
\\
 & \quad\; - \:
\left(\:
   X_y \:.\: \frac{\partial \,dx}{\partial y}
   \;+\; 
   Y_y \:.\: \frac{\partial \,dy}{\partial y} 
   \;+\; 
   Z_y \:.\: \frac{\partial \,dz}{\partial y} 
\:\right)
\\
 & \quad\; - \:
\left(\:
   X_z \:.\: \frac{\partial \,dx}{\partial z}
   \;+\; 
   Y_z \:.\: \frac{\partial \,dy}{\partial z} 
   \;+\; 
   Z_z \:.\: \frac{\partial \,dz}{\partial z} 
\:\right) \; .
\end{aligned}
\;\;\;\;
\right.
\nonumber 
\end{equation}

\vspace*{-3mm} 
\begin{center}
--------------------------------------------------- 
\end{center}
\vspace*{-3mm}

For a frictionless fluid we would have: 
$$
  k \;=\; 0 \; ,
       \;\;\;\;\;\;\;\;\;\;\;
  X_x \;=\; Y_y \;=\; Z_z \;=\; p \; ,
       \;\;\;\;\;\;\;\;\;\;\;
  X_y \;=\; Y_z \;=\; .\:.\:. \;=\; 0 \; ,
$$ 
which actually satisfies the equation because of the incompressibility condition. 

However, if $k$ is different from $0$, then terms are added to the values of the pressure components, which are obviously most easily determined from the equation in the following way: 
\vspace*{1mm} 
\begin{equation}
\left.
\begin{aligned}
 X_x \;=\; p 
     \:-\: 2 \; k \; \frac{\partial u}{\partial x} \; ,
 \;\;\;\;
 Y_y \;=\; p 
     \:-\: 2 \; k \; \frac{\partial v}{\partial y} \; ,
 \;\;\;\;
 Z_z \;=\; p 
     \:-\: 2 \; k \; \frac{\partial w}{\partial z} \; ,
 \\
 Y_z \;=\; Z_y \;=\;
 -\:k\:
 \left(
   \frac{\partial w}{\partial y}
   \;+\;
   \frac{\partial v}{\partial z}
 \right) \; ,
 \;\;\;\;
 Z_x \;=\; X_z \;=\;
 -\:k\:
 \left(
   \frac{\partial u}{\partial z}
   \;+\;
   \frac{\partial w}{\partial x}
 \right) \; ,
 \\
 X_y \;=\; Y_x \;=\;
 -\:k\:
 \left(
   \frac{\partial v}{\partial x}
   \;+\;
   \frac{\partial u}{\partial y}
 \right) \; ,
 \;\;\;\;\;\;\;\;\;\;\;\;\;\;\;\;
 \;\;\;\;\;\;\;\;\;\;\;\;\;\;\;\;
\end{aligned}
\;\;\;\;
\right.
\nonumber 
\end{equation}
whereby in 
a known manner$\,$\footnote{$\:$G. Kirchhoff: Mechanik {\it\color{red}(Mechanics)}, 1877, p.370.}
the values of compressive forces acting in a frictional incompressible fluid are given. This determines the entire movement.

\vspace*{-3mm} 
\begin{center}
=========
{\it\color{red}Conclusions} 
=========
\end{center}
\vspace*{-3mm}

In our \dashuline{remarks on mechanical energy} we have tried to show that \dashuline{the laws of mechanics}, including Newton's axioms, \dashuline{can be completely deduced from the principle of the conservation of energy} (not, however, by strict deduction, because mechanics, like no other branch of physics, can be constructed by purely deductive means) but \dashuline{by the repeated use of certain inductive conclusions} (which, however, are all essentially based on the unique and same idea). 

\vspace*{2mm} 
\begin{center}
{\it\color{red}\bf
---------
(p.187) 
---------
}
\end{center}
\vspace*{-3mm}

{\it\color{red}Indeed,} if we had established for a point (or an element of mass) the equation which \dashuline{measures the change of its energy} by the \dashuline{equivalent of the external effects}, we divided this equation into $2$ or more individual equations, whereby each time the proposition was expressed that \dashuline{the total energy decomposes into a sum of individual kinds of energy} which change independently of each other, each according to the measure of the external effects corresponding precisely to its peculiarity. 

Thus \dashuline{the kinetic energy} of a point breaks down into the $3$ individual types which correspond to the $3$ dimensions of space. 
These can never transform directly into each other, but each changes independently of the others only as a result of the external work corresponding to it. 

It is similar with the \dashuline{division of the energy} of an element into \dashuline{external} (molar) and \dashuline{internal} (molecular) \dashuline{energy}. How the division of energy is to be made in each case, however, must be taught by experience in each particular case. 

This \dashuline{principle of superposition} plays a most important part \dashuline{in all physics}, as we have already repeatedly emphasised, {\it\color{red}because} without it all phenomena would become mixed up with each other, and it would no longer be possible to establish the dependence of the individual phenomena on each other (since if each effect interferes with the other, the possibility of recognising the causal connection naturally ceases). 

We can therefore never get over this principle {\it\color{red}(of superposition)}, whether we emphasise it expressly or use it tacitly.
It is contained in the \dashuline{law of inertia} just as well as in the \dashuline{parallelogram of forces} and the \dashuline{law of action and counteraction}.

\vspace*{-3mm} 
\begin{center}
--------------------------------------------------- 
\end{center}
\vspace*{-3mm}

But what gives an essential advantage to \dashuline{the method here adopted}, is, on the one hand, \dashuline{the clearness with which the connection of the propositions deduced}, and \dashuline{the nature of their dependence on the principle of the conservation of energy}, is brought out: how much of it is to be regarded as a necessary consequence of the principle itself, and how much as established by particular experience independent of it.

\vspace*{-6mm} 
\begin{center}
{\it\color{red}\bf
---------
(p.188) 
---------
}
\end{center}
\vspace*{-3mm}

{\it\color{red}However,} on the other hand, the circumstance is of decisive importance that \dashuline{all the parts of physics can be treated in a completely uniform and consistent manner} by the method adopted. 

\dashuline{Energy} is a concept which finds its measure and its meaning \dashuline{in every natural phenomenon}, and \dashuline{the principle of superposition} likewise \dashuline{governs all natural effects}.

We may therefore be sure that, in case any as yet unknown agent should be discovered, \dashuline{the principle of superposition of effects could not serve to deduce the laws of the new force from the energy principle}, but would \dashuline{help us to pose those questions} the answering of which by experience affords the only means of discovering \dashuline{the laws of the phenomenal world}.

\vspace*{2mm}
\subsection{\underline{Thermal and chemical energy} (p.188-200)}
\label{Subsection-3-2}
\vspace*{0mm}

The \dashuline{discovery of the energy principle} had the most direct and powerful influence on the development of the \dashuline{theory of heat}, which is why even today people are sometimes inclined to regard this part of physics as the actual field of application of the principle, although apart from the historical facts mentioned, there is absolutely no evidence of this there is a special reason. 

Yes, one can say that \dashuline{the heat theory} owes its development and the successes it has achieved in modern times not even mainly to the \dashuline{discovery of the energy principle}, even though this was the first impetus for its transformation, but to the same extent perhaps \dashuline{even more so the application of Carnot's principle}, which is completely independent of this, which \dashuline{Clausius} introduced as the \dashuline{second law into the heat theory}, and also of the mechanical conception of heat founded by \dashuline{Joule}, \dashuline{Krönig} and \dashuline{Clausius}, which is also \dashuline{completely independent of the energy principle}.
Through the latter assumption, which has so far been very well proven, \dashuline{the heat theory} has been made \dashuline{a part of mechanics}, and \dashuline{the thermal energy therefore leads back to the mechanical energy}, which we have just discussed. 

\vspace*{-2mm} 
\begin{center} 
{\it\color{red}\bf
---------
(p.189) 
---------
}
\end{center}
\vspace*{-3mm}

Therefore, \dashuline{our task} here should be \dashuline{to develop the conclusions of the energy principle using only} the closest \dashuline{empirical facts}, \dashuline{completely independent of the mechanical conception of heat}, i.e. \dashuline{of any hypothesis about the molecular structure of bodies, and also of Carnot's principle}. 

After all, the field that presents itself for application under these restrictive conditions is still general. 
 
%

\vspace*{-2mm} 
\begin{center}
--------------------------------------------------- 
\end{center}
\vspace*{-3mm}

In the following we will again use the sentence that \dashuline{the increase in energy of a material system} is equal to the corresponding \dashuline{work done outside the system}, or \dashuline{its equivalent}. 
The external effects can consist of both \dashuline{mechanical and thermal changes}, and in the latter case we have to determine the mechanical equivalent of such a change first of all by special consideration. 

{\it\color{red}I have previously shown/recalled that\:}
the work value of any change is equal to the amount of mechanical work which either produces that change, or which is produced by it, regardless of the way in which this occurs. 
Although we cannot convert heat into work without further secondary changes (which also includes an increase in volume), we can, conversely, convert work completely into heat, and experiments have shown that heating can always be brought about by expending work whose amount corresponds to the quantity the heat generated is in a constant ratio, independent of materials, temperature, etc.
\dashuline{Every amount of heat} (more precisely: every heating) \dashuline{is equivalent to a certain amount of work}.

%

This fact could only be established through \dashuline{experience}, and only on the basis of this special experience is \dashuline{it possible to apply the principle of conservation of energy to thermal processes}. 
The \dashuline{equivalence of heat and work} cannot in any way be deduced from the energy principle.
The latter also leaves room for completely different views, such as \dashuline{Carnot}'s theory, according to which mechanical work would not be equivalent to a quantity of heat, but rather to the product of the quantity of heat and temperature (cf. p.13). 

\vspace*{-6mm} 
\begin{center}
{\it\color{red}\bf
---------
(p.190) 
---------
}
\end{center}
\vspace*{-3mm}

The number which indicates the ratio of a quantity of heat to its equivalent work was known 
from Joule in the metric system and referred to Cels.Grade set at $423.55$~g \,{\it\color{red}(\,i.e.\:4.155\:J~cal${}^{\:-1}$)}.
However, \dashuline{recent repeated measurements} have shown that this number is too small and that approximately \dashuline{$428$~g}\,{\it\color{red}(\,i.e.\:4.20\:J~cal${}^{\:-1}$)} must be substituted (after all, the third digit may still be inaccurate by $1$ or $2$ units{\it\color{red}, \,and indeed it should be 1.5 unit lower only to give the modern values of $\:426.5$~Kgr-m corresponding to \:4.184\:J~cal${}^{\:-1}$}). 
Of the compilations of all the numerous calculations of these constants carried out in a variety of ways, we mention, in addition to those already mentioned in the first section, 
those by 
\dashuline{Sacchetti}$\,$\footnote{$\:$G. Sacchetti: Considerazioni intorno all' origine della teoria meccanica del calore {\it\color{red}(Considerations regarding the origin of the mechanical theory of heat)} Memor. dell' Acc. di Bologna VIII (2) p.149, 1869.}
and especially those by 
\dashuline{Rowland}$\,$\footnote{$\:$H. A. Rowland: On the mechanical equivalent of heat, with subsidiary researches on the variation of the mercurial from the air thermometer and on the variation of the specific heat of water. Proc. Amer. Acad. (2) VII, p.75, 1880.}

%

If one sets the heat unit as the heat which is equivalent to work of $1$, then \dashuline{the mechanical equivalent of heat becomes $=1$}, and the expressions are simplified accordingly.
\dashuline{We will make use of this measurement system in the following}.

Now that we are able to measure the value of the work of the external effects that correspond to a specific change in the state of the material system under consideration, we can use this to \dashuline{calculate the} resulting \dashuline{change in the energy of the system}. 
However, here again, as with the investigation of elastic forces above, the \dashuline{superposition principle} makes it possible \dashuline{to resolve the energy equation} into two individual equations and thereby simplify the task. 
Firstly, the energy of the material system breaks down into 2 parts: \dashuline{the molar-{\it\color{red}macroscopic} energy} (living force-{\it\color{red}kinetic energy} of mass movement, potential of gravity, etc.) and the \dashuline{molecular-{\it\color{red}internal} energy} (heat, chemical energy).

\vspace*{-2mm} 
\begin{center}
{\it\color{red}\bf
---------
(p.191) 
---------
}
\end{center}
\vspace*{-4mm}

On the other hand, you can always divide the external effects into two parts, one of which includes all those effects that influence the molar-{\it\color{red}(macroscopic)} energy, the other those that change the molecular-{\it\color{red}(internal)} energy.

%

\dashuline{By equating the changes in the two types of energy} individually with the external effects relating to them, \dashuline{two equations are obtained}, only one of which is of particular interest here, since the other belongs to mechanics (see the previous section). 
It is obvious that this separation can generally be carried out with greater or lesser ease 
whether 
the molar-{\it\color{red}(macroscopic)} energy is already known in advance. 
One has only to think of the whole process as being divided into a part relating to molar-{\it\color{red}(macroscopic)} movement and a part relating to molecular-{\it\color{red}(microscopic)} movement. 
The heat supplied from outside by \dashuline{conduction or radiation} 
only affects the molecular-{\it\color{red}(internal)} energy, while the mechanical external effects (pressure, shock, friction) generally cause both molar and molecular changes.

%

We therefore want to simplify the task from now on by examining only those states of a (finite or infinitely small) body system in which the molar-{\it\color{red}(macroscopic)} energy is not taken into account at all. 
Let us then denote the molecular-{\it\color{red}(internal)} energy by $U$, also the heat that disappeared during any (finite or infinitely small) change of state in the environment (transferred into the system through conduction or radiation) by $Q$ (measured mechanically), and finally the heat expended in the environment mechanical work with $A$, then the work value of the applied external effects is $Q + A$, therefore: 
\begin{align}\tag{1}
 U' \;-\; U \;=\; Q \;+\; A \; , 
\label{label_eq_thermal_1}
\end{align} 
where \dashuline{$U$ and $U'$ are the values of the {\it\color{red}(internal)} energy of the system in the initial state and in the final state} of the process under consideration. 
The type {\it\color{red}(kind)} of transition is irrelevant. 
If we take a closer look at the mechanical nature of heat, we can divide $U$ into actual energy (free heat) and potential energy (internal work, chemical energy).
However, since these two types of energy are not linked to special external effects, we will not make use of this distinction in the following.

\vspace*{-2mm} 
\begin{center}
{\it\color{red}\bf
---------
(p.192) 
---------
}
\end{center}
\vspace*{-3mm}

Depending on the nature of the change in state, various statements about the thermal or chemical behavior of the system can be derived from the above equation. 
If we first assume a body whose state is determined by a single variable, then {\it\color{red}(the internal energy)} $U$ can be understood as a function of this unique variable. 
In general, the choice of independent variables will be arbitrary, therefore $U$ can appear in different forms, but they all represent the same value. 
For example, do we have a liquid or a gas that is heated infinitely little by the supply of heat from outside at a constant volume, then $Q= c_v \:.\: d\vartheta$ and $A=0$, where \dashuline{$c_v$ is the heat capacity at a constant volume} and \dashuline{$\vartheta$ means the absolute temperature} measured by the air thermometer (the exact definition of the temperature is, however, only possible with the help of Carnot's principle, although experience shows that it agrees within fairly wide limits with that obtained by the expansion of the air). 
It follows from equation (\ref{label_eq_thermal_1}) that: 
$$ d\,U \;=\; c_v \:.\: d\vartheta \; . $$
If we now carry out a finite process while maintaining the assumed conditions, we can consider $\vartheta$ as the only independent variable and  
get$\,${\color{red}\footnote{$\:${\it\color{red}Note that an integration constant $U_0$ should appear here to give: $U({\vartheta})= \int_{\vartheta_0}^{\vartheta} c_v(\vartheta')\,.\, d\vartheta' \,+\, U_0({\vartheta_0})$ (P. Marquet)}.}}: 
$U= \int c_v\,.\, d\vartheta$, where $c_v$ of course only depends on $\vartheta$.
But we can easily introduce another quantity (such as the pressure $p$) determined by $\vartheta$ and $v$ as an independent variable, and express both $\vartheta$ and $U$ in terms of the same (because since $v$ is constant, the value of $p$ also determines the entire state). 

Let us now assume the more general case that when the state changes, the volume $v$ is also changed, but in such a way that a certain relationship exists between volume and temperature via $f(v, \vartheta) = 0$ (such as in adiabatic changes of state), where $f$ means any function of two variables.
Then the state again depends on only one variable ($\vartheta$ or $v$ or $p$), of which $3$ variables one is always co-determined by the other two.

\vspace*{-6mm} 
\begin{center}
{\it\color{red}\bf
---------
(p.193) 
---------
}
\end{center}
\vspace*{-3mm}

Accordingly, the {\it\color{red}(internal)} energy $U$ can be represented at will as a function of $\vartheta$ or $v$ or $p$, and as long as one limits oneself to the introduced condition, it is quite indifferent which of these 3 different forms one wants to consider as the primary expression of the {\it\color{red}(internal)} energy and to take as a basis for the view.

But if we move on to processes whose course is determined by two independent variables, the arbitrariness in the representation of $U$ diminishes by one degree. 
If a body changes its temperature by $d\vartheta$ and at the same time, independently of this, its volume by $dv$, which also determines the change in pressure $dp$, then its energy changes by a certain amount $d\,U$. 
As always, this also gives the mechanical equivalent of the corresponding external effects: the sum of the work expended and the heat supplied, and in general one of these two quantities (e.g. the external work) is still arbitrary, 
while the other is determined by it.
Let us now  
assume that the external work {\it\color{red}($\,A\,$)} consists precisely in overcoming the pressure $p$.
Then we have: 
$$ A \;=\; - \: p \:.\: dv \; , $$ 
and thus according to (\ref{label_eq_thermal_1}): 
$$ d\,U \;=\;  Q \;-\; p \:.\: dv \; .  $$
The heat supplied $Q$ can be expressed by the {\it\color{red}(specific)} heat capacities at constant volume $c_v$ and at constant pressure $c_p$ by assuming, for example, that the body is first brought from pressure $p$ to pressure $p + dp$ at constant volume by the supply of heat, and then from volume $v$ to volume $v + dv$ at constant pressure ($p + dp$). 
If we now use $v$ and $p$ as independent variables for the sake of convenience, we obtain by simple calculation for the total heat supplied from outside: 
$$ Q \;=\; 
 c_v \:.\: \frac{\partial\,\vartheta}{\partial\,p} 
 \:.\: dp
 \;+\;
 c_p \:.\: \frac{\partial\,\vartheta}{\partial\,v}  
 \:.\: dv
 \; , $$ 
and therefore for the change in energy: 
\begin{align}\tag{2}
 d\,U \;=\; 
 c_v \:.\: \frac{\partial\,\vartheta}{\partial\,p} 
 \:.\: dp
 \;+\;
 \left(
 c_p \:.\: \frac{\partial\,\vartheta}{\partial\,v} 
 \;-\; p
 \right) 
 .\: dv 
 \; , 
\label{label_eq_thermal_2}
\end{align} 
\vspace*{-6mm} 
\begin{center}
{\it\color{red}\bf
---------
(p.194) 
---------
}
\end{center}
\vspace*{-3mm}
from which it immediately follows: 
$$ 
 \frac{\partial}{\partial\,v} 
 \left(
 c_v \:.\: \frac{\partial\,\vartheta}{\partial\,p} 
 \right) 
 \;=\;
 \frac{\partial}{\partial\,p} 
 \left(
 c_p \:.\: \frac{\partial\,\vartheta}{\partial\,v} 
 \;-\; p
 \right) 
 \; .
$$
If $c_v$ and $c_p$ are constant, as with perfect gases, then we have: 
$$
\left( c_p \:-\: c_v \right) \:.\:\:
 \frac{\partial^2\,\vartheta}{\partial\,p \,.\,\partial\,v}
 \;=\; 1 
 \; .
$$
In fact then:
$$ 
 \vartheta \;=\; \frac{p\:.\:v}{c_p\;-\;c_v}
 \; .
$$
Furthermore, it follows from (\ref{label_eq_thermal_2}):
$$ 
 U \;=\; \frac{c_v}{c_p\;-\;c_v} \:.\: p\:v \; , 
 \;\;\mbox{or} \;=\; c_v\:.\:\vartheta
 \; .
$$ 

Each of these two forms of {\it\color{red}(internal)} energy is equally justified as long as the relation expressed by the combined Mariotte's and Gay Lussac's (Boyle's and Charles') law is allowed. 
The \dashuline{internal energy of a gas} can then be imagined at will either as a \dashuline{tension force} (which would give the expansion endeavour of the gas approximately the same character as that of a system of resting repelling points) or as a \dashuline{living force {\it\color{red}(kinetic energy)}} (in which case the pressure force is provided by the impact of freely flying molecules). 

However, if Gay Lussac-Mariotte's law is not understood as an identity that applies from the outset, but rather (as is the case in the mechanical theory of heat) as the special property of a state that has become stationary (in that the pressure $p$ is only defined at all when the velocities of the molecules are equal), if the velocities of the molecules of a gas element (taken relative to that of their centre of gravity) are balanced in a certain way in terms of magnitude and direction, then the primary, generally valid form of energy can no longer depend on the pressure, and it is necessary to recur to the expression $c_v\: . \:\vartheta$ which (with a suitable mechanical definition of temperature) can be retained under all circumstances.

The same applies to the {\it\color{red}(internal)} energy $U$ of any body whose state depends on two variables. 
Since the so-called equation of state always gives a relationship between pressure, volume and temperature, $U$ can always be expressed and interpreted using any two of these three quantities. 

\vspace*{-2mm} 
\begin{center}
{\it\color{red}\bf
---------
(p.195) 
---------
}
\end{center}
\vspace*{-3mm}

In the mechanical view, however, the form that relates to the independent variables $v$ and $\vartheta$ will always appear as primary, because pressure represents a concept that is only derived from these two quantities under special circumstances.


The internal state of a homogeneous isotropic body that is deformed in any way and in elastic equilibrium depends on 7 variables, namely the 6 deformation quantities (p. 173), and the temperature. Accordingly, the internal energy of the body is also a function of these 7 variables, and only under special conditions (e.g. if, as was assumed on page 173, the deformation takes place when heat is prevented) can the number of variables be reduced. The application of the \dashuline{principle of conservation of energy} here again follows equation (\ref{label_eq_thermal_1}). 

The external work $A$ can generally be easily calculated with the aid of the formulae of the theory of elasticity, whereas the heat $Q$ supplied from outside cannot be expressed directly by the known constants that depend on the nature of the body, since knowledge of the heat capacities at constant pressure and constant volume is not sufficient for this purpose. 
The more complicated the properties of the bodies to be analysed become, the more variables the value of the energy will depend on, and the less arbitrary will be the primary expression that can be generally established for the {\it\color{red}(internal?)} energy of a body.

A certain type of molecular energy of a body that has not yet been taken into account deserves special attention, which, in contrast to the types discussed so far, is referred to as surface 
energy$\,$\footnote{$\:$See: J. Cl. Maxwell: Theory of heat. Deutsch von F. Neesen, Braunschw. 1878, p. 318.}. 
Experience shows that the energy of a body is generally dependent not only on its internal state, but also on the shape of its surface, so that a link must be added to the expression of energy considered so far, which depends on the nature of the body surface and the nature of the adjacent body. 

\vspace*{-3mm} 
\begin{center}
{\it\color{red}\bf
---------
(p.196) 
---------
}
\end{center}
\vspace*{-4mm}

Essentially, however, this type of energy is not at all different in nature from the types discussed so far.
Indeed, one must never think of the contact of two bodies in such a way that the inner state of the bodies is the same at all points up to the immediate surface and then suddenly the jump into the neighbouring medium takes place, but one always has to do with a physical boundary layer, of small but finite thickness, in which forces are active which depend on the nature of the two media at the same time, and it is the internal energy of this boundary layer which one is inclined 
to introduce as surface energy. 
This view also resolves the question of whether it is necessary to include the amount of surface energy in the body's energy, since it does not only depend on the state of the body itself, but also on that of its surroundings. If the boundary layer is included in the system under consideration, the surface energy must be included in the expression of the energy of the system.
Otherwise it will be left out, and all surface changes must then be taken into account as external effects.

Surface energy plays the most important role in bodies whose surface changes slightly, i.e. in liquids. In relation to a single surface element, the energy will of course be proportional to the size of this element.
Therefore, the energy of a finite piece of the surface, if the state of the boundary layer is the same at all points on it, is equal to the size of this piece of surface, multiplied by a constant which depends on the nature of the two adjacent media, their temperature, etc. 
The conditions of capillary equilibrium follow easily from this theorem. 

If movement occurs through the action of the forces acting on the surface, surface energy is transformed into the living force {\it\color{red}(kinetic energy)} of molar movement, and the surface energy must always decrease because the living force {\it\color{red}(kinetic energy)} necessarily increases from the value $0$ onwards. 

\vspace*{-2mm} 
\begin{center}
{\it\color{red}\bf
---------
(p.197) 
---------
}
\end{center}
\vspace*{-3mm}

If the surface energy is at a minimum in a state of rest, no movement can occur because the surface energy can no longer decrease, i.e. in this case there is stable equilibrium. 
In a similar way, the task is completed if, in addition to the surface forces, gravity is also assumed to be effective.
 Let us now proceed to the consideration of processes which are formed not only by physical changes in the bodies concerned, but also by \dashuline{chemical effects}, whether they are chemical compounds in the strict sense (i.e. reactions according to fixed weight ratios) or also so-called chemical mixtures, to which the processes of absorption, dissolution, etc., are also to be reckoned.
In each such case, the application of \dashuline{the energy principle} proceeds according to equation (\ref{label_eq_thermal_1}), and the increase in energy of the entire changed system caused by the process is thus equal to the sum of the heat supplied from outside and the work expended from outside. 
 In very many cases, the value of the latter is negligibly small compared to the other amounts of energy taken into account here, and it disappears completely if the effect takes place at constant total volume, but can also often be neglected in other reactions that take place under constant pressure (atmospheric pressure), namely in the case of solid and liquid bodies, which generally change their volume so little that the external work performed in the process is omitted.
Even when it comes to the expansion of gases (like in combustion or explosion phenomena), the value of the external work often forms such a small fraction of the amounts of energy that are otherwise converted that this work can be neglected even with significant changes in volume. 
If you leave it out, the equation reduces to: 
$$ U'\;-\; U \;=\; 0 \; .$$
\vspace*{-11mm} 
\begin{center}
{\it\color{red}\bf
---------
(p.198) 
---------
}
\end{center}
\vspace*{-3mm} 
The heat supplied from outside alone is equal to the \dashuline{increase in energy}.
Its amount therefore \dashuline{only depends on the initial and final state of the system}, but not on the type of {\it\color{red}(the kind of)} transition. This expresses the well-known statement (see p.20) that the heat generated by a series of chemical reactions is independent of the order in which the individual reactions are carried out. Of course, this theorem only applies if one either ignores the external work entirely, as we do here, or takes particular care to ensure that it has a value that is independent of the order of state changes (as in reactions under constant pressure).

Since the heat $Q$ can be observed directly, the equation established makes it possible to \dashuline{measure the energy change}. 
Of outstanding importance is the case where the system returns to the initial temperature (and pressure) after the reaction has ended. 
Then $Q$ is very often negative (exothermic reaction), i.e. the system has given off heat to the outside, and in this sense the amount of $Q$ is referred to as the \dashuline{heat of reaction}, \dashuline{heat tone}, \dashuline{heat value}, \dashuline{heat of formation}, etc. --a term that (as is well known) can be separated from that of \dashuline{chemical relationship} or \dashuline{affinity}.
A special circumstance must be taken into account when forming the value $U$ of the energy of a chemically composed body in any state. 
It goes without saying that this value {\bf\dashuline{must always be related to the same zero state}} (see p. 102), \dashuline{which can be chosen at random at the beginning}, but  {\bf\dashuline{must be retained afterwards during the transition to other}}, also chemically different, {\bf\dashuline{states}}. 

Since the energy corresponding to any given state is always defined by the work value of the external effects that appear {\bf\dashuline{during the transition to the zero state}}, {\bf\dashuline{the choice of the zero state}} is best made with a view \dashuline{to making the transition to it as easy as possible}. 

As long as \dashuline{only physical changes} of the body are involved, \dashuline{the zero state} will be any conveniently situated state of the same body.
Then the \dashuline{transition to the zero state\;} is only a purely physical process.

\vspace*{-2mm} 
\begin{center}
{\it\color{red}\bf
---------
(p.199) 
---------
}
\end{center}
\vspace*{-3mm}

But if chemical changes come into consideration, then in general that transition will have to be associated with chemical effects. 
 Assuming, for example, that we are dealing with the complete or partial decomposition (or formation) of a quantity of water, the most convenient way to {\bf\dashuline{find the value of the energy as a zero state}} is to assume the \dashuline{corresponding quantities of oxygen and hydrogen completely separately next to each other}, for example at 0° Celsius and 1 atmosphere.
Then, in any state of the system, with any advanced decomposition, the energy is easy to define, because for any separated quantity of oxygen or hydrogen it is given by the physical transition to 0° Celsius and 1 atmosphere of pressure, whereas for a quantum of water the chemical decomposition into the two elements must be added (it has been generally explained above on page 101 that the reverse process; i.e. \dashuline{the formation of water}, \dashuline{can be used instead to define the energy}).
This shows that as soon as changes in the chemical composition of a body are involved, \dashuline{an additive constant must be added to the physical energy} of the body: the \dashuline{chemical energy}, which is always \dashuline{equal and opposite to the heat value of the formation of the body from its constituent parts}. 
During the {\bf\dashuline{transition to the zero state (decomposition of the body)}}, the \dashuline{chemical energy} then appears as (positive or negative) heat. 
This also applies to solutions, mixtures, etc.: the energy of any salt solution is equal to the negative heat of solution of the salt. 
This value is sufficient if we confine ourselves to the investigation of changes in the per cent content of the solution.
But if we extend the consideration to processes in which the salt or the solvent itself undergoes decomposition, a further element must be included in the expression of energy, which is equal to the heat released in the decomposition of the substance in question into its other constituents.

\vspace*{-2mm} 
\begin{center}
{\it\color{red}\bf
---------
(p.200) 
---------
}
\end{center}
\vspace*{-3mm}

When \dashuline{applying the energy principle to chemical processes}, it is of course just no more necessary, as in other cases, to regard all the bodies taking part in the reaction as a single system (Basic System p. 121).
Rather, one can pick out a particular (finite or infinitely small) material complex at will, and then analyse its energy change. 
This is, as always, equal to the \dashuline{mechanical equivalent of the external effects} that are expended in the change of state in question. 
Moreover, for the practical determination of \dashuline{chemical energy}, the theorems we discussed in the previous section (on page 100 and following) are used with favour.

\vspace*{2mm}

\subsection{\underline{Electrical and magnetic energy} (p.200-247)}
\label{Subsection-3-3}
\vspace*{-1mm}

Of all the conclusions which \dashuline{the principle of conservation of energy} allows us to draw on \dashuline{the laws of action of the various forces active in nature}, those relating to \dashuline{electricity and magnetism} claim \dashuline{a particularly outstanding interest}, because in no other part of physics does the \dashuline{fruitfulness of this principle} come into its own \dashuline{in such a pure, direct way}. 

First of all, \dashuline{as far as mechanics is concerned}, it had already reached a high degree of development a considerable time \dashuline{before the discovery of the general energy principle}, so that all that remained for the \dashuline{application of the principle} to the doctrine of the phenomena of motion \dashuline{was to prove its agreement with the laws already established} with sufficient certainty by other means, and thus to verify one known theorem on another known theorem.

\dashuline{Things were somewhat different in the theory of heat}: here the newly discovered principle had a powerful influence on the development of the ideas that people tried to form about the nature of thermal processes.
However, even in this area of research it did not remain the only recognized and proven leader for long. 
Since the high fertility of the \dashuline{mechanical conception of heat} has become apparent, people are generally \dashuline{much more inclined} to allow themselves \dashuline{to be guided by the idea of purely mechanical processes} when considering thermal processes \dashuline{than to go back to the general energy principle} that is independent of these ideas.

\vspace*{-5mm} 
\begin{center}
{\it\color{red}\bf
---------
(p.201) 
---------
}
\end{center}
\vspace*{-3mm}

Indeed, it is immediately obvious that consideration of the latter alone does not provide the specialization of ideas necessary to achieve a clear view, which makes the mechanical imagination so valuable to us.


\vspace*{-5mm} 
\begin{center}
--------------------------------------------------- 
\end{center}
\vspace*{-3mm}


However, in the field of electricity and magnetism we now encounter very different conditions from those just described. 


 
So far we have by no means succeeded in \dashuline{uniting the variety of phenomena} that confront us here \dashuline{under a uniform conception}, namely in tracing them back to a \dashuline{thorough analogy with phenomena that are known to us from other sources} and have become familiar to us through habit (i.e. of a mechanical nature), and it is perhaps doubtful whether this will ever succeed.

At all events, however, in undertaking the task of discovering the laws of electricity, we find ourselves from the first in possession of \dashuline{no other reliable aid} to research \dashuline{than the principle of the conservation of energy} alone.  

This is where \dashuline{the meaning of the principle {\it\color{red}(of the conservation of energy)}} becomes clearest: detached from all secondary ideas, it \dashuline{forms the only safe starting point for the investigation}, as it indicates the guiding train of ideas, which  \dashuline{is an essential prerequisite} for the rational use of the results of experiment and observation.

\vspace*{-5mm} 
\begin{center}
--------------------------------------------------- 
\end{center}
\vspace*{-3mm}

However, the remark must be added here that, in the interest of a convenient and understandable form of expression, it often seems unavoidable to use terminology in the application of the principle to certain individual processes that is reminiscent of \dashuline{certain special ideas about the operation of electricity}. 

We speak, for instance, of \dashuline{electricity as if it were a special substance} that moves and exerts forces. 
However, these expressions in no way involve any judgment about the nature of electrical effects.
Rather, the nature of the effects remains completely open. 

\vspace*{-2mm} 
\begin{center}
{\it\color{red}\bf
---------
(p.202) 
---------
}
\end{center}
\vspace*{-4mm}

Yes, we can even leave undecided for the time being the question (which is already somewhat near to being finally decided) whether \dashuline{electricity and magnetism act directly into the distance}, or whether the \dashuline{effects are brought about by corresponding changes in the intermediate medium}, although we shall sometimes be compelled, in order to fix the conception to some extent, to base our designations on one or the other mode of conception.

\vspace*{-3mm} 
\begin{center}
--------------------------------------------------- 
\end{center}
\vspace*{-3mm}


All electrical and magnetic effects can be divided into two large groups, depending on whether they cause movements of the ponderable matter itself (current conductors, magnetized bodies, etc.) or changes in the internal electrical or magnetic state of the bodies. 

We want to characterize these two types of effects, as is usually done, as \dashuline{ponderomotive} (mechanical) and \dashuline{induction} (electromotive, magnetomotive) effects.

We must take both into account in every application of our principle.

\vspace*{-3mm} 
\begin{center}
--------------------------------------------------- 
\end{center}
\vspace*{-3mm}

Let us now deal first with processes which are caused by the \dashuline{effects of quantities of electricity} that are \dashuline{at rest in the bodies}. 

It is assumed that: a number of material points, each charged with a certain quantum of electricity, are in an insulating medium (air); then the effects of the electricity will be limited to the ponderomotive ones, and the points will begin to move under their influence, whereby we will simply assume that the medium does not offer any noticeable resistance to the movement. 

For the sake of completeness, however, we must include another assumption here (which we will maintain in the following), namely that \dashuline{the relative velocities} of moving bodies \dashuline{are always infinitesimally small} compared to \dashuline{the so-called critical speed (300,000 km per second)} because otherwise certain electrodynamic effects will occur in addition to the electrostatic ones (which we will only have to take into account later). 

Then we are given the most convenient view of the entire course of the movement through the idea that the \dashuline{electricities act on one another at a distance}, the like ones \dashuline{repelling}, the unlike ones \dashuline{attractive}, according to \dashuline{Newton's general law of gravitation}. (Polarization of the dielectrics see below p. 206).

\vspace*{-3mm} 
\begin{center}
{\it\color{red}\bf
---------
(p.203) 
---------
}
\end{center}
\vspace*{-3mm}


If no external effects are exerted on the system, its energy is constant, but it consists of two parts: the \dashuline{living force {\it\color{red}(kinetic energy)}} of the moving points, and the \dashuline{potential of the active (central) forces} (p.165). 
This potential 
$$  P \;=\; \sum \: \frac{e \:.\: e'}{r} \;  $$ 
is therefore to be regarded in this case as an electrostatic energy $U$, 
where $e$ and $e'$ (with corresponding signs) denote the electrical charges of two points (each combination taken once in the sum $\sum$) and $r$ their distance (always positive).

Their change {\it\color{red}(namely of the kinetic energy and of the potential $P$)} in a certain time is equal and opposite to the work done during this time by the electric-powderomotive forces.

From the expression of $P$ it is easy to see that the electrical potential of several systems is equal to the sum of the potentials of the systems on themselves, increased by the sum of the potentials of two systems on each other.

\vspace*{-5mm} 
\begin{center}
--------------------------------------------------- 
\end{center}
\vspace*{-3mm}

Since $P$ (as energy) has the dimension of a quantity of work, the above equation also represent the \dashuline{electrostatic measure of electricity}.

This also solves a question that one occasionally hears from students: {\it what is the {\it\color{red}value of the} mechanical equivalent of electricity} (more precisely: the electrical potential). 

This {\it\color{red}value} is $=1$, both in the electrostatic and in the magnetic dimension (for which, however, the above equation {\it\color{red}for $P$} no longer applies).

\vspace*{-3mm} 
\begin{center}
--------------------------------------------------- 
\end{center}
\vspace*{-3mm}

If one selects a limited number from the entire set of points, and considers them as a basic system (p.121), \dashuline{the energy of this system} changes by the amount of the \dashuline{external effects applied}. 
  And the same is obviously true for  
the work of the ponderomotor forces 
(which is carried out by the points located outside the points of the basic system).

\vspace*{-3mm} 
\begin{center}
{\it\color{red}\bf
---------
(p.204) 
---------
}
\end{center}
\vspace*{-3mm}


The process is somewhat different when the carriers of electricity are not individual points, but rather spatially extended conductors, because in this case induction effects are added to the ponderomotive effects. 
We can assume that the displacement of electricity required to establish electrical equilibrium within a conductor occurs at a speed infinitely greater than the movements of the conductor.

Then each constellation of the conductor system corresponds to a specific arrangement of the electricity in the conductors, which is given by the condition that the value of the electrical potential function  
$$  \varphi \;=\; \sum \: \frac{e}{r} \;  $$ 
is constant for all points of 
one and the same 
conductor, while in the surrounding insulator the condition: $\Delta \varphi=0$ applies everywhere. 

This and the given charge quantities of the conductors determine the function $\varphi$, while the electrical density on the surface of a conductor has the value $-1/(4\:\pi)\:.\:\partial \varphi /\partial n$ 
 (if $n$ 
  means the normal drawn to the inside of the insulator.

\vspace*{-5mm} 
\begin{center}
--------------------------------------------------- 
\end{center}
\vspace*{-3mm}

Let us now assume that the conductors move under the influence of the forces emanating from their charges, and consider the change which the electrical potential of the entire system on itself: 
$$  P \;=\; \sum \: \frac{e \:.\: e'}{r} \; ,
\;\;\;\;\mbox{or} \;
\;=\; \frac{1}{2} \: \sum \: e \:.\: \varphi \; , $$ 
suffered in a time element. 

For this purpose we break down the entire infinitely small change that the system undergoes into two parts: 
1) the conductors change their position in space, while the electricity in them is fixed; 
  2) caused by the changed in the constellation, the electricities assume a new equilibrium position in the conductors, while these themselves are at rest.

By each of these two processes $P$ will be changed, and it is easy to see that the first change is nothing other than the negative work of the ponderomotive forces, just as in the case previously discussed, while the second change compared to the first is infinitely small. 

\vspace*{-3mm} 
\begin{center}
{\it\color{red}\bf
---------
(p.205) 
---------
}
\end{center}
\vspace*{-3mm}

Because the new electrical distribution differs from the initial one in that an infinitely small (positive or negative) particle $\delta\,e$ has been added to each originally existing electrical particle $e$, and the resulting change in the entire electrical potential is, according to p. 203, the potential of all newly added electricity $\delta\,e$ on the entire originally existing charge $e$ (because the potential of the newly added electricity on itself is small a,d of a higher order). 

The desired change is therefore $\sum(\varphi\:.\:\delta\,e)$, and this sum is equal to $0$ for each individual conductor, since $\varphi$ has the same value everywhere in it, and its total charge $ \sum\,e$ remains unchanged.

\vspace*{-3mm} 
\begin{center}
--------------------------------------------------- 
\end{center}
\vspace*{-3mm}


Therefore, the temporal change in the potential $\delta P$ is fully represented by the negative ponderomotive work, i.e. by the decrease in the live force {\it\color{red}(kinetic energy)} of the moving conductor.
\dashuline{The sum of the potential and the living force {\it\color{red}(kinetic energy)} is therefore constant}. 

Since, on the other hand, \dashuline{the total energy of the system remains constant}, it follows that the kinetic energy is supplemented by an electrostatic {\it\color{red}(energy)} $U$, which in this case is again measured by the positive electrical potential $P$. 

Therefore, the effects caused by a movement process (such as the one just considered) consist only in the \dashuline{mutual conversion of electrostatic and kinetic energy}. 

The induction effects that take place in the conductors have no finite work value, and therefore \dashuline{cannot deliver any electrical heat} because, although a finite amount of electricity generally flows through a cross-section of a conductor in a finite time, the work it does is infinitely small, since the current does not have a finite potential gradient.

\vspace*{-3mm} 
\begin{center}
--------------------------------------------------- 
\end{center}
\vspace*{-3mm}


A thermal effect only occurs when two conductors come close enough that they can equalize each other's charges. 
In this case there is a rapid reduction in the value of the electrostatic energy, which does not benefit the living force {\it\color{red}(kinetic energy)} of the conductor movement, but rather the molecular energy.

\vspace*{-6mm} 
\begin{center}
{\it\color{red}\bf
---------
(p.206) 
---------
}
\end{center}
\vspace*{-3mm}
If it is a matter of \dashuline{heat generation} (when discharging electrical batteries), this directly gives the amount of discharge heat in mechanical measure. 

However, the \dashuline{transformation of electricity into heat} does not occur instantaneously, but rather more or less complicated electrodynamic processes usually take place (e.g. oscillating discharge), and in the final state all types of energy are reduced again to \dashuline{thermal and electrostatic energy}.

\vspace*{-3mm} 
\begin{center}
--------------------------------------------------- 
\end{center}
\vspace*{-3mm}


Let's now move on to the case where, in addition to the conductors, there are also dielectrics in the system. 
For the sake of simplicity, we will initially assume that the dielectrics, like the conductors, are rigid, and only the latter will be charged with certain given amounts of electricity, and both types of bodies may move freely in a completely insulating non-polarizable medium. 

According to Faraday, we define the constitution of a dielectric as the fact that a very large number of very small conducting bodies are inserted into an absolutely insulating medium, in which, under the influence of electrical forces, an electrical distribution is induced just as in finitely extended conductors. 

In mathematical treatment, this assumption leads to the theory of dielectric (or magnetic) polarization founded by Poisson, according to which the dielectric moment relating to the unit volume at any point of a dielectric is the same in direction and proportional in magnitude to that at this point acting electrical force, and also a constant that depends on the nature of the dielectric. So the components of this moment satisfy the equations:
\vspace*{1mm} 
\begin{equation} \tag{1}
\left.
\begin{aligned}
  \lambda \;=\; -\:\varkappa\:.\:\frac{\partial\,\varphi}{\partial x}
  \; , \;\;\;\;\;\;\;\;\;\;\;\;
  \mu     \;=\; -\:\varkappa\:.\:\frac{\partial\,\varphi}{\partial y}
  \; , \;\;\;\;\;\;\;\;\;\;\;\;
  \nu     \;=\; -\:\varkappa\:.\:\frac{\partial\,\varphi}{\partial z}
  \;\; .
\end{aligned}
\;\;\;\;
\right.
\label{Eq_Planck_1887_Elec_Mag_1} 
\end{equation}
The larger the values of the (positive) constant $\varkappa$, the more the behavior of the dielectric approaches that of a conductor, and (however)  for $\varkappa = 0$ the polarization disappears completely.

\vspace*{-3mm} 
\begin{center}
{\it\color{red}\bf
---------
(p.207) 
---------
}
\end{center}
\vspace*{-4mm}

Since dielectric polarization is completely traced back to the phenomena presented by perfect conductors through the idea on which we are based, we can easily transfer the results obtained for the \dashuline{application of the energy principle} to this case. 

Also, the value of the electrostatic energy $U$ is given here {\it\color{red}(for dielectrics)} by the electrical potential of the entire system on itself, i.e. by the expression
\vspace*{1mm} 
\begin{equation}
\left.
\begin{aligned}
 P \;=\; \frac{1}{2} \: \sum \: e \:.\: \varphi \;\; .
\end{aligned}
\;\;\;\;
\right.
\nonumber 
\end{equation}
Since $\varphi$ has a constant value in each of the small conducting bodies of which a dielectric is composed, and in addition the quantity of charge in question is equal to $0$, this sum vanishes for all quantities of electricity $e$ that are in the dielectrics, and needs to be extended only to the conductors with finite charges. 
The presence of a dielectric only influences the value of the electrostatic energy insofar as the potential function $\varphi$ is modified by it.

\vspace*{-3mm} 
\begin{center}
--------------------------------------------------- 
\end{center}
\vspace*{-3mm}

The characteristics of the mode of operation of a dielectric are brought much closer to the view and made more accessible to calculation by the well-known theorem that the electricity contained in a dielectric can be completely replaced with regard to all of its physical effects by a simple electrical layer which is on the surface of the dielectric Dielectric is spread out and at any point has the density: 
\vspace*{1mm} 
\begin{equation}
\left.
\begin{aligned}
-\:\left\{\:
        \lambda \: \cos\,(\,n\:x\,) 
  \:+\:   \mu   \: \cos\,(\,n\:y\,)
  \:+\:   \nu   \: \cos\,(\,n\:z\,)
\:\right\}
\;=\; \varkappa\:.\:\frac{\partial\,\varphi}{\partial n}
 \;\; , 
\end{aligned}
\;\;\;\;
\right.
\nonumber 
\end{equation}
where $n$ means the direction of the normal drawn to the interior of the dielectric. The total mass of this layer is $0$.

\vspace*{-5mm} 
\begin{center}
--------------------------------------------------- 
\end{center}
\vspace*{-3mm}

Therefore, while the potential function $\varphi$ (with its differential quotients, inside and outside the dielectric) retains the same value at all points in space if the electrical distribution in a dielectric is replaced by the fictitious surface layer mentioned, it should be noted that the same is not the case with the value of the potential of all the electricity present in the system on itself.

\vspace*{-2mm} 
\begin{center}
{\it\color{red}\bf
---------
(p.208) 
---------
}
\end{center}
\vspace*{-3mm}

Indeed, the general expression of this potential 
$P = (1/2) \: \sum \: e \:.\: \varphi $ 
delivers a value that is different from the one above, as you can immediately see when you look at it, and consider the part of the sum that comes from the electricity $e$ of the fictitious surface layers. 
And since the potential function $\varphi$ is not constant everywhere on the surface of a dielectric, the corresponding sum will generally assume a (negative) value other than $0$, whereas in the case considered above (on the previous page) and based on the electrics $e$ of the dielectric, $0$ became. 

On the other hand, we get exactly the same numbers as above for the electricity $e$ contained in the conductors.
Therefore, the resulting value of the potential $P$ here will in any case be different (smaller) than in the case previously considered. 

From all what has been said, it follows that as soon as the electrical distribution in a dielectric is thought to be replaced by the physically equivalent surface layer, one has to distinguish between the concepts: \dashuline{the potential} $P$ and \dashuline{the energy} $U$ (because the latter quantity naturally remains the same in both cases, due to its physical meaning).

\vspace*{-3mm} 
\begin{center}
--------------------------------------------------- 
\end{center}
\vspace*{-3mm}


This difference also arises immediately if we use an earlier consideration to \dashuline{derive the energy from the potential}. 
If we consider the infinitesimal change $\delta P$, which the potential $P$ undergoes when the conductors and dielectrics move in a time element, we can imagine it broken down into two parts: 
\begin{enumerate}[leftmargin=8mm,parsep=0mm,itemsep=1mm,topsep=-1mm,rightmargin=2mm]
\item 
\!\!\!) the change due to the spatial displacement of the conductors and dielectrics, while the electrics are fixed in them;
\item 
\!\!\!) the change due to the new arrangement of the electrics in the bodies, while they themselves are fixed in space.
\end{enumerate}
The first change represents the negative work of the ponderomotive forces, i.e. the decrease in the living force {\it\color{red}(kinetic energy)} $T$, but the second is equal to the potential of all newly added electricity $\delta e$ to the originally existing $e$ (p. 205).
\vspace*{2mm} 
\begin{center}
{\it\color{red}\bf
---------
(p.209) 
---------
}
\end{center}
\vspace*{-3mm}
Therefore this is
$= \sum \: (\,\varphi \:.\:\delta e\,)$, 
and: 
$$ \delta P 
   \;=\; -\:\delta\,T 
   \:+\: \sum \: (\,\varphi \:.\:\delta e\,) \; ,  $$ 
or, since the total energy: $T + U = const.$: 
$$ \delta P 
   \;=\; \delta\,u 
   \:+\: \sum \: (\,\varphi \:.\:\delta e\,) \; .  $$ 
The second term on the right side of the equation is by no means $=0$ if the corresponding surface coverage is substituted for the electrical state of the dielectric, because $\varphi$ is not constant on the surface of a dielectric, but that term can be determined with the help of of the conditions established for the equilibrium of electricity prove to be a complete differential of a certain function of the state by converting it into a space integral using (\ref{Eq_Planck_1887_Elec_Mag_1}) by integrating with time and obtaining: 
\vspace*{1mm} 
\begin{equation} \tag{2}
\left.
\begin{aligned}
 P \;=\; U 
 -\:\bigintsss 
  \: \left(\:\lambda^2\:+\:\mu^2\:+\:\nu^2\:\right)
  \:.\;\, \frac{d\,\tau}{2\:\varkappa}
\;\; , 
\end{aligned}
\;\;\;\;
\right.
\label{Eq_Planck_1887_Elec_Mag_2} 
\end{equation}
where the integral over $d\tau$ is to extend over the volume elements of all dielectrics. 
Of course, the difference between the quantities $P$ and $U$ is the same as that already stated above. 
For $\varkappa=\infty$ (dielectric completely conductive) energy and potential merge into one another, just as for $\varkappa = 0$ (dielectric completely unpolarizable), because then $\lambda,\:\mu,\:\nu = 0$ (See magnetic induction p. 225).

\vspace*{-3mm} 
\begin{center}
--------------------------------------------------- 
\end{center}
\vspace*{-3mm}

In the same way, the case can be treated in which the dielectrics are charged from the outset with certain amounts of electricity that have been supplied to them from outside, either on the surface or in the interior. Here the principle of superimposing the effects simply leads to the goal. Furthermore, if the dielectrics are represented by liquid or deformable solid bodies instead of rigid ones, which can also continuously fill the spaces between the conductors, then similar considerations can also be made successfully, although it must also be taken into account that the Dielectric constant $\varkappa$ generally varies with the deformation state of the dielectric 
substance$\,$\footnote{$\:$H. v . Helmholtz: Über die auf das Innere magnetisch oder dielektrisch polarisierter Körper wirkenden Kräfte {\it\color{red}(On the forces acting on the interior of magnetically or dielectrically polarized bodies)}. Wied. Ann. 13, p.385, 1881. Wiss. Abh. I, p.798.}.

\vspace*{-2mm} 
\begin{center}
{\it\color{red}\bf
---------
(p.210) 
---------
}
\end{center}
\vspace*{-3mm}

We can summarise the previous considerations relating to electrostatics in the following theorem: 
\\
\vspace*{-8mm}
\begin{quote}
in all cases where electrically charged conductors and dielectrics move under mutual influence, as long as the equations established above for the electrical distribution (p. 206) apply, \dashuline{the electrical energy} is only ever converted into \dashuline{the living force {\it\color{red}(kinetic energy)} of the molar movement}, and the magnitude of \dashuline{the electrical energy $U$} is given by \dashuline{the value of the potential $P$} of the total electricity present in the system on itself, with a modification given by equation (\ref{Eq_Planck_1887_Elec_Mag_2}) for the case where the corresponding surface layer is used instead of the arrangement of the electricity in a dielectric.

This sentence only has an exception when two conductors come so close to each other that their charges equalize, and in this case \dashuline{molecular energy} is created.
\end{quote}
\vspace*{-4mm}

\vspace*{-3mm} 
\begin{center}
--------------------------------------------------- 
\end{center}
\vspace*{-3mm}


It is understood that by means of the \dashuline{conditional equations} which apply to electrical equilibrium, the \dashuline{expression of electrical energy} can be expressed in \dashuline{very diverse forms} (which, however, differ only in relation to mathematical usefulness, while in physical respect they are quite equally justified, as long as those equations exist). 

Therefore, for the \dashuline{application of the energy principle}, it is completely irrelevant whether one puts together the \dashuline{expression of energy} from the combinations of two finitely distant electrical particles, starting from the assumption of an \dashuline{immediate action at a distance}, or whether one calculates the energy by summation, assuming \dashuline{purely molecular effects} over all space elements that are in an electrically constrained state: \dashuline{the numerical value of the energy is the same in both cases}. 

\dashuline{A decision} in favor of a particular of \dashuline{the two theories mentioned} can only be obtained by \dashuline{observing phenomena} in which the conditions assumed above for electrical equilibrium do not exist.

\vspace*{-2mm} 
\begin{center} 
{\it\color{red}\bf
---------
(p.211) 
---------
}
\end{center}
\vspace*{-3mm}
We will return to this question later.

\vspace*{-3mm} 
\begin{center}
--------------------------------------------------- 
\end{center}
\vspace*{-3mm}

 Let us now take a closer look at a case that we had to exclude from the discussion above: the contact of several conductors.

First, we imagine two rigid metallic conductors touching each other along any surface, located in an insulating, non-polarizable medium, which contains any other electrified conductors, and charged with a certain quantity of electricity. 

As experience shows, the electricity in both conductors arranges itself in such a way that its potential function $\varphi$ at the interface of the conductors undergoes a jump that depends only on the molecular state of the bodies (substance, temperature), while inside each conductor is constant in accordance with what has been said so far. 

  From this it is clear that there is an electrical double layer at the interface, consisting of two simple electrical layers of opposite and equal density that are very close to each other (since there is no discontinuity in the values --all $= 0$ {\it\color{red}(null in average)}-- 
of the differential quotients of $\varphi$).

If the product of the density of the positive layer in the distance between the layers (the moment of the double layer) is denoted by $\omega$, then $\varphi$ grows in the positive direction as it passes through the double layer (i.e. from the negative to the positive layer) around 
$4\:\pi\:.\:\omega=\phi_0$ (electrical voltage).

\vspace*{-3mm} 
\begin{center}
--------------------------------------------------- 
\end{center}
\vspace*{-3mm}


This provides a means of measuring $\omega$ electrostatically, while the size of the two factors from which $\omega$ is formed can only be approximately estimated. 

Helmholtz has shown that the formation of such a double layer can be attributed to the effectiveness of the specific attractive forces of the ponderable molecules of the conductor substances on the electricity.
Then the electrical voltage is represented as the work that these forces do when a positive unit of electricity passes through the layer in the above-mentioned direction. 

\vspace*{-2mm} 
\begin{center}
{\it\color{red}\bf
---------
(p.212) 
---------
}
\end{center}
\vspace*{-3mm}

If one also adds the conditions that the potential function $\varphi$ (and its differential quotients) must fulfill (inside the conductors, in the external space and at the boundary of both), as well as those {\it\color{red}(conditions)} which are given by the quantity of charges on the pair of conductor (and the other conductors), then, for each constellation of the conductor system in space, only a single value of $\varphi$ (and thus only a single arrangement of the electricity) is possible.

\vspace*{-3mm} 
\begin{center}
--------------------------------------------------- 
\end{center}
\vspace*{-3mm}

Let us now again form the expression of the electrostatic energy of the system $U$, i.e. the potential of the entire electricity on itself: 
$P = (1/2)\:\sum\:(e\:.\:\,\varphi)$ 
and examine the change that this quantity undergoes in a time element when the two (firmly connected) conductors move simultaneously with the other conductors in the electrical field under the influence of the electrical forces. 

This change $\delta P$ can again be thought of as being brought about successively: 
\begin{enumerate}[leftmargin=8mm,parsep=0mm,itemsep=1mm,topsep=-1mm,rightmargin=2mm]
\item 
\!\!\!) by a displacement of the conductors, while the electricity is fixed in them (negative ponderomotive work or decrease in living force: $-\delta T)$;
\item 
\!\!\!) by changing the electrical arrangement in the stationary conductors, whereby one can imagine that a positive or a negative value $\delta e$ is added to each electrical particle $e$ of the system.
\end{enumerate}
Then the corresponding change in the total potential is equal to the potential of the electricity $\delta e$ on the originally existing $e$ (p. 205), i.e.: $= \sum \: (\varphi\:.\:\delta e)$.  

For an isolated conductor, this sum is equal to $0$, since $\varphi$ is constant in it, but for the permanently connected pair of conductors, as one can easily see, it is equal to $\varphi_0\:.\:\delta E$, if $\varphi_0$ is the (positive) potential difference of the conductors and $\delta E$ denotes the entire amount of positive electricity transferred in a positive direction from one metal to the other (during the element of time under consideration). 

For the sake of simplicity, we only consider positive electricity as moving. So we have: 
$$
 \delta\,U \;=\; \delta\,P \;=\;
   -\:\delta\,T \;+\; \varphi_0\:.\:\delta E
 \; ,
$$ 
or 
$$
 \delta\,U \;+\; \delta\,T \;-\; \varphi_0\:.\:\delta E 
 \;=\; 0 \; .
$$

\vspace*{-3mm} 
\begin{center}
--------------------------------------------------- 
\end{center}
\vspace*{-3mm}


From this it becomes clear the important result that, in the case under consideration, \dashuline{the sum} of the \dashuline{living force {\it\color{red}(kinetic energy)}} $T$ and the \dashuline{electrostatic energy} $U$ is by no means constant.

\vspace*{-2mm} 
\begin{center}
{\it\color{red}\bf
---------
(p.213) 
---------
}
\end{center}
\vspace*{-4mm}

Rather, both values can even grow or decrease at the same time, and their sum therefore \dashuline{does not represent the total energy} of the system, but \dashuline{a third type of energy} must come into play here, the change in which just compensates for the first two changes mentioned. 
The amount of this is known immediately: it is \dashuline{the product of the electrical voltage} of the metals \dashuline{into the positive quantity of electricity} that passes from the negative to the positive metal.

However, no final decision has been made as to what physical form this type of energy should take.

\vspace*{-3mm} 
\begin{center}
--------------------------------------------------- 
\end{center}
\vspace*{-3mm}

There are two main views that come into consideration here, which will be discussed below. 

According to one view, in addition to the \dashuline{usual electrostatic energy} (which arises from the interactions of electricity with one another), there is a special type of energy (of a potential nature) which has its origin in the effects which are exerted on the electricity by the ponderable molecules.
One could therefore call it ``\,\dashuline{electromolecular}\,'' \dashuline{energy}. 

According to this, every electrical particle $e$ always has a special \dashuline{electromolecular energy}, and its amount is equal to the product of the quantity of electricity $e$ into a constant $G$ (which is dependent on the molecular nature --substance, temperature-- of the conductor in which it is currently located, but not on its electrical state). 

It is clear that this assumption easily fulfills its purpose, because 
as long as a quantity of electricity $\delta E$ moves between molecules of the same conductor, its electromolecular energy $G\:.\:\delta E$ (a very definite quantity) remains constant, whereas as soon as it passes into another conductor, its magnitude changes by $(G'-G)\:.\:\delta E$.

Therefore, one only has to assume the difference $G'-G$ to be equal and opposite to the known electrical voltage $\varphi_0$, in order to obtain the change in energy which is required by \dashuline{the principle of conservation of energy}.
{\it\color{red}However,} \dashuline{an {\bf additive constant} remains {\bf arbitrary} in the value of $G$}. 

Then the principle {\it\color{red}(of conservation of energy)} is satisfied by the fact that in the entire system under consideration \dashuline{the sum} of the \dashuline{living force} {\it\color{red}(kinetic energy)}, the \dashuline{electrostatic energy} and the \dashuline{electromolecular energy} is a quantity \dashuline{independent of time}.

\vspace*{-3mm} 
\begin{center}
{\it\color{red}\bf
---------
(p.214) 
---------
}
\end{center}
\vspace*{-4mm}


According to a remark made above (p.211), $G$ is also equal (and opposite) to the work that the electromolecular forces do when the unit of electricity transfers from an arbitrarily fixed normal metal to the relevant (arbitrarily electrified) conductor. 
Helmholtz referred to this work as the ``\,galvanic value\,'' of the 
conductor$\,$\footnote{$\:$H. v. Helmholtz: Wiss. Abh. I, p.910 ($G$ ist hier mit entgegengesetztem Vorzeichen definiert) {\it\color{red}(\,$G$ is defined here with the opposite sign)}.}. 

The value of $G$ increases as one progresses from basic to more noble metals. 
Since the hypothesis described above describes the value of the electrical voltage of two conductors as the difference between two quantities $G$ and $G'$ (which depend only on the nature of the conductors), it obviously has the advantage that Volta's law of voltage necessarily follows from it ({\it\color{red}even if} for the same reason, however, it fails to serve for second-class conductors).

\vspace*{-3mm} 
\begin{center}
--------------------------------------------------- 
\end{center}
\vspace*{-3mm}

  In contrast to this, however, a different view can be put forward, which assumes that such a special type of energy (as \dashuline{the electromolecular energy} just described) \dashuline{does not exist at all}, but that the lack of energy stated above is \dashuline{covered by} a corresponding \dashuline{change in the usual molecular (thermal, chemical) energy}.

Accordingly, the sum of the living force {\it\color{red}(kinetic energy)}, the electrostatic energy and the molecular energy would be unchangeable, and in the motion process examined above (in connection with the proven change in the known types of energy) \dashuline{there would have to be molecular (here thermal) energy} of the amount $-\:\varphi_0 \:.\:\delta\,E$ appear. 

More generally, this requirement would be: {\it every time a positive quantity of electricity passes through an electrical double layer in the positive direction, a reduction in molecular energy occurs by the amount of the product of the quantity of electricity in the potential difference caused by the double layer}. 
On the other hand, the transfer of electricity from one conductor to another would, in itself (i.e. without the work of forces of electrical origin), not involving any change in energy.

\vspace*{-3mm} 
\begin{center}
{\it\color{red}\bf
---------
(p.215) 
---------
}
\end{center}
\vspace*{-4mm}


The experimental testing of this theorem actually leads to phenomena which can be understood as confirmations of it.

For conductors of the first class, it is the thermal effects discovered by Peltier at the soldering points of two metals, which, however, are only directly perceptible when considerable amounts of electricity are conducted in a continuous current through the separating surface of the conductors. The amount of heat generated or absorbed is actually proportional to the quantity and sign of the amount of electricity passed through.

On the other hand, as is well known, the values of the electrical voltages calculated from this generally do not agree at all with the numbers which indicate the voltaic voltages measured by direct electrostatic means. 

At least one can still help oneself here by assuming that, with the so-called voltaic voltage, we are not dealing with the real contact difference between the two metals, but that the presence of minimal amounts of gases or vapors affects the properties of the surfaces, and thus significantly modify the value of the voltage, as this can actually be 
proven$\,$\footnote{$\:$Über diesen Punkt vgl. {\it\color{red}/ On this point cf.} J. Cl. Maxwell: A treatise on electricity and magnetism. Oxford 1873. I. Ch.249 ; and R. Clausius: Die mechanische Behandlung der Elektricität {\it\color{red}(The mechanical treatment of electricity)}. Braunschweig 1878, p.172 ff.}.

\vspace*{-3mm} 
\begin{center}
--------------------------------------------------- 
\end{center}
\vspace*{-3mm}

To be on the safe side, the most advisable thing would be to combine both hypotheses by distributing the amount of energy in question 
$(-\:\varphi_0\:.\:\delta E)$ 
in indefinite parts between the electromolecular and the ordinary molecular energy.

Then it remains up to future experience whether one of these parts disappears completely, or in what proportion the two types of energy participate in the amount mentioned.

\vspace*{-3mm} 
\begin{center}
--------------------------------------------------- 
\end{center}
\vspace*{-3mm}


{\it\color{red}Differently}, since it is the task of natural science to describe the phenomena in the simplest possible way, it makes sense, before one decides to accept \dashuline{a completely new type of energy}, to try to see whether it works with the 
\dashuline{already known types of energy}. 

\vspace*{-2mm} 
\begin{center}
{\it\color{red}\bf
---------
(p.216) 
---------
}
\end{center}
\vspace*{-3mm}

In fact, given the current state of our experience, such an attempt not only does not seem hopeless, but even promises success to a high degree, because in addition to the heat phenomena occurring in first-class conductors (Peltier's, Thomson's heat), the chemical processes observed in second-class conductors can also be satisfactorily explained in this way. 

However, one can then no longer represent the voltaic voltage law applicable to first-class conductors as a consequence of the energy principle (it then appears rather as a consequence of the Carnot-Clausius principle {\it\color{red}(i.e. of the 2nd law and definition of the entropy 22 years before)}).

On the other hand, the advantage is gained that the second class conductors submit to exactly the same treatment as those of the first class, and are only distinguished from them in that chemical effects also take place in them.

\vspace*{-3mm} 
\begin{center}
--------------------------------------------------- 
\end{center}
\vspace*{-3mm}

In the following, according the views expressed by \dashuline{W. Thomson} and \dashuline{Cl. Maxwell}, we will therefore start from the assumption that \dashuline{there is no electromolecular energy}, but that: ``\,{\it as often as a positive quantity of electricity passes through an electric double layer in the positive direction, the molecular (thermal or chemical) energy there is reduced by the product of the quantity of electricity in the potential difference of the double layer}.\,'' 

We can leave the details of this process completely undiscussed here.

\vspace*{-3mm} 
\begin{center}
--------------------------------------------------- 
\end{center}
\vspace*{-3mm}

Accordingly, the application of the \dashuline{principle of the conservation of energy} to the case serving as the starting point for our considerations is as follows: 
{\it if several electrostatically charged metallic conductors of the same temperature }(some of which are isolated and some of which are conductively connected to each other in any number, the generalisation is immediate) {\it move in an electric field, not only does a conversion take place between the living force of the movement and the electrostatic energy, but heat effects also occur at the contact points of the conductors} (by immediately dissipating the generated heat into the environment, the temperature and thus the electrical voltage at the contact points remaining  unchanged).

\vspace*{2mm} 
\begin{center} 
{\it\color{red}\bf
---------
(p.217) 
---------
}
\end{center}
\vspace*{-3mm}

Against this Peltier heat, the Joule heat, which is generated inside the conductor by the flowing electricity, naturally disappears as infinitesimally small (cf. above p. 205).

What is strange here is that this transformation of living force {\it\color{red}(kinetic energy)} and electricity into heat is mediated by a reversible process, from which interesting conclusions arise for the Carnot-Clausius principle {\it\color{red}(the second law)}, but which do not belong to the scope of our current investigation.

\vspace*{-3mm} 
\begin{center}
--------------------------------------------------- 
\end{center}
\vspace*{-3mm}

If in a system of contacting conductors (first and second class) the fulfilment of all the conditions necessary for equilibrium becomes impossible (as is generally the case, for example, when several conductors of any temperature connected in series are united by contact of the first with the last to form a closed chain), a flow of electricity takes place which continues until the possibility of satisfying all the conditions of equilibrium is created (by the occurrence of certain chemical, thermal or mechanical changes).

At first we only want to deal with the case where the flow has become stationary, i.e. when all variables relating to the electrical state of the system (such as current intensity, electrostatic charges on the conductors, potential function, etc.) are independent of time. 

Then the types of electrical energy (electrostatic, electrodynamic energy) are also constant and can therefore be completely ignored, and any thermal and chemical processes may take place in the chain (with or without polarization).

\vspace*{-3mm} 
\begin{center}
--------------------------------------------------- 
\end{center}
\vspace*{-3mm}


First, we apply \dashuline{the energy principle} by imagining the entirety of the conductors forming the galvanic chain as a single system.

Since it is not subject to any external effects (induction effects are excluded), its \dashuline{total energy} is constant and therefore, according to what was said above, the \dashuline{sum of the thermal and chemical} (possibly also the mechanical) \dashuline{energy} remains unchanged. 

\vspace*{-2mm} 
\begin{center} 
{\it\color{red}\bf
---------
(p.218) 
---------
}
\end{center}
\vspace*{-3mm}

From this follows \dashuline{the well-known theorem}: {\it the thermal effects produced in a stationary galvanic current are equivalent to the chemical effects} (we want to exclude mechanical effects).

\vspace*{-3mm} 
\begin{center}
--------------------------------------------------- 
\end{center}
\vspace*{-3mm}

If by this proposition we obtain information about the magnitude of the effects produced by a current in its whole circumference, we do not yet know anything about the changes which take place in the individual parts of the chain.
But we can also gain knowledge of this if we subject not the whole conductor system, but rather some part of it as a basic system, to the \dashuline{application of the energy principle}.

Let us first take a case we have already discussed: the point of contact between two conductors. 
We pick out a part of the whole conductor system by imagining a separating surface on both sides of a contact point between two metallic conductors (and in such close proximity to the contact point that: essentially only the processes in the boundary layer itself come into question).

  The material complex thus delimited, consisting of two thin metal lamellae, considered as a basic system, will suffer certain external effects from the current passing through it, which \dashuline{change its energy}, and the \dashuline{mechanical work} value of the external effects, as always, gives the \dashuline{amount of the increase in energy}.

Now, if we do not know this work value directly, we can calculate it by using the reverse procedure, since our previous investigations have enabled us to state the \dashuline{change in energy} of the system under consideration. 

According to the assumption, the \dashuline{electrical energy} remains unchanged, the \dashuline{chemical energy} too, but the \dashuline{thermal energy} changes (according to the principle stated on page 216) in the time unit by $\:-\:\varphi_0\:.\:i\:$, where $i$ denotes the intensity of the current (always thought to be positive), and where $\varphi_0$ (the potential difference of the conductors) is positive if $\varphi$ increases in the direction of the current. 

\vspace*{-2mm} 
\begin{center} 
{\it\color{red}\bf
---------
(p.219) 
---------
}
\end{center}
\vspace*{-3mm}

Then, if $\varphi_0$ is measured electrostatically and by observing the heat generated, $i$ can be expressed in electrostatic measure. 
And since the \dashuline{change in energy} is now known, it follows that the \dashuline{mechanical work} value of the applied external effects is also expressed by $\:-\:\varphi_0\:.\:i\:$.

\vspace*{-3mm} 
\begin{center}
--------------------------------------------------- 
\end{center}
\vspace*{-3mm}



Let us now move on to another basic system, 
made of 
the total number of current conductors that remains if we exclude the lamellar part (just discussed) from the entire circuit. 
In this new system, the increase in \dashuline{molecular (thermal and chemical) energy} (and therefore also the \dashuline{work value} of the external effects expended) is $=\:+\:\varphi_0\:.\:i\:$, because the \dashuline{molecular energy} remains unchanged in the entire closing circuit.

\vspace*{-3mm} 
\begin{center}
--------------------------------------------------- 
\end{center}
\vspace*{-3mm}


We see from this that the amount of work done by the external effects depends only on the current intensity, and then on the difference in the values of the potential function at the two boundaries of the system. 


We can therefore state the following proposition: ``\,{\it in an (unclosed)  conductor system through which a stationary current flows (and in which any molecular changes take place), the molecular (thermal and chemical) energy is increased (in the unit of time) by an amount which is given by the product of: 
1) the ``current intensity'' and 
2) the ``decrease which the value of the potential function suffers'' (from the point of entry of the current to the point of exit, both assumed to be metallic)}\,'' 
(and it does not matter whether and which jumps the potential function makes inside the system.

\vspace*{-3mm} 
\begin{center}
--------------------------------------------------- 
\end{center}
\vspace*{-3mm}

  This theorem, which can just as easily be expressed for arbitrary distribution of current, as well as for spatially variable current densities in physical conductors, has a number of important applications.

\vspace*{-3mm} 
\begin{center}
--------------------------------------------------- 
\end{center}
\vspace*{-3mm}

Let's consider, for example, the effects exerted inside a metallic conductor by considering an infinitely small piece of an infinitely thin current thread with the cross section $q$ and the length $dn$. 
If $j$ denotes the current density, the intensity of the current flowing through the piece of conductor is $(\,j\:.\:q\,)$, and the decrease in the potential function is $\:-\:(\partial \varphi/\partial n) \:dn\,$, i.e. the change in molecular energy (here heat generation) $-\:j\:.\:(\partial \varphi/\partial n)\:.\:q \:dn$. 


\vspace*{-5mm} 
\begin{center}
{\it\color{red}\bf
---------
(p.220) 
---------
}
\end{center}
\vspace*{-3mm}

Since, according to Joule's experiments, the heat $j^2\:w\:.\:q \:dn$ appears, whereby $w$ (the specific resistance of the conductor) is defined, the Ohm's law (for conductors of the first class) follows by equating the two expressions, leading to: 
$$ j\:w \;=\; -\:\frac{\partial \varphi}{\partial n} \; . $$

  If there are temperature differences inside the conductor, this generally also causes electrical voltages, and the potential gradient then changes,  and therefore the heat generated inside (due to Thomson's effect, this is obviously equal to the product of the current intensity by the relevant electrical voltage).

Conversely, from Joule's law, any observed deviation in the value of the current heat generated in a conductor is an evidence of the presence of electrical voltages (electromotive forces) inside the conductor.

\vspace*{-3mm} 
\begin{center}
--------------------------------------------------- 
\end{center}
\vspace*{-3mm}

We have already discussed Peltier's effects above.
Let us now turn to what is happening in the second class conductors. 

If a stationary current passes between two metallic electrodes through an electrolyte (in which it produces any desired effects) then, according to the developed theorem, the growth of the molecular energy within (and on) the surface of the electrolyte is equal to the product of 1) the current intensity in 2) the potential difference between the positive and the negative electrode, regardless of whether local chemical reactions take place inside, or not.

Although the change in the total molecular energy of an electrolyte is easy to specify, it is much more difficult to separate the effects taking place on the two electrode surfaces from one another. 

  For this purpose we have to consider a basic system which is bounded (on the one hand) by a metallic conductor and (on the other) by an electrolyte. Then, our theorem does not apply without further explanation to the entry or exit of the current through a second-class conductor, because the external effects transmitted into the system by an electrolyte are not only mediated by flowing electricity, but also by flowing matter.

\vspace*{-5mm} 
\begin{center}
{\it\color{red}\bf
---------
(p.221) 
---------
}
\end{center}
\vspace*{-4mm}

Indeed, the transfer of the ions has a special effect on the system, simply because the ions carry molecular energy with them.
In this case, the principle stated on page 146 (under item 2) applies.

\vspace*{-3mm} 
\begin{center}
--------------------------------------------------- 
\end{center}
\vspace*{-3mm}

For the interior of an electrolyte, the \dashuline{application of the energy principle} is again simple, since due to the immutability of the chemical composition one can assume that in a space that is located entirely within an electrolyte, the ions carry in \dashuline{as much energy} on one side as on the other taken out by others. 
We can therefore completely neglect the external effects of this kind, and then get Joule's heat exactly as above, as it has also been proven experimentally, in connection with Ohm's law. 

This result now makes it easy to calculate the effects at the interfaces of an electrolyte separately from those inside. 
Suppose that the current $i$ flows through an electrolyte between two metallic electrodes in which the potential function has the values $\varphi$ and $\varphi'$. 
Furthermore, the potential change from an electrode to the interior of the initially adjacent undecomposed liquid layer ($\varepsilon$, or $\varepsilon'$) is positive if the potential function increases in the direction of the current. 
Then, according to the above, the change in the \dashuline{total molecular energy} inside and on the surface of the electrolyte (in the unit of time) is: $i\:.\:(\varphi-\varphi')\,$. 


On the other hand, the \dashuline{molecular energy} (Joule heat) developed inside the electrolyte is given (just as with first-class conductors) by the product of the current intensity in the decrease of the potential function, i.e. by: 
$$ i\:.\: \left(\: 
          [\: \varphi  \:-\: \varepsilon \:]
    \;-\; [\: \varphi' \:-\: \varepsilon'\:]
         \:\right) 
\; . $$ 
And so, by subtracting the last value from the first, the quantity $- i\:.\:( \varepsilon + \varepsilon')$ remains for the development of molecular energy at the two interfaces of the electrolyte, namely the product of the current intensity in the sum of the electromotive voltages acting on both electrodes (including any polarisation, of course).

\vspace*{-2mm} 
\begin{center}
{\it\color{red}\bf
---------
(p.222) 
---------
}
\end{center}
\vspace*{-3mm}

If the temperature at both electrodes remains constant, the molecular effect on the generation of chemical energy is reduced, and one can therefore set the electromotive force $( \varepsilon + \varepsilon' )$ acting on both electrode surfaces directly equal to that developed by the current unit chemical energy; i.e. the change of heat that would accompany the course of the same chemical processes if they proceeded in the usual way, without electrical excitation. 

In general, thermal effects (direct or secondary) will also occur on the electrode surfaces, and therefore this conclusion must of course be modified. The separate calculation of the thermal and chemical effects is a task for which the \dashuline{principle of conservation of energy} alone is not sufficient.

\vspace*{-3mm} 
\begin{center}
--------------------------------------------------- 
\end{center}
\vspace*{-3mm}

We see from this that the theorem (according to which the \dashuline{molecular energy} generated at the point of contact between two metallic conductors is given by the product of the current intensity in the electrical voltage prevailing there), only applies to electrolytes insofar as the \dashuline{molecular energy} developed together at both electrode surfaces is measured by the product of the current intensity into the sum of the voltages existing at both interfaces. 
This is precisely because the migration of the ions in the electrolyte transfers a certain \dashuline{amount of energy} from one electrode to the other, which cannot be determined from the outset. 

With regard to the processes that take place individually at the electrodes, no general rule can currently be derived from the \dashuline{energy principle}, as investigations into this are still in full swing. 
In any case, according to the explanations we have given, these processes are closely related to the migration of the ions, since the question of the \dashuline{amount of energy} that is transferred to or from the electrodes plays a decisive role here.
The answer to them is likely to be of essential importance for the assessment of these complicated processes.

Of course, it still remains to carry out the separation of the chemical from the thermal effects, which has already been touched on above, and which has recently been successfully based on the Carnot-Clausius principle {\it\color{red}(i.e. the second law of thermodynamics)}.

\vspace*{-3mm} 
\begin{center}
{\it\color{red}\bf
---------
(p.223) 
---------
}
\end{center}
\vspace*{-4mm}

Let us now return to the more general case that a stationary current passes  through an arbitrary system of conductors and between two metallic electrodes (with the potential functions $\varphi$ and $\varphi'$). 
According to the above, the increase in \dashuline{molecular energy} in the time unit is: $(\varphi-\varphi')\:.\:i\,$. 

But we can also express the same value, on the other hand, through the effects that take place inside and at the interfaces of the individual conductors. 
If $W$ denotes the sum of all resistances of the first and second class conductors (including any transition or secondary resistances), then the total Joule heat is: $i^2\:.\:W$. 

Furthermore, if $\varepsilon$ is the sum of all electromotive forces acting in the conductor system (including any voltages caused by differences in temperature, or structure, or by polarization inside the conductors), then according to the foregoing the sum of all molecular effects caused by this is: $-\, i\:.\:\varepsilon$. 

Therefore, for the \dashuline{total change in molecular energy} per unit time we have the expression: 
\vspace*{1mm} 
\begin{equation} \tag{3}
\left.
\begin{aligned}
 i^2\:.\:W  \;-\; i\:.\:\varepsilon  \;\; , 
\end{aligned}
\;\;\;\;
\right.
\label{Eq_Planck_1887_Elec_Mag_3} 
\end{equation}
which, identified with the above, gives the equation:  
\vspace*{1mm} 
\begin{equation}
\left.
\begin{aligned}
 i\:.\:W \;=\; \varepsilon 
         \;+\; \varphi \;-\; \varphi' \;\; , 
\end{aligned}
\;\;\;\;
\right.
\nonumber 
\end{equation}
in accordance with Ohm's law. 

If the beginning and end of the conductor system coincide so that the conductors form a closed circle, then $\varphi=\varphi'$, and we get Ohm's formula applied to the entire chain.

\vspace*{-3mm} 
\begin{center}
--------------------------------------------------- 
\end{center}
\vspace*{-3mm}

It is easy to extend the considerations made to branched{\it\color{red}(/arborescent?/ramified?)} currents and to the mutual contact of second-class conductors.
We shall therefore only take a brief look at the processes in inconstant currents.

If a conductor system is connected to form a closed loop, and then left to its own way, the result is generally a more or less rapidly changing current. 
But by considering it only for a very short time, we can treat it entirely from the point of view of the stationary currents just examined.

\vspace*{-2mm} 
\begin{center}
{\it\color{red}\bf
---------
(p.224) 
---------
}
\end{center}
\vspace*{-3mm}

However, it should be taken into account that not only the \dashuline{molecular energy} but also the \dashuline{electrical (electrostatic and electrodynamic) energy} changes, which makes the process somewhat more complicated, as the phenomena of self-induction come into play. 

However, if the current fluctuations only occur relatively slowly, induction can generally be ignored and the situation is the same as above (only with the difference that, here, as a result of the molecular effects, both the electromotive forces and the resistances generally change every moment and therefore, according to Ohm's formula, give rise to ever new current intensities and potential functions).

\vspace*{-3mm} 
\begin{center}
--------------------------------------------------- 
\end{center}
\vspace*{-3mm}


The interaction of magnetic or magnetized bodies can essentially be derived from the same assumptions as that of electrostatically charged conductors and dielectrics.
When discussing these phenomena we can therefore refer to the results obtained there (pp. 206 to 210). 

We will again base our analysis on a system of magnetic or magnetized bodies that move in a non-resisting, unpolarizable medium under the influence of their mutual ponderomotive and magnetomotive forces. 

Here we must also maintain a condition that we set earlier (p.202) for the electrostatic effects: namely, that the relative velocities do not exceed a certain upper limit, otherwise certain electrical forces are awakened in the moving bodies, which can modify their interaction.

\vspace*{-3mm} 
\begin{center}
--------------------------------------------------- 
\end{center}
\vspace*{-3mm}


The effects of each magnet can be replaced by those of a simple magnetic layer spread out in a certain way on its surface (generally combined with a certain spatial distribution of magnetic mass inside). 
The following definitions refer to this fictitious magnetic charge. 

The magnetic potential of the system on itself is 
$P=\sum \: m \:.\: m'\,/\,r$, in easy-to-understand symbols, or also 
$P = (1/2)\:\sum \:m\:.\:\varphi$ if the magnetic potential function is 
$\varphi = \sum \:m\,/\,r\,$. 

\vspace*{-2mm} 
\begin{center}
{\it\color{red}\bf
---------
(p.225) 
---------
}
\end{center}
\vspace*{-3mm}

The expression of $P$, which has the dimension of a work, also gives the unity of the quantity of magnetism in magnetic measure.

\vspace*{-3mm} 
\begin{center}
--------------------------------------------------- 
\end{center}
\vspace*{-3mm}


If only permanent magnets are involved, only ponderomotive effects come into consideration, and the magnetic energy $U$ is represented by the potential $P$. 

If, on the other hand, there are also magnetizable (paramagnetic or diamagnetic) substances in the magnetic field, the changeability of the temporary magnetism causes the potential $P$ of all magnetic and magnetized bodies to change, which does not benefit the ponderomotive effect. 

If we make the assumption that the magnetic equilibrium is currently established for each configuration of the system according to the equations of Poisson (see p.206), as can generally be assumed to be correct for weak magnetizations, then the \dashuline{energy conversion} takes place in the same way as in previous case, only between the living force {\it\color{red}(kinetic energy)} of the molar movement and the magnetic energy $U$. 

However, here $U$ is no longer measured by the potential $P$, but by the expression: 
\vspace*{0mm} 
\begin{equation}
\left.
\begin{aligned}
 U \;=\; P 
  \; + \bigintsss 
  \: \left(\:\lambda^2\:+\:\mu^2\:+\:\nu^2\:\right)
  \:.\;\, \frac{d\,\tau}{2\:\varkappa}
\;\; , 
\end{aligned}
\;\;\;\;
\right.
\nonumber 
\end{equation}
which can be obtained by the same considerations as the corresponding p. 209 {\it\color{red}and from Equation (\ref{Eq_Planck_1887_Elec_Mag_2})}.

\vspace*{-3mm} 
\begin{center}
--------------------------------------------------- 
\end{center}
\vspace*{-3mm}

In exactly the same way, the value of the magnetic energy $U$ is determined in the case that a body contains permanent and temporary magnetism at the same time. 
Only then the quantities $(\lambda, \:\mu, \:\nu)$ do not refer to the entire magnetic moment of the volume $d\tau$, but only to the induced (temporary) moment. 
If this disappears, $U$ is reduced again to $P$. 

It can be seen from this that, as long as the coercive force is effective only in the direction of maintaining permanent magnetism (but otherwise Poisson's induction conditions apply), any loss of \dashuline{magnetic energy} is completely replaced by a gain of \dashuline{living force {\it\color{red}(kinetic energy)}}, similar to the movement of points that are subject to fixed conditions independent of time.

\vspace*{-3mm} 
\begin{center}
{\it\color{red}\bf
---------
(p.226) 
---------
}
\end{center}
\vspace*{-4mm}

This still applies if other magnetization equations are assumed instead of Poisson's magnetization equations, if only temporary magnetism occurs as a specific function of the magnetizing force. 

Therefore, if the system passes from one configuration into a specific other (no matter how), then, since the magnetic energy suffers a specific change as a result, the increase in \dashuline{living force {\it\color{red}(kinetic energy)}} (and thus the ponderomotive work) must also be a very specific change (but the induction in the magnetizable substances can take place in very different ways in different ways).

\vspace*{-3mm} 
\begin{center}
--------------------------------------------------- 
\end{center}
\vspace*{-3mm}


However, as soon as the coercive force manifests itself in a time delay in the establishment of the internal magnetic equilibrium, there is generally a loss of \dashuline{magnetic energy} that is not covered by a corresponding increase in \dashuline{living force {\it\color{red}(kinetic energy)}}, so that the \dashuline{energy principle} here indicates the appearance of a \dashuline{third type of energy (heat)} demands, similar to the friction of moving bodies. 

Important conclusions from the energy principle can also be linked to the variability of magnetism with temperature; However, since these only become essentially fruitful through the use of the second law of heat theory, we will not go into them in more detail here.

\vspace*{-3mm} 
\begin{center}
--------------------------------------------------- 
\end{center}
\vspace*{-3mm}


So far we have treated the electrical and magnetic phenomena as two completely separate areas, considering only the interactions between electricity and magnetism. 
This also made it possible to use the electrostatic and magnetic measuring systems side by side, at the same time. 

  The relationship is different, however, when we move on to the consideration of electromagnetic effects, since they create a bridge, so to speak, between the two areas mentioned, which makes it possible to transfer either the electrostatic ground system into the magnetic realm, or vice versa, the magnetic ground system into the electrical realm, at will.
  
\vspace*{-3mm} 
\begin{center}
{\it\color{red}\bf
---------
(p.227) 
---------
}
\end{center}
\vspace*{-4mm}  

This transition is carried out by the so-called Ampère's theorem, which is to be regarded as the basic law of all electromagnetic phenomena, and which says that: ``\,{\it a linear electric current of intensity $i$ exerts exactly the same (ponderomotive, electromotive and magnetomotive) outward effects 
like 
a magnetic double layer of the moment $\omega$ bounded by the current curve and occupied in the corresponding known sense, and with $\omega = i\,$}.\,''

\vspace*{-3mm} 
\begin{center}
--------------------------------------------------- 
\end{center}
\vspace*{-3mm}


This equates the dimension of the \dashuline{moment of a magnetic double layer} {\it\color{red}``\,$\omega$\,''} (magnetic area density times length) to that of a \dashuline{current intensity} {\it\color{red}``\,$i$\,''} (amount of electricity divided by time), and one can therefore use this equation arbitrarily to \dashuline{either measure magnetism by electricity or electricity by magnetism} (but is in any case forced to decide on one of these two procedures). 

In the following we want to record the magnetic measurement system {\it\color{red}(namely ``\,$\omega$\,'')} .

\vspace*{-3mm} 
\begin{center}
--------------------------------------------------- 
\end{center}
\vspace*{-3mm}


  Let us now imagine the case that (in an infinitely extended field) any number of permanent magnets and linear (closed) unbranched current conductors (without sliding points), in which constant galvanic elements are active, move freely under mutual influence (whereby we disregard the influences of gravity).
Electrostatically charged bodies (conductors and dielectrics) can also be in the field, as the current conductors already contain resting free electricity. 
However, in the following we want to completely ignore the forces that emanate from the stationary electricity and which are otherwise exerted on it, on the assumption that they are simply superimposed on the interactions between currents and magnets.


  However, this assumption is not strictly correct, since it is highly likely that certain ponderomotive effects take place between electrostatically charged (resting) bodies and moving magnets or 
variable currents$\,$\footnote{$\:$H. Hertz: Über die Beziehungen zwischen den Maxwell'schen elektrodynamischen Grundgleichungen und den Grundgleichungen der gegnerischen Elektrodynamik {\it\color{red}(On the relationships between Maxwell's fundamental electrodynamic equations and the fundamental equations of opposing electrodynamics)}. Wied. Ann. 23, p.84, 1884.}, 
as well as between stationary magnets and moving static electricity (i.e. electricity at rest in the 
conductors$\,$\footnote{$\:$H. v. Helmholtz: Bericht betr. Versuche über die elektromagnetische Wirkung elektrischer Convection, ausgeführt von H. A. Rowland {\it\color{red}(Report regarding experiments on the electromagnetic effects of electrical convection, carried out by H. A. Rowland)}. Pogg. Ann. 158, p.487, 1876. Wiss. Abh. I, p.791.}).
%

\vspace*{-6mm} 
\begin{center}
{\it\color{red}\bf
---------
(p.228) 
---------
}
\end{center}
\vspace*{-4mm}

  {\it However, we can avoid the conclusions that can be drawn from these effects} (which are very weak in themselves, and which are directly related to the question of the propagation speed of the electromagnetic effects) {\it by assuming that the speed of the changes in the electromagnetic field} (as caused by movements of the magnets and conductors, or by fluctuations in the current intensities) {\it is so small that it does not come into consideration against this propagation speed} (a condition that is usually fulfilled). 

  Under this assumption, the effects mentioned disappear completely, and the theory of electromagnetic and electrodynamic effects founded by Ampère and F. Neumann forms a self-contained, complete and objection-free entity.

\vspace*{-5mm} 
\begin{center}
--------------------------------------------------- 
\end{center}
\vspace*{-3mm}


The course of the process taking place in the given system is completely determined if the state of the system is known at any point in time (i.e. the entirety of the positions, velocities, temperatures, current intensities, etc., see p. 107).
All of these quantities, including the last ones mentioned, must be given individually and independently of one another.
Then, the temporal changes are determined from this, and thus the entire process in a finite time. 

Since the \dashuline{total energy of the system} remains constant, its change in a time interval $dt$ is equal to $0$.

Let us form the corresponding expression. 

We have to differentiate between \dashuline{different types of energy}: the \dashuline{living force {\it\color{red}(kinetic energy)}} of conductors and magnets, the \dashuline{molecular energy} of conductors, and finally the \dashuline{electrodynamic and magnetic energy} caused by the presence of currents and magnets.

\vspace*{-3mm} 
\begin{center}
{\it\color{red}\bf
---------
(p.229) 
---------
}
\end{center}
\vspace*{-3mm}

The growth of \dashuline{living force {\it\color{red}(kinetic energy)}} is given by the work of all ponderomotive forces emanating from currents and magnets. 

According to Ampère's principle, a current $i$ acts on a magnet like a magnetic double layer of moment $i$, and consequently, according to the \dashuline{mechanical principle of action and counteraction}, a magnet acts on a current with a corresponding ponderomotive force, and finally, since all the effects of a magnet can be replaced by those of a current, one current on another in the same way. 

The ponderomotive work of two magnets is measured by the decrease in the magnetic potential if the magnetisms are considered constant.

Therefore, {\it by analogy with the magnetic potential $U$} (of the magnets on each other), {\it we now form the electromagnetic potential $V$} (of the magnets on the currents $i$, if these are thought to be replaced by magnetic double layers $i$) {\it and the electrodynamic potential $W$} (the current on each other, under the same assumption; here the potential of a current on itself must also be included, since a current conductor can also have a ponderomotive effect on its own parts). 

Then: 
  $$ V \;=\; i_1\:v_1 \;+\; i_2\:v_2 \;+\; .\:.\:. $$ 
where the sizes $v$ depend on the positions of the relevant conductors and all magnets, and where we always take the current intensities $i$ to be positive. 

Furthermore: 
\vspace*{0mm} 
\begin{equation}
\left.
\begin{aligned}
 W &\:=\; i_1\:i_2 \:.\: w_{12} \;+\; i_1\:i_3 \:.\: w_{13} 
   \;+\; .\:.\:. 
   \;\;\;
   \mbox{(each combination 
   {\it\color{red}--\:like $12$ and $21$--} calculated once)}
 \nonumber \\  
  & \quad
    \:+\; \frac{(i_1)^2}{2} \:.\: w_{11}
    \;+\; \frac{(i_2)^2}{2} \:.\: w_{22}
    \;+\; .\:.\:. 
\;\; , 
\end{aligned}
\;\;\;\;
\right.
\nonumber 
\end{equation}
where the sizes $w$ only depend on the positions of the conductors.  
It is easy to put it in the following form: 
\vspace*{0mm} 
\begin{equation}
\left.
\begin{aligned}
 w_{12} &\:=\; 
  -\:\bigintssss \bigintssss 
      \: \frac{\cos(d\,s_1 \, , \:  d\,s_2)}{r}
      \;\: d\,s_1 \; d\,s_2 
  \;\;\left(\:=\;  w_{21} \:\right)
 \nonumber 
\;\; , \\
 w_{11} &\:=\; 
  -\:\bigintssss \bigintssss 
      \: \frac{\cos(d\,s_1 \, , \:  d\,s'_1)}{r}
      \;\: d\,s_1 \; d\,s'_1 
 \nonumber 
\;\; , 
\end{aligned}
\;\;\;\;
\right.
\nonumber 
\end{equation}
in an easy-to-understand notation. 

The arc elements $ds$ are to be taken as positive in the direction in which the current flows through them.
In the integral of the conductor on itself {\it\color{red}(\,i.e. like  $w_{11}$\,)}, each arc element appears both as $ds$ and as $ds'$ because the factor $1/2$ is already included in the expression of $W$.






\vspace*{-2mm} 
\begin{center}
{\it\color{red}\bf
---------
(p.230) 
---------
}
\end{center}
\vspace*{-3mm}

The required work of the ponderomotive forces, i.e. the \dashuline{total increase in the living force {\it\color{red}(kinetic energy)} of conductors and magnets}, is therefore: 
\vspace*{0mm} 
\begin{equation} \tag{4}
\left.
\begin{aligned}
 -\:d\,U 
 & \:-\; 
 \left(\:
  i_1\:.\:dv_1 \;+\; i_2\:.\:dv_2 \;+\; .\:.\:. 
 \:\right)
 \nonumber \\  
 & \:-\; 
  \left(
          i_1 \; i_2 \:.\: dw_{12}
    \;+\; i_1 \; i_3 \:.\: dw_{13}
    \;+\; .\:.\:. 
    \;+\; \frac{(i_1)^2}{2} \:.\: dw_{11}
    \;+\; \frac{(i_2)^2}{2} \:.\: dw_{22}
    \;+\; .\:.\:. 
  \:\right)
\;\; .
\end{aligned}
\;\;\;\;
\right\}
\label{Eq_Planck_1887_Elec_Mag_4} 
\end{equation}


In the interest of later application to ponderomotive work in systems with physical current conductors and sliding points, it is not unimportant to emphasize here that this expression does not represent the complete decrease in the total potential $U + V + W$ in the time element $dt$, but rather only the partial decrease that arises from the movement of the bodies when the current intensities (and the magnetisms) are all considered constant.



The second type of energy to be considered is the molecular (thermal and chemical) energy of the conductor. 
If $w$ denotes the resistance, $E$ the (galvanic) electromotive force of a conductor circuit (positive if it acts in the sense of $i$), then according to (\ref{Eq_Planck_1887_Elec_Mag_3}), p.223: \dashuline{the molecular energy generated in it during the time $dt$} is $(i^2 \:.\: w - i \:.\: E) \:.\: dt$. 
For a stationary current left to its own devices, it is $= 0$. 

Now let's call $i \:.\: w - E = e$ the electromotive force induced in the conductor circuit (which makes Ohm's law also applicable to induced currents), then the \dashuline{increase in molecular energy} in all conductors is represented by: 
\vspace*{0mm} 
\begin{equation} \tag{5}
\left.
\begin{aligned}
 \left(\:
  i_1\;e_1 \:+\; i_2\;e_2 \:+\;  .\:.\:.  
 \:\right) \,.\: dt
 \nonumber 
\;\; .
\end{aligned}
\;\;\;\;
\right.
\label{Eq_Planck_1887_Elec_Mag_5} 
\end{equation}



Finally, there is the increase in the type of energy caused by the presence of the currents and magnets. If we set it $= Q$, we obtain the corresponding \dashuline{increase in energy}: 
\vspace*{0mm} 
\begin{equation} \tag{6}
\left.
\begin{aligned}
 d\,Q
 \nonumber 
\;\; .
\end{aligned}
\;\;\;\;
\right.
\label{Eq_Planck_1887_Elec_Mag_6} 
\end{equation}


\vspace*{-4mm} 
\begin{center}
{\it\color{red}\bf
---------
(p.231) 
---------
}
\end{center}
\vspace*{-4mm}


According to the \dashuline{principle of conservation of energy}, the sum of the expressions 
(\ref{Eq_Planck_1887_Elec_Mag_4}), 
(\ref{Eq_Planck_1887_Elec_Mag_5}) and 
(\ref{Eq_Planck_1887_Elec_Mag_6}) 
is equal to 0, {\it\color{red}which can be} symbolised by: 
\vspace*{0mm} 
\begin{equation}
\left.
\begin{aligned}
 \mbox{(\ref{Eq_Planck_1887_Elec_Mag_4})}
 \;+\;
 \mbox{(\ref{Eq_Planck_1887_Elec_Mag_5})}
 \;+\;
 \mbox{(\ref{Eq_Planck_1887_Elec_Mag_6})}
 \;=\; 0
 \nonumber 
\;\; .
\end{aligned}
\;\;\;\;
\right.
\nonumber 
\end{equation}




This equation can be used in two ways: 
either you can use it (if $Q$ is known) to calculate the values of the induced electromotive forces $e$ (each one individually); 
or you can (if the $e$ are given) find the value of $Q$.
However, \dashuline{the energy principle} does not achieve both at the same time (cf. p.46). 


We shall take the latter path here by applying the 
law of induction$\,$\footnote{$\:$Franz Ernst Neumann: Allgemeine Gesetze der inducierten Ströme {\it\color{red}(General laws of induced currents)}. Abh. d. kgl. Akad. d. Wiss. Berlin 1815. Pogg. Ann. 67, p.31, 1846.}
established by F. Neumann independently of our principle. 
Even without this, the values of $e$ are not completely arbitrary, because  for example we can see from the beginning how they depend on time, since the differential $dt$ only occurs explicitly in (\ref{Eq_Planck_1887_Elec_Mag_5}).

\vspace*{-5mm} 
\begin{center}
--------------------------------------------------- 
\end{center}
\vspace*{-3mm}

According to F. Neumann, the electromotive force induced in a (closed) conductor in the positive direction is equal to the time-based differential quotient of the potential of all magnets and currents on the imaginary conductor with the current $1$ flowing through it in the positive direction, multiplied by a positive absolute constant $\varepsilon$ (which we can leave undefined here), i.e.: 
\vspace*{0mm} 
\begin{equation}
\left.
\begin{aligned}
 e_1 \;=\;
 \varepsilon \:.\:
 \frac{d}{dt}\left(\:
 v_1 \;+\; i_1\:.\:w_{11} 
 \;+\; i_2\:.\:w_{12} \;+\; i_3\:.\:w_{13} 
 \:+\;  .\:.\:. 
 \:\right)
 \nonumber 
\;\; .
\end{aligned}
\;\;\;\;
\right.
\nonumber 
\end{equation}
Here the factor $1/2$ has been omitted for $w_{11}$ (coefficient of self-induction) because in this expression every combination of two conductor elements must appear twice, depending on whether current $i_1$ or current $1$ flows through one of them.


\vspace*{-4mm} 
\begin{center} 
{\it\color{red}\bf
---------
(p.232) 
---------
}
\end{center}
\vspace*{-4mm}

Substituting these values of $e_1, e_2, ...$ into expression (\ref{Eq_Planck_1887_Elec_Mag_5}) we obtain from the equation of conservation of energy:
\vspace*{0mm} 
\begin{equation}
\left.
\begin{aligned}
 dQ \;=\: d\,U 
 & \:+\; 
 \left(\:
  i_1\:.\:dv_1 \;+\; i_2\:.\:dv_2 \;+\; .\:.\:. 
 \:\right)
 \nonumber \\  
 & \:+\; 
  \left(\:
          i_1 \; i_2 \:.\: dw_{12}
    \;+\; i_1 \; i_3 \:.\: dw_{13}
    \;+\; .\:.\:. 
  \:\right)
 \nonumber \\  
 & \:+\; 
  \left(\:
          \frac{(i_1)^2}{2} \:.\: dw_{11}
    \;+\; \frac{(i_2)^2}{2} \:.\: dw_{22}
    \;+\; .\:.\:. 
  \:\right)
\;\; 
 \nonumber \\  
 & \:-\; 
 \varepsilon \:.\: i_1 \:.\: 
 d\left(\:
 v_1 \;+\; i_1\:.\:w_{11} 
 \;+\; i_2\:.\:w_{12} \;+\; i_3\:.\:w_{13} 
 \:+\;  .\:.\:. 
 \:\right)
 \nonumber \\  
 & \:-\; 
 \varepsilon \:.\: i_2 \:.\: 
 d\left(\:
 v_2 \;+\; i_1\:.\:w_{21} 
 \;+\; i_2\:.\:w_{22} \;+\; i_3\:.\:w_{23} 
 \:+\;  .\:.\:. 
 \:\right)
 \nonumber 
 \nonumber \\  
 & \:-\; 
    .\:.\:. \; .\:.\:. \; .\:.\:. \; .\:.\:. \; .\:.\:.
 \; .\:.\:. \; .\:.\:. \; .\:.\:. \; .\:.\:. \; .\:.\:.
 \; .\:.\:. \; .\:.\:. \; .\:.\:. \; .\:.\:. \; .\:.\:.
 \nonumber 
\end{aligned}
\right.
\nonumber 
\end{equation}
Since $dQ$ is a complete differential, it follows: 
\vspace*{0mm} 
\begin{equation}
\left.
\begin{aligned}
 & \varepsilon \;=\: 1 
 \;\;\;\;\;\;\;\;\;
 \mbox{(see p.45) \;\;\; and}
 \nonumber \\  
 Q \;=\: U 
 & \:-\; 
  \left(\:
          i_1 \; i_2 \:.\: dw_{12}
    \;+\; i_1 \; i_3 \:.\: dw_{13}
    \;+\; .\:.\:. 
    \;+\; \frac{(i_1)^2}{2} \:.\: dw_{11}
    \;+\; \frac{(i_2)^2}{2} \:.\: dw_{22}
    \;+\; .\:.\:. 
  \:\right)
\;\; 
 \nonumber \\  
\end{aligned}
\right.
\nonumber 
\end{equation}
or, taking into account the value of $W$: 
\vspace*{-3mm} 
$$ Q \;=\; U \;-\; W \;\; . $$

\vspace*{0mm} 
\begin{center}
--------------------------------------------------- 
\end{center}
\vspace*{-3mm}


Accordingly, the energy caused by the presence of magnets and currents consists of two types: the \dashuline{magnetic energy} (measured by the positive magnetic potential $U$, as before) and the \dashuline{electrodynamic (electrokinetic) energy} (measured by the negative electrodynamic potential $W$), while the \dashuline{electromagnetic potential} $V$ makes no contribution to the energy at all.

We will have to talk about this circumstance in more detail below (p.234). 

First, the equations formulated should be briefly applied to a few simple cases.

\vspace*{-3mm} 
\begin{center}
--------------------------------------------------- 
\end{center}
\vspace*{-3mm}

If there is no magnet in the field and only a single circuit, the energy $Q$ is reduced to $(i_1^2/2)\:.\:w_{11}$, and the equation for the conservation of the sum of the three types of energy is:
\vspace*{0mm} 
\begin{equation}
\left.
\begin{aligned}
  \left[\:
    -\; \frac{(i_1)^2}{2} \:.\: dw_{11}
  \:\right]
  \;+\;
  \left[\:
    i_1 \:.\: e_1 \:.\: dt 
  \:\right]
  \;-\; 
  d\left[\:
    \frac{(i_1)^2}{2} \:.\: w_{11}
  \:\right]
 \;=\; 0
 \;\; .
 \nonumber
\end{aligned}
\right.
\nonumber 
\end{equation}

\vspace*{-3mm} 
\begin{center}
--------------------------------------------------- 
\end{center}
\vspace*{-3mm}


For a rigid conductor, $w_{11}$ is unchangeable, i.e. the ponderomotive work $=0$, and the self-induction remains 
$e_1 = w_{11} \:.\: di_1/dt$. 

\vspace*{-4mm} 
\begin{center}
{\it\color{red}\bf
---------
(p.233) 
---------
}
\end{center}
\vspace*{-4mm}

From this, the current intensity $i_1$ is calculated according to Ohm's law (p. 230) if it is given at any given time. 

So here we only have \dashuline{conversions of different types of energy} within the circuit, while the \dashuline{total energy} of the conductor remains constant.

For a conductor and a magnet, there is also the \dashuline{living force {\it\color{red}(kinetic energy)}} generated by the ponderomotive effects.
Then we get the corresponding equation: 
\vspace*{0mm} 
\begin{equation}
\left.
\begin{aligned}
  \left[\:
    -\; i_1 \:.\:  dv_1
    \;-\; \frac{(i_1)^2}{2} \:.\: dw_{11}
  \:\right]
  \;+\;
  \left[\:
    i_1 \:.\: e_1 \:.\: dt 
  \:\right]
  \;-\; 
  d\left[\:
    \frac{(i_1)^2}{2} \:.\: w_{11}
  \:\right]
 \;=\; 0
 \;\; ,
 \nonumber
\end{aligned}
\right.
\nonumber 
\end{equation}
where: 
$$ e_1 \;=\; \frac{d(\, v_1 \:+\: i_1\:w_{11} \,)}{dt} \; . $$

\vspace*{-3mm} 
\begin{center}
--------------------------------------------------- 
\end{center}
\vspace*{-3mm}

If you ignore the self-induction of the current (i.e. if you set $w_{11}=0$), you can no longer choose $i_1$ arbitrarily at the beginning of the movement, but the current strength is then given immediately because the link with $di_1/dt$ is omitted from the equation, and therefore a condition between ``\,state variables\,'' (p.108) is created using Ohm’s formula.

\vspace*{-3mm} 
\begin{center}
--------------------------------------------------- 
\end{center}
\vspace*{-3mm}


If we look at the energy (\dashuline{living force {\it\color{red}(kinetic energy)}, molecular energy, electrokinetic energy}) of the conductor on its own, it only undergoes a change due to external influences. 

These consist firstly in the ponderomotive work that the magnet performs on the conductor, and through which the \dashuline{living force {\it\color{red}(kinetic energy)}} of the conductor is increased, secondly in the induction effect of the magnet on the conductor, the work value of which ($i_1 \:.\: dv_1/dt  \:.\: dt$) to increase the \dashuline{molecular energy} of the conductor is used, as can be seen directly from the above equation. 

Conversely, if we make the magnet the basic system, its energy consists only of the \dashuline{living force {\it\color{red}(kinetic energy)}} of its movement, and the only effect exerted on it by the current is the ponderomotive work through which its \dashuline{living force {\it\color{red}(kinetic energy)}} is changed. 

Therefore, the interactions between magnets and currents are not completely mutual, which we will come back to very soon.

\vspace*{-3mm}
\vspace*{-2mm} 
\begin{center}
{\it\color{red}\bf
---------
(p.234) 
---------
}
\end{center}
\vspace*{-3mm}


If one imagines that the magnet is replaced by a second current, the ponderomotor and induction effects take place in a corresponding manner on both sides. 

Then the energy of each individual conductor is increased in the time $dt$:  1) by the ponderomotive work done on it from outside, which turns into \dashuline{living force {\it\color{red}(kinetic energy)}}; and 
2) by the external induction work (product of the externally induced electromotive force in its current strength multiplied by $dt$), which contributes to the \dashuline{molecular energy}. 

The remaining changes in 
\dashuline{living force {\it\color{red}(kinetic energy)}} and 
\dashuline{molecular energy} 
are provided by the conductor circuit's own 
\dashuline{electrokinetic energy}. 

What is also noteworthy here is the fact that, due to the special form in which the \dashuline{electrokinetic energy} occurs, the \dashuline{total energy} of the two current conductors is not the simple \dashuline{sum of the energies} of the individual conductors, while in the case of a conductor and a magnet the corresponding theorem is valid.

\vspace*{-5mm} 
\begin{center}
--------------------------------------------------- 
\end{center}
\vspace*{-3mm}

Let us now go back to the general case of any number of permanent magnets and currents and take a closer look at the expression of the resulting \dashuline{energy $Q=U-W$}. 
It contains no member which, analogous to the \dashuline{living force {\it\color{red}(kinetic energy)}} of moving ponderable masses, allows one to conclude that the moving electricity is 
inertia$\,$\footnote{$\:$H. R. Hertz: Versuche zur Feststellung einer oberen Grenze für die kinetische Energie der elektrischen Strömung {\it\color{red}(Experiments to determine an upper limit for the kinetic energy of electrical flow)}. Wied. Ann. 10, p.414, 1880 , Wied. Ann. 14, p.581, 1881.}, 
nor does it contain one which indicates a direct interaction between electricity and 
ponderable matter$\,$\footnote{$\:$Vgl. R. Colley: Nachweis der Existenz der Maxwell'schen elektrom. Kraft $Y_{me}$ {\it\color{red}(Proof of the existence of Maxwell's electrom. Strength $Y_{me}$)}. Wied. Ann. 17, p.55, 1882.}, 
but consists only of the magnetic element and the electrodynamic potential. 

At first glance it might seem like a kind of inconsistency that the magnetic potential $U$ is included in the \dashuline{value of the energy} with a positive sign, the electromagnetic $V$ not at all, while the electrodynamic $W$ is included in the \dashuline{value of the energy} with a negative sign. 

\vspace*{-2mm} 
\begin{center}
{\it\color{red}\bf
---------
(p.235) 
---------
}
\end{center}
\vspace*{-3mm}

Indeed, since we have derived the electromagnetic and electrodynamic effects directly from the purely magnetic ones, it is reasonable to assume that the three potentials mentioned are viewed as completely similar and have equal rights, and in fact: as soon as we identify a magnet with a current, the conclusion is inevitable that the magnetic potential plays exactly the same role in the \dashuline{formation of energy} as the electromagnetic and electrodynamic.

\vspace*{-3mm} 
\begin{center}
--------------------------------------------------- 
\end{center}
\vspace*{-3mm}


However, here the matter is different. We based the derivation of the electromagnetic and electrodynamic effects from the magnetic ones solely on the generally proven empirical principle (p.227) that a current exerts exactly the same (ponderomotive and induction) effects as a magnetic double layer with corresponding properties. 

Moreover, it does not follow from this that current and magnets behave identically, because otherwise they would have to suffer the same effects (\,``\,ceteris paribus\,'' 
{\it\color{red}/ all other things being equal}\,). 

But only with regard to the ponderomotor forces can an equality of the passive effects also be derived from the mechanical principle of action and counteraction (p.229), whereas this identity is by no means present for the induction effects.

While certain \dashuline{changes in energy} are caused in a current by external electromotive forces, the internal state of a corresponding permanent magnet placed in the same place (which can always be imagined as non-conducting for electricity) remains completely unchanged according to the \dashuline{equation of conservation of energy} that we have established.

\vspace*{0mm} 
\begin{center}
--------------------------------------------------- 
\end{center}
\vspace*{-3mm}


This fact is well illustrated by the following example. 

Let us imagine a permanent magnet in the form of a uniformly magnetized double layer, and with it a linear constant current (e.g. hydroelectric)  that flows through the boundary line of the magnetic surface, in such a way that the effects emanating from the magnet are just canceled out.

This system, whether it is moving or not, will not exert any forces (neither ponderomotive nor inducing) in the environment. 

If another permanent magnet (or current) moves somehow nearby, it will behave in exactly the same way as if it were all alone in space (its speed remains unchanged, etc). 

But that is not the case with our combined system of magnet and electricity.

\vspace*{-2mm} 
\begin{center}
{\it\color{red}\bf
---------
(p.236) 
---------
}
\end{center}
\vspace*{-3mm}


While nothing changes in the magnet, induction effects are caused in the current by the movement of the outer magnet, i.e. molecular energy is generated (and without altering the work value of these effects, one can assume the resistance of the power line to be so high that the induced current against the already existing constants can be neglected). 

The energy generated in this way obviously does not arise at the expense of the \dashuline{living force} {\it\color{red}\dashuline{(kinetic} \dashuline{energy)}} of the inducing magnet, but rather at the expense of the \dashuline{magnetic energy} $U$, which is caused by the presence of the two magnets. 
And if the current is interrupted (which does not involve work), this \dashuline{energy becomes ponderomotive}.

The situation is similar if a current is placed in place of the outer magnet, only then it is not the \dashuline{magnetic energy} but the \dashuline{electrokinetic energy} that comes into play: the negative electrodynamic potential $W$. From this we see how necessary it is to carry out the separation in the behavior of permanent magnets and that of their equivalent currents.

\vspace*{-3mm} 
\begin{center}
--------------------------------------------------- 
\end{center}
\vspace*{-3mm}


The above application of the \dashuline{principle of conservation of energy} to the interactions of magnets and currents is based solely on facts of experience.
It is, in particular (what we attach particular importance to) independent of any more specific idea about the nature of magnetism. 

But if we now accept Ampère's assumption (which was very plausible from the beginning) that the magnets are nothing other than systems of correspondingly oriented molecular currents, then the application of the \dashuline{energy principle} suffers a significant modification. 

All magnetic and electromagnetic effects then transition into electrodynamic ones, and must be treated as such. 

There is only one electrodynamic potential, the size of which is $( U + V + W )$, and accordingly only one \dashuline{electrokinetic energy}, which is equal and opposite to the potential.

\vspace*{-2mm} 
\begin{center}
{\it\color{red}\bf
---------
(p.237) 
---------
}
\end{center}
\vspace*{-3mm}


Furthermore, the \dashuline{internal energy} of a permanent magnet is no longer constant, but it changes in a finite way with a finite change in the magnetic field in which it is located (and no hypothesis about the nature of molecular currents can help with this conclusion). 

How one should imagine this changeability in detail is a question that does not belong here, where we are only concerned with the immediate results of experience (and which can only be answered in connection with more general problems that will be briefly touched on below).

\vspace*{-3mm} 
\begin{center}
--------------------------------------------------- 
\end{center}
\vspace*{-3mm}



As in the case of the interaction of linear unbranched {\it\color{red}(direct)} currents and permanent magnets, the laws of Ampère and F. Neumann can also be taken as a basis for the application of the \dashuline{energy principle} in the more general case of physically extended (closed) conductors, and any magnetisable bodies with equally satisfactory success, only then one must again distinguish between \dashuline{magnetic energy} $U$ and \dashuline{magnetic potential} $P$ as in the case of purely magnetic effects (p. 225).
Otherwise the propositions presented retain their wording in full.

\vspace*{-3mm} 
\begin{center}
--------------------------------------------------- 
\end{center}
\vspace*{-4mm}


The effects caused by the presence of sliding points in physically extended conductors require a special remark, simply because they play an interesting role in the historical development of the theory of current effects. 

  We have already pointed out above (p.230) that: {\it \dashuline{the ponderomotive work} performed by the currents and magnets during any change in the electromagnetic field is \dashuline{not measured by the real, complete temporal decrease of the total magnetic, electromagnetic and electrodynamic potential} (namely $U + V + W$), \dashuline{but by the partial decrease of this quantity}} (which is caused by the movements of the bodies with invariable current intensities and magnetisms).

  If physically extended conductors, which can also suffer mechanical deformations or are subject to slippage, move in the field, the invariability of the current intensities must be understood to mean that the current intensity remains the same in each ponderable particle. 

 Therefore, the current filaments can bend or stretch during movement (in the case of deformations or slippage), but must not be torn off.

\vspace*{-5mm} 
\begin{center}
{\it\color{red}\bf
---------
(p.238) 
---------
}
\end{center}
\vspace*{-4mm}


In any given case, one can imagine the complete temporal change in the potential as being composed of two partial changes that occur one after the other, which correspond to the following two processes: 
1) the ponderable conductor parts move into their new position, while the current threads remain in them, 
2) the current threads and current intensities take on the size and direction determined by the new constellation, while the conductors are at rest. 

Only the first partial change is taken into account when calculating the ponderomotive work. 
Applying this rule always leads to correct results, because all objections that have been raised against this theorem (first expressed in the present generalized form by 
Helmholtz$\,$\footnote{$\:$H. v. Helmholtz : Wiss. Abh. I, p.692.}) 
are based on an unjustified confusion of the described partial change in potential with the complete one (which for certain electrodynamic rotations  is $= 0$).

\vspace*{-5mm} 
\begin{center}
--------------------------------------------------- 
\end{center}
\vspace*{-3mm}


Let us think, for example, to the simple case that a linear current conductor $L$ slides with one end along the surface of a conductive body $K$ (such as mercury). 
The current that flows through $L$ to $K$ spreads in all directions through $K$ when it leaves $L$. 

The ponderomotive work done during the assumed movement can be found by considering an infinitesimal displacement of the conductor along the conducting surface, for example from point $A$ of the surface to a neighboring point $B$, and by calculating that (partial) change in the total potential of the current on itself, which results if the current intensity is thought to be constant in all parts of the conductor. 

\vspace*{-5mm} 
\begin{center}
{\it\color{red}\bf
---------
(p.239) 
---------
}
\end{center}
\vspace*{-4mm}

After the shift has been made, the current is not to be imagined flowing from $L$ through $B$ directly to $K$, but from $L$ through $B$ first linearly to $A$, and from there in the same way as before through $K$. Here we particularly emphasize that in the expression of the potential (as always with self-potentials) each combination of two current elements only occurs once (cf. the general expression of ponderomotor work, p.230).
\vspace*{0mm} 
\begin{center}
--------------------------------------------------- 
\end{center}
\vspace*{-3mm}


The application of F. Neumann's principle of induction to the present case is very similar, although somewhat more complicated. 

When calculating the change over time in the potential of all currents (and magnets) on a line through which current 1 flows, one must differentiate precisely between the changes in the currents and magnets (which have an inducing effect) and the changes in the power line (in which an electromotive one force is induced). 

In the former, the real, complete change in the current intensities (regardless of whether sliding points are present or not) and the magnetisms must always be taken into account.

In the latter, however, the current 1 must be thought of as flowing invariably in the same way as shown above, i.e.~so that the current threads retain their positions in the ponderable conductor parts (even when deformation occurs and are not torn off even in the event of sliding points).

\vspace*{-3mm} 
\begin{center}
--------------------------------------------------- 
\end{center}
\vspace*{-3mm}


Therefore, in order for example to find the self-induction of the previously considered current (which flows through the linear conductor $L$ into the physical conductor $K$) along whose surface $L$ slides, one first has to form the potential (at time $t$) of the current $L -A - K$ to the same imaginary line $L - A - K$ through which current 1 flows.

{\it\color{red}It is then needed} to subtract this expression from the potential (at time $t + dt$) of the completely changed current, which with its new intensity $[\: i + (di/dt) \:dt\:]$ flows through $L$ (and spreads from the new point of contact $B$ directly into the conductor $K$) onto the imaginary line $L - B - A - K$ flowed through by current 1, where the current threads in $K$ are just as they are at time $t$. 

{\it\color{red}Then,} if you divide the difference found by $dt$, you get the induced electromotive force.

\vspace*{-1mm} 
\begin{center}
{\it\color{red}\bf
---------
(p.240) 
---------
}
\end{center}
\vspace*{-3mm}



Here, of course, we are not dealing with a self-potential, because every combination of two elements occurs twice, depending on whether one of them is thought to be flowed through by the current $i$ or $i + (di/dt) \:dt$, or by the current 1 (cf. p.231). 
Neumann's induction principle only leads to correct results for physical conductors if the stated rule is observed.

\vspace*{-3mm} 
\begin{center}
--------------------------------------------------- 
\end{center}
\vspace*{-3mm}


In the case of uniform electrodynamic rotation, the current strength $i$ and the \dashuline{electrokinetic energy} are constant.
In fact, one easily finds that the ponderomotive work is equal to the work value of the electromotive effects, i.e. that the work required to maintain rotation (overcoming frictional resistance) is provided by the \dashuline{molecular energy} of the electrical conductors. 

The phenomena of magneto-electric (including the so-called unipolar) induction also find their direct explanation in the manner indicated. 

However, the question of the ponderomotive or electromotive forces acting in the individual conductor parts still remains completely 
open$\,$\footnote{$\:$ Vgl. hierüber E. Riecke: Zur Theorie der unipolaren Induction und der Plücker'schen Versuche {\it\color{red}(On the theory of unipolar induction and Plücker's experiments)}. Gött. Nachr. 1876, p.332. (Wied. Ann. 1 , p.110, 1877). Wied. Ann. 11 , p.413, 1880. Ferner F. Koch : Untersuchungen über magnetelektrische Rotationserscheinungen {\it\color{red}(Investigations into magneto-electric rotation phenomena)}. Wied. Ann. 19 , p.143 , 1883.}.

\vspace*{-3mm} 
\begin{center}
--------------------------------------------------- 
\end{center}
\vspace*{-3mm}


Thus, the \dashuline{Ampère-Neumann theory} of the interactions between closed currents and magnets appears to us as a complete system that is internally \dashuline{connected by the energy principle} and which forms the solid basis for further investigations, secured by the results of numerous experiments. 

Whether you want to think of the effects of the individual current elements according to Ampère's or 
Grassmann's$\,$\footnote{$\:$H. Grassmann: Neue Theorie der Elektrodynamik {\it\color{red}(New theory of electrodynamics)}. Pogg. Ann. 64, p.1, 1845.} 
elementary laws (or even according to Helmholtz's potential law) is up to the individual's taste.

\vspace*{-5mm} 
\begin{center}
{\it\color{red}\bf
---------
(p.241) 
---------
}
\end{center}
\vspace*{-4mm}


This theory only shows an internal 
gap$\,$\footnote{$\:$H. Hertz: Über die Beziehungen zwischen den Maxwell'schen elektrodynamischen Grundgleichungen und den Grundgleichungen der gegnerischen Elektrodynamik {\it\color{red}(On the relationships between Maxwell's fundamental electrodynamic equations and the fundamental equations of opposing electrodynamics)}. Wied. Ann. 23, p.84, 1884.}
if one drops the assumption we made at the beginning (p.228), namely that the speed of the changes in the electromagnetic field compared to the critical velocity is not taken into account. 

In this case, certain phenomena emerge which are to be understood as manifestations of the fact that the electromagnetic needs.

\vspace*{-5mm} 
\begin{center}
--------------------------------------------------- 
\end{center}
\vspace*{-4mm}


Significantly, all theories put forward about the nature of electricity (no matter how fundamentally different they may be in terms of origin and ideas --even those which are based on the assumption of an immediate long-distance effect) have reached this precise point in their further development. 

Gauss$\,$\footnote{$\:$C. F. Gauss: Brief an W. Weber. Werke V, p.627. Vgl. R. Clausius: Über die von Gauss angeregte neue Auffassung der elektrodynamischen Erscheinungen {\it\color{red}(Letter to W. Weber. Works~V, p.627. Cf. R. Clausius: On the new conception of electrodynamic phenomena suggested by Gauss)}, Pogg. Ann. 135, p.606, 1868.}
already viewed the derivation of the forces caused by the movement of electricity from an effect that is not instantaneous, but propagates at a finite speed as the ``\,keystone\,'' of electrodynamics and described the failure of his efforts in this direction as the reason why he did not consider the basic electrical law he had established to be ready for the public. 

B. Riemann$\,$\footnote{$\:$B. Riemann: Ein Beitrag zur Elektrodynamik {\it\color{red}(A contribution to electrodynamics)}. Pogg. Ann. 131, p.237, 1867.}
pursued similar ideas, and 
C. Neumann$\,$\footnote{$\:$C. Neumann: Die Principien der Elektrodynamik {\it\color{red}(The principles of electrodynamics)}. Gött. Nachr. 1868, p.223. Ferner: Math. Aunal. I, p.317, 1868. VIII, p.555, 1875.}
succeeded in an excellent way in reducing Weber's basic law to the assumption that the usual electrostatic potential spreads evenly in all directions at a certain velocity, and that this spread is the sole reason why the electrical forces also appear to be dependent on the velocities and accelerations of the acting electricity particles.

\vspace*{-2mm} 
\begin{center}
{\it\color{red}\bf
---------
(p.242) 
---------
}
\end{center}
\vspace*{-3mm}


  The question now arises, however, whether such an idea is at all compatible with the assumption of an unmediated action at a distance, and whether the assumption of a finite velocity of propagation of the electrical effects does not directly compel us to assume (with Faraday, Maxwell and many other physicists) a change in the intermediate medium that accompanies and mediates the propagation.

Because the state of an electrical system cannot be explicitly dependent on time, but only on the physical changes that the material parts of the system have suffered at the relevant point in time (including for the ether). 

Even Clausius' fundamental  
law$\,$\footnote{$\:$R. Clausius: Über ein neues Grundgesetz der Elektrodynamik {\it\color{red}(On a new basic law of electrodynamics)}. Pogg. Ann. 156, p.657, 1875. Crelle J. 82, p.85, 1876. Die mechanische Behandlung der Elektricität {\it\color{red}(The mechanical treatment of electricity)}. Braunschweig, 1879, p.277.}
(which was derived without any consideration of an existing intermediate medium) cannot do without the participation of such a medium, since a truly ``\,absolute\,'' velocity cannot be physically defined at 
all$\,${\it\color{red}\footnote{$\:${\it\color{red}Note that Max Planck (in 1887) was not aware of (or did not trust in) the observations of Michelson (in 1881) and Michelson and Morley (1887), and did not anticipate the heuristic formulas of Lorentz (1895-1904) and Poincaré (1900-1905-1906), and then above all the theory of relativity by Einstein in 1905, where $c\approx 300,000$~km/s will become a true ``\,absolute\,'' velocity / P. marquet)}.}}.

\vspace*{0mm} 
\begin{center}
--------------------------------------------------- 
\end{center}
\vspace*{-3mm}


If the essential importance of the intermediate medium for the emergence of the electrical-magnetic effects is now recognized, the idea is to give up the pure long-distance effect completely and to transfer the complete mediation of these effects to the intermediate medium, or in other words  (after a phrase used by C. Neumann): all ``telescopic'' effects can be attributed to ``microscopic'' ones.

\vspace*{-4mm} 
\begin{center}
--------------------------------------------------- 
\end{center}
\vspace*{-4mm}


Before this fundamental question all others must (in my opinion) take a back seat, such as: whether one has to distinguish between two different kinds of electricity? (whether there are unclosed currents) and if so: how the interactions of two current elements can be deduced from the interactions of closed currents ? and furthermore: how one has to conceive of the molecular currents in the magnets (which is finally the fundamental law of electrical effects) . . . and so on. 

Indeed, depending on the answer to the main question, the series of ideas (and thus the course of speculation) is directed into completely different ways. And this may mean a formal revolution of all our views of the nature of the forces at work in nature (which have been handed down to us by Newton and have become habitual).

\vspace*{-6mm} 
\begin{center}
{\it\color{red}\bf
---------
(p.243) 
---------
}
\end{center}
\vspace*{-4mm}

Because even if, according to Newton's own process, we regard only the appearance as the given (and leave entirely untouched the question of processes which may take place somewhere else), but which for the time being elude perception, our present view of nature is nevertheless (on the whole) permeated and dominated by the idea of \dashuline{direct action at a distance}, in the cosmic as well as in the molecular world.


Namely, we believe that, between the celestial bodies and between the atoms, nothing further takes place which is necessarily connected with the movements of these bodies (an opinion which has its good reason in the fact that we have in fact no perception of such processes in the movement of the celestial bodies, whereas in the case of the atoms it is based only on a conclusion by analogy).

\vspace*{-3mm} 
\begin{center}
--------------------------------------------------- 
\end{center}
\vspace*{-4mm}


And yet should it definitely succeed (and there is currently a high degree of probability for this) to trace the entirety of electrical phenomena to forces that only act at infinitely small distances, then there can hardly be any doubt that we will also have to get used to looking at the effects of gravity (which follow so much simpler laws) and as a result also the chemical phenomena (from the same point of view).

Indeed, the simplification that the new view brings to all our ideas about nature cannot easily be valued highly enough, as we will endeavor to show in more detail below. 

  The inconvenience of having to \dashuline{renounce a connection of ideas that} has been firmly rooted over a long period of time will not be able to change this task, because just as it took the arduous work of many centuries to make the idea of \dashuline{a direct remote effect at a distance} a living habit, it must be possible to get rid of this habit, once it has actually been established that the idea has done its job.

\vspace*{-3mm} 
\begin{center}
{\it\color{red}\bf
---------
(p.244) 
---------
}
\end{center}
\vspace*{-4mm}


Without prematurely anticipating the final decision on this fundamental question, we would like to conclude our investigations by highlighting the main consequences that arise from the general implementation of \dashuline{the new theory} for the application of the \dashuline{principle of conservation of energy}. 

In the absence of a short, suitable name for this theory, I will now take the liberty of calling it the ``\,\dashuline{infinitesimal theory}.\,''

\vspace*{-5mm} 
\begin{center}
--------------------------------------------------- 
\end{center}
\vspace*{-3mm}


First of all, it is important to emphasize that the two opposing theories are by no means to be viewed as coordinated, but that the \dashuline{theory of action at a distance} turns out to be the more general, just as a finite quantity contains an infinitely small one as a special case. 
Indeed, according to the infinitesimal theory, the forces which act on the parts of a body depend only on its own state, while according to the other theory, they also depend on all the bodies which fill the entire universe. 

This circumstance is also the reason why we have adhered more to the \dashuline{more general idea of direct action at a distance} in our presentation and expression so far. 

If the \dashuline{infinitesimal theory} is confirmed, then at the same time a \dashuline{new general law of nature} is proven, namely: {\it the law that \dashuline{all changes} that take place in and on any material element \dashuline{are completely determined by the current processes within and at the limit of the elements}}. 

It goes without saying that this sentence goes deep into the nature and functioning of all natural forces.

\vspace*{-7mm} 
\begin{center}
--------------------------------------------------- 
\end{center}
\vspace*{-3mm}


As a result, \dashuline{the concept of energy} now gains a much simpler meaning, as the behavior of \dashuline{energy} follows that of \dashuline{matter} even more closely. 

The amount of matter in the world can neither be increased nor decreased, but what is even more important is that \dashuline{matter cannot disappear in one place and} at the same time \dashuline{reappear in another place at a finite distance from it}, but it can only do so steadily over time change their place. 

The amount of \dashuline{matter in a closed space can only be changed by matter entering or exiting through the interface of the space}, and the magnitude of the change is measured precisely by the quantum passing through the surface.

\vspace*{-5mm} 
\begin{center}
{\it\color{red}\bf
---------
(p.245) 
---------
}
\end{center}
\vspace*{-3mm}

\dashuline{Things are different with energy as long as the theory of action at a distance is maintained}. 

Although \dashuline{the sum of energies in nature remains unchanged}, the energy can suddenly pass from one body to one at a distance from it, a planet can transfer its living force directly to another, a magnet creates instantaneous heat in a body through the energy of its movement induced current line etc. (i.e.~according to the infinitesimal theory, energy, like matter, can only change its location continuously over time). 

The energy in a closed space can only be increased or decreased by external effects that are mediated by physical processes in the boundary surface of the space, so here too one can speak of the \dashuline{energy passing through this surface}. 

Then the energy of a material system can always be broken down into elements, each of which belongs to a specific material element and finds its place in it (while, for example, the potential energy of two bodies acting on each other from a distance always only appears as an indivisible whole). 

Therefore, if several material systems are combined into a single one, \dashuline{the energy of the entire system is equal to the sum of the energies of the individual systems} --a sentence that is peculiar to \dashuline{infinitesimal theory} (cf. p.123).

\vspace*{-3mm} 
\begin{center}
--------------------------------------------------- 
\end{center}
\vspace*{-3mm}


With this great \dashuline{simplification of the view of nature}, as offered by the \dashuline{infinitesimal theory}, it is even more urgently suggested to physical research to examine the justification of this theory in detail by revealing its consequences in detail (because only in this way can one obtain the means of either confirming or refuting them). 

It is obviously of the utmost importance to completely separate the essence of this theory from all hypotheses with which one comes to the aid of the view, but which have nothing to do with the theory in and of itself.

\vspace*{-5mm} 
\begin{center}
{\it\color{red}\bf
---------
(p.246) 
---------
}
\end{center}
\vspace*{-3mm}


The difficulties that can arise for our imagination are completely out of the question, that e.g. \dashuline{the ether does not behave like} one of the \dashuline{solid, liquid or gaseous bodies} known to us, is a circumstance that does not cause the \dashuline{infinitesimal theory} the least embarrassment. 

Over time we will be able to get used to \dashuline{the specific mode of action of the ether} as well as to \dashuline{the properties that any other body shows us}, and will then soon include it in the series of phenomena that we are familiar with through many experiences.

\vspace*{-3mm} 
\begin{center}
--------------------------------------------------- 
\end{center}
\vspace*{-4mm}


However, it cannot be denied that \dashuline{the assumption of a special body} (that is so essentially different from all known ones) \dashuline{does not serve} our striving for \dashuline{the simplest possible description of nature}. 

However, the simplification that the uniform implementation of  \dashuline{the infinitesimal theory} affords in the entire realm of nature is infinitely greater than the disadvantage that arises from the introduction of a new body, which is indispensable in the theory of light anyway (already there, due to its high degree of elasticity and its minimal density, it occupies a very exceptional position in the series of solid bodies). 

In any case, the final decision on this question can be described as one of the most valuable achievements that are in prospect for the next period of scientific research.

\vspace*{-3mm} 
\begin{center}
{\it\color{red}\bf
---------
(p.246-247) 
---------
}
\end{center}
\vspace*{-4mm}


Finally, I would like to point out a remarkable analogy here. 

  It used to be believed that all events in nature (both spiritual and physical) find their cause and sufficient explanation not only in simultaneously acting circumstances, but that in general both past and future events (teleology) have a direct influence on the course of events, and thus influence the law of causality.

  Modern natural science (and this is precisely the basis of the powerful advantage it has over ancient science) has destroyed this belief, and it assumes that ultimately for the present state: {\it what is happening in the world as a whole, at a moment, is the completely determining cause of what will happen in the next moment, and that in the continuous chain of changes each link is determined independently and in its entirety by the immediately preceding one}.

\vspace*{-3mm} 
\begin{center}
{\it\color{red}\bf
---------
(p.247) 
---------
}
\end{center}
\vspace*{-4mm}


In other words: with regard to \dashuline{temporal effects}, the \dashuline{infinitesimal theory} has achieved \dashuline{thorough recognition}. 

It will probably be reserved for the next few decades to do the same \dashuline{for the spatial effects} by showing that there is \dashuline{no more direct influence} from a spatial distance than from a temporal distance but that all spatial effects, just \dashuline{like the temporal ones}, appear ultimately composite from such effects that spread from element to 
element$\,${\it\color{red}\footnote{$\:${\it\color{red}Note that the future revolutionary theories of Relativities and Quantum Mechanics will be indeed based on ``\,local\,'' theories and without any more ``\,interaction at a distance\,'' and in particular without any need to imagine the strange and useless ``\,ether medium\,'' / P. Marquet)}.}}.

Then \dashuline{every phenomenon} finds its complete \dashuline{explanation} in the \dashuline{spatially and temporally immediately {\bf adjacent} circumstances}, and \dashuline{all finite processes are composed of {\bf infinitesimal} effects}. 

  This second step seems to me to be on a par with the first, to which we owe \dashuline{the successes of today's natural science} to such an outstanding degree, and one is entitled to expect that it too will prove to be of similarly \dashuline{far-reaching importance for its further development}.

\vspace*{8mm} 
\begin{center}
{\bf- - - - - - - - - - - - - - - - - - - - - - - - - - - - - - - - - - - - - - - - - - - - - -} 
\end{center}
\vspace*{-2mm}

\noindent
Kgl. Hof- und UniversitätsLuchdruckerei von Dr. C. Wolf \& Sohn.

\noindent
{\it\color{red}Royal Court and University Printing House of Dr C. Wolf \& Sohn.}

\end{document}